# First-principles modeling of electromagnetic scattering by discrete and discretely heterogeneous random media

Michael I. Mishchenko[a,*], Janna M. Dlugach[b], Maxim A. Yurkin[c,d], Lei Bi[e], Brian Cairns[a], Li Liu[a,f], R. Lee Panetta[e], Larry D. Travis[a], Ping Yang[e], Nadezhda T. Zakharova[g]

[a]*NASA Goddard Institute for Space Studies, 2880 Broadway, New York, NY 10025, USA*
[b]*Main Astronomical Observatory of the National Academy of Sciences of Ukraine, 27 Zabolotny Str., 03680, Kyiv, Ukraine*
[c]*Voevodsky Institute of Chemical Kinetics and Combustion, SB RAS, Institutskaya str. 3, 630090 Novosibirsk, Russia*
[d]*Novosibirsk State University, Pirogova 2, 630090 Novosibirsk, Russia*
[e]*Department of Atmospheric Sciences, Texas A&M University, College Station, TX 77843, USA*
[f]*Columbia University, 2880 Broadway, New York, NY 10025, USA*
[g]*Trinnovim LLC, 2880 Broadway, New York, NY 10025, USA*

[*]Corresponding author. Fax: +1 212 678 5622.
*E-mail address:* michael.i.mishchenko@nasa.gov (M. I. Mishchenko).

**Abstract**

A discrete random medium is an object in the form of a finite volume of a vacuum or a homogeneous material medium filled with quasi-randomly and quasi-uniformly distributed discrete macroscopic impurities called small particles. Such objects are ubiquitous in natural and artificial environments. They are often characterized by analyzing theoretically the results of laboratory, *in situ*, or remote-sensing measurements of the scattering of light and other electromagnetic radiation. Electromagnetic scattering and absorption by particles can also affect the energy budget of a discrete random medium and hence various ambient physical and chemical processes. In either case electromagnetic scattering must be modeled in terms of appropriate optical observables, i.e., quadratic or bilinear forms in the field that quantify the reading of a relevant optical instrument or the electromagnetic energy budget. It is generally believed that time-harmonic Maxwell's equations can accurately describe elastic electromagnetic scattering by macroscopic particulate media that change in time much more slowly than the incident electromagnetic field. However, direct solutions of these equations for discrete random media had been impracticable until quite recently. This has led to a widespread use of various phenomenological approaches in situations when their very applicability can be questioned. Recently, however, a new branch of physical optics has emerged wherein electromagnetic scattering by discrete and discretely heterogeneous random media is modeled directly by using analytical or numerically exact computer solutions of the Maxwell equations. Therefore, the main objective of this Report is to formulate the general theoretical framework of electromagnetic scattering by discrete random media rooted in the Maxwell–Lorentz electromagnetics and discuss its immediate analytical and numerical consequences. Starting from the microscopic Maxwell–Lorentz equations, we trace the development of the first-principles formalism enabling accurate calculations of monochromatic and quasi-monochromatic scattering by static and randomly varying multiparticle groups. We illustrate how this general framework can be coupled with state-of-the-art computer solvers of the Maxwell equations and applied to direct modeling of electromagnetic scattering by representative random multi-particle groups with arbitrary packing densities. This first-principles modeling yields general physical insights unavailable with phenomenological approaches. We discuss how the first-order-scattering approximation, the radiative transfer theory, and the theory of weak localization of electromagnetic waves can be derived as immediate corollaries of the Maxwell equations for very specific and well-defined kinds of particulate medium. These recent developments confirm the mesoscopic origin of the radiative transfer, weak localization, and effective-medium regimes and help evaluate the numerical accuracy of widely used approximate modeling methodologies.

# 1. Introduction

In this Report we discuss fundamental aspects of the scattering of electromagnetic radiation by a discrete random medium (DRM), i.e., an object in the form of a distinct finite volume of a vacuum or a homogeneous material medium filled with quasi-randomly and quasi-uniformly distributed discrete macroscopic impurities called small particles. The general subject of electromagnetic scattering is extremely broad and can hardly be fully covered in a single review, which necessitates exercising proper selectivity and a careful delineation of the overall scope of the discussion. Therefore, the main purpose of the opening section is to introduce, in a somewhat *ad hoc* and qualitative manner, several basic definitions and notions and to explain the main focus of this Report.

## *1.1. Small particles*

The term "small particles" or just "particles" is ubiquitous in the discipline of light (or, more generally, electromagnetic) scattering [1–35] and even enters the very titles of several specialized monographs [1,8,11,13,16,18,20,24–26,28–30,34,35]. However, it may not be straightforward to give a universal and unambiguous physical definition of this term. For the purposes of this Report, a small particle is defined as a *small yet optically macroscopic body*. More specifically, a small particle is a finite discrete physical body that is "small" (or "tiny" or "minute") and yet consists of a number of atoms large enough that the body can be characterized by bulk optical constants such as the electric permittivity, magnetic permeability, and conductivity. The adjective "discrete" means that the body can be thought of as having a distinct macroscopic surface separating it from the surrounding host medium and acting as an optical interface between the interior and exterior materials with different refractive indices. The distribution of the refractive index inside the particle does not need to be homogeneous.

We will see later that the requirement of being characterized by optical constants appropriate to bulk matter allows one to define rather unambiguously the minimal permissible size of a small particle. It is not as straightforward to define the requisite "smallness" of the particle. More often than not, the characterization of being small follows from the human visual perception or from the need for an optical or electronic microscope to even see the particle. Often however it is more appropriate to refer to the *optical size* of the particle (or its *size parameter*), defined as the ratio of the circumference of the particle's smallest circumscribed sphere to the wavelength of the incident electromagnetic wave. Then the smallness of a particle may be defined by restricting the dimensionless particle size parameter to be a few orders of magnitude or less. For example, leaves, birds, or decimeter-sized clumps of ice forming Saturn's rings do not appear to be small particles when looked at by a human eye, but they are small particles from the perspective of probing them electromagnetically with a remote decimeter-wavelength radar. The factor that makes it convenient and possible to define the smallness of a particle in terms of its size parameter is the fundamental so-called scale invariance rule of electromagnetic scattering. This rule states that all dimensionless scattering and absorption characteristics of a finite object depend only on the ratio of the object's size and the wavelength of radiation but not on their individual values [36].

Fig. 1 shows that in many cases the above definition of a small particle can be rather unequivocal. Fig. 2 illustrates however that a degree of ambiguity can remain in some cases. Indeed, on one hand a fractal soot aggregate with touching components can be considered an individual particle when it is suspended in the atmosphere and is widely separated from other atmospheric particulates. On the other hand, it can also be considered a composite object consisting of individ-ual particles in the form of soot spherules. The possibility of this and similar ambiguities should always be kept in mind.

### *1.2. Discrete random media*

Using the above definition of a small particle, a Type-1 discrete random medium (DRM; see Fig. 3a) can be defined as a morphologically complex object in the form of an imaginary volume $V$ populated by a large number $N$ of small particles in such a way that

- the spatial distribution of the particles throughout the volume is quasi-random and quasi-uniform, and
- the physical states of the individual particles are statistically independent of each other and of the particle positions

(e.g., [7,12,19]). The physical state of a particle is defined as the combination of the particle's size, morphology (including the spatial distribution of the refractive index), and orientation. A defining trait of electromagnetic scattering by a DRM is the absence of speckles in scattering patterns.

It is imperative to recognize that at any moment in time, the spatial distribution of particles in a multi-particle group is definite rather than random. Therefore, if the group is illuminated by a monochromatic or quasi-monochromatic[1] parallel beam of light then statistical randomness and spatial uniformity of a DRM and the requisite absence of speckles can be achieved only over a sufficiently long period of time owing to random temporal changes of particle positions. This is precisely what happens naturally in a multi-particle group suspended in a gas or a liquid and causes smooth, speckle-free patterns of electromagnetic scattering.

In some cases (e.g., in a particulate surface) the particles do not move relative to each other and yet, from the perspective of electromagnetic scattering, can often be thought of as forming a DRM. This can happen, for example, when particle positions in a group are "maximally random" and "maximally uniform", while the entire multi-particle group is moving relative to the source of light and/or the detector. Then even small changes of the source-of-light → multi-particle group → detector configuration during the measurement are equivalent to multi-wavelength shifts in particle positions and can, in essence, result in a random particulate sample generating speckle-free scattering patterns [44,45] (see also Section 1.4 of [25]). Another way to achieve a speckle-free regime is to illuminate a fixed quasi-random multi-particle group by incoherent polychromatic and/or uncollimated light. Such scattering scenarios help broaden the notion of a DRM and often lead to useful practical applications.

The volume $V$ hosting the $N$-particle group can also have a distinct physical boundary $S$ separating the finite interior and the infinite exterior space with different refractive indices (Fig. 3b). Such a heterogeneous object can be classified as a morphologically complex DRM as well, provided that the distribution of particle positions throughout the actual physical volume $V$ is sufficiently random and uniform. We will refer to such an object as a Type-2 DRM.

In reality, the spatial distribution of the constituent particles can never be completely random and statistically uniform because the particles are not allowed to overlap and because their cumulative volume $V_{\text{part}}$ (defined as the union of the individual particle volumes) is nonzero. For the same reason the orientations of nonspherical particles cannot be completely independent of each other and of the particle coordinates. We will assume however that the particles are distributed throughout the (physical or imaginary) volume $V$ as randomly and uniformly as the volume packing density $\rho = V_{\text{part}}/V < 1$ permits.[2] In this regard the morphology of a DRM is fundamentally different from that of fractal-like multi-particle clusters such as those studied, e.g., in [41–43,46–49] and illustrated in Fig. 2. On a somewhat pedestrian level, the spatial distribution of the particles can be considered statistically quasi-uniform if the average number of particles per unit volume $n(\mathbf{r})$ is independent of the position vector $\mathbf{r}$ over distances of the order of several times the average separation between two neighbouring particles. Of course, $n(\mathbf{r})$ is allowed to change over much greater distances. An instructive discussion of various mathematical parameterizations of physical disorder can be found in [50].

[1] According to conventional terminology, the qualifier "monochromatic" refers to a purely time-harmonic electromagnetic field, while the qualifier "quasi-monochromatic" refers to a time-harmonic electromagnetic field subjected to relatively slow quasi-random fluctuations.

### *1.3. Why to study electromagnetic scattering by discrete random media?*

Discrete random media are ubiquitous in natural and artificial environments. Typical examples are clouds of interstellar dust; the cloud of interplanetary dust in the solar system; dusty atmospheres of comets; particulate planetary rings; clouds in planetary atmospheres; geophysical, biomedical, and technical particle suspensions; aerosol particles with numerous inclusions; heterogeneous polymeric materials; and particulate surfaces (cf. Figs. 4 and 5a–d). Another important class of DRMs is represented by technical coatings such as layers of paint [54] (Figs. 5e,f).

The extreme morphological complexity of the majority of natural and artificial DRMs makes their characterization a daunting task. More often than not, one has to infer the micro- and macrophysical parameters of a DRM by analyzing theoretically the results of laboratory, *in situ*, or remote-sensing measurements of light and other electromagnetic radiation scattered by the medium. Thus the use of electromagnetic scattering as a potent noninvasive characterization technique represents a major reason to study this phenomenon. Another major reason has to do with the fact that scattering and absorption of electromagnetic radiation by particles can affect the energy budget of a volume of DRM and hence various ambient physical and chemical processes. In either case electromagnetic scattering must be described in terms of appropriate *optical observables*, i.e., quadratic or bilinear forms in the electromagnetic field that quantify the reading of a relevant optical instrument and/or the electromagnetic energy budget.

### *1.4. The general scope of this Report*

The practical solution of optical-characterization and energy-budget problems has the following four main ingredients:

- Formulation of appropriate optical observables for a given DRM and a specific type of illumination.
- Theoretical modeling of these observables for a specific DRM (the so-called direct scattering problem).
- Practical measurement of these observables.
- Solution of the so-called inverse scattering problem, i.e., finding the physical model of a

[2] Note that the volume packing density $\rho$ of a DRM can vary from essentially zero for a cloud to more than 0.5 for a particulate surface.

DRM that provides the best fit of theoretical simulations of electromagnetic scattering to the measurement data.

To keep the size of this Report manageable, we will focus only on the first two ingredients. Specifically, our primary objective is to outline the *first-principles* theoretical framework of electromagnetic scattering by DRMs rooted in the microscopic Maxwell–Lorentz equations and discuss its immediate analytical and numerical consequences.

### *1.5. The need for first-principles approach*

Until quite recently, theoretical calculations of electromagnetic scattering by a DRM had typically been based on *ad hoc* approaches with poorly known or undefined accuracies and ranges of applicability. Perhaps the most notable examples are the phenomenological[3] radiative transfer theory [7,12,19,55–71] originally developed for sparse turbid media such as clouds, effective-medium rules [32,72–78], and the geometric-optics ray-tracing method [79–83]. Even simplistic phenomenological approaches such as the Gershun theory of the light field [84] or the Kubelka–Munk [85–89] and Hapke [90] theories, have found frequent – even though questionable – use (cf. [91,92]). The underlying principles of some of these methodologies can be traced all the way back to the classical yet thoroughly outdated work by Pierre Bouguer [93,94], Johann Lambert [95], August Beer [96], Eugen von Lommel [97], Orest Khvolson [98], and Arthur Schuster [99] (see [100] for an account of the history of the phenomenological radiative transfer theory). The basic "physically obvious" premise in many studies of electromagnetic scattering by DRMs has been the belief that if the individual far-field scattering properties of each constituent particle are known then all scattering properties of the entire DRM can somehow be constructed from those of the constituent particles. This generally incorrect assumption is based on the lack of recognition that from the perspective of electromagnetics, the entire DRM is a unified scattering target, while the only essential consequence of the complex object's morphology (e.g., being composed of what appears to the human eye as discrete units, called particles) is to make the corresponding electromagnetic boundary conditions more complicated.

The main objective of this Report is to expose the fundamental physical nature of the phenomenon of electromagnetic scattering by a DRM and introduce the general theoretical formalism enabling first-principles modeling of relevant optical observables. We demonstrate how recent advances in the development of computer solvers of the macroscopic Maxwell equations and the availability of powerful computers and computer clusters have made possible direct modeling of electromagnetic scattering by representative random multi-particle groups with arbitrary packing densities. Furthermore, we discuss how the first-order-scattering approximation, the radiative transfer theory and the theory of weak localization of electromagnetic waves can be derived as immediate corollaries of Maxwell's electromagnetics for very specific and well-defined kinds of DRM. These recent developments have decisively brought the subject of electromagnetic energy transport in macroscopic DRMs and their optical characterization into the realm of physical optics. In particular, they have helped establish a mesoscopic link between the macrophysical regime of radiative transfer, weak localization, and effective-medium approximations on one hand and the microscopic Maxwell–Lorenz equations on the other. We make a spe-cial effort to state explicitly what results have been established definitively and what aspects of this research discipline necessitate further analysis.

[3] A physical theory is called phenomenological if it expresses mathematically the results of observed phenomena without tracing and clarifying their fundamental origin and significance. Typically, the development of a phenomenological theory is based on heuristic (i.e., experience-based) shortcuts lacking rigorous justification.

The unquestionable advantage of the first-principles approach is that it yields the definitive physical understanding of the phenomenon of electromagnetic scattering by a DRM and its corollaries. However, technical complexities of solving the Maxwell equations directly (both analytically and numerically) often diminish the applicability of the first-principles approach to real physical systems encountered in practice. As a consequence, it is reasonable to expect that various analytical, phenomenological, and heuristic approximations such as those mentioned above will still be widely used in the foreseeable future. Hence an important function of the first-principles approach is to characterize the accuracy and range of applicability of these approximations. To illustrate this function, we include a discussion of how direct computer solutions of the macroscopic Maxwell equations can be used to quantify the errors of such popular modeling tools as the first-order-scattering approximation, the radiative transfer equation, the theory of weak localization, and the effective-medium approach.

### *1.6. Further guidelines*

To make the scope of this Report manageable, we will discuss only elastic scattering of electromagnetic waves. In other words, nonlinear optics effects will be excluded by assuming that the optical constants of the scattering object as well as of the surrounding medium are independent of the electric and magnetic fields.

We will also exclude from specific consideration the small Doppler shift of frequency of the scattered light relative to that of the incident light due to the movement of particles with respect to the source of illumination. Furthermore, we will not discuss the scattering of transient electromagnetic fields such as ultra-short laser pulses (cf. [101]) and will discuss only frequency-domain electromagnetic scattering by assuming that all "quasi-instantaneous" fields and sources are time-harmonic and satisfy the frequency-domain Maxwell equations. In other words, we focus on the scattering of a monochromatic or quasi-monochromatic electromagnetic field and assume that the scattering object varies in time much more slowly than the field.

In the majority of this Report we will assume that the randomness of a particulate medium is ensured by its temporal variability. The extension of the concept of a DRM to a fixed particulate medium illuminated by an incoherent source will be discussed in Section 12.

## 2. Electromagnetics, optical observables, and averaging

The most advanced theory of light–matter interactions available today is quantum electrodynamics (QED) [102–108] followed, in the hierarchy of generality and complexity, by the semi-classical approach [109–111] and the Maxwell–Lorentz microscopic electromagnetics [112–116]. Since the specific subject of this Report is elastic (i.e., not involving changes in frequency) electromagnetic scattering, we will assume that from the standpoint of a wide range of practical applications, all relevant physics can be adequately captured by the classical microscopic Maxwell–Lorentz equations.

Despite this simplification, the actual quantification of electromagnetic scattering by a DRM is still highly problematic because solving the Maxwell–Lorentz equations either analytically or numerically is essentially impossible given the enormous number of elementary electric charges forming macroscopic objects. This makes it imperative to derive a theoretical formalism that is much simpler than the microscopic Maxwell–Lorentz electromagnetics and bypasses the unnecessarily detailed computation of the actual electromagnetic field. It turns out that doing this is feasible by exploiting the two-layer structure of electromagnetics along with making hierarchal use of volume, time, and/or ensemble averaging.

Indeed, in the words of Freeman Dyson [117],

> The modern view of the world that emerged from Maxwell's theory is a world with two layers. The first layer, the layer of the fundamental constituents of the world, consists of fields satisfying simple linear equations. The second layer, the layer of the things that we can directly touch and measure, consists of mechanical stresses and energies and forces. The two layers are connected, because the quantities in the second layer are quadratic or bilinear combinations of the quantities in the first layer. To calculate energies or stresses, you take the square of the electric field-strength or multiply one component of the field by another… The objects on the first layer, the objects that are truly fundamental, are abstractions not directly accessible to our senses. The objects that we can feel and touch are on the second layer, and their behavior is only determined indirectly by the equations that operate on the first layer.

Owing to this two-layer structure, the framework of the simplified theoretical formalism can be formulated as the following two-stage procedure:

- first, define relevant *optical observables* as quadratic and bilinear forms in the electromagnetic field that (i) can be directly measured with suitably designed instruments, and/or (ii) quantify the energy budget of a macroscopic object[4]; and
- second, develop an efficient way to directly calculate appropriate *averages* of these observables even if the detailed computation of the exact (microscopic) electromagnetic field itself is sacrificed.

Indeed, the majority of applications do not require the knowledge of instantaneous (or quasi-instantaneous) local values of the optical observables but rather deal with averages taken over extended time intervals and/or finite (rather than infinitesimal) volume elements. Simple examples of the experimental use of time averages are the exposure time of a camera, the integrating time of the rod cells in our eyes, and the signal integration over the sensitive face of a detector. Moreover, in many situations time averaging can be replaced by ensemble averaging, thereby resulting in further dramatic simplifications.

In what follows, we will discuss how the two-layer structure of electromagnetics in combination with appropriate averaging procedures yields an important effective-field approximation called macroscopic electromagnetics. This approximation is based on the introduction of a mathematical entity called the macroscopic electromagnetic field and can be used to quantify time-harmonic electromagnetic scattering by a fixed macroscopic object. Although the macroscopic electromagnetic field in and of itself is not an actual physical field, it can yield suitably averaged optical observables directly, i.e., without the prior computation of the exact (i.e., microscopic) electromagnetic field. A straightforward generalization makes this approach applicable to quasi-monochromatic macroscopic fields and/or time-variable macroscopic objects. The resulting formalism enables the computation of relevant time-averaged optical observables for a DRM by using analytical expressions and equations completely devoid of the actual electromagnetic field scattered by the medium.

[4] Note that this definition of optical observables does not exclude quadratic and bilinear forms in the field that cannot be measured directly. The prime example of a bilinear form that often is not measurable is the Poynting vector [34].

All three types of averaging mentioned so far (i.e., volume, time, and ensemble averaging) have been used in various publications on electromagnetic scattering. Some of these publications may give the impression that different types of averaging may be used interchangeably or that a type of averaging can be selected almost at will. It is imperative to keep in mind however that each type of averaging has its own conditions of applicability and that indiscriminate use of any one of them can lead to physically meaningless results. Therefore, in what follows we will be very explicit in justifying the use of a specific type of averaging and explaining why an alternative choice can be inappropriate.

## 3. Macroscopic Maxwell equations

*War es ein Gott, der diese Zeichen schrieb (?) (Was it a God who wrote these signs (?))*
[Ludwig Boltzmann [118], from Goethe's *Faust*]

*From a long view of the history of mankind – seen from, say, ten thousand years from now – there can be little doubt that the most significant event of the 19th century will be judged as Maxwell's discovery of the laws of electrodynamics. The American Civil War will pale into provincial insignificance in comparison with this important scientific event of the same decade.*
[Richard P. Feynman [119]]

The macroscopic Maxwell equations (MMEs) were postulated by James Clerk Maxwell 150 years ago [120] as the most fundamental laws of electromagnetics consistent with the totality of experimental data accumulated by that time. Maxwell's ideas, summarized in his famous *Treatise* [121], were picked up, systematized, and reworked mathematically by his immediate followers [122,123], most notably by Oliver Heaviside [124]. The subsequent notion that the MMEs must be a corollary of the more fundamental microscopic Maxwell–Lorentz equations [125] was put forth by Hendrik Lorentz in [112] and has been further developed by many authors [113–116,126–142].

In the framework of classical electromagnetics, the microscopic electromagnetic field is the only actual physical field which, in the vast majority of situations, is an extremely intricate function of the position vector $\mathbf{r}$ and time $t$. The basic idea of macroscopic electromagnetics is that the detailed knowledge of the exceedingly complex dependence of the microscopic field on $\mathbf{r}$ and $t$ is often not required in practice. Instead, this dependence is artificially simplified by averaging the microscopic field over either $\mathbf{r}$ or $t$, thereby yielding contrived macroscopic field vectors. It is imperative to recognize that these fictitious mathematical entities can only be useful to the extent to which they simplify the computation of macroscopic optical observables. In this respect macroscopic electromagnetics is the prime example of an effective-field approximation.

The temporal variability of the microscopic electromagnetic field inside a macroscopic object is caused by the incessant microscopic movements of the constituent elementary charges, by macroscopic temporal changes of the object, and by time-harmonic oscillations and quasi-random fluctuations induced by the external sources. Throughout this Report, we will assume that field variations caused by macroscopic changes of the object occur much more slowly than those caused by the microscopic movements of the constituent elementary charges as well as much more slowly than the externally induced time-harmonic oscillations and, potentially, quasi-random fluctuations. Furthermore, we will assume that the quasi-random oscillations of the field occur much more slowly than its time-harmonic oscillations. A fundamental corollary of these assumptions is that monochromatic and quasi-monochromatic electromagnetic scattering by the slowly varying macroscopic object can be described at any moment in time by assuming that the object is fixed and solving the corresponding quasi-instantaneous boundary-value problem for the frequency-domain MMEs.

### *3.1. Averaging over physically small volume elements*

The fundamental equations governing electromagnetic phenomena for point charges serving as building blocks of a macroscopic material medium are the four microscopic Maxwell–Lorentz equations:

$$\nabla \cdot \mathbf{e}(\mathbf{r},t) = \frac{\eta(\mathbf{r},t)}{\varepsilon_0}, \tag{1}$$

$$\nabla \times \mathbf{e}(\mathbf{r},t) + \frac{\partial \mathbf{b}(\mathbf{r},t)}{\partial t} = \mathbf{0}, \tag{2}$$

$$\nabla \cdot \mathbf{b}(\mathbf{r},t) = 0, \tag{3}$$

$$\nabla \times \mathbf{b}(\mathbf{r},t) - \varepsilon_0\mu_0 \frac{\partial \mathbf{e}(\mathbf{r},t)}{\partial t} = \mu_0 \mathbf{j}(\mathbf{r},t), \tag{4}$$

where $\mathbf{e}$ and $\mathbf{b}$ are the microscopic electric and magnetic fields; $\eta$ and $\mathbf{j}$ are the microscopic charge and current densities; $\varepsilon_0$ and $\mu_0$ are the electric permittivity and the magnetic permeability of a vacuum; and $\mathbf{0}$ is a zero vector. Fundamentally, the microscopic fields are functions of $\mathbf{r}$ and $t$ only. This implies that *the position vector and time are the only parameters over which* $\mathbf{e}(\mathbf{r},t)$ *and* $\mathbf{b}(\mathbf{r},t)$ *can in principle be averaged.*

According to the first averaging approach dating back to Lorentz [112], the microscopic field is homogenized, at any moment in time, over "physically small" volume elements $\delta V$ centered at $\mathbf{r}$ in order to smooth out drastic variations of $\mathbf{e}(\mathbf{r},t)$ and $\mathbf{b}(\mathbf{r},t)$ over interatomic distances [114,126–128,133]:

$$\mathbf{E}^{\delta V}(\mathbf{r},t) = \langle \mathbf{e}(\mathbf{r},t)\rangle_{\delta V}, \tag{5}$$

$$\mathbf{B}^{\delta V}(\mathbf{r},t) = \langle \mathbf{b}(\mathbf{r},t)\rangle_{\delta V}. \tag{6}$$

The well-known result of this approach is the system of the four MMEs:

$$\nabla \cdot \mathbf{D}^{\delta V}(\mathbf{r},t) = \rho^{\delta V}(\mathbf{r},t), \tag{7}$$

$$\nabla \times \mathbf{E}^{\delta V}(\mathbf{r},t) + \frac{\partial \mathbf{B}^{\delta V}(\mathbf{r},t)}{\partial t} = \mathbf{0}, \tag{8}$$

$$\nabla \cdot \mathbf{B}^{\delta V}(\mathbf{r},t) = 0, \tag{9}$$

$$\nabla \times \mathbf{H}^{\delta V}(\mathbf{r},t) - \frac{\partial \mathbf{D}^{\delta V}(\mathbf{r},t)}{\partial t} = \mathbf{J}^{\delta V}(\mathbf{r},t), \tag{10}$$

where $\mathbf{H}^{\delta V}(\mathbf{r},t)$ is the macroscopic magnetic intensity vector; $\mathbf{D}^{\delta V}(\mathbf{r},t)$ is the macroscopic electric displacement vector; $\rho^{\delta V}(\mathbf{r},t)$ and $\mathbf{J}^{\delta V}(\mathbf{r},t)$ are the macroscopic free charge density and current density, respectively. In the case of a macroscopically isotropic and time-dispersive material, the macroscopic field vectors entering the MMEs (7)–(10) are further related by the con-stitutive relations [116]

$$\mathbf{D}^{\delta V}(\mathbf{r},t) = \int_{-\infty}^{t} \mathrm{d}t' \varepsilon^{\delta V}(\mathbf{r},t-t')\mathbf{E}^{\delta V}(\mathbf{r},t'), \tag{11}$$

$$\mathbf{H}^{\delta V}(\mathbf{r},t) = \int_{-\infty}^{t} \mathrm{d}t' \frac{1}{\mu^{\delta V}(\mathbf{r},t-t')}\mathbf{B}^{\delta V}(\mathbf{r},t'), \tag{12}$$

$$\mathbf{J}^{\delta V}(\mathbf{r},t) = \int_{-\infty}^{t} \mathrm{d}t' \sigma^{\delta V}(\mathbf{r},t-t')\mathbf{E}^{\delta V}(\mathbf{r},t'). \tag{13}$$

One must recognize that this averaging procedure yields artificial field vectors formally satisfying the MMEs rather than an exact physical electromagnetic field. Furthermore, the problem of actual practical significance is to compute macroscopic averages of specific optical observables, including quantities describing electromagnetic energy budget. In other words, one needs volume averages of *quadratic and bilinear forms in the electromagnetic field*, such, for example, as the Poynting vector [143,144]

$$\mathbf{s}(\mathbf{r},t) = \frac{1}{\mu_0}\mathbf{e}(\mathbf{r},t)\times\mathbf{b}(\mathbf{r},t). \tag{14}$$

This implies that for a mathematical solution of the MMEs to be physically significant and practically useful, it must enable the computation of relevant macroscopic optical observables, including the macroscopic Poynting vector

$$\mathbf{S}^{\delta V}(\mathbf{r},t) = \langle \mathbf{s}(\mathbf{r},t)\rangle_{\delta V} = \frac{1}{\mu_0}\langle \mathbf{e}(\mathbf{r},t)\times\mathbf{b}(\mathbf{r},t)\rangle_{\delta V}, \tag{15}$$

by a simple substitution of the macroscopic field vectors for the microscopic ones, e.g.,

$$\mathbf{S}^{\delta V}(\mathbf{r},t) = \frac{1}{\mu_0}\mathbf{E}^{\delta V}(\mathbf{r},t)\times\mathbf{B}^{\delta V}(\mathbf{r},t). \tag{16}$$

Since the average of the vector product of two vectors in Eq. (15) is not necessarily equal to the vector product of the individual averages, Eq. (16) is highly nontrivial and by no means obvious.

To the best of our knowledge, Eq. (16) has been derived rigorously only for structured periodic nonmagnetic materials and only for the case of time-harmonic fields [141,142], while in all other situations (e.g., in the case of amorphous solids and liquids), it still has to be *postulated*. It must be recognized however that without Eq. (16) and similar formulas for other relevant second moments of the electromagnetic field the MMEs would largely lose their physical significance and become irrelevant or not helpful if one wishes to make useful predictions.

Let us now assume that the macroscopic field vectors are monochromatic, while the medium is non-magnetic. It is convenient to represent the real field variables as real parts of the respective complex time-harmonic variables:

$$\mathbf{E}^{\delta V}(\mathbf{r},t) = \mathrm{Re}[\widetilde{\mathbf{E}}^{\delta V}(\mathbf{r})\exp(-\mathrm{i}\omega t)], \tag{17}$$

$$\mathbf{B}^{\delta V}(\mathbf{r},t) = \mathrm{Re}[\widetilde{\mathbf{B}}^{\delta V}(\mathbf{r})\exp(-\mathrm{i}\omega t)], \tag{18}$$

$$\mathbf{D}^{\delta V}(\mathbf{r},t) = \mathrm{Re}[\widetilde{\mathbf{D}}^{\delta V}(\mathbf{r})\exp(-\mathrm{i}\omega t)], \tag{19}$$

$$\mathbf{H}^{\delta V}(\mathbf{r},t) = \mathrm{Re}[\widetilde{\mathbf{H}}^{\delta V}(\mathbf{r})\exp(-\mathrm{i}\omega t)], \tag{20}$$

$$\rho^{\delta V}(\mathbf{r},t) = \mathrm{Re}[\tilde{\rho}^{\delta V}(\mathbf{r})\exp(-\mathrm{i}\omega t)], \tag{21}$$

$$\mathbf{J}^{\delta V}(\mathbf{r},t) = \mathrm{Re}[\tilde{\mathbf{J}}^{\delta V}(\mathbf{r})\exp(-\mathrm{i}\omega t)], \tag{22}$$

where $\omega$ is the angular frequency, $\mathrm{i} = (-1)^{1/2}$, and "Re" stands for "the real part of". It is then straightforward to show that the macroscopic field vectors satisfy the standard frequency-domain MMEs

$$\nabla \cdot \tilde{\mathbf{D}}^{\delta V}(\mathbf{r}) = \tilde{\rho}^{\delta V}(\mathbf{r}), \tag{23}$$

$$\nabla \times \tilde{\mathbf{E}}^{\delta V}(\mathbf{r}) - \mathrm{i}\omega\tilde{\mathbf{B}}^{\delta V}(\mathbf{r}) = \mathbf{0}, \tag{24}$$

$$\nabla \cdot \tilde{\mathbf{B}}^{\delta V}(\mathbf{r}) = 0, \tag{25}$$

$$\nabla \times \tilde{\mathbf{H}}^{\delta V}(\mathbf{r}) + \mathrm{i}\omega\tilde{\mathbf{D}}^{\delta V}(\mathbf{r}) = \tilde{\mathbf{J}}^{\delta V}(\mathbf{r}) \tag{26}$$

supplemented by the constitutive relations

$$\tilde{\mathbf{D}}^{\delta V}(\mathbf{r}) = \tilde{\varepsilon}^{\delta V}(\mathbf{r},\omega)\tilde{\mathbf{E}}^{\delta V}(\mathbf{r}), \tag{27}$$

$$\tilde{\mathbf{H}}^{\delta V}(\mathbf{r}) = \frac{1}{\mu_0}\tilde{\mathbf{B}}^{\delta V}(\mathbf{r}), \tag{28}$$

$$\tilde{\mathbf{J}}^{\delta V}(\mathbf{r}) = \tilde{\sigma}^{\delta V}(\mathbf{r},\omega)\tilde{\mathbf{E}}^{\delta V}(\mathbf{r}), \tag{29}$$

where the frequency-dependent electric permittivity $\tilde{\varepsilon}^{\delta V}(\mathbf{r},\omega)$ and conductivity $\tilde{\sigma}^{\delta V}(\mathbf{r},\omega)$ are, in general, complex valued:

$$\tilde{\varepsilon}^{\delta V}(\mathbf{r},\omega) = \int_0^\infty \mathrm{d}t\, \varepsilon^{\delta V}(\mathbf{r},t)\exp(\mathrm{i}\omega t), \tag{30}$$

$$\tilde{\sigma}^{\delta V}(\mathbf{r},\omega) = \int_0^\infty \mathrm{d}t\, \sigma^{\delta V}(\mathbf{r},t)\exp(\mathrm{i}\omega t). \tag{31}$$

According to Eq. (16), the near-instantaneous time-independent macroscopic Poynting vector is now given by the time average

$$\begin{aligned}\langle \mathbf{S}^{\delta V}(\mathbf{r},t)\rangle &= \frac{1}{T}\int_{t-T/2}^{t+T/2} \mathrm{d}t'\, \mathbf{E}^{\delta V}(\mathbf{r},t') \times \mathbf{H}^{\delta V}(\mathbf{r},t') \\ &\approx \frac{1}{2}\mathrm{Re}\{\tilde{\mathbf{E}}^{\delta V}(\mathbf{r}) \times [\tilde{\mathbf{H}}^{\delta V}(\mathbf{r})]^*\}, \qquad T \gg T_\mathrm{o},\end{aligned} \tag{32}$$

where the asterisk denotes a complex-conjugate value and

$$T_\mathrm{o} = \frac{2\pi}{\omega} \tag{33}$$

is the period of time-harmonic oscillations.

### *3.2. Ensemble averaging*

The above methodology is intended to yield optical observables homogenized over physically small volume elements *at each moment in time*. In the case of a high-frequency time-harmonic electromagnetic field, one could think of a different averaging approach intended to yield time-averaged optical observables *at each point in space*. Specifically, the microscopic electric and magnetic fields as well as the microscopic charge and current densities are factorized according to

$$\mathbf{e}(\mathbf{r},t) = \mathrm{Re}[\tilde{\mathbf{e}}(\mathbf{r},t)\exp(-\mathrm{i}\omega t)], \tag{34}$$

$$\mathbf{b}(\mathbf{r},t) = \mathrm{Re}[\tilde{\mathbf{b}}(\mathbf{r},t)\exp(-\mathrm{i}\omega t)], \tag{35}$$

$$\eta(\mathbf{r},t) = \mathrm{Re}[\tilde{\eta}(\mathbf{r},t)\exp(-\mathrm{i}\omega t)], \tag{36}$$

$$\mathbf{j}(\mathbf{r},t) = \mathrm{Re}[\tilde{\mathbf{j}}(\mathbf{r},t)\exp(-\mathrm{i}\omega t)] \tag{37}$$

[131,133]. The dependence of the complex amplitudes $\tilde{\mathbf{e}}(\mathbf{r},t)$, $\tilde{\mathbf{b}}(\mathbf{r},t)$, $\tilde{\eta}(\mathbf{r},t)$, and $\tilde{\mathbf{j}}(\mathbf{r},t)$ on time is *implicit* in that it is caused by relatively slow random movements of the microscopic charges occurring independently of the rapid oscillatory motions described by the time-harmonic factor $\exp(-\mathrm{i}\omega t)$. This means that at any moment in time, the complex amplitudes satisfy the frequency-domain microscopic Maxwell–Lorentz equations

$$\nabla \cdot \tilde{\mathbf{e}}(\mathbf{r},t) = \frac{\tilde{\eta}(\mathbf{r},t)}{\varepsilon_0}, \tag{38}$$

$$\nabla \times \tilde{\mathbf{e}}(\mathbf{r},t) - \mathrm{i}\omega\tilde{\mathbf{b}}(\mathbf{r},t) = \mathbf{0}, \tag{39}$$

$$\nabla \cdot \tilde{\mathbf{b}}(\mathbf{r},t) = 0, \tag{40}$$

$$\nabla \times \tilde{\mathbf{b}}(\mathbf{r},t) + \mathrm{i}\omega\varepsilon_0\mu_0\tilde{\mathbf{e}}(\mathbf{r},t) = \mu_0\tilde{\mathbf{j}}(\mathbf{r},t) \tag{41}$$

provided that

$$\left|\frac{\partial\tilde{\mathbf{e}}(\mathbf{r},t)}{\partial t}\right| \ll \omega\left|\tilde{\mathbf{e}}(\mathbf{r},t)\right|, \tag{42}$$

$$\left|\frac{\partial\tilde{\mathbf{b}}(\mathbf{r},t)}{\partial t}\right| \ll \omega\left|\tilde{\mathbf{b}}(\mathbf{r},t)\right|. \tag{43}$$

Note that whether the inequalities (42) and (43) are satisfied can be expected to depend on several factors, including the angular frequency, the material in question, and the material temperature.

Averaging the fields (34) and (35) over time is meaningless since the rapidly oscillating factor $\exp(-\mathrm{i}\omega t)$ causes both averages to vanish:

$$\frac{1}{T}\int_{t-T/2}^{t+T/2} \mathrm{d}t' \exp(-\mathrm{i}\omega t') \underset{T \gg T_\mathrm{o}}{=} 0. \tag{44}$$

However, the vector product of the complex electric field and the complex conjugate of the magnetic field varies with time much more slowly since the factors $\exp(-\mathrm{i}\omega t)$ and $[\exp(-\mathrm{i}\omega t)]^*$ cancel each other. Therefore, averaging $\mathbf{e}(\mathbf{r},t)\times\mathbf{b}(\mathbf{r},t)$ over a period of time $T$ much longer than $T_\mathrm{o}$ but much shorter than the typical temporal scale $T'$ of variability of the complex amplitudes $\tilde{\mathbf{e}}(\mathbf{r},t)$ and $\tilde{\mathbf{b}}(\mathbf{r},t)$ is meaningful and yields a quasi-instantaneous Poynting vector slowly varying in time:

$$\langle\langle \mathbf{s}(\mathbf{r},t)\rangle\rangle = \frac{1}{\mu_0}\frac{1}{T}\int_{t-T/2}^{t+T/2} \mathrm{d}t'\mathbf{e}(\mathbf{r},t')\times\mathbf{b}(\mathbf{r},t')$$

$$\approx \frac{1}{2\mu_0} \mathrm{Re}\{\tilde{\mathbf{e}}(\mathbf{r},t) \times [\tilde{\mathbf{b}}(\mathbf{r},t)]^*\}, \qquad T_o \ll T \ll T', \tag{45}$$

where the symbol $\langle\langle \cdots \rangle\rangle$ hereinafter denotes averaging over a sufficiently long time interval, the actual length of the time interval being clear from the context. The time-independent macroscopic Poynting vector is now defined as the average over a time interval $T$ much longer than $T'$ :

$$\langle\langle \mathbf{S}(\mathbf{r},t) \rangle\rangle = \frac{1}{2\mu_0} \frac{1}{T} \mathrm{Re} \int_{t-T/2}^{t+T/2} \mathrm{d}t' \, \tilde{\mathbf{e}}(\mathbf{r},t') \times [\tilde{\mathbf{b}}(\mathbf{r},t')]^*, \qquad T \gg T'. \tag{46}$$

The computation of $\langle\langle \mathbf{S}(\mathbf{r},t) \rangle\rangle$ is usually simplified by assuming *ergodicity* of the ensemble of elementary charges (see, e.g., Section 10.4 of Ref. [34][5]) and replacing the temporal average in Eq. (46) by the statistical average over the ensemble $\psi$ of configurations of all the microscopic charges:

$$\langle\langle \mathbf{S}(\mathbf{r},t) \rangle\rangle = \frac{1}{2\mu_0} \mathrm{Re} \int \mathrm{d}\psi \, \tilde{\mathbf{e}}(\mathbf{r},\psi) \times [\tilde{\mathbf{b}}(\mathbf{r},\psi)]^* \, p(\psi), \tag{47}$$

where $p(\psi)$ is a suitable time-independent probability density function. Similar expressions can be written for macroscopic versions of other quadratic and bilinear forms in the microscopic electromagnetic field.

In general, the computation of the ensemble average (47) is still quite involved. A major simplification is based on the as yet unproven *assumption* according to which

$$\langle\langle \mathbf{S}(\mathbf{r},t) \rangle\rangle = \frac{1}{2\mu_0} \mathrm{Re}\{\tilde{\mathbf{E}}^{\psi}(\mathbf{r}) \times [\tilde{\mathbf{B}}^{\psi}(\mathbf{r})]^*\} \tag{48}$$

(and similarly for other second moments), where all materials are assumed to be non-magnetic and the corresponding macroscopic complex field vectors are given by the individual time averages replaced by ensemble averages:

$$\tilde{\mathbf{E}}^{\psi}(\mathbf{r}) = \frac{1}{T} \int_{t-T/2}^{t+T/2} \mathrm{d}t' \, \tilde{\mathbf{e}}(\mathbf{r},t') = \int \mathrm{d}\psi \, \tilde{\mathbf{e}}(\mathbf{r},\psi) p(\psi), \qquad T \gg T', \tag{49}$$

$$\tilde{\mathbf{B}}^{\psi}(\mathbf{r}) = \frac{1}{T} \int_{t-T/2}^{t+T/2} \mathrm{d}t' \, \tilde{\mathbf{b}}(\mathbf{r},t') = \int \mathrm{d}\psi \, \tilde{\mathbf{b}}(\mathbf{r},\psi) p(\psi), \qquad T \gg T'. \tag{50}$$

It can then be shown [131] that the macroscopic field vectors satisfy the standard frequency-domain MMEs

$$\nabla \cdot \tilde{\mathbf{D}}^{\psi}(\mathbf{r}) = \tilde{\rho}^{\psi}(\mathbf{r}), \tag{51}$$

$$\nabla \times \tilde{\mathbf{E}}^{\psi}(\mathbf{r}) - \mathrm{i}\omega \tilde{\mathbf{B}}^{\psi}(\mathbf{r}) = \mathbf{0}, \tag{52}$$

$$\nabla \cdot \tilde{\mathbf{B}}^{\psi}(\mathbf{r}) = 0, \tag{53}$$

$$\nabla \times \tilde{\mathbf{H}}^{\psi}(\mathbf{r}) + \mathrm{i}\omega \tilde{\mathbf{D}}^{\psi}(\mathbf{r}) = \tilde{\mathbf{J}}^{\psi}(\mathbf{r}) \tag{54}$$

---

[5] Instructive discussions of the ergodic hypothesis as a basic underlying principle of classical and quantum statistical mechanics and kinetic theory can be found in [145–149].

supplemented by the constitutive relations

$$\widetilde{\mathbf{D}}^{\psi}(\mathbf{r}) = \widetilde{\varepsilon}^{\psi}(\mathbf{r},\omega)\widetilde{\mathbf{E}}^{\psi}(\mathbf{r}), \tag{55}$$

$$\widetilde{\mathbf{H}}^{\psi}(\mathbf{r}) = \frac{1}{\mu_0}\widetilde{\mathbf{B}}^{\psi}(\mathbf{r}), \tag{56}$$

$$\widetilde{\mathbf{J}}^{\psi}(\mathbf{r}) = \widetilde{\sigma}^{\psi}(\mathbf{r},\omega)\widetilde{\mathbf{E}}^{\psi}(\mathbf{r}), \tag{57}$$

where, again, the frequency-dependent macroscopic electric permittivity $\widetilde{\varepsilon}^{\psi}(\mathbf{r},\omega)$ and conductivity $\widetilde{\sigma}^{\psi}(\mathbf{r},\omega)$ are, in general, complex valued.

We have seen in the preceding subsection that equations having the same mathematical structure as Eqs. (51)–(57) can also be obtained using the volume-averaging approach. As a consequence, it is often believed that a formal re-multiplication of the vectors $\widetilde{\mathbf{E}}^{\psi}(\mathbf{r})$ and $\widetilde{\mathbf{B}}^{\psi}(\mathbf{r})$ given by Eqs. (49) and (50) by the time-harmonic factor $\exp(-\mathrm{i}\omega t)$ yields the *ensemble-averaged* time-dependent electromagnetic field. This belief is questionable since ensemble averaging is not a primordial physical concept and can only be used as a substitute for time averaging (see, e.g., [146] and pages 1–6 of [150]). Averaging the right-hand sides of Eqs. (34) and (35) over time yields a zero result provided that $T_o \ll T'$. Therefore, the ensemble-averaged time-dependent electromagnetic field must also be zero. The reader should recall that the vectors $\widetilde{\mathbf{E}}^{\psi}(\mathbf{r})$ and $\widetilde{\mathbf{B}}^{\psi}(\mathbf{r})$ are obtained by:

- artificially omitting the time-harmonic factor $\exp(-\mathrm{i}\omega t)$ in Eqs. (34) and (35);
- taking the time average of the remaining factors; and
- replacing this time average by the ensemble average based on the assumption of ergodicity according to Eqs. (49) and (50).

It is thus obvious that the subsequent re-multiplication of $\widetilde{\mathbf{E}}^{\psi}(\mathbf{r})$ and $\widetilde{\mathbf{B}}^{\psi}(\mathbf{r})$ by $\exp(-\mathrm{i}\omega t)$ yields quantities of questionable veracity rather than actual time or ensemble averages.

### *3.3. Further discussion*

The ensemble averaging approach described in Subsection 3.2 bypasses the introduction of time-dependent macroscopic vector fields altogether, whereas the volume averaging approach outlined in Subsection 3.1 does yield macroscopic field vectors explicitly dependent on time as well as on coordinates. As a consequence, certain solutions of the time-domain MMEs (7)–(10) do describe vector waves unfolding in space and time, which may seem to provide substance to the widespread belief that $\mathbf{E}^{\delta V}(\mathbf{r},t)$ and $\mathbf{H}^{\delta V}(\mathbf{r},t)$ represent an actual physical field in a "homogenized macroscopic medium" rather than a purely mathematical entity. It is imperative to recognize however that (i) any type of averaging is a purely human intervention resulting in an artificial rather than an actual physical field; (ii) the computation of the macroscopic field vectors $\mathbf{E}^{\delta V}(\mathbf{r},t)$ and $\mathbf{H}^{\delta V}(\mathbf{r},t)$ (or $\widetilde{\mathbf{E}}^{\psi}(\mathbf{r})$ and $\widetilde{\mathbf{B}}^{\psi}(\mathbf{r})$) is almost never an end in itself; and (iii) irrespective of the averaging approach chosen, the ultimate purpose of macroscopic electromagnetics is the computation of time- and/or volume-averaged optical observables and the average electromagnetic energy budget. The MMEs are useful and meaningful only to the extent to which they help achieve this objective by eliminating the need to solve explicitly the microscopic

Maxwell–Lorentz equations. Despite substantial recent progress in the microphysical justification of the MMEs, this problem still awaits a definitive solution.

Although the formal mathematical structure of Eqs. (23)–(29) and (51)–(57) is the same, their solutions can, in principle, be different. Indeed, the specific procedures for the computation of the corresponding macroscopic electric permittivities and conductivities are not necessarily the same, and it is far from being obvious that $\tilde{\varepsilon}^{\delta V}(\mathbf{r},\omega) \equiv \tilde{\varepsilon}^{\psi}(\mathbf{r},\omega)$ and $\tilde{\sigma}^{\delta V}(\mathbf{r},\omega) \equiv \tilde{\sigma}^{\psi}(\mathbf{r},\omega)$. In what follows, we will usually omit the superscripts $\delta V$ and $\psi$ for the sake of brevity, but it should be kept in mind that the actual values of $\tilde{\varepsilon}(\mathbf{r},\omega)$ and $\tilde{\sigma}(\mathbf{r},\omega)$ may depend on the averaging approach chosen.

The relative merits of either homogenizing optical observables over physically small volume elements or averaging them over a time interval $T \gg T'$ remain somewhat unclear. The two averaging strategies are likely to yield similar results if they are applied to the calculation of the time-averaged radiation budget of a macroscopic volume with dimensions greatly exceeding the wavelength, provided that $\tilde{\varepsilon}^{\delta V}(\mathbf{r},\omega) \equiv \tilde{\varepsilon}^{\psi}(\mathbf{r},\omega)$ and $\tilde{\sigma}^{\delta V}(\mathbf{r},\omega) \equiv \tilde{\sigma}^{\psi}(\mathbf{r},\omega)$. However, the modeling of the interaction of the electromagnetic field with a photodetector may depend on whether one uses the macroscopic field vectors homogenized over physically small volume elements at a given moment in time or those averaged over time at a given point in space. This issue obviously needs to be further clarified.

On the more fundamental level, the MMEs must be derived in the framework of the QED by quantizing the microscopic electromagnetic field. There has been notable progress in this direction [151–154], but definitive studies are still needed.

## 4. Monochromatic and quasi-monochromatic scattering by a fixed macroscopic object

Consistent with the preceding discussion, the foundation of the theory of electromagnetic scattering by a DRM can be built of the following four major building blocks:

- the theory of monochromatic scattering by a fixed finite object;
- the theory of quasi-monochromatic scattering by a fixed finite object;
- the theory of monochromatic scattering by a randomly changing object; and
- the theory of quasi-monochromatic scattering by a randomly changing object.

The main purpose of this section is to give an explicit formulation of the electromagnetic scattering problem for a fixed object in maximally general terms and discuss its immediate implications. We start with monochromatic scattering and then, in Subsection 4.12, generalize the formalism to encompass the case of quasi-monochromatic radiation. Monochromatic and quasi-monochromatic scattering by a time-variable object such as a DRM will be considered in the following section.

Let us define the characteristic length $l$ according to

$$d \ll l \ll \lambda, \tag{58}$$

where $d$ is the typical distance between a molecule and its closest neighbors and $\lambda$ is the wavelength. A major premise of the previous discussion is that frequency-domain macroscopic electromagnetics can be used to

- compute the value of any second moment of the microscopic electromagnetic field homogenized over physically small volume elements with dimensions of the order of $l$ and averaged over a time interval $T \gg T_0$ (hereinafter averaging strategy 1; Subsection 3.1), or
- compute the value of any second moment of the microscopic electromagnetic field averaged over a time interval $T \gg T'$ at any fixed point in space (hereinafter averaging strategy 2; Subsection 3.2).

This is done by first solving the frequency-domain MMEs

$$\nabla \cdot \tilde{\mathbf{D}}(\mathbf{r}) = \tilde{\rho}(\mathbf{r}), \tag{59}$$

$$\nabla \times \tilde{\mathbf{E}}(\mathbf{r}) - \mathrm{i}\omega\tilde{\mathbf{B}}(\mathbf{r}) = \mathbf{0}, \tag{60}$$

$$\nabla \cdot \tilde{\mathbf{B}}(\mathbf{r}) = 0, \tag{61}$$

$$\nabla \times \tilde{\mathbf{H}}(\mathbf{r}) + \mathrm{i}\omega\tilde{\mathbf{D}}(\mathbf{r}) = \tilde{\mathbf{J}}(\mathbf{r}) \tag{62}$$

supplemented by the constitutive relations

$$\tilde{\mathbf{D}}(\mathbf{r}) = \tilde{\varepsilon}(\mathbf{r}, \omega)\tilde{\mathbf{E}}(\mathbf{r}), \tag{63}$$

$$\tilde{\mathbf{H}}(\mathbf{r}) = \frac{1}{\mu_0}\tilde{\mathbf{B}}(\mathbf{r}), \tag{64}$$

$$\tilde{\mathbf{J}}(\mathbf{r}) = \tilde{\sigma}(\mathbf{r}, \omega)\tilde{\mathbf{E}}(\mathbf{r}) \tag{65}$$

and then substituting the resulting macroscopic field vectors $\tilde{\mathbf{E}}(\mathbf{r})$ and $\tilde{\mathbf{B}}(\mathbf{r})$ in the appropriately modified formula for the requisite second moment of the microscopic field. Typically the modification amounts to applying the operation "$(1/2)\,\mathrm{Re}$". For example, the microscopic Poynting vector is defined by Eq. (14), while its time-independent average macroscopic counterpart is given by

$$\bar{\mathbf{S}}(\mathbf{r}) = \frac{1}{2}\mathrm{Re}\{\tilde{\mathbf{E}}(\mathbf{r}) \times [\tilde{\mathbf{H}}(\mathbf{r})]^*\}, \tag{66}$$

where $\bar{\mathbf{S}}(\mathbf{r})$ stands for either $\langle \mathbf{S}^{\delta V}(\mathbf{r}, t)\rangle$ or $\langle\langle \mathbf{S}(\mathbf{r}, t)\rangle\rangle$. According to [155], the frequency-domain macroscopic formalism can be expected to work well as long as the smallest homogeneous element of the scattering object exceeds ~50Å. And even for smaller elements, it may produce meaningful results if combined with empirical corrections for $\tilde{\varepsilon}(\mathbf{r}, \omega)$ and $\tilde{\sigma}(\mathbf{r}, \omega)$ [156].

The three basic ingredients of some phenomenological approaches to electromagnetic scattering by a DRM have been:

(i) the visual perception of the DRM as being assembled of separate "building blocks" in the form of particles;

(ii) the presumed knowledge of how to compute specific optical observables for each individual building block in the absence of all the other blocks; and

(iii) the belief that the optical observables for the assembly of the building blocks can somehow be expressed in terms of the optical observables computed for the separate building blocks.

The latter belief has usually been justified by verbal "simple physical considerations" and accepted as needing no rigorous mathematical proof. However, this approach is generally incorrect since unlike the human eye, the electromagnetic field perceives the DRM in its entirety rather than one particle at a time. Therefore, any first-principles approach to electromagnetic scattering by a DRM must originate in the explicit formulation of the MMEs and appropriate boundary conditions *for the entire DRM* rather than in the set of separate formulations for the individual particles.

### *4.1. The standard scattering problem*

Consider a fixed finite scattering object imbedded in an infinite medium; the latter is assumed to be homogeneous, linear, isotropic, nonmagnetic, and nonabsorbing. In general, the object is a cluster consisting of a finite number $N \ge 1$ of separated or touching distinct components. It occupies a finite interior region $V_{\mathrm{INT}}$ given by

$$V_{\mathrm{INT}} = \bigcup_{i=1}^{N} V_i, \tag{67}$$

where $V_i$ is the interior volume of the $i$th component (Fig. 6). The object is surrounded by the infinite exterior region $V_{\mathrm{EXT}}$ defined such that $V_{\mathrm{INT}} \cup V_{\mathrm{EXT}} = \Re^3$, where $\Re^3$ denotes the entire three-dimensional space. It is further assumed that the interior volume $V_{\mathrm{INT}}$ is filled with isotropic, linear, nonmagnetic, and possibly inhomogeneous material. Point $O$ serves as both the common origin of all position vectors and the origin of the laboratory coordinate system.

Unlike the general microscopic Maxwell–Lorentz equations (1)–(4), the four frequency-domain MMEs (59)–(62) are not independent since Eqs. (59) and (61) follow from Eqs. (60) and (62) [34]. This allows one to consider only the Maxwell curl equations, re-written as

$$\left.\begin{aligned} \nabla \times \tilde{\mathbf{E}}(\mathbf{r}) &= \mathrm{i}\omega\mu_0 \tilde{\mathbf{H}}(\mathbf{r}) \\ \nabla \times \tilde{\mathbf{H}}(\mathbf{r}) &= -\mathrm{i}\omega\varepsilon_1 \tilde{\mathbf{E}}(\mathbf{r}) \end{aligned}\right\} \quad \mathbf{r} \in V_{\mathrm{EXT}}, \tag{68}$$

$$\left.\begin{aligned} \nabla \times \tilde{\mathbf{E}}(\mathbf{r}) &= \mathrm{i}\omega\mu_0 \tilde{\mathbf{H}}(\mathbf{r}) \\ \nabla \times \tilde{\mathbf{H}}(\mathbf{r}) &= -\mathrm{i}\omega\varepsilon_2(\mathbf{r},\omega) \tilde{\mathbf{E}}(\mathbf{r}) \end{aligned}\right\} \quad \mathbf{r} \in V_{\mathrm{INT}}, \tag{69}$$

where $\varepsilon_1$ is the real-valued electric permittivity of the infinite host medium and

$$\varepsilon_2(\mathbf{r},\omega) = \tilde{\varepsilon}_2(\mathbf{r},\omega) + \mathrm{i}\,\frac{\tilde{\sigma}_2(\mathbf{r},\omega)}{\omega} \tag{70}$$

is the (potentially co-ordinate-dependent) so-called complex permittivity of the scattering object. The corresponding boundary conditions read:

$$\left.\begin{aligned} \hat{\mathbf{n}} \times [\tilde{\mathbf{E}}_1(\mathbf{r}) - \tilde{\mathbf{E}}_2(\mathbf{r})] &= \mathbf{0} \\ \hat{\mathbf{n}} \times [\tilde{\mathbf{H}}_1(\mathbf{r}) - \tilde{\mathbf{H}}_2(\mathbf{r})] &= \mathbf{0} \end{aligned}\right\} \quad \mathbf{r} \in S_{\mathrm{INT}}, \tag{71}$$

where the subscripts 1 and 2 correspond to the exterior and interior sides of the boundary $S_{\mathrm{INT}}$ of the object, respectively, $\hat{\mathbf{n}}$ is the local *outward* normal to $S_{\mathrm{INT}}$, and $S_{\mathrm{INT}}$ is the union of the closed surfaces of the $N$ components of the object:

$$S_{\mathrm{INT}} = \bigcup_{i=1}^{N} S_i. \tag{72}$$

Let us now assume that the complex amplitudes $\tilde{\mathbf{E}}(\mathbf{r})$ and $\tilde{\mathbf{H}}(\mathbf{r})$ everywhere in $\Re^3$ can be written as a superposition of the complex amplitudes of a plane-wave "incident field" (superscript "inc") propagating in the direction of the unit vector $\hat{\mathbf{n}}^{\text{inc}}$ and those of a "scattered field" (superscript "sca"):

$$\tilde{\mathbf{E}}(\mathbf{r}) = \tilde{\mathbf{E}}^{\text{inc}}(\mathbf{r}) + \tilde{\mathbf{E}}^{\text{sca}}(\mathbf{r}), \tag{73}$$

$$\tilde{\mathbf{H}}(\mathbf{r}) = \tilde{\mathbf{H}}^{\text{inc}}(\mathbf{r}) + \tilde{\mathbf{H}}^{\text{sca}}(\mathbf{r}), \tag{74}$$

where

$$\tilde{\mathbf{E}}^{\text{inc}}(\mathbf{r}) = \tilde{\mathbf{E}}_0^{\text{inc}} \exp(\mathrm{i}k_1\hat{\mathbf{n}}^{\text{inc}} \cdot \mathbf{r}), \tag{75}$$

$$\tilde{\mathbf{H}}^{\text{inc}}(\mathbf{r}) = \tilde{\mathbf{H}}_0^{\text{inc}} \exp(\mathrm{i}k_1\hat{\mathbf{n}}^{\text{inc}} \cdot \mathbf{r}) = \sqrt{\frac{\varepsilon_1}{\mu_0}}\, \hat{\mathbf{n}}^{\text{inc}} \times \tilde{\mathbf{E}}_0^{\text{inc}} \exp(\mathrm{i}k_1\hat{\mathbf{n}}^{\text{inc}} \cdot \mathbf{r}), \tag{76}$$

and

$$k_1 = \omega\sqrt{\varepsilon_1\mu_0} \tag{77}$$

is the wave number in the exterior region $V_{\text{EXT}}$. Furthermore, we require the scattered field amplitudes to satisfy the following asymptotic condition at infinity:

$$\lim_{r\to\infty} \{\sqrt{\mu_0}\, \mathbf{r} \times \tilde{\mathbf{H}}^{\text{sca}}(\mathbf{r}) + r\sqrt{\varepsilon_1}\, \tilde{\mathbf{E}}^{\text{sca}}(\mathbf{r})\} = \mathbf{0}, \tag{78}$$

where $r = |\mathbf{r}|$ is the distance from $O$ to the observation point (Fig. 6). The limit (78) is traditionally called the Silver–Müller radiation condition at infinity [157,158] and holds uniformly over all outgoing directions $\hat{\mathbf{r}} = \mathbf{r}/r$.

The combination of the Maxwell curl equations (68) and (69), the boundary conditions (71), the decomposition (73)–(76), and the asymptotic condition (78) represents the so-called *standard electromagnetic scattering problem for plane-wave illumination*.

The mathematical decomposition (73)–(76) of the "total" macroscopic frequency-domain field vectors makes it clear that the scattered field is defined as the difference between the total fields corresponding to the situations when the object is and is not present. This is consistent with the point of view that the incident field is not transformed into or replaced by the scattered field. In other words, the physical cause of frequency-domain scattering by the object is not the incident field, but rather the very presence of an object with optical properties different from that of the exterior medium [159,160].

Note that to the best of our knowledge, the boundary conditions (71) have not been derived explicitly from the microscopic Maxwell–Lorentz equations. *Ad hoc* ways of introducing these conditions in the framework of macroscopic electromagnetics are discussed in Section 2.8 of [161].

Since the first relations in Eqs. (68) and (69) yield the magnetic field vector provided that the electric field vector is known everywhere, the solution of the standard scattering problem is often sought in terms of only the electric field vector.

### *4.2. Existence and uniqueness of solution of the standard problem*

The statement of the standard scattering problem would be of little practical use if this problem had no solution and/or if the solution was not unique. Fortunately, both the existence and the uniqueness of solution have recently been demonstrated for particles with smooth surfaces (see [162,163] as well as Section 9.1 of [115]). Certain results for particles with edges do exist, but this case is fundamentally more difficult since the formulation of the boundary condition becomes ambiguous unless appropriately modified (see Chapter 9.2 of [115] and the discussion in Subsection 4.3).

### *4.3. Volume integral equation*

Although the standard scattering problem is formulated for the incident field in the form of a plane electromagnetic wave, the range of its actual applicability is much wider. Indeed, the linearity of both the MMEs, the boundary conditions, and the radiation condition at infinity implies that solving the standard problem yields the solution of a more general scattering problem as long as the corresponding incident electromagnetic field can be expanded in plane electromagnetic waves. This profound fact becomes especially evident if we consider a *mathematically equivalent formulation* [162,163] of electromagnetic scattering in terms of the so-called volume integral equation (VIE) [34,164] (see also [165]):

$$\begin{aligned}\tilde{\mathbf{E}}(\mathbf{r}) &= \tilde{\mathbf{E}}^{\mathrm{inc}}(\mathbf{r}) + k_1^2 \int_{V_{\mathrm{INT}}} \mathrm{d}^3\mathbf{r}' \overleftrightarrow{G}(\mathbf{r},\mathbf{r}') \cdot \tilde{\mathbf{E}}(\mathbf{r}')[m^2(\mathbf{r}') - 1] \\ &= \tilde{\mathbf{E}}^{\mathrm{inc}}(\mathbf{r}) + k_1^2 \left( \overleftrightarrow{I} + \frac{1}{k_1^2} \nabla \otimes \nabla \right) \cdot \int_{V_{\mathrm{INT}}} \mathrm{d}^3\mathbf{r}' \, \tilde{\mathbf{E}}(\mathbf{r}') \frac{\exp(\mathrm{i}k_1|\mathbf{r}-\mathbf{r}'|)}{4\pi|\mathbf{r}-\mathbf{r}'|}[m^2(\mathbf{r}') - 1], \quad \mathbf{r} \in \Re^3, \end{aligned} \tag{79}$$

where

$$\overleftrightarrow{G}(\mathbf{r},\mathbf{r}') = \left( \overleftrightarrow{I} + \frac{1}{k_1^2} \nabla \otimes \nabla \right) \frac{\exp(\mathrm{i}k_1|\mathbf{r}-\mathbf{r}'|)}{4\pi|\mathbf{r}-\mathbf{r}'|} \tag{80}$$

is the free-space dyadic Green's function,

$$m(\mathbf{r}') = \sqrt{\frac{\varepsilon_2(\mathbf{r}',\omega)}{\varepsilon_1}} \tag{81}$$

is the (complex) refractive index of the object's interior relative to that of the host exterior medium, $\overleftrightarrow{I}$ is the identity dyadic, and $\otimes$ is the dyadic product sign.

The second equality of Eq. (79) is a mathematically rigorous expression which has been used in [162,163,166] to deduce several useful corollaries. By contrast, the first equality is a shorter expression, but contains a strong singularity (strictly speaking, a non-integrable one) when $\mathbf{r} \in V_{\mathrm{INT}}$. Then the integration must be carried in the following specific principal-value sense to ensure that it is equivalent to the rigorous expression [167,168]:

$$\int_{V_{\mathrm{INT}}} \mathrm{d}^3\mathbf{r}' \overleftrightarrow{G}(\mathbf{r},\mathbf{r}') F(\mathbf{r}') = \lim_{V_0 \to 0} \int_{V_{\mathrm{INT}} \setminus V_0} \mathrm{d}^3\mathbf{r}' \overleftrightarrow{G}(\mathbf{r},\mathbf{r}') F(\mathbf{r}') - \frac{\mathrm{i}4\pi}{3} F(\mathbf{r}), \tag{82}$$

where $V_0$ is a spherical exclusion volume around $\mathbf{r}$. In what follows, we use the compact version of Eq. (79) and similar ones, but always assume that it implies the abbreviation (82).

The VIE incorporates the boundary and radiation conditions and expresses the total field everywhere in space in terms of the total internal field. It can even be considered to be more inclusive since it is well behaved (has a unique solution) even for particles with sharp edges [166].

Therefore, in the following we do not impose any limitations on the object's boundary and assume, in a somewhat *ad hoc* fashion, that in the presence of sharp edges the scattering problem is formulated through its VIE representation.

Owing to its specific mathematical form, the VIE serves as the very embodiment of the concept of frequency-domain electromagnetic scattering [159,160]. Indeed, it shows that if the scattering object is absent $(m(\mathbf{r}') \equiv 1)$, then the total field is identically equal to the incident field. The presence of the object $(m(\mathbf{r}') \neq 1)$ changes the total field, thereby allowing the definition of the scattered field as the difference between the total fields in the presence and in the absence of the object. Furthermore, the VIE implies that the incident field is not modified by the presence of the object and, thus, is not transformed into the scattered field.

The linearity of the VIE suggests that it should be convenient in many cases to express, purely mathematically, the scattered electric field in terms of the incident field:

$$\tilde{\mathbf{E}}^{\text{sca}}(\mathbf{r}) = \int_{V_{\text{INT}}} \mathrm{d}^3\mathbf{r}' \overleftrightarrow{G}(\mathbf{r},\mathbf{r}') \cdot \int_{V_{\text{INT}}} \mathrm{d}^3\mathbf{r}'' \overleftrightarrow{T}(\mathbf{r}',\mathbf{r}'') \cdot \tilde{\mathbf{E}}^{\text{inc}}(\mathbf{r}''), \qquad \mathbf{r} \in \Re^3, \tag{83}$$

where $\overleftrightarrow{T}$ is the so-called dyadic transition operator of the scattering object. Eqs. (79) and (83) imply the following integral equation for $\overleftrightarrow{T}$ traditionally called the Lippmann–Schwinger equation (cf. [10,169,170]):

$$\overleftrightarrow{T}(\mathbf{r},\mathbf{r}') = k_1^2[m^2(\mathbf{r}) - 1]\,\delta(\mathbf{r}-\mathbf{r}')\overleftrightarrow{I} + k_1^2[m^2(\mathbf{r}) - 1]\int_{V_{\text{INT}}} \mathrm{d}^3\mathbf{r}'' \overleftrightarrow{G}(\mathbf{r},\mathbf{r}'') \cdot \overleftrightarrow{T}(\mathbf{r}'',\mathbf{r}'), \qquad \mathbf{r},\mathbf{r}' \in V_{\text{INT}}, \tag{84}$$

where $\delta(\mathbf{r})$ is the three-dimensional delta function. Note that $\overleftrightarrow{T}$ is independent of the electromagnetic field and is defined only by the spatial distribution of the relative refractive index throughout $V_{\text{INT}}$. As such, it can be viewed as a complete "optical identifier" of the scattering object.

### 4.4. *Scattering in the far zone of the entire object*

The spatial distribution of $\tilde{\mathbf{E}}(\mathbf{r})$ and $\tilde{\mathbf{H}}(\mathbf{r})$ inside the scattering object as well as in its immediate vicinity can be quite complex. However, there is a drastic simplification as the distance from the object increases since, irrespective of the specific nature of the object, the scattered field ultimately becomes a spherical outgoing electromagnetic wave. Indeed, a key property of the dyadic Green's function is the asymptotic behavior

$$\overleftrightarrow{G}(\mathbf{r},\mathbf{r}') \underset{r\to\infty}{\to} (\overleftrightarrow{I} - \hat{\mathbf{r}} \otimes \hat{\mathbf{r}}) \frac{\exp(\mathrm{i}k_1 r)}{4\pi r} \exp(-\mathrm{i}k_1 \hat{\mathbf{r}} \cdot \mathbf{r}'), \tag{85}$$

where, as before, $\hat{\mathbf{r}} = \mathbf{r}/r$. As a consequence, placing the origin of the laboratory coordinate system $O$ at the geometrical center of the scattering object, as shown in Fig. 7, and substituting Eqs. (75) and (85) in Eq. (83) yields [34]

$$\tilde{\mathbf{E}}^{\text{sca}}(\mathbf{r}) \underset{r\to\infty}{\to} \frac{\exp(\mathrm{i}k_1 r)}{r} \tilde{\mathbf{E}}_1^{\text{sca}}(\hat{\mathbf{n}}^{\text{sca}}) = \frac{\exp(\mathrm{i}k_1 r)}{r} \overleftrightarrow{A}(\hat{\mathbf{n}}^{\text{sca}}, \hat{\mathbf{n}}^{\text{inc}}) \cdot \tilde{\mathbf{E}}_0^{\text{inc}}, \qquad \hat{\mathbf{n}}^{\text{sca}} \cdot \tilde{\mathbf{E}}_1^{\text{sca}}(\hat{\mathbf{n}}^{\text{sca}}) = 0. \tag{86}$$

Here, $\hat{\mathbf{n}}^{\text{sca}} = \hat{\mathbf{r}}$ is a unit vector in the scattering direction and $\overleftrightarrow{A}(\hat{\mathbf{n}}^{\text{sca}}, \hat{\mathbf{n}}^{\text{inc}})$ is the scattering dyad-ic having the dimension of length and possessing the properties

$$\hat{\mathbf{n}}^{\text{sca}} \cdot \overleftrightarrow{A}(\hat{\mathbf{n}}^{\text{sca}}, \hat{\mathbf{n}}^{\text{inc}}) = \overleftrightarrow{A}(\hat{\mathbf{n}}^{\text{sca}}, \hat{\mathbf{n}}^{\text{inc}}) \cdot \hat{\mathbf{n}}^{\text{inc}} = \mathbf{0}. \tag{87}$$

The explicit expression for the scattering dyadic in terms of the dyadic transition operator is as follows:

$$\overleftrightarrow{A}(\hat{\mathbf{n}}^{\text{sca}}, \hat{\mathbf{n}}^{\text{inc}}) = \frac{1}{4\pi}(\overleftrightarrow{I} - \hat{\mathbf{n}}^{\text{sca}} \otimes \hat{\mathbf{n}}^{\text{sca}}) \cdot \int_{V_{\text{INT}}} \mathrm{d}^3\mathbf{r}' \exp(-\mathrm{i}k_1 \hat{\mathbf{n}}^{\text{sca}} \cdot \mathbf{r}')$$
$$\times \int_{V_{\text{INT}}} \mathrm{d}^3\mathbf{r}'' \overleftrightarrow{T}(\mathbf{r}', \mathbf{r}'') \exp(\mathrm{i}k_1 \hat{\mathbf{n}}^{\text{inc}} \cdot \mathbf{r}'') \cdot (\overleftrightarrow{I} - \hat{\mathbf{n}}^{\text{inc}} \otimes \hat{\mathbf{n}}^{\text{inc}}). \tag{88}$$

Eq. (86) implies that the electric and magnetic field vectors of the scattered electromagnetic field vibrate in the plane perpendicular to the propagation direction and decay inversely with distance from the object.

The formal mathematical conditions of applicability of Eq. (86) are as follows:

$$k_1(r - a) \gg 1, \tag{89}$$

$$r \gg a, \tag{90}$$

$$r \gg \frac{k_1 a^2}{2}, \tag{91}$$

where $a$ is the radius of the smallest circumscribing sphere of the entire scattering object centered at $O$. The physical meaning of these inequalities is discussed in [34,171].

The main attraction of the far-zone approximation is that the entire object is implicitly treated as a point source of scattered radiation, while the scattered field is reduced to a simple outgoing spherical wave. Furthermore, Eq. (87) implies that out of the nine components of the scattering dyadic in spherical coordinates centered at the origin, only four are independent. It is thus convenient to introduce a $2 \times 2$ amplitude scattering matrix $\mathbf{S}$ having the dimension of length and expressing the $\theta$- and $\varphi$-components of the scattered spherical wave in the $\theta$- and $\varphi$-components of the incident plane wave:

$$\begin{bmatrix} \tilde{E}_\theta^{\text{sca}}(r\hat{\mathbf{n}}^{\text{sca}}) \\ \tilde{E}_\varphi^{\text{sca}}(r\hat{\mathbf{n}}^{\text{sca}}) \end{bmatrix} = \frac{\exp(\mathrm{i}k_1 r)}{r} \mathbf{S}(\hat{\mathbf{n}}^{\text{sca}}, \hat{\mathbf{n}}^{\text{inc}}) \begin{bmatrix} \tilde{E}_{0\theta}^{\text{inc}} \\ \tilde{E}_{0\varphi}^{\text{inc}} \end{bmatrix}. \tag{92}$$

Here, $\theta \in [0, \pi]$ is the polar (zenith) angle measured from the positive $z$-axis and $\varphi \in [0, 2\pi)$ is the azimuth angle measured from the positive $x$-axis in the clockwise direction when looking in the direction of the positive $z$-axis.

A fundamental property of the scattering dyadic and the amplitude scattering matrix is the following symmetry with respect to reversing and interchanging the incidence and scattering directions [172]:

$$\overleftrightarrow{A}(-\hat{\mathbf{n}}^{\text{inc}}, -\hat{\mathbf{n}}^{\text{sca}}) = [\overleftrightarrow{A}(\hat{\mathbf{n}}^{\text{sca}}, \hat{\mathbf{n}}^{\text{inc}})]^{\text{T}}, \tag{93a}$$

$$\mathbf{S}(-\hat{\mathbf{n}}^{\text{inc}}, -\hat{\mathbf{n}}^{\text{sca}}) = \begin{bmatrix} S_{11}(\hat{\mathbf{n}}^{\text{sca}}, \hat{\mathbf{n}}^{\text{inc}}) & -S_{21}(\hat{\mathbf{n}}^{\text{sca}}, \hat{\mathbf{n}}^{\text{inc}}) \\ -S_{12}(\hat{\mathbf{n}}^{\text{sca}}, \hat{\mathbf{n}}^{\text{inc}}) & S_{22}(\hat{\mathbf{n}}^{\text{sca}}, \hat{\mathbf{n}}^{\text{inc}}) \end{bmatrix}, \tag{93b}$$

where T stands for "transposed." Eqs. (93a) and (93b) are traditionally called reciprocity rela-tions.

### 4.5. *Well-collimated radiometers*

By solving the MMEs either analytically or numerically, one can model a wide range of optical observables, including those that can be measured with actual optical instruments. Some of these instruments are expressly intended for near-field applications [173], while some can measure both near- and far-field observables. As explained in [100,174], the overwhelming majority of laboratory, *in situ*, and remote-sensing instruments measuring specific manifestations of electromagnetic energy transport in particulate media belong to the category of well-collimated radiometers (WCRs). Depending on the measurement setting, these instruments can work in the near as well as in the far zone of the particulate scattering object, but in either case it is assumed that the total electromagnetic field at the observation point is a superposition of plane or near-plane wavefronts.

The principal functional elements of a WCR are the objective and relay lenses, the pinhole diaphragm, and the photoelectric detector, as shown schematically in Fig. 8a. The physical nature of the measurement afforded by the WCR can be illustrated by considering the response of the instrument to the field formed by superposing two plane electromagnetic waves propagating in the directions of the unit vectors $\hat{\mathbf{q}}_1$ and $\hat{\mathbf{q}}_2$, respectively. The effect of the objective lens on the total field is a superposition of its effects on each plane-wave component. According to the paraxial approximation (see Section 5.1 of [175]), in the near zone of the objective lens either plane wavefront is transformed into a converging spherical wavefront (Fig. 8b) with its focal point located in the plane of the diaphragm. The first spherical wavefront passes freely through the pinhole, is converted back into a plane wavefront by the relay lens, and impinges on the sensitive surface of the photodetector, thereby defining the signal generated by the WCR. The second spherical wavefront becomes extinguished by the diaphragm and never reaches the photodetector. Thus the combination {objective lens, diaphragm} serves to filter out only plane (or near-plane) wavefronts propagating in directions very close to the optical axis of the WCR and falling within its small acceptance solid angle

$$\Delta\Omega = \frac{\pi d^2}{4 f^2}, \tag{94}$$

where $d$ is the diameter of the pinhole and $f$ is the focal length of the objective lens.

Typically a photodetector reacts only to the intensity of the beam impinging on its sensitive surface. However, by inserting special optical elements (such as polarizers and retarders) between the relay lens and the detector in Fig. 8a, it is possible to modify the resulting beam impinging on the detector in such a way that its new intensity contains information about the polarization state of the original superposition of the plane or near-plane wavefronts filtered out by the {objective lens, diaphragm} combination. The result is a photopolarimetric WCR.

Despite having quite different appearances, the one natural and six manmade devices in Fig. 9 have the same basic physical function: they filter out electromagnetic wavefronts rather than electromagnetic energy currents. In a radio telescope (Fig. 9b) or a reflecting optical telescope (Fig. 9c), the functional role of the objective lens is played by the radio antenna or the primary mirror, respectively. In the final analysis, however, all these devices are WCRs, possibly with an added panoramic (or imaging) capability[6]. The basic functionality of a WCR makes it quite useful in cases when the total electromagnetic field at the observation point can naturally be represented as a superposition of plane or locally near-plane wavefronts, the electromagnetic field in the far zone of a finite object being a prime example. Since the measurement enabled by a WCR is well defined in terms of basic concepts of light–matter interactions, it should be amenable to theoretical modeling. This explains why the combination of a WCR and an appropriate theoretical analysis tool often serves as an efficient means of optical characterization.

### *4.6. Far-zone optical observables*

The formalism embodied by Eqs. (75) and (92) helps illustrate how to define specific far-field optical observables measurable with WCRs. The main results of the following analysis will be straightforward consequences of the total field in the far zone being a superposition of two transverse wavefronts, i.e., the incident plane-wave field and the scattered spherical-wave field.

As we have already mentioned, interposing one or more optical elements between the relay lens and the photodetector of a WCR can enable it to measure the power corresponding to particular polarization components of the superposition of plane or near-plane wavefront passed by the {objective lens, diaphragm} angular filter. Similarly, interposing one or more such optical elements before the scattering object, we can generate the incident field with a specific state of polarization. Repeating the measurement for a number of different combinations and/or orientations of these optical elements enables us to determine the specific mathematical relationship between a complete set of polarization characteristics of the incident field and that of the field impinging on the objective lens of a WCR. This relationship is traditionally formulated in terms of 4-element columns formed by the Stokes parameters and $4\times4$ so-called phase and extinction matrices.

According to the preceding discussion, a WCR that is not facing the incident wave and is not centered at the scattering object will not generate any signal. Therefore, let us first consider the situation when the instrument has its optical axis centered at the object in the scattering direction away from the incidence direction, i.e., $\hat{\mathbf{r}} \neq \hat{\mathbf{n}}^{\text{inc}}$ (WCR 1 in Fig. 10). It is clear that in this case the instrument filters out only the quasi-plane part of the outgoing spherical wave cut out by its objective lens, as shown schematically by the dashed curve. Therefore, the average polarization response of WCR 1 per unit time can be expressed in terms of the so-called Stokes column vector of the scattered wave as follows:

$$\overline{\textbf{Signal 1}} = S_{\text{ol}}\mathbf{I}^{\text{sca}}(r\hat{\mathbf{n}}^{\text{sca}}), \tag{95}$$

where the overbar has the same meaning as in Eq. (66), $r$ is the distance from the scattering object to WCR 1, $\hat{\mathbf{n}}^{\text{sca}} = \hat{\mathbf{r}}_1$, and $S_{\text{ol}}$ is the surface area of the objective lens. Recalling the definition of the real-valued Stokes parameters of a transverse electromagnetic wave [20,34] and Eq. (92), we have

[6] In this case each pixel of a charge-coupled device or each photoreceptor cell of the retina has a dual role of the diaphragm and the detector of electromagnetic energy flow [34].

$$\mathbf{I}^{\text{sca}}(r\hat{\mathbf{n}}^{\text{sca}}) = \begin{bmatrix} I^{\text{sca}}(r\hat{\mathbf{n}}^{\text{sca}}) \\ Q^{\text{sca}}(r\hat{\mathbf{n}}^{\text{sca}}) \\ U^{\text{sca}}(r\hat{\mathbf{n}}^{\text{sca}}) \\ V^{\text{sca}}(r\hat{\mathbf{n}}^{\text{sca}}) \end{bmatrix}$$

$$= \frac{1}{2r^2}\sqrt{\frac{\varepsilon_1}{\mu_0}} \begin{bmatrix} \widetilde{E}_{1\theta}^{\text{sca}}(\hat{\mathbf{n}}^{\text{sca}})[\widetilde{E}_{1\theta}^{\text{sca}}(\hat{\mathbf{n}}^{\text{sca}})]^* + \widetilde{E}_{1\varphi}^{\text{sca}}(\hat{\mathbf{n}}^{\text{sca}})[\widetilde{E}_{1\varphi}^{\text{sca}}(\hat{\mathbf{n}}^{\text{sca}})]^* \\ \widetilde{E}_{1\theta}^{\text{sca}}(\hat{\mathbf{n}}^{\text{sca}})[\widetilde{E}_{1\theta}^{\text{sca}}(\hat{\mathbf{n}}^{\text{sca}})]^* - \widetilde{E}_{1\varphi}^{\text{sca}}(\hat{\mathbf{n}}^{\text{sca}})[\widetilde{E}_{1\varphi}^{\text{sca}}(\hat{\mathbf{n}}^{\text{sca}})]^* \\ -\widetilde{E}_{1\theta}^{\text{sca}}(\hat{\mathbf{n}}^{\text{sca}})[\widetilde{E}_{1\varphi}^{\text{sca}}(\hat{\mathbf{n}}^{\text{sca}})]^* - \widetilde{E}_{1\varphi}^{\text{sca}}(\hat{\mathbf{n}}^{\text{sca}})[\widetilde{E}_{1\theta}^{\text{sca}}(\hat{\mathbf{n}}^{\text{sca}})]^* \\ \mathrm{i}\{\widetilde{E}_{1\varphi}^{\text{sca}}(\hat{\mathbf{n}}^{\text{sca}})[\widetilde{E}_{1\theta}^{\text{sca}}(\hat{\mathbf{n}}^{\text{sca}})]^* - \widetilde{E}_{1\theta}^{\text{sca}}(\hat{\mathbf{n}}^{\text{sca}})[\widetilde{E}_{1\varphi}^{\text{sca}}(\hat{\mathbf{n}}^{\text{sca}})]^*\} \end{bmatrix}. \tag{96}$$

Analogously, the polarization state of the plane incident wave (75) can be characterized in terms of the Stokes column vector

$$\mathbf{I}^{\text{inc}} = \begin{bmatrix} I^{\text{inc}} \\ Q^{\text{inc}} \\ U^{\text{inc}} \\ V^{\text{inc}} \end{bmatrix} = \frac{1}{2}\sqrt{\frac{\varepsilon_1}{\mu_0}} \begin{bmatrix} \widetilde{E}_{0\theta}^{\text{inc}}(\widetilde{E}_{0\theta}^{\text{inc}})^* + \widetilde{E}_{0\varphi}^{\text{inc}}(\widetilde{E}_{0\varphi}^{\text{inc}})^* \\ \widetilde{E}_{0\theta}^{\text{inc}}(\widetilde{E}_{0\theta}^{\text{inc}})^* - \widetilde{E}_{0\varphi}^{\text{inc}}(\widetilde{E}_{0\varphi}^{\text{inc}})^* \\ -\widetilde{E}_{0\theta}^{\text{inc}}(\widetilde{E}_{0\varphi}^{\text{inc}})^* - \widetilde{E}_{0\varphi}^{\text{inc}}(\widetilde{E}_{0\theta}^{\text{inc}})^* \\ \mathrm{i}[\widetilde{E}_{0\varphi}^{\text{inc}}(\widetilde{E}_{0\theta}^{\text{inc}})^* - \widetilde{E}_{0\theta}^{\text{inc}}(\widetilde{E}_{0\varphi}^{\text{inc}})^*] \end{bmatrix}. \tag{97}$$

The corresponding transformation law reads:

$$\mathbf{I}^{\text{sca}}(r\hat{\mathbf{n}}^{\text{sca}}) = \frac{1}{r^2}\mathbf{Z}(\hat{\mathbf{n}}^{\text{sca}}, \hat{\mathbf{n}}^{\text{inc}})\mathbf{I}^{\text{inc}}, \tag{98}$$

where $\mathbf{Z}(\hat{\mathbf{n}}^{\text{sca}}, \hat{\mathbf{n}}^{\text{inc}})$ is the $4 \times 4$ Stokes phase matrix with real-valued elements given by the following quadratic and bilinear combinations of the elements of the amplitude scattering matrix $\mathbf{S}(\hat{\mathbf{n}}^{\text{sca}}, \hat{\mathbf{n}}^{\text{inc}})$ [20,25,34]:

$$Z_{11} = \tfrac{1}{2}(|S_{11}|^2 + |S_{12}|^2 + |S_{21}|^2 + |S_{22}|^2), \tag{99}$$

$$Z_{12} = \tfrac{1}{2}(|S_{11}|^2 - |S_{12}|^2 + |S_{21}|^2 - |S_{22}|^2), \tag{100}$$

$$Z_{13} = -\mathrm{Re}(S_{11}S_{12}^* + S_{22}S_{21}^*), \tag{101}$$

$$Z_{14} = -\mathrm{Im}(S_{11}S_{12}^* - S_{22}S_{21}^*), \tag{102}$$

$$Z_{21} = \tfrac{1}{2}(|S_{11}|^2 + |S_{12}|^2 - |S_{21}|^2 - |S_{22}|^2), \tag{103}$$

$$Z_{22} = \tfrac{1}{2}(|S_{11}|^2 - |S_{12}|^2 - |S_{21}|^2 + |S_{22}|^2), \tag{104}$$

$$Z_{23} = \mathrm{Re}(S_{11}S_{12}^* - S_{22}S_{21}^*), \tag{105}$$

$$Z_{24} = -\mathrm{Im}(S_{11}S_{12}^* + S_{22}S_{21}^*), \tag{106}$$

$$Z_{31} = -\mathrm{Re}(S_{11}S_{21}^* + S_{22}S_{12}^*), \tag{107}$$

$$Z_{32} = -\mathrm{Re}(S_{11}S_{21}^* - S_{22}S_{12}^*), \tag{108}$$

$$Z_{33} = \mathrm{Re}(S_{11}S_{22}^* + S_{12}S_{21}^*), \tag{109}$$

$$Z_{34} = \mathrm{Im}(S_{11}S_{22}^* + S_{21}S_{12}^*), \tag{110}$$

$$Z_{41} = -\mathrm{Im}(S_{21}S_{11}^* + S_{22}S_{12}^*), \tag{111}$$

$$Z_{42} = -\mathrm{Im}(S_{21}S_{11}^* - S_{22}S_{12}^*), \tag{112}$$

$$Z_{43} = \mathrm{Im}(S_{22}S_{11}^* - S_{12}S_{21}^*), \tag{113}$$

$$Z_{44} = \mathrm{Re}(S_{22}S_{11}^* - S_{12}S_{21}^*). \tag{114}$$

The Stokes phase matrix has the dimension of area.

Let us now consider a polarimetric WCR with its axis centered at the scattering object in the exact forward-scattering direction $\hat{\mathbf{r}} = \hat{\mathbf{n}}^{\mathrm{inc}}$, i.e., WCR 2 in Fig. 10. Now the {objective lens, diaphragm} angular filter of the instrument accepts both the incident plane wave and the part of the outgoing spherical wave propagating in the forward-scattering direction and cut out by the objective lens. As a consequence, we can define the Stokes column vector of the total field for propagation directions $\hat{\mathbf{r}}'$ very close to $\hat{\mathbf{n}}^{\mathrm{inc}}$:

$$\mathbf{I}(r'\hat{\mathbf{r}}') = \frac{1}{2}\sqrt{\frac{\varepsilon_1}{\mu_0}} \begin{bmatrix} \tilde{E}_\theta(r'\hat{\mathbf{r}}')[\tilde{E}_\theta(r'\hat{\mathbf{r}}')]^* + \tilde{E}_\varphi(r'\hat{\mathbf{r}}')[\tilde{E}_\varphi(r'\hat{\mathbf{r}}')]^* \\ \tilde{E}_\theta(r'\hat{\mathbf{r}}')[\tilde{E}_\theta(r'\hat{\mathbf{r}}')]^* - \tilde{E}_\varphi(r'\hat{\mathbf{r}}')[\tilde{E}_\varphi(r'\hat{\mathbf{r}}')]^* \\ -\tilde{E}_\theta(r'\hat{\mathbf{r}}')[\tilde{E}_\varphi(r'\hat{\mathbf{r}}')]^* - \tilde{E}_\varphi(r'\hat{\mathbf{r}}')[\tilde{E}_\theta(r'\hat{\mathbf{r}}')]^* \\ \mathrm{i}\{\tilde{E}_\varphi(r'\hat{\mathbf{r}}')[\tilde{E}_\theta(r'\hat{\mathbf{r}}')]^* - \tilde{E}_\theta(r'\hat{\mathbf{r}}')[\tilde{E}_\varphi(r'\hat{\mathbf{r}}')]^*\} \end{bmatrix}, \quad \hat{\mathbf{r}}' \in \Delta\Omega_2, \tag{115}$$

where $\Delta\Omega_2$ is the acceptance solid angle of WCR 2 and the total electric field is given by

$$\tilde{\mathbf{E}}(r'\hat{\mathbf{r}}') = \tilde{\mathbf{E}}^{\mathrm{inc}}(r'\hat{\mathbf{r}}') + \tilde{\mathbf{E}}^{\mathrm{sca}}(r'\hat{\mathbf{r}}'). \tag{116}$$

Integrating the elements of $\mathbf{I}(r'\hat{\mathbf{r}}')$ over the objective lens of WCR 2 yields the following expression for the average recorded polarized signal per unit time [20,25,34]:

$$\overline{\mathbf{Signal\ 2}} = S_{\mathrm{ol}}\mathbf{I}^{\mathrm{inc}} - \mathbf{K}(\hat{\mathbf{n}}^{\mathrm{inc}})\mathbf{I}^{\mathrm{inc}} + \frac{S_{\mathrm{ol}}}{r^2}\mathbf{Z}(\hat{\mathbf{n}}^{\mathrm{inc}}, \hat{\mathbf{n}}^{\mathrm{inc}})\mathbf{I}^{\mathrm{inc}}, \tag{117}$$

where $\mathbf{Z}(\hat{\mathbf{n}}^{\mathrm{inc}}, \hat{\mathbf{n}}^{\mathrm{inc}})$ is the forward-scattering Stokes phase matrix and $\mathbf{K}(\hat{\mathbf{n}}^{\mathrm{inc}})$ is the real $4 \times 4$ Stokes extinction matrix. The elements of the latter are linear combinations of the elements of the forward-scattering amplitude matrix $\mathbf{S}(\hat{\mathbf{n}}^{\mathrm{inc}}, \hat{\mathbf{n}}^{\mathrm{inc}})$:

$$K_{jj} = \frac{2\pi}{k_1}\mathrm{Im}(S_{11} + S_{22}), \quad j = 1, \ldots, 4, \tag{118}$$

$$K_{12} = K_{21} = \frac{2\pi}{k_1}\mathrm{Im}(S_{11} - S_{22}), \tag{119}$$

$$K_{13} = K_{31} = -\frac{2\pi}{k_1}\mathrm{Im}(S_{12} + S_{21}), \tag{120}$$

$$K_{14} = K_{41} = \frac{2\pi}{k_1}\mathrm{Re}(S_{21} - S_{12}), \tag{121}$$

$$K_{23} = -K_{32} = \frac{2\pi}{k_1}\,\mathrm{Im}(S_{21} - S_{12}), \tag{122}$$

$$K_{24} = -K_{42} = -\frac{2\pi}{k_1}\,\mathrm{Re}(S_{12} + S_{21}), \tag{123}$$

$$K_{34} = -K_{43} = \frac{2\pi}{k_1}\,\mathrm{Re}(S_{22} - S_{11}), \tag{124}$$

where "Im" stands for "imaginary part of". Like the phase matrix, the extinction matrix has the dimension of area.

Eq. (117) is the most general form of the so-called optical theorem. It demonstrates that the presence of the scattering object not only changes the total power of the electromagnetic radiation recorded by the WCR facing the incident wave (WCR 2 in Fig. 10), but also can change its state of polarization. The latter phenomenon is called dichroism and is caused by different attenuation rates for different polarization components of the incident wave in the case of an object lacking perfect spherical symmetry. Moving WCR 2 sufficiently far from the scattering object can render the contribution of the third term on the right-hand side of Eq. (117) negligibly small,

$$\overline{\textbf{Signal 2}} \underset{r\to\infty}{=} S_{\mathrm{ol}}\mathbf{I}^{\mathrm{inc}} - \mathbf{K}(\hat{\mathbf{n}}^{\mathrm{inc}})\mathbf{I}^{\mathrm{inc}}, \tag{125}$$

and thereby make the extinction matrix a directly measurable quantity.

Among the most general properties of the phase and extinction matrices [18,20,178] are the inequalities

$$0 \le Z_{11}, \qquad |Z_{ij}| \le Z_{11} \quad (i, j = 1, \ldots, 4); \tag{126}$$

the reciprocity relations

$$\mathbf{Z}(-\hat{\mathbf{n}}^{\mathrm{inc}}, -\hat{\mathbf{n}}^{\mathrm{sca}}) = \mathbf{\Delta}_3[\mathbf{Z}(\hat{\mathbf{n}}^{\mathrm{sca}}, \hat{\mathbf{n}}^{\mathrm{inc}})]^{\mathrm{T}}\mathbf{\Delta}_3 \tag{127}$$

and

$$\mathbf{K}(-\hat{\mathbf{n}}^{\mathrm{inc}}) = \mathbf{\Delta}_3[\mathbf{K}(\hat{\mathbf{n}}^{\mathrm{inc}})]^{\mathrm{T}}\mathbf{\Delta}_3;$$

the backscattering theorem

$$Z_{11}(-\hat{\mathbf{n}}, \hat{\mathbf{n}}) - Z_{22}(-\hat{\mathbf{n}}, \hat{\mathbf{n}}) + Z_{33}(-\hat{\mathbf{n}}, \hat{\mathbf{n}}) - Z_{44}(-\hat{\mathbf{n}}, \hat{\mathbf{n}}) = 0; \tag{128}$$

and the symmetry relation

$$\mathbf{K}(-\hat{\mathbf{n}}^{\mathrm{inc}}) = \begin{bmatrix} K_{11}(\hat{\mathbf{n}}^{\mathrm{inc}}) & K_{12}(\hat{\mathbf{n}}^{\mathrm{inc}}) & -K_{13}(\hat{\mathbf{n}}^{\mathrm{inc}}) & K_{14}(\hat{\mathbf{n}}^{\mathrm{inc}}) \\ K_{21}(\hat{\mathbf{n}}^{\mathrm{inc}}) & K_{22}(\hat{\mathbf{n}}^{\mathrm{inc}}) & K_{23}(\hat{\mathbf{n}}^{\mathrm{inc}}) & -K_{24}(\hat{\mathbf{n}}^{\mathrm{inc}}) \\ -K_{31}(\hat{\mathbf{n}}^{\mathrm{inc}}) & K_{32}(\hat{\mathbf{n}}^{\mathrm{inc}}) & K_{33}(\hat{\mathbf{n}}^{\mathrm{inc}}) & K_{34}(\hat{\mathbf{n}}^{\mathrm{inc}}) \\ K_{41}(\hat{\mathbf{n}}^{\mathrm{inc}}) & -K_{42}(\hat{\mathbf{n}}^{\mathrm{inc}}) & K_{43}(\hat{\mathbf{n}}^{\mathrm{inc}}) & K_{44}(\hat{\mathbf{n}}^{\mathrm{inc}}) \end{bmatrix}, \tag{129}$$

where

$$\mathbf{\Delta}_3 = \mathbf{\Delta}_3^{\mathrm{T}} = \mathbf{\Delta}_3^{-1} = \begin{bmatrix} 1 & 0 & 0 & 0 \\ 0 & 1 & 0 & 0 \\ 0 & 0 & -1 & 0 \\ 0 & 0 & 0 & 1 \end{bmatrix}. \tag{130}$$

The properties (127)–(130) follow directly from the reciprocity relation (93) combined with Eqs. (99)–(114) and (118)–(124).

### 4.7. *Energy budget*

In the preceding subsection we explained how to quantify the electromagnetic response of a far-field WCR. In this subsection we discuss the theoretical solution of the energy-budget problem for an arbitrary volume $V$ enclosing the entire scattering object (Fig. 11a). Since the host medium is assumed to be nonabsorbing, the net average rate at which the electromagnetic energy crosses the closed boundary $S$ of the volume is always nonnegative and is equal to the power absorbed by the object:

$$\overline{W}^{\mathrm{abs}} = -\int_S \mathrm{d}^2\mathbf{r}\, \overline{\mathbf{S}}(\mathbf{r}) \cdot \hat{\mathbf{n}}, \tag{131}$$

where, as before,

$$\overline{\mathbf{S}}(\mathbf{r}) = \frac{1}{2}\,\mathrm{Re}\{\widetilde{\mathbf{E}}(\mathbf{r}) \times [\widetilde{\mathbf{H}}(\mathbf{r})]^*\} \tag{132}$$

is the average macroscopic Poynting vector and $\hat{\mathbf{n}}$ is the unit vector in the direction of the local outward normal to $S$. According to Eqs. (73) and (74), $\overline{W}^{\mathrm{abs}}$ can be written as a combination of three terms:

$$\overline{W}^{\mathrm{abs}} = \overline{W}^{\mathrm{inc}} - \overline{W}^{\mathrm{sca}} + \overline{W}^{\mathrm{ext}}, \tag{133}$$

where

$$\overline{W}^{\mathrm{inc}} = -\frac{1}{2}\,\mathrm{Re}\int_S \mathrm{d}^2\mathbf{r}\, \{\widetilde{\mathbf{E}}^{\mathrm{inc}}(\mathbf{r}) \times [\widetilde{\mathbf{H}}^{\mathrm{inc}}(\mathbf{r})]^*\} \cdot \hat{\mathbf{n}}, \tag{134}$$

$$\overline{W}^{\mathrm{sca}} = \frac{1}{2}\,\mathrm{Re}\int_S \mathrm{d}^2\mathbf{r}\, \{\widetilde{\mathbf{E}}^{\mathrm{sca}}(\mathbf{r}) \times [\widetilde{\mathbf{H}}^{\mathrm{sca}}(\mathbf{r})]^*\} \cdot \hat{\mathbf{n}}, \tag{135}$$

$$\overline{W}^{\mathrm{ext}} = -\frac{1}{2}\,\mathrm{Re}\int_S \mathrm{d}^2\mathbf{r}\, \{\widetilde{\mathbf{E}}^{\mathrm{inc}}(\mathbf{r}) \times [\widetilde{\mathbf{H}}^{\mathrm{sca}}(\mathbf{r})]^* + \widetilde{\mathbf{E}}^{\mathrm{sca}}(\mathbf{r}) \times [\widetilde{\mathbf{H}}^{\mathrm{inc}}(\mathbf{r})]^*\} \cdot \hat{\mathbf{n}}. \tag{136}$$

It can easily be seen that $\overline{W}^{\mathrm{inc}}$ vanishes identically because $\{\widetilde{\mathbf{E}}^{\mathrm{inc}}(\mathbf{r}) \times [\widetilde{\mathbf{H}}^{\mathrm{inc}}(\mathbf{r})]^*\}/2$ is a constant vector independent of $\mathbf{r}$, which is a trivial consequence of the surrounding medium being lossless. Therefore, the absorption rate is equal to the difference between the energy extinction rate and the energy scattering rate:

$$\overline{W}^{\mathrm{abs}} = \overline{W}^{\mathrm{ext}} - \overline{W}^{\mathrm{sca}}. \tag{137}$$

Again, one can exploit the assumption that the infinite host medium surrounding the object is nonabsorbing to show that the values of $\overline{W}^{\mathrm{ext}}$ and $\overline{W}^{\mathrm{sca}}$ would not change if $V$ were chosen to be a spherical volume with its boundary $S$ residing in the far zone of the entire object. Then it is straightforward to derive that

$$\overline{W}^{\text{ext}} = K_{11}(\hat{\mathbf{n}}^{\text{inc}})I^{\text{inc}} + K_{12}(\hat{\mathbf{n}}^{\text{inc}})Q^{\text{inc}} + K_{13}(\hat{\mathbf{n}}^{\text{inc}})U^{\text{inc}} + K_{14}(\hat{\mathbf{n}}^{\text{inc}})V^{\text{inc}}, \tag{138}$$

$$\overline{W}^{\text{sca}} = \int_{4\pi} \mathrm{d}\hat{\mathbf{r}}\,[Z_{11}(\hat{\mathbf{r}},\hat{\mathbf{n}}^{\text{inc}})I^{\text{inc}} + Z_{12}(\hat{\mathbf{r}},\hat{\mathbf{n}}^{\text{inc}})Q^{\text{inc}} + Z_{13}(\hat{\mathbf{r}},\hat{\mathbf{n}}^{\text{inc}})U^{\text{inc}} + Z_{14}(\hat{\mathbf{r}},\hat{\mathbf{n}}^{\text{inc}})V^{\text{inc}}] \tag{139}$$

(see Sec. 13.4 of [34]).

It is important to recognize that although the extinction and phase matrices are inherently far-field quantities, Eqs. (137)–(139) are valid for any volume enclosing the entire scattering object even if its boundary lies in the object's near field. Of course, a trivial modification of Eqs. (137)–(139) would not work for a volume enclosing only part of the scattering object, as illustrated in Fig. 11b, since in this case one would need to know the specific near-field spatial distribution of the Poynting vector over $S$. Eqs. (137)–(139) would also not apply if the host medium were absorbing [179].

### *4.8. Foldy equations*

The general scattering formalism described in Subsections 4.1–4.3 applies equally to an object in the form of a single body and to a fixed multi-particle group. However, when the object is a group of touching and/or separated distinct components then it can sometimes be advantageous to modify the formalism by expressing the total scattered field as a vector superposition of the partial fields contributed by the individual components. Specifically, let us consider the scattering by a fixed configuration of $N \geq 2$ distinct finite particles collectively occupying the interior region $V_{\text{INT}}$ according to Eq. (67) (see Fig. 6). It has been shown in [180,181] (see also Section 11.3 of [170]) that the solution of the VIE everywhere in space can be expressed as follows:

$$\tilde{\mathbf{E}}(\mathbf{r}) = \tilde{\mathbf{E}}^{\text{inc}}(\mathbf{r}) + \sum_{i=1}^{N} \int_{V_i} \mathrm{d}^3\mathbf{r}' \overleftrightarrow{G}(\mathbf{r},\mathbf{r}') \cdot \int_{V_i} \mathrm{d}^3\mathbf{r}'' \overleftrightarrow{T}_i(\mathbf{r}',\mathbf{r}'') \cdot \tilde{\mathbf{E}}_i(\mathbf{r}''), \qquad \mathbf{r} \in \Re^3, \tag{140}$$

where the electric field vector $\tilde{\mathbf{E}}_i(\mathbf{r})$ "exciting" particle $i$ is given by

$$\tilde{\mathbf{E}}_i(\mathbf{r}) = \tilde{\mathbf{E}}^{\text{inc}}(\mathbf{r}) + \sum_{j(\neq i)=1}^{N} \tilde{\mathbf{E}}_{ij}^{\text{exc}}(\mathbf{r}), \tag{141}$$

the $\tilde{\mathbf{E}}_{ij}^{\text{exc}}(\mathbf{r})$ are "particle–particle exciting field vectors" given by

$$\tilde{\mathbf{E}}_{ij}^{\text{exc}}(\mathbf{r}) = \int_{V_j} \mathrm{d}^3\mathbf{r}' \overleftrightarrow{G}(\mathbf{r},\mathbf{r}') \cdot \int_{V_j} \mathrm{d}^3\mathbf{r}'' \overleftrightarrow{T}_j(\mathbf{r}',\mathbf{r}'') \cdot \tilde{\mathbf{E}}_j(\mathbf{r}''), \qquad \mathbf{r} \in V_i, \tag{142}$$

and $\overleftrightarrow{T}_i$ is the $i$th-particle dyadic transition operator with respect to the common laboratory coordinate system. The $\overleftrightarrow{T}_i$ satisfies the integral equation

$$\overleftrightarrow{T}_i(\mathbf{r},\mathbf{r}') = U_i(\mathbf{r})\delta(\mathbf{r}-\mathbf{r}')\overleftrightarrow{I} + U_i(\mathbf{r})\int_{V_i} \mathrm{d}^3\mathbf{r}'' \overleftrightarrow{G}(\mathbf{r},\mathbf{r}'') \cdot \overleftrightarrow{T}_i(\mathbf{r}'',\mathbf{r}'), \qquad \mathbf{r},\mathbf{r}' \in V_i, \tag{143}$$

where the $U_i(\mathbf{r})$ is the $i$th-particle potential function given by

$$U_i(\mathbf{r}) = \begin{cases} 0, & \mathbf{r} \notin V_i, \\ k_1^2[m_i^2(\mathbf{r}) - 1], & \mathbf{r} \in V_i \end{cases} \tag{144}$$

and $m_i(\mathbf{r})$ is the refractive index of particle $i$ relative to that of the host medium. All position vectors originate at the common origin $O$ of the laboratory coordinate system. Eqs. (140)–(143) form the system of integral so-called Foldy equations (FEs). They automatically incorporate all boundary conditions at individual-particle surfaces as well as the radiation condition at infinity and rigorously describe the scattered field everywhere in space. Comparison of Eqs. (84) and (143) reveals that, quite conveniently, $\overleftrightarrow{T}_i$ is the dyadic transition operator of the $i$th particle in the absence of all the other particles. As such, it can be considered an individual optical identifier of the $i$th component of the group.

### *4.9. Frequency-domain "multiple" scattering*

Ever since Heaviside's *Electromagnetic Theory* [182], the concept of "multiple" scattering has been quite popular in discussions of electromagnetic scattering by multi-particle groups (see [183,184] and references therein). To demonstrate the actual nature of this concept in the context of frequency-domain electromagnetics [92], let us introduce the $i$th potential dyadic centered at the origin of the laboratory reference frame according to

$$\overleftrightarrow{U}_i(\mathbf{r}, \mathbf{r}') = U_i(\mathbf{r})\delta(\mathbf{r} - \mathbf{r}')\overleftrightarrow{I} \tag{145}$$

and introduce the following operator notation:

$$\hat{B}E = \int \mathrm{d}^3\mathbf{r}' \overleftrightarrow{B}(\mathbf{r}, \mathbf{r}') \cdot \widetilde{\mathbf{E}}(\mathbf{r}'). \tag{146}$$

Iterating Eqs. (141) and (142), we have

$$E_i = E^{\mathrm{inc}} + \sum_{j(\neq i)=1}^{N} \hat{G}\hat{T}_j E^{\mathrm{inc}} + \sum_{j(\neq i)=1}^{N} \sum_{l(\neq j)=1}^{N} \hat{G}\hat{T}_j \hat{G}\hat{T}_l E^{\mathrm{inc}} + \sum_{j(\neq i)=1}^{N} \sum_{l(\neq j)=1}^{N} \sum_{m(\neq l)=1}^{N} \hat{G}\hat{T}_j \hat{G}\hat{T}_l \hat{G}\hat{T}_m E^{\mathrm{inc}} + \cdots, \tag{147}$$

whereas substituting Eq. (147) in Eq. (140) yields the following Neumann expansion of the total field:

$$E = E^{\mathrm{inc}} + \sum_{i=1}^{N} \hat{G}\hat{T}_i E^{\mathrm{inc}} + \sum_{i=1}^{N} \sum_{j(\neq i)=1}^{N} \hat{G}\hat{T}_i \hat{G}\hat{T}_j E^{\mathrm{inc}} + \sum_{i=1}^{N} \sum_{j(\neq i)=1}^{N} \sum_{l(\neq j)=1}^{N} \hat{G}\hat{T}_i \hat{G}\hat{T}_j \hat{G}\hat{T}_l E^{\mathrm{inc}} + \cdots. \tag{148}$$

It is clear that the Neumann series is fundamentally based on the previously mentioned fact that $\overleftrightarrow{T}_i$ for each $i$ is an individual property of the $i$th particle computed as if this particle were alone rather than a member of the group. As a consequence, it has been rather common to characterize

Eq. (148) as the "order-of-scattering expansion" for the $N$-particle group. It should be remembered however that the FEs have a solution even when the corresponding Neumann series (148) does not converge. Numerical examples of possible divergence can be found in [185,186].

### 4.10. Far-field Foldy equations

In principle, the FEs can be solved numerically in order to compute the field scattered by a fixed finite configuration of arbitrarily positioned particles. However, the solution becomes impracticable quite rapidly with increasing $N$. One way to simplify the problem and make it tractable is to consider a very sparse configuration by assuming that:

- the $N$ particles are widely separated so that each of them resides in the far zones of all the other particles; and
- the observation point is located in the far zone of any particle (but not necessarily in the far zone of the entire group).

Specifically, if the incident electric field vector is given by Eq. (75) then the FEs imply that the total electric field vector is still given by the superposition (73), where the scattered electric field vector is now given by [34,181]

$$\tilde{\mathbf{E}}^{\text{sca}}(\mathbf{r}) = \sum_{i=1}^{N} g(r_i)\left[\overleftrightarrow{A}_i(\hat{\mathbf{r}}_i, \hat{\mathbf{n}}^{\text{inc}}) \cdot \tilde{\mathbf{E}}^{\text{inc}}(\mathbf{R}_i) + \sum_{j(\neq i)=1}^{N} \overleftrightarrow{A}_i(\hat{\mathbf{r}}_i, \hat{\mathbf{R}}_{ij}) \cdot \tilde{\mathbf{E}}_{ij}\right]. \tag{149}$$

Here, $\overleftrightarrow{A}_i(\hat{\mathbf{n}}', \hat{\mathbf{n}})$ is the far-zone scattering dyadic of particle $i$ centered at the particle's own origin $O_i$ (Subsection 4.4);

$$g(r) = \frac{\exp(\mathrm{i}k_1 r)}{r}; \tag{150}$$

and the vectors $\tilde{\mathbf{E}}_{ij}$ satisfy the following system of equations:

$$\tilde{\mathbf{E}}_{ij} = g(R_{ij})\overleftrightarrow{A}_j(\hat{\mathbf{R}}_{ij}, \hat{\mathbf{n}}^{\text{inc}}) \cdot \tilde{\mathbf{E}}^{\text{inc}}(\mathbf{R}_j) + g(R_{ij}) \sum_{l(\neq j)=1}^{N} \overleftrightarrow{A}_j(\hat{\mathbf{R}}_{ij}, \hat{\mathbf{R}}_{jl}) \cdot \tilde{\mathbf{E}}_{jl}, \quad i, j = 1, \ldots, N, \quad j \neq i. \tag{151}$$

The vector notation used in Eqs. (149) and (151) is explained in Fig. 12; a hat denotes a unit vector in the respective direction. Eqs. (149)–(151) are called the far-field FEs. It is evident indeed that the linear algebraic system (151) is much simpler than the initial system of integral equations (141)–(142).

Eqs. (73) and (149) show that the total field at any observation point located sufficiently far from any particle in the sparse multi-particle configuration is a superposition of the incident plane wave and $N$ partial spherical wavelets centered at the $N$ particles. Importantly, the observation point $\mathbf{r}$ does not have to be in the far zone of the entire $N$-particle group: it can be anywhere in space (e.g., between particles $i$ and $j$ in Fig. 12) as long as it is in the far zone of any particle entering the group. The total scattered field (149) is not, in general, a transverse electromagnetic wave. It becomes a transverse wave only in the far zone of the entire $N$-particle configuration defined by the criteria (89)–(91), where $a$ is the radius of the smallest sphere encompassing all $N$ particles.

*4.11. Dyadic correlation function and Poynting–Stokes tensor*

We have already mentioned in Section 2 that it can be possible in some cases to derive an analytical expression for an optical observable that is explicitly devoid of the electromagnetic field. Sometimes this expression is a closed-form equation solving which can serve as a highly efficient means of calculating the optical observable directly, without the prior detailed computation of the electromagnetic field itself.

The general methodology enabling one to bypass an explicit use of the electromagnetic field is well exemplified by the far-field formulas (95), (98), and (117), in which case the observable in question is the 4-element Stokes column vector. However, this observable can be defined only for a transverse (i.e., plane or spherical) electromagnetic wave, whereas the total electromagnetic field in the near zone of any object (e.g., at any observation point inside a cloud of particles) is never a transverse wave. Furthermore, the Stokes column vector contains no explicit information on the direction of the Poynting vector and cannot be used in situations when this direction is not known *a priori*.

The Poynting vector is another optical observable extensively discussed in the preceding sections. Its obvious analytical limitation is that different pairs of electric and magnetic field vectors can yield the same Poynting vector. This implies that the Poynting vector cannot be used to describe the phenomenon of electromagnetic scattering by, for example, expressing the Poynting vector of the scattered field in that of the incident field. In other words, the Poynting vector does not carry sufficient information about the participating electric and magnetic fields and, in particular, carries no information about the polarization state of a transverse electromagnetic field.

It is therefore highly desirable to introduce an alternative quantity that:

- can be defined for any electromagnetic field;
- has the dimension of electromagnetic energy flux; and
- enables a complete and self-contained description of electromagnetic scattering in the context of practical optical analysis.

It has been shown in [34,187] that a rather general quantity satisfying all these requirements is the so-called dyadic correlation function involving electric field vectors at two different points in space:

$$\overleftrightarrow{C}(\mathbf{r}', \mathbf{r}) = \mathbf{E}(\mathbf{r}') \otimes [\mathbf{E}(\mathbf{r})]^{*}. \tag{152}$$

This quantity along with the Maxwell curl equations (68) and (69) can be used to compute other observables, including those involving both the electric and the magnetic field vector. An important example is the so-called Poynting–Stokes tensor defined as the dyadic product of the magnetic and complex-conjugated electric field vectors taken at the same point in space:

$$\overleftrightarrow{P}(\mathbf{r}) = \frac{1}{2} \widetilde{\mathbf{H}}(\mathbf{r}) \otimes \widetilde{\mathbf{E}}^{*}(\mathbf{r}). \tag{153}$$

Indeed, it is easily verified that

$$\overleftrightarrow{P}(\mathbf{r}) = \frac{1}{2\mathrm{i}\omega\mu_0} [\nabla_{\mathbf{r}'} \times \overleftrightarrow{C}(\mathbf{r}', \mathbf{r})] \big|_{\mathbf{r}'=\mathbf{r}}, \tag{154}$$

where the subscript $\mathbf{r}'$ means that the $\nabla$ operator acts only on $\mathbf{E}(\mathbf{r}')$. Unlike the Stokes parameters, this quantity is applicable to any electromagnetic field (e.g., the near field of a scattering object) and not just to a transverse electromagnetic wave. Furthermore, unlike the Poynting vector, the Poynting–Stokes tensor preserves all the information about the participating electric and magnetic fields that gets lost upon taking the vector product of these fields. We will see in the following sections that owing to their generality, the dyadic correlation function and the Poynting–Stokes tensor enable the derivation of compact closed-form analytical formulas and equations directly describing electromagnetic scattering in terms of optical observables.

It is straightforward to see that with respect to the Poynting–Stokes tensor, the Poynting vector and the Stokes parameters are derivative quantities. Indeed, we have in general

$$\overline{\mathbf{S}}(\mathbf{r}) = \mathrm{Re}\{[P^*_{zy}(\mathbf{r}) - P^*_{yz}(\mathbf{r})]\,\hat{\mathbf{x}} + [P^*_{xz}(\mathbf{r}) - P^*_{zx}(\mathbf{r})]\,\hat{\mathbf{y}} + [P^*_{yx}(\mathbf{r}) - P^*_{xy}(\mathbf{r})]\,\hat{\mathbf{z}}\}, \tag{155}$$

where $\hat{\mathbf{x}}$, $\hat{\mathbf{y}}$, and $\hat{\mathbf{z}}$ are the unit vectors of a right-handed Cartesian coordinate system. In the case of a transverse electromagnetic wave, the Stokes column vector is given by

$$\mathbf{I}(\mathbf{r}) = \begin{bmatrix} P^*_{\varphi\theta}(\mathbf{r}) - P^*_{\theta\varphi}(\mathbf{r}) \\ P^*_{\varphi\theta}(\mathbf{r}) + P^*_{\theta\varphi}(\mathbf{r}) \\ P^*_{\theta\theta}(\mathbf{r}) - P^*_{\varphi\varphi}(\mathbf{r}) \\ \mathrm{i}[P^*_{\theta\theta}(\mathbf{r}) + P^*_{\varphi\varphi}(\mathbf{r})] \end{bmatrix}, \tag{156}$$

where $\theta$ and $\varphi$ are the zenith and azimuth angles defining the local direction $\hat{\mathbf{n}}$ of wave propagation such that $\hat{\mathbf{n}} = \hat{\boldsymbol{\theta}} \times \hat{\boldsymbol{\varphi}}$.

According to Eqs. (73), (75), and (83), the total field can be expressed as

$$\tilde{\mathbf{E}}(\mathbf{r}) = \overset{\leftrightarrow}{\mathfrak{I}}_E(\mathbf{r}, \hat{\mathbf{n}}^{\mathrm{inc}}) \cdot \tilde{\mathbf{E}}_0^{\mathrm{inc}}, \qquad \mathbf{r} \in \mathfrak{R}^3, \tag{157}$$

where $\overset{\leftrightarrow}{\mathfrak{I}}_E(\mathbf{r}, \hat{\mathbf{n}}^{\mathrm{inc}})$ is a transformation dyadic independent of $\tilde{\mathbf{E}}_0^{\mathrm{inc}}$ and given by

$$\overset{\leftrightarrow}{\mathfrak{I}}_E(\mathbf{r}, \hat{\mathbf{n}}^{\mathrm{inc}}) = \exp(\mathrm{i}k_1\hat{\mathbf{n}}^{\mathrm{inc}} \cdot \mathbf{r})\overset{\leftrightarrow}{I} + \int_{V_{\mathrm{INT}}} \mathrm{d}^3\mathbf{r}'\overset{\leftrightarrow}{G}(\mathbf{r}, \mathbf{r}') \cdot \int_{V_{\mathrm{INT}}} \mathrm{d}^3\mathbf{r}''\overset{\leftrightarrow}{T}(\mathbf{r}', \mathbf{r}'')\exp(\mathrm{i}k_1\hat{\mathbf{n}}^{\mathrm{inc}} \cdot \mathbf{r}''). \tag{158}$$

It follows from Eqs. (68), (69), (74), and (76) that a relationship similar to Eq. (157) must exist for the magnetic field as well:

$$\tilde{\mathbf{H}}(\mathbf{r}) = \overset{\leftrightarrow}{\mathfrak{I}}_H(\mathbf{r}, \hat{\mathbf{n}}^{\mathrm{inc}}) \cdot \tilde{\mathbf{H}}_0^{\mathrm{inc}}, \qquad \mathbf{r} \in \mathfrak{R}^3, \tag{159}$$

where the transformation dyadic $\hat{\mathfrak{I}}_H(\mathbf{r}, \hat{\mathbf{n}}^{\mathrm{inc}})$ is independent of $\tilde{\mathbf{H}}_0^{\mathrm{inc}}$ and is given by

$$\overset{\leftrightarrow}{\mathfrak{I}}_H(\mathbf{r}, \hat{\mathbf{n}}^{\mathrm{inc}}) = \frac{\mathrm{i}}{k_1}\nabla \times \overset{\leftrightarrow}{\mathfrak{I}}_E(\mathbf{r}, \hat{\mathbf{n}}^{\mathrm{inc}}) \times \hat{\mathbf{n}}^{\mathrm{inc}}. \tag{160}$$

Then we have for the Poynting–Stokes tensor of the total field [34]:

$$\overset{\leftrightarrow}{P}(\mathbf{r}) = \overset{\leftrightarrow}{\mathfrak{I}}_H(\mathbf{r}, \hat{\mathbf{n}}^{\mathrm{inc}}) \cdot \overset{\leftrightarrow}{P}^{\mathrm{inc}} \cdot [\overset{\leftrightarrow}{\mathfrak{I}}_E(\mathbf{r}, \hat{\mathbf{n}}^{\mathrm{inc}})]^{\mathrm{T}*}, \qquad \mathbf{r} \in \mathfrak{R}^3, \tag{161}$$

where

$$\overset{\leftrightarrow}{P}^{\mathrm{inc}} = \frac{1}{2}\tilde{\mathbf{H}}_0^{\mathrm{inc}} \otimes [\tilde{\mathbf{E}}_0^{\mathrm{inc}}]^* \tag{162}$$

is the Poynting–Stokes tensor calculated separately for the plane-wave incident field.

We will see below that Eq. (161) is a general template for many closed-form relationships between *observable* characteristics of the incident and total fields. Importantly, this formula demonstrates that the elements of the tensor $\vec{\vec{P}}(\mathbf{r})$ generally depend on all the elements of the tensor $\vec{\vec{P}}^{\rm inc}$. In other words, as we have already mentioned, the complex Poynting vector of the total field cannot be uniquely expressed in that of the incident field. This implies that the widespread characterization of electromagnetic scattering as causing the transformation of the intensity of the incident light into that of the scattered light is fundamentally wrong.

Eq. (161) implies the existence of a linear (but not necessarily tensorial) operator expressing the Poynting–Stokes tensor of the total field in that of the incident plane-wave field [34]. We will denote this operator by $\hat{\mathfrak{I}}$ and write symbolically:

$$\vec{\vec{P}}(\mathbf{r}) = \hat{\mathfrak{I}}(\mathbf{r}, \hat{\mathbf{n}}^{\rm inc})\vec{\vec{P}}^{\rm inc}. \tag{163}$$

The reader may find it instructive to rewrite Eq. (163) in the matrix form with respect to the Cartesian laboratory coordinate system and thereby express the elements of the $9 \times 9$ matrix representing the operator $\hat{\mathfrak{I}}$ in terms of the elements of the $3 \times 3$ matrices representing the dyadics $\vec{\vec{\mathfrak{I}}}_E$ and $\vec{\vec{\mathfrak{I}}}_H$. Analogously, we can write

$$\vec{\vec{P}}^{\rm sca}(\mathbf{r}) = \hat{\mathfrak{I}}^{\rm sca}(\mathbf{r}, \hat{\mathbf{n}}^{\rm inc})\vec{\vec{P}}^{\rm inc}, \tag{164}$$

where

$$\vec{\vec{P}}^{\rm sca} = \frac{1}{2}\widetilde{\mathbf{H}}^{\rm sca}(\mathbf{r}) \otimes [\widetilde{\mathbf{E}}^{\rm sca}(\mathbf{r})]^* \tag{165}$$

is the Poynting–Stokes tensor calculated separately for the scattered field.

Eqs. (163) and (164) represent a remarkably compact yet general way of describing electromagnetic scattering in terms of optical observables rather than macroscopic field vectors. As such, they will be central to the following discussion, especially when it comes to the scattering of quasi-monochromatic fields by temporally variable objects. The reader can verify that Eqs. (98) and (117) are but specific coordinate-dependent manifestations of these formulas.

Formulas analogous to Eqs. (163) and (164) can be derived for optical observables other than the Poynting–Stokes tensor. Each such formula serves as a linear transformer with an optical observable of the incident electromagnetic field as the input and an optical observable of the total or scattered electromagnetic field as the output. Such linear transformers are essential in practice because of the two-layer structure of electromagnetics discussed in Section 2.

### *4.12. Quasi-monochromatic scattering by a fixed object*

The formalism summarized above provides an efficient means of computing time-averaged macroscopic optical observables without solving explicitly the microscopic Maxwell–Lorentz equations. It is based, in particular, on the assumption that the complex amplitudes $\widetilde{\mathbf{E}}_0^{\rm inc}$ and $\widetilde{\mathbf{H}}_0^{\rm inc} = \sqrt{\varepsilon_1/\mu_0}\,\hat{\mathbf{n}}^{\rm inc} \times \widetilde{\mathbf{E}}_0^{\rm inc}$ entering the solution of the standard scattering problem are independent of time. Let us now imagine a situation wherein these amplitudes remain constant over periods of time $T_{\rm f}$ such that

$$T_{\rm f} \gg T_{\rm o} \text{ (averaging strategy 1)} \tag{166}$$

and

$$T_{\mathrm{f}} \gg T' \text{ (averaging strategy 2)}, \tag{167}$$

but fluctuate over longer time scales. In other words,

$$\widetilde{\mathbf{E}}^{\mathrm{inc}}(\mathbf{r},t) = \widetilde{\mathbf{E}}_0^{\mathrm{inc}}(t)\exp(\mathrm{i}k_1\hat{\mathbf{n}}^{\mathrm{inc}}\cdot\mathbf{r}), \tag{168}$$

$$\widetilde{\mathbf{H}}^{\mathrm{inc}}(\mathbf{r},t) = \widetilde{\mathbf{H}}_0^{\mathrm{inc}}(t)\exp(\mathrm{i}k_1\hat{\mathbf{n}}^{\mathrm{inc}}\cdot\mathbf{r}) = \sqrt{\frac{\varepsilon_1}{\mu_0}}\,\hat{\mathbf{n}}^{\mathrm{inc}}\times\widetilde{\mathbf{E}}_0^{\mathrm{inc}}(t)\exp(\mathrm{i}k_1\hat{\mathbf{n}}^{\mathrm{inc}}\cdot\mathbf{r}), \tag{169}$$

where significant random changes of the complex amplitude $\widetilde{\mathbf{E}}_0^{\mathrm{inc}}(t)$ occur over periods of time longer than $T_{\mathrm{f}}$. The solution of the standard scattering problem for a temporal succession of $\widetilde{\mathbf{E}}_0^{\mathrm{inc}}(t)$-values then yields a temporal succession of the total field vector values that can be thought of as defining time-dependent macroscopic field vectors fluctuating randomly on time scales longer than $T_{\mathrm{f}}$: $\{\widetilde{\mathbf{E}}(\mathbf{r}),\widetilde{\mathbf{H}}(\mathbf{r})\}\to\{\widetilde{\mathbf{E}}(\mathbf{r},t),\widetilde{\mathbf{H}}(\mathbf{r},t)\}$.

According to the above discussion, the quasi-instantaneous values of $\widetilde{\mathbf{E}}(\mathbf{r},t)$ and $\widetilde{\mathbf{H}}(\mathbf{r},t)$ are postulated to define optical observables averaged over time intervals of the order of $T_{\mathrm{f}}$. For example,

$$\overline{\mathbf{S}}(\mathbf{r},t) = \frac{1}{2}\mathrm{Re}\{\widetilde{\mathbf{E}}(\mathbf{r},t)\times[\widetilde{\mathbf{H}}(\mathbf{r},t)]^*\} \tag{170}$$

and

$$\vec{\vec{P}}(\mathbf{r},t) = \frac{1}{2}\widetilde{\mathbf{H}}(\mathbf{r},t)\otimes\widetilde{\mathbf{E}}^*(\mathbf{r},t). \tag{171}$$

The corresponding averages over much longer time intervals are then calculated according to

$$\langle\langle\overline{\mathbf{S}}(\mathbf{r},t)\rangle\rangle = \frac{1}{T}\int_{t-T/2}^{t+T/2}\mathrm{d}t'\,\overline{\mathbf{S}}(\mathbf{r},t'), \qquad T\gg T_{\mathrm{f}}, \tag{172}$$

$$\langle\langle\vec{\vec{P}}(\mathbf{r},t)\rangle\rangle = \frac{1}{T}\int_{t-T/2}^{t+T/2}\mathrm{d}t'\,\vec{\vec{P}}(\mathbf{r},t'), \qquad T\gg T_{\mathrm{f}}. \tag{173}$$

These averages are time independent provided that $\overline{\mathbf{S}}(\mathbf{r},t)$ and $\vec{\vec{P}}(\mathbf{r},t)$ are stationary random processes (see, e.g., [188]).

The random macroscopic field vectors $\widetilde{\mathbf{E}}(\mathbf{r},t)$ and $\widetilde{\mathbf{H}}(\mathbf{r},t)$ are traditionally said to represent a quasi-monochromatic macroscopic electromagnetic field. In particular, Eqs. (168) and (169) are said to describe a quasi-monochromatic plane electromagnetic wave (or a quasi-monochromatic parallel beam of light).

Despite the inequalities (166) and (167), typical fluctuations of quasi-instantaneous optical observables still occur too rapidly to be traced by many optical instruments. It is therefore *postulated* that the intrinsic functionality of such instruments is to record the integral of an optical observable over an extended period of time without resolving the quasi-instantaneous values of this observable explicitly.[7] Thus the practical usefulness of the notion of a quasi-monochromatic electromagnetic field turns out to be two-fold. First, it helps combine the simplicity of the frequency-domain scattering formalism with a more realistic representation of the majority of artificial and natural sources of the electromagnetic field. Second, it allows one to account for inherent limitations of typical optical devices.

The generalization of Eqs. (161), (163), and (164) to the case of quasi-monochromatic scattering by a fixed object is quite straightforward:

$$\langle\langle \vec{\vec{P}}(\mathbf{r},t)\rangle\rangle = \vec{\vec{\mathfrak{I}}}_H(\mathbf{r},\hat{\mathbf{n}}^{\rm inc})\cdot\langle\langle \vec{\vec{P}}^{\rm inc}(t)\rangle\rangle\cdot[\vec{\vec{\mathfrak{I}}}_E(\mathbf{r},\hat{\mathbf{n}}^{\rm inc})]^{\rm T*}, \qquad \mathbf{r}\in\Re^3, \tag{174}$$

$$\langle\langle \vec{\vec{P}}(\mathbf{r},t)\rangle\rangle = \hat{\mathfrak{I}}(\mathbf{r},\hat{\mathbf{n}}^{\rm inc})\langle\langle \vec{\vec{P}}^{\rm inc}(t)\rangle\rangle, \tag{175}$$

$$\langle\langle \vec{\vec{P}}^{\rm sca}(\mathbf{r},t)\rangle\rangle = \hat{\mathfrak{I}}^{\rm sca}(\mathbf{r},\hat{\mathbf{n}}^{\rm inc})\langle\langle \vec{\vec{P}}^{\rm inc}(t)\rangle\rangle. \tag{176}$$

The quasi-monochromatic versions of the main formulas of Subsections 4.6 and 4.7 are again coordinate-specific manifestations of Eqs. (175) and (176):

$$\langle\langle \overline{\mathbf{Signal\ 1}}(t)\rangle\rangle = S_{\rm ol}\langle\langle \mathbf{I}^{\rm sca}(r\hat{\mathbf{n}}^{\rm sca},t)\rangle\rangle = \frac{S_{\rm ol}}{r^2}\mathbf{Z}(\hat{\mathbf{n}}^{\rm sca},\hat{\mathbf{n}}^{\rm inc})\langle\langle \mathbf{I}^{\rm inc}(t)\rangle\rangle, \tag{177}$$

$$\langle\langle \overline{\mathbf{Signal\ 2}}(t)\rangle\rangle = S_{\rm ol}\langle\langle \mathbf{I}^{\rm inc}(t)\rangle\rangle - \mathbf{K}(\hat{\mathbf{n}}^{\rm inc})\langle\langle \mathbf{I}^{\rm inc}(t)\rangle\rangle + \frac{S_{\rm ol}}{r^2}\mathbf{Z}(\hat{\mathbf{n}}^{\rm inc},\hat{\mathbf{n}}^{\rm inc})\langle\langle \mathbf{I}^{\rm inc}(t)\rangle\rangle, \tag{178}$$

$$\langle\langle \overline{W}^{\rm abs}(t)\rangle\rangle = \langle\langle \overline{W}^{\rm ext}(t)\rangle\rangle - \langle\langle \overline{W}^{\rm sca}(t)\rangle\rangle, \tag{179}$$

$$\begin{aligned}\langle\langle \overline{W}^{\rm ext}(t)\rangle\rangle = {} & K_{11}(\hat{\mathbf{n}}^{\rm inc})\langle\langle I^{\rm inc}(t)\rangle\rangle + K_{12}(\hat{\mathbf{n}}^{\rm inc})\langle\langle Q^{\rm inc}(t)\rangle\rangle \\ & + K_{13}(\hat{\mathbf{n}}^{\rm inc})\langle\langle U^{\rm inc}(t)\rangle\rangle + K_{14}(\hat{\mathbf{n}}^{\rm inc})\langle\langle V^{\rm inc}(t)\rangle\rangle,\end{aligned} \tag{180}$$

$$\begin{aligned}\langle\langle \overline{W}^{\rm sca}(t)\rangle\rangle = \int_{4\pi} {\rm d}\hat{\mathbf{r}}\,[ & Z_{11}(\hat{\mathbf{r}},\hat{\mathbf{n}}^{\rm inc})\langle\langle I^{\rm inc}(t)\rangle\rangle + Z_{12}(\hat{\mathbf{r}},\hat{\mathbf{n}}^{\rm inc})\langle\langle Q^{\rm inc}(t)\rangle\rangle \\ & + Z_{13}(\hat{\mathbf{r}},\hat{\mathbf{n}}^{\rm inc})\langle\langle U^{\rm inc}(t)\rangle\rangle + Z_{14}(\hat{\mathbf{r}},\hat{\mathbf{n}}^{\rm inc})\langle\langle V^{\rm inc}(t)\rangle\rangle],\end{aligned} \tag{181}$$

where

$$\langle\langle \mathbf{I}^{\rm inc}(t)\rangle\rangle = \frac{1}{2}\sqrt{\frac{\varepsilon_1}{\mu_0}}\begin{bmatrix} \langle\langle \widetilde{E}_{0\theta}^{\rm inc}(t)[\widetilde{E}_{0\theta}^{\rm inc}(t)]^* + \widetilde{E}_{0\varphi}^{\rm inc}(t)[\widetilde{E}_{0\varphi}^{\rm inc}(t)]^*\rangle\rangle \\ \langle\langle \widetilde{E}_{0\theta}^{\rm inc}(t)[\widetilde{E}_{0\theta}^{\rm inc}(t)]^* - \widetilde{E}_{0\varphi}^{\rm inc}(t)[\widetilde{E}_{0\varphi}^{\rm inc}(t)]^*\rangle\rangle \\ \langle\langle -\widetilde{E}_{0\theta}^{\rm inc}(t)[\widetilde{E}_{0\varphi}^{\rm inc}(t)]^* - \widetilde{E}_{0\varphi}^{\rm inc}(t)[\widetilde{E}_{0\theta}^{\rm inc}(t)]^*\rangle\rangle \\ {\rm i}\langle\langle \widetilde{E}_{0\varphi}^{\rm inc}(t)[\widetilde{E}_{0\theta}^{\rm inc}(t)]^* - \widetilde{E}_{0\theta}^{\rm inc}(t)[\widetilde{E}_{0\varphi}^{\rm inc}(t)]^*\rangle\rangle \end{bmatrix}. \tag{182}$$

All time averages in Eqs. (174)–(182) are independent of time.

All results of this subsection can easily be generalized to the case of a polychromatic incident field with quasi-monochromatic components [34].

---

[7] The ultimate justification of this postulate must be based, in particular, on the explicit QED treatment of light–matter interactions [110,189].

## 5. Electromagnetic scattering by a randomly changing macroscopic object

### 5.1. *Dynamic and static scattering*

So far we have been discussing electromagnetic scattering by a fixed macroscopic object. In the case of a randomly changing macroscopic object such as a DRM, temporal changes in particle positions and/or physical states result in significant variations in the solution of the standard scattering problem even if the incident field is monochromatic. The typical time interval over which macroscopic quadratic and bilinear forms in the field vary significantly will be denoted by $T_\mathrm{v}$. We will assume hereinafter that $T_\mathrm{v} \gg T_\mathrm{o}$, $T_\mathrm{v} \gg T'$, and $T_\mathrm{v} \gg T_\mathrm{f}$.

In some cases, the temporal resolution of optical measurements is finer than $T_\mathrm{v}$, i.e., is sufficient to trace the random variations in macroscopic optical observables. Such measurements and their theoretical simulations constitute the subject of *dynamic light scattering* [190,191]. In other cases such random variations occur too rapidly to be captured by an actual optical device accumulating the signal over an extended period of time. This type of measurements and their theoretical modeling belong to the discipline of *static light scattering* [34]. If $T$ is the integration time of an optical measurement defining its temporal resolution, then

$$\left\{\begin{matrix} T_\mathrm{o} \\ T' \end{matrix}\right\} \ll T_\mathrm{f} \ll T \ll T_\mathrm{v} \tag{183}$$

for dynamic scattering and

$$\left\{\begin{matrix} T_\mathrm{o} \\ T' \end{matrix}\right\} \ll T_\mathrm{f} \ll T_\mathrm{v} \ll T \tag{184}$$

for static scattering. In either case the practical quantification of electromagnetic scattering by a stochastic macroscopic object requires, strictly speaking, repeated solutions of the standard scattering problem for temporally evolving instantaneous states of the object.

In what follows, we will mostly discuss static scattering of monochromatic and quasi-monochromatic electromagnetic fields.

### 5.2. *Monochromatic static scattering by a randomly changing macroscopic object*

To represent an actual static measurement, quadratic and bilinear forms in the field must be averaged over a sufficiently long period of time $T \gg T_\mathrm{v}$. In the case of monochromatic scattering, Eqs. (170) and (171) become

$$\langle\langle \bar{\mathbf{S}}(\mathbf{r},t) \rangle\rangle = \frac{1}{2} \mathrm{Re} \langle\langle \tilde{\mathbf{E}}(\mathbf{r},t) \times [\tilde{\mathbf{H}}(\mathbf{r},t)]^* \rangle\rangle, \qquad \mathbf{r} \in V_\mathrm{EXT}, \tag{185}$$

$$\langle\langle \vec{P}(\mathbf{r},t) \rangle\rangle = \frac{1}{2} \langle\langle \tilde{\mathbf{H}}(\mathbf{r},t) \otimes [\tilde{\mathbf{E}}(\mathbf{r},t)]^* \rangle\rangle, \qquad \mathbf{r} \in V_\mathrm{EXT}, \tag{186}$$

where the macroscopic field vectors $\tilde{\mathbf{E}}(\mathbf{r},t)$ and $\tilde{\mathbf{H}}(\mathbf{r},t)$ depend on time owing to the temporal changes of the scattering object, while $\langle\langle \bar{\mathbf{S}}(\mathbf{r},t) \rangle\rangle$ and $\langle\langle \vec{P}(\mathbf{r},t) \rangle\rangle$ are time independent provided that $\bar{\mathbf{S}}(\mathbf{r},t)$ and $\vec{P}(\mathbf{r},t)$ are stationary random processes. Now the temporal average on the right-hand side of Eq. (185) or (186) cannot, in general, be expressed as a product of the individual averages,

$$\langle\langle \overline{\mathbf{S}}(\mathbf{r},t)\rangle\rangle \neq \frac{1}{2}\,\mathrm{Re}[\langle\langle \widetilde{\mathbf{E}}(\mathbf{r},t)\rangle\rangle \times \langle\langle \widetilde{\mathbf{H}}(\mathbf{r},t)\rangle\rangle^{*}], \tag{187}$$

$$\langle\langle \vec{P}(\mathbf{r},t)\rangle\rangle \neq \frac{1}{2}\,\langle\langle \widetilde{\mathbf{H}}(\mathbf{r},t)\rangle\rangle \otimes \langle\langle \widetilde{\mathbf{E}}(\mathbf{r},t)\rangle\rangle^{*}, \tag{188}$$

and must be calculated explicitly. As usual, this computation is drastically simplified by assuming ergodicity of the scattering object and the resulting ergodicity of the random processes $\overline{\mathbf{S}}(\mathbf{r},t)$ and $\vec{P}(\mathbf{r},t)$ :

$$\begin{aligned}\langle\langle \overline{\mathbf{S}}(\mathbf{r},t)\rangle\rangle &= \langle \overline{\mathbf{S}}(\mathbf{r},\Psi)\rangle_{\Psi} = \int \mathrm{d}\Psi\, \overline{\mathbf{S}}(\mathbf{r},\Psi)P(\Psi)\\ &= \frac{1}{2}\,\mathrm{Re}\int \mathrm{d}\Psi\, \widetilde{\mathbf{E}}(\mathbf{r},\Psi)\times[\widetilde{\mathbf{H}}(\mathbf{r},\Psi)]^{*}P(\Psi),\end{aligned} \tag{189}$$

$$\begin{aligned}\langle\langle \vec{P}(\mathbf{r},t)\rangle\rangle &= \langle \vec{P}(\mathbf{r},\Psi)\rangle_{\Psi} = \int \mathrm{d}\Psi\, \vec{P}(\mathbf{r},\Psi)P(\Psi)\\ &= \frac{1}{2}\int \mathrm{d}\Psi\, \widetilde{\mathbf{H}}(\mathbf{r},\Psi)\otimes[\widetilde{\mathbf{E}}(\mathbf{r},\Psi)]^{*}P(\Psi),\end{aligned} \tag{190}$$

where $\Psi$ defines the state of the macroscopic object (rather than that of the constituent molecules) and $P(\Psi)$ is a suitable time-independent probability density function.

Eqs. (161), (163), and (164) now become

$$\begin{aligned}\langle\langle \vec{P}(\mathbf{r},t)\rangle\rangle &= \langle\langle \vec{\mathfrak{I}}_{H}(\mathbf{r},\hat{\mathbf{n}}^{\mathrm{inc}};t)\cdot \vec{P}^{\mathrm{inc}}\cdot[\vec{\mathfrak{I}}_{E}(\mathbf{r},\hat{\mathbf{n}}^{\mathrm{inc}};t)]^{\mathrm{T}*}\rangle\rangle\\ &= \langle \vec{\mathfrak{I}}_{H}(\mathbf{r},\hat{\mathbf{n}}^{\mathrm{inc}};\Psi)\cdot \vec{P}^{\mathrm{inc}}\cdot[\vec{\mathfrak{I}}_{E}(\mathbf{r},\hat{\mathbf{n}}^{\mathrm{inc}};\Psi)]^{\mathrm{T}*}\rangle_{\Psi}, \qquad \mathbf{r}\in\Re^{3},\end{aligned} \tag{191}$$

$$\langle\langle \vec{P}(\mathbf{r},t)\rangle\rangle = \langle\langle \hat{\mathfrak{I}}(\mathbf{r},\hat{\mathbf{n}}^{\mathrm{inc}};t)\rangle\rangle \vec{P}^{\mathrm{inc}} = \langle \hat{\mathfrak{I}}(\mathbf{r},\hat{\mathbf{n}}^{\mathrm{inc}};\Psi)\rangle_{\Psi}\vec{P}^{\mathrm{inc}}, \tag{192}$$

$$\langle\langle \vec{P}^{\mathrm{sca}}(\mathbf{r},t)\rangle\rangle = \langle\langle \hat{\mathfrak{I}}^{\mathrm{sca}}(\mathbf{r},\hat{\mathbf{n}}^{\mathrm{inc}};t)\rangle\rangle \vec{P}^{\mathrm{inc}} = \langle \hat{\mathfrak{I}}^{\mathrm{sca}}(\mathbf{r},\hat{\mathbf{n}}^{\mathrm{inc}};\Psi)\rangle_{\Psi}\vec{P}^{\mathrm{inc}}, \tag{193}$$

while the main formulas of Subsections 4.6 and 4.7 take the form

$$\langle\langle \overline{\mathbf{Signal\,1}(t)}\rangle\rangle = \frac{S_{\mathrm{ol}}}{r^{2}}\langle \mathbf{Z}(\hat{\mathbf{n}}^{\mathrm{sca}},\hat{\mathbf{n}}^{\mathrm{inc}};\Psi)\rangle_{\Psi}\mathbf{I}^{\mathrm{inc}}, \tag{194}$$

$$\langle\langle \overline{\mathbf{Signal\,2}(t)}\rangle\rangle = S_{\mathrm{ol}}\mathbf{I}^{\mathrm{inc}} - \langle \mathbf{K}(\hat{\mathbf{n}}^{\mathrm{inc}};\Psi)\rangle_{\Psi}\mathbf{I}^{\mathrm{inc}} + \frac{S_{\mathrm{ol}}}{r^{2}}\langle \mathbf{Z}(\hat{\mathbf{n}}^{\mathrm{inc}},\hat{\mathbf{n}}^{\mathrm{inc}};\Psi)\rangle_{\Psi}\mathbf{I}^{\mathrm{inc}}, \tag{195}$$

$$\langle\langle \overline{W}^{\mathrm{abs}}(t)\rangle\rangle = \langle\langle \overline{W}^{\mathrm{ext}}(t)\rangle\rangle - \langle\langle \overline{W}^{\mathrm{sca}}(t)\rangle\rangle, \tag{196}$$

$$\begin{aligned}\langle\langle \overline{W}^{\mathrm{ext}}(t)\rangle\rangle &= \langle K_{11}(\hat{\mathbf{n}}^{\mathrm{inc}};\Psi)\rangle_{\Psi}I^{\mathrm{inc}} + \langle K_{12}(\hat{\mathbf{n}}^{\mathrm{inc}};\Psi)\rangle_{\Psi}Q^{\mathrm{inc}}\\ &\quad + \langle K_{13}(\hat{\mathbf{n}}^{\mathrm{inc}};\Psi)\rangle_{\Psi}U^{\mathrm{inc}} + \langle K_{14}(\hat{\mathbf{n}}^{\mathrm{inc}};\Psi)\rangle_{\Psi}V^{\mathrm{inc}},\end{aligned} \tag{197}$$

$$\begin{aligned}\langle\langle \overline{W}^{\mathrm{sca}}(t)\rangle\rangle &= \int_{4\pi}\mathrm{d}\hat{\mathbf{r}}\,[\langle Z_{11}(\hat{\mathbf{r}},\hat{\mathbf{n}}^{\mathrm{inc}};\Psi)\rangle_{\Psi}I^{\mathrm{inc}} + \langle Z_{12}(\hat{\mathbf{r}},\hat{\mathbf{n}}^{\mathrm{inc}};\Psi)\rangle_{\Psi}Q^{\mathrm{inc}}\\ &\quad + \langle Z_{13}(\hat{\mathbf{r}},\hat{\mathbf{n}}^{\mathrm{inc}};\Psi)\rangle_{\Psi}U^{\mathrm{inc}} + \langle Z_{14}(\hat{\mathbf{r}},\hat{\mathbf{n}}^{\mathrm{inc}};\Psi)\rangle_{\Psi}V^{\mathrm{inc}}].\end{aligned} \tag{198}$$

Again, all time averages in Eqs. (191)–(198) are independent of time.

### 5.3. *Quasi-monochromatic static scattering by a randomly changing macroscopic object*

Let us now consider the situation wherein the solution of the standard scattering problem fluctuates in time owing to random temporal variations of both the incident field and the macroscopic object. Eqs. (163) and (164) now become

$$\langle\langle \vec{\vec{P}}(\mathbf{r},t)\rangle\rangle = \langle\langle \hat{\mathfrak{I}}(\mathbf{r},\hat{\mathbf{n}}^{\mathrm{inc}};t)\vec{\vec{P}}^{\mathrm{inc}}(t)\rangle\rangle, \tag{199}$$

$$\langle\langle \vec{\vec{P}}^{\mathrm{sca}}(\mathbf{r},t)\rangle\rangle = \langle\langle \hat{\mathfrak{I}}^{\mathrm{sca}}(\mathbf{r},\hat{\mathbf{n}}^{\mathrm{inc}};t)\vec{\vec{P}}^{\mathrm{inc}}(t)\rangle\rangle, \tag{200}$$

where the averages are taken over a period of time much longer than both $T_{\mathrm{f}}$ and $T_{\mathrm{v}}$. It is reasonable to assume that morphological changes of the scattering object are completely independent of the temporal fluctuations of the externally generated incident field. More specifically, we assume that $\hat{\mathfrak{I}}(\mathbf{r},\hat{\mathbf{n}}^{\mathrm{inc}};t)$ and $\vec{\vec{P}}^{\mathrm{inc}}(t)$ as well as $\hat{\mathfrak{I}}^{\mathrm{sca}}(\mathbf{r},\hat{\mathbf{n}}^{\mathrm{inc}};t)$ and $\vec{\vec{P}}^{\mathrm{inc}}(t)$ are pairs of independent stationary random processes, which implies that both $\langle\langle \vec{\vec{P}}(\mathbf{r},t)\rangle\rangle$ and $\langle\langle \vec{\vec{P}}^{\mathrm{sca}}(\mathbf{r},t)\rangle\rangle$ are independent of time and are given by

$$\langle\langle \vec{\vec{P}}(\mathbf{r},t)\rangle\rangle = \langle\langle \hat{\mathfrak{I}}(\mathbf{r},\hat{\mathbf{n}}^{\mathrm{inc}};t)\rangle\rangle\langle\langle\vec{\vec{P}}^{\mathrm{inc}}(t)\rangle\rangle, \tag{201}$$

$$\langle\langle \vec{\vec{P}}^{\mathrm{sca}}(\mathbf{r},t)\rangle\rangle = \langle\langle \hat{\mathfrak{I}}^{\mathrm{sca}}(\mathbf{r},\hat{\mathbf{n}}^{\mathrm{inc}};t)\rangle\rangle\langle\langle\vec{\vec{P}}^{\mathrm{inc}}(t)\rangle\rangle. \tag{202}$$

Finally, assuming ergodicity of the scattering object, we obtain

$$\langle\langle \vec{\vec{P}}(\mathbf{r},t)\rangle\rangle = \langle \hat{\mathfrak{I}}(\mathbf{r},\hat{\mathbf{n}}^{\mathrm{inc}};\Psi)\rangle_{\Psi}\langle\langle\vec{\vec{P}}^{\mathrm{inc}}(t)\rangle\rangle, \tag{203}$$

$$\langle\langle \vec{\vec{P}}^{\mathrm{sca}}(\mathbf{r},t)\rangle\rangle = \langle \hat{\mathfrak{I}}^{\mathrm{sca}}(\mathbf{r},\hat{\mathbf{n}}^{\mathrm{inc}};\Psi)\rangle_{\Psi}\langle\langle\vec{\vec{P}}^{\mathrm{inc}}(t)\rangle\rangle. \tag{204}$$

In other words, the time averaging of the Poynting–Stokes tensor of the incident quasi-monochromatic beam and the ensemble averaging of the transformation operators $\hat{\mathfrak{I}}$ and $\hat{\mathfrak{I}}^{\mathrm{sca}}$ are completely separated. The corresponding generalization of the main formulas of Subsections 4.6 and 4.7 reads

$$\langle\langle \overline{\mathbf{Signal\,1}}(t)\rangle\rangle = \frac{S_{\mathrm{ol}}}{r^2}\langle \mathbf{Z}(\hat{\mathbf{n}}^{\mathrm{sca}},\hat{\mathbf{n}}^{\mathrm{inc}};\Psi)\rangle_{\Psi}\langle\langle \mathbf{I}^{\mathrm{inc}}(t)\rangle\rangle, \tag{205}$$

$$\langle\langle \overline{\mathbf{Signal\,2}}(t)\rangle\rangle = S_{\mathrm{ol}}\langle\langle \mathbf{I}^{\mathrm{inc}}(t)\rangle\rangle - \langle \mathbf{K}(\hat{\mathbf{n}}^{\mathrm{inc}};\Psi)\rangle_{\Psi}\langle\langle \mathbf{I}^{\mathrm{inc}}(t)\rangle\rangle + \frac{S_{\mathrm{ol}}}{r^2}\langle \mathbf{Z}(\hat{\mathbf{n}}^{\mathrm{inc}},\hat{\mathbf{n}}^{\mathrm{inc}};\Psi)\rangle_{\Psi}\langle\langle \mathbf{I}^{\mathrm{inc}}(t)\rangle\rangle, \tag{206}$$

$$\langle\langle \overline{W}^{\mathrm{abs}}(t)\rangle\rangle = \langle\langle \overline{W}^{\mathrm{ext}}(t)\rangle\rangle - \langle\langle \overline{W}^{\mathrm{sca}}(t)\rangle\rangle, \tag{207}$$

$$\begin{aligned}\langle\langle \overline{W}^{\mathrm{ext}}(t)\rangle\rangle = {}& \langle K_{11}(\hat{\mathbf{n}}^{\mathrm{inc}};\Psi)\rangle_{\Psi}\langle\langle I^{\mathrm{inc}}(t)\rangle\rangle + \langle K_{12}(\hat{\mathbf{n}}^{\mathrm{inc}};\Psi)\rangle_{\Psi}\langle\langle Q^{\mathrm{inc}}(t)\rangle\rangle \\ & + \langle K_{13}(\hat{\mathbf{n}}^{\mathrm{inc}};\Psi)\rangle_{\Psi}\langle\langle U^{\mathrm{inc}}(t)\rangle\rangle + \langle K_{14}(\hat{\mathbf{n}}^{\mathrm{inc}};\Psi)\rangle_{\Psi}\langle\langle V^{\mathrm{inc}}(t)\rangle\rangle,\end{aligned} \tag{208}$$

$$\begin{aligned}\langle\langle \overline{W}^{\mathrm{sca}}(t)\rangle\rangle = \int_{4\pi} \mathrm{d}\hat{\mathbf{r}}\,[&\langle Z_{11}(\hat{\mathbf{r}},\hat{\mathbf{n}}^{\mathrm{inc}};\Psi)\rangle_{\Psi}\langle\langle I^{\mathrm{inc}}(t)\rangle\rangle + \langle Z_{12}(\hat{\mathbf{r}},\hat{\mathbf{n}}^{\mathrm{inc}};\Psi)\rangle_{\Psi}\langle\langle Q^{\mathrm{inc}}(t)\rangle\rangle \\ & + \langle Z_{13}(\hat{\mathbf{r}},\hat{\mathbf{n}}^{\mathrm{inc}};\Psi)\rangle_{\Psi}\langle\langle U^{\mathrm{inc}}(t)\rangle\rangle + \langle Z_{14}(\hat{\mathbf{r}},\hat{\mathbf{n}}^{\mathrm{inc}};\Psi)\rangle_{\Psi}\langle\langle V^{\mathrm{inc}}(t)\rangle\rangle].\end{aligned} \tag{209}$$

It is straightforward to generalize all results of this subsection to the case of a polychromatic incident field with quasi-monochromatic components [34].

Let us now assume that the transformation dyadics $\ddot{\Im}_E$ and $\ddot{\Im}_H$ as well as the electric and magnetic field amplitudes of the quasi-monochromatic plane-wave incident field can be represented as sums of average and fluctuating (subscript "f") components:

$$\ddot{\Im}_E(\mathbf{r},\hat{\mathbf{n}}^{\rm inc};t) = \langle\langle\ddot{\Im}_E(\mathbf{r},\hat{\mathbf{n}}^{\rm inc};t)\rangle\rangle + \ddot{\Im}_E^{\rm f}(\mathbf{r},\hat{\mathbf{n}}^{\rm inc};t) = \langle\ddot{\Im}_E(\mathbf{r},\hat{\mathbf{n}}^{\rm inc};\Psi)\rangle_\Psi + \ddot{\Im}_E^{\rm f}(\mathbf{r},\hat{\mathbf{n}}^{\rm inc};t), \tag{210}$$

$$\ddot{\Im}_H(\mathbf{r},\hat{\mathbf{n}}^{\rm inc};t) = \langle\langle\ddot{\Im}_H(\mathbf{r},\hat{\mathbf{n}}^{\rm inc};t)\rangle\rangle + \ddot{\Im}_H^{\rm f}(\mathbf{r},\hat{\mathbf{n}}^{\rm inc};t) = \langle\ddot{\Im}_H(\mathbf{r},\hat{\mathbf{n}}^{\rm inc};\Psi)\rangle_\Psi + \ddot{\Im}_H^{\rm f}(\mathbf{r},\hat{\mathbf{n}}^{\rm inc};t), \tag{211}$$

$$\widetilde{\mathbf{E}}_0^{\rm inc}(t) = \langle\langle\widetilde{\mathbf{E}}_0^{\rm inc}(t)\rangle\rangle + \widetilde{\mathbf{E}}_{0\rm f}^{\rm inc}(t) = \widetilde{\mathbf{E}}_{0\rm c}^{\rm inc} + \widetilde{\mathbf{E}}_{0\rm f}^{\rm inc}(t), \tag{212}$$

$$\widetilde{\mathbf{H}}_0^{\rm inc}(t) = \langle\langle\widetilde{\mathbf{H}}_0^{\rm inc}(t)\rangle\rangle + \widetilde{\mathbf{H}}_{0\rm f}^{\rm inc}(t) = \widetilde{\mathbf{H}}_{0\rm c}^{\rm inc} + \widetilde{\mathbf{H}}_{0\rm f}^{\rm inc}(t), \tag{213}$$

where $\mathbf{r} \in \Re^3$, the subscript "c" stands for "coherent", and, by definition,

$$\langle\langle\ddot{\Im}_E^{\rm f}(\mathbf{r},\hat{\mathbf{n}}^{\rm inc};t)\rangle\rangle = \ddot{0}, \tag{214}$$

$$\langle\langle\ddot{\Im}_H^{\rm f}(\mathbf{r},\hat{\mathbf{n}}^{\rm inc};t)\rangle\rangle = \ddot{0}, \tag{215}$$

$$\langle\langle\widetilde{\mathbf{E}}_{0\rm f}^{\rm inc}(t)\rangle\rangle = \mathbf{0}, \tag{216}$$

$$\langle\langle\widetilde{\mathbf{H}}_{0\rm f}^{\rm inc}(t)\rangle\rangle = \mathbf{0}, \tag{217}$$

where $\ddot{0}$ is a zero dyad. Then averaging Eq. (161) over a time interval much longer than both $T_{\rm f}$ and $T_{\rm v}$ while assuming statistical independence of the random incident field and the random scattering object yields

$$\langle\langle\ddot{P}(\mathbf{r},t)\rangle\rangle = \langle\langle\ddot{\Im}_H(\mathbf{r},\hat{\mathbf{n}}^{\rm inc};t)\cdot\ddot{P}^{\rm inc}(t)\cdot[\ddot{\Im}_E(\mathbf{r},\hat{\mathbf{n}}^{\rm inc};t)]^{\rm T*}\rangle\rangle \tag{218a}$$

$$\begin{aligned}&= \frac{1}{2}\langle\ddot{\Im}_H(\mathbf{r},\hat{\mathbf{n}}^{\rm inc};\Psi)\rangle_\Psi\cdot\{\widetilde{\mathbf{H}}_{0\rm c}^{\rm inc}\otimes[\widetilde{\mathbf{E}}_{0\rm c}^{\rm inc}]^*\}\cdot\langle[\ddot{\Im}_E(\mathbf{r},\hat{\mathbf{n}}^{\rm inc};\Psi)]^{\rm T*}\rangle_\Psi\\
&+\frac{1}{2}\langle\ddot{\Im}_H(\mathbf{r},\hat{\mathbf{n}}^{\rm inc};\Psi)\rangle_\Psi\cdot\langle\langle\widetilde{\mathbf{H}}_{0\rm f}^{\rm inc}(t)\otimes[\widetilde{\mathbf{E}}_{0\rm f}^{\rm inc}(t)]^*\rangle\rangle\cdot\langle[\ddot{\Im}_E(\mathbf{r},\hat{\mathbf{n}}^{\rm inc};\Psi)]^{\rm T*}\rangle_\Psi\\
&+\frac{1}{2}\langle\ddot{\Im}_H^{\rm f}(\mathbf{r},\hat{\mathbf{n}}^{\rm inc};\Psi)\cdot\{\widetilde{\mathbf{H}}_{0\rm c}^{\rm inc}\otimes[\widetilde{\mathbf{E}}_{0\rm c}^{\rm inc}]^*\}\cdot[\ddot{\Im}_E^{\rm f}(\mathbf{r},\hat{\mathbf{n}}^{\rm inc};\Psi)]^{\rm T*}\rangle_\Psi\\
&+\frac{1}{2}\langle\ddot{\Im}_H^{\rm f}(\mathbf{r},\hat{\mathbf{n}}^{\rm inc};\Psi)\cdot\langle\langle\widetilde{\mathbf{H}}_{0\rm f}^{\rm inc}(t)\otimes[\widetilde{\mathbf{E}}_{0\rm f}^{\rm inc}(t)]^*\rangle\rangle\cdot[\ddot{\Im}_E^{\rm f}(\mathbf{r},\hat{\mathbf{n}}^{\rm inc};\Psi)]^{\rm T*}\rangle_\Psi.\end{aligned} \tag{218b}$$

This formula can alternatively be written as

$$\begin{aligned}\langle\langle\ddot{P}(\mathbf{r},t)\rangle\rangle &= \langle\ddot{\Im}_H(\mathbf{r},\hat{\mathbf{n}}^{\rm inc};\Psi)\rangle_\Psi\cdot\langle\langle\ddot{P}^{\rm inc}(t)\rangle\rangle\cdot\langle[\ddot{\Im}_E(\mathbf{r},\hat{\mathbf{n}}^{\rm inc};\Psi)]^{\rm T*}\rangle_\Psi\\
&+\langle\ddot{\Im}_H^{\rm f}(\mathbf{r},\hat{\mathbf{n}}^{\rm inc};\Psi)\cdot\langle\langle\ddot{P}^{\rm inc}(t)\rangle\rangle\cdot[\ddot{\Im}_E^{\rm f}(\mathbf{r},\hat{\mathbf{n}}^{\rm inc};\Psi)]^{\rm T*}\rangle_\Psi\end{aligned} \tag{219a}$$

$$= \langle\ddot{\Im}_H(\mathbf{r},\hat{\mathbf{n}}^{\rm inc};\Psi)\cdot\langle\langle\ddot{P}^{\rm inc}(t)\rangle\rangle\cdot[\ddot{\Im}_E(\mathbf{r},\hat{\mathbf{n}}^{\rm inc};\Psi)]^{\rm T*}\rangle_\Psi. \tag{219b}$$

We see again that averaging the Poynting–Stokes tensor of the incident quasi-monochromatic field over time is completely decoupled from the ensemble averaging. This implies that to solve the quasi-monochromatic scattering problem, one can solve the monochromatic scattering prob-lem and then make the formal substitution $\vec{\vec{P}}^{\text{inc}} \to \langle\langle \vec{\vec{P}}^{\text{inc}}(t) \rangle\rangle$.

## 6. Effective-object methodology

Direct computer solutions of the MMEs for morphologically complex objects can be quite time-consuming and in many cases impracticable. As a consequence, there has been a widespread use of phenomenological so-called effective-medium rules intended to drastically simplify the computation (see [72–78,192–199] and references therein). Implicitly, the main idea of an effective-object approximation (EOA) (more commonly known as an effective-medium approximation, or EMA) is to replace a morphologically complex object, either fixed or randomly varying in time, by a much simpler "effective" object possessing essentially the same scattering properties. For example, one could think of replacing the Type-1 and -2 DRMs shown in Figs. 13a,b by homogeneous scattering objects with the same overall shape defined by the surface $S$, as shown in Figs. 13c,d.

In terms of the transformation operators $\hat{\mathfrak{I}}$ and $\hat{\mathfrak{I}}^{\text{sca}}$, one can think of the following hierarchy of EOAs:

- A deterministic EOA amounts to replacing a fixed morphologically complex scattering object by a fixed simple "effective" object such that

$$\hat{\mathfrak{I}}(\mathbf{r}, \hat{\mathbf{n}}^{\text{inc}}) \approx \hat{\mathfrak{I}}_{\text{eff}}(\mathbf{r}, \hat{\mathbf{n}}^{\text{inc}}), \tag{220}$$

$$\hat{\mathfrak{I}}^{\text{sca}}(\mathbf{r}, \hat{\mathbf{n}}^{\text{inc}}) \approx \hat{\mathfrak{I}}^{\text{sca}}_{\text{eff}}(\mathbf{r}, \hat{\mathbf{n}}^{\text{inc}}). \tag{221}$$

- A semi-stochastic EOA amounts to replacing a stochastic morphologically complex scattering object by a fixed simple "effective" object such that

$$\langle \hat{\mathfrak{I}}(\mathbf{r}, \hat{\mathbf{n}}^{\text{inc}}; \Psi) \rangle_{\Psi} \approx \hat{\mathfrak{I}}_{\text{eff}}(\mathbf{r}, \hat{\mathbf{n}}^{\text{inc}}), \tag{222}$$

$$\langle \hat{\mathfrak{I}}^{\text{sca}}(\mathbf{r}, \hat{\mathbf{n}}^{\text{inc}}; \Psi) \rangle_{\Psi} \approx \hat{\mathfrak{I}}^{\text{sca}}_{\text{eff}}(\mathbf{r}, \hat{\mathbf{n}}^{\text{inc}}). \tag{223}$$

- A stochastic EOA amounts to replacing a stochastic morphologically complex scattering object by a stochastic simple "effective" object such that

$$\langle \hat{\mathfrak{I}}(\mathbf{r}, \hat{\mathbf{n}}^{\text{inc}}; \Psi) \rangle_{\Psi} \approx \langle \hat{\mathfrak{I}}_{\text{eff}}(\mathbf{r}, \hat{\mathbf{n}}^{\text{inc}}; \Psi_{\text{eff}}) \rangle_{\Psi_{\text{eff}}}, \tag{224}$$

$$\langle \hat{\mathfrak{I}}^{\text{sca}}(\mathbf{r}, \hat{\mathbf{n}}^{\text{inc}}; \Psi) \rangle_{\Psi} \approx \langle \hat{\mathfrak{I}}^{\text{sca}}_{\text{eff}}(\mathbf{r}, \hat{\mathbf{n}}^{\text{inc}}; \Psi_{\text{eff}}) \rangle_{\Psi_{\text{eff}}}. \tag{225}$$

Note that we intentionally defined the three EOAs in terms of the linear operators $\hat{\mathfrak{I}}$ and $\hat{\mathfrak{I}}^{\text{sca}}$ acting on an optical observable rather than on the macroscopic field vectors (of course these definitions can be generalized to include types of optical observables other than the Poynting–Stokes tensor). Traditionally, however, EMAs have been introduced with the purpose of replicating the average macroscopic field vectors rather than specific optical observables [72–78,192–199]. In other words, a semi-stochastic EOA would normally be introduced as a recipe for replacing a stochastic morphologically complex scattering object by a fixed simple "effective" object such that

$$\langle \vec{\vec{\mathfrak{I}}}_E(\mathbf{r}, \hat{\mathbf{n}}^{\text{inc}}; \Psi) \rangle_{\Psi} \approx \vec{\vec{\mathfrak{I}}}^{\text{eff}}_E(\mathbf{r}, \hat{\mathbf{n}}^{\text{inc}}), \tag{226}$$

$$\langle \overleftrightarrow{\Im}_H(\mathbf{r}, \hat{\mathbf{n}}^{\text{inc}}; \Psi) \rangle_\Psi \approx \overleftrightarrow{\Im}_H^{\text{eff}}(\mathbf{r}, \hat{\mathbf{n}}^{\text{inc}}) \tag{227}$$

in Eq. (218b). This is equivalent to defining the EMA as that replicating the "coherent field" created by the object. This explains why an EMA recipe would typically be formulated in terms of replacing an actual heterogeneous object by that made of a homogeneous material with an "effective refractive index".

Unfortunately, the traditional way of formulating an EMA is somewhat limited since it helps determine only two out of four terms on the right-hand side of Eq. (218b). The three EOAs defined by Eqs. (220)–(225) are more general and useful. Furthermore, they yield automatically the traditional field-based EMAs in cases when the last two terms on the right-hand side of Eq. (218b) can be neglected. In addition, they do not rely on the contrived notion of the coherent field.

To the best of our knowledge, EOAs remain unproven hypotheses since none of them has been derived directly from the time-domain or frequency-domain MMEs under well-defined and reproducible conditions. In the words of Chýlek et al. written in 2000 [77], EMAs

> are not approximations in a strict mathematical sense. It is not generally possible to estimate the accuracy of a given approximation by considering the magnitude of neglected terms with respect to those that are kept. EMAs are often based on an *ad hoc* assumption that leads to a simplified, solvable model of a real, complicated, and usually unsolvable situation. As a result, one is able to derive a simple or only moderately complicated prescription (e.g., the mixing rule) of how to calculate the average optical properties of a heterogeneous composite material from the known properties and amounts of its individual components. Because there are no specific algebraic terms neglected and because the exact solution of the problem is usually unknown, the accuracy of such derived effective material constants (effective dielectric constants or effective refractive indices of material) and the precise conditions for their permissible use are not easy to assess.

Fortunately, the current availability of efficient computer solvers of the MMEs and powerful computer clusters makes it possible to validate EOAs *numerically*, at least in special cases. Recent progress in this direction will be discussed in Subsection 8.2 and Section 9.

## 7. Direct computer solvers of the macroscopic Maxwell equations

According to the preceding discussion, the main objective of the discipline of electromagnetic scattering by particulate objects is the computation of optical observables that can be used to quantify the energy budget of a macroscopic volume or the results of measurements with actual optical instruments. Alternatively, this objective can be formulated as the computation of quantities such as the transformation dyadics $\overleftrightarrow{\Im}_E(\mathbf{r}, \hat{\mathbf{n}}^{\text{inc}})$ and $\overleftrightarrow{\Im}_H(\mathbf{r}, \hat{\mathbf{n}}^{\text{inc}})$ entering Eq. (161); the transformation operators $\hat{\Im}(\mathbf{r}, \hat{\mathbf{n}}^{\text{inc}})$ and $\hat{\Im}^{\text{sca}}(\mathbf{r}, \hat{\mathbf{n}}^{\text{inc}})$ entering Eqs. (163) and (164); the ensemble-averaged transformation operators $\langle \hat{\Im}(\mathbf{r}, \hat{\mathbf{n}}^{\text{inc}}; \Psi) \rangle_\Psi$ and $\langle \hat{\Im}^{\text{sca}}(\mathbf{r}, \hat{\mathbf{n}}^{\text{inc}}; \Psi) \rangle_\Psi$ entering Eqs. (192), (193), (203), and (204); and their various coordinate-specific representations. Far-field examples of the latter are the (ensemble-averaged) extinction and phase matrices.

Whenever possible, all these quantities should be calculated by using a direct, numerically exact computer solver of the MMEs[8] in combination with a suitable ensemble-averaging procedure. The majority of direct solvers of the MMEs belong to one of two broad categories. Differential-equation techniques yield the scattered field by solving the differential MMEs or the vector wave equation in the frequency or in the time domain. Classical examples of the frequency-domain differential-equation techniques are the Lorenz–Mie theory for a homogeneous or radially inhomogeneous spherical particle [1,22,200–204] and the separation of variables technique for a homogeneous or layered spheroid [205–211]. General differential-equation techniques applicable to an object with essentially any morphology are the frequency-domain finite-difference method [212,213], the finite-difference time-domain method (FDTDM) [214–218], and the pseudo-spectral time-domain method (PSTDM) [219,220]. In both FDTDM and PSTDM, it is necessary to truncate the computational domain by introducing an artificial outer boundary and then ensure that spurious reflections from this boundary are suppressed. This is effectively accomplished by using the perfectly matched layer proposed in [221,222] and its refinements. Integral equation methods are based on the volume or surface integral counterparts of the MMEs, the prime examples being the frequency-domain volume integral equation method and the closely related discrete-dipole approximation (DDA) [168,223–228]. There are also hybrid techniques or methods that can be derived using different approaches. Furthermore, there are general formulations, such as the frequency-domain $T$-matrix method (TMM) [20,24,28,35,229–235], based on expanding relevant electric field vectors in special mathematical functions possessing desirable analytical properties and then using various *ad hoc* techniques to compute the transition matrices relating the resulting columns of the expansion coefficients.

The FDTDM and the DDA are examples of general direct solvers of the MMEs that are rather insensitive to the object's morphology and thus can be applied to a multi-particle group as well as to a compact single-body object using the same basic computer program. Some approaches, such as the TMM, can be made much more efficient by explicitly accounting for the specific object's morphology, for example, its aggregate structure. Each direct numerical solver of the MMEs has its own advantages and drawbacks in terms of computer memory and execution time requirements, convergence rate, accuracy, and range of applicability. For example, the more traditional versions of the TMM can be less flexible than the FDTDM and the DDA in terms of the scattering object's morphology, but appear to be the fastest and most accurate techniques within the range of their convergence. Further information on direct computer solvers of the MMEs can be found in the reviews [236,237].

By definition, running a direct computer solver of the MMEs yields the monochromatic scattering properties of a fixed object. However, the angular scattering patterns typical of a fixed object with a size comparable to or greater than the wavelength are typically burdened by numerous sharp maxima and minima (called speckles) that must be smoothed out to yield representative static-scattering results [34,92,238,239]. The necessity of repeating computations for a large number of realizations of a random object for the purpose of ensemble averaging still represents a great practical challenge. For example, the computation of electromagnetic scattering by a DRM can require averaging over an excessive number of fixed multi-particle configurations. In this respect the advantage of the superposition TMM (STMM) [20,230,240] is the ex-tremely efficient quasi-analytical procedure which allows one to create a fixed quasi-random $N$-particle configuration and then average relevant optical observables over all possible orientations of this configuration with respect to the laboratory coordinate system [241–244]. This procedure captures in effect an infinite continuous set of random realizations of the $N$-particle group, eliminates completely the notorious speckle "noise", and yields exceedingly accurate results.

[8] By definition, a direct computer solver of the MMEs is called numerically exact if it can generate numerical results with a guaranteed number of correct decimals. The number of correct decimals may vary depending on the available computer resources and practical accuracy requirements. However, all reported decimals can, in principle, be validated by modifying computer program settings in order to accommodate a more stringent accuracy requirement.

The first studies of electromagnetic scattering by random three-dimensional multi-particle groups based on direct computer solutions of the MMEs [245–248] exploited the frequency-domain multi-sphere method [249–253] which can be considered a particular case of the STMM. More recently, other numerical solvers of the MMEs have been used, such as the PSTDM and its variations [254–258], the DDA [259–274], the FDTDM [275–277], and the hybrid finite element–boundary integral–characteristic basis function method [278–281]. However, the STMM appears to have been the most frequently used technique [268,282–328]. Studies of two-dimensional DRMs composed of parallel infinite cylinders have been based on the multi-cylinder solution of the MMEs [329–331], the PSTDM [332,333] and the FDTDM [334].

Given the extreme complexity of direct computer calculations of electromagnetic scattering by a DRM, it is imperative to characterize the accuracy of the various numerical techniques and certify that internal (subjective) convergence of a technique (if achieved) ensures objectively converged results. This can be done by comparing benchmark data generated for the same scattering object with software implementations of completely independent methods.

As an example, let us compare far-field results obtained with five totally independent computer programs based on the STMM [244], DDA [228], invariant-imbedding TMM (II-TMM) [234,335], FDTDM [336], and PSTDM [337] for the same compound scatterer in the form of a spherical particle hosting 10 identical non-overlapping spherical inclusions (Fig. 14a). The size parameters of the host and the inclusions are $k_1 R = 10$ and $k_1 r = 2.5$, respectively, where $R$ is the radius of the host and $r$ is that of the inclusions. The corresponding refractive indices relative to that of the infinite surrounding medium are 1.33 and 1.55 + i0.003. The coordinates of the 10 inclusions (in units of size parameter) are listed in Table 1. It is assumed that the compound particle is illuminated by a quasi-monochromatic plane electromagnetic wave incident in the direction of the positive $z$-axis, as shown in Fig. 14b. For demonstration purposes, we define the $4\times 4$ dimensionless scattering matrix $\widetilde{\mathbf{F}}(\Theta)$ according to

$$\widetilde{\mathbf{F}}(\Theta) = \frac{4\pi}{C_{\text{sca}}} \langle \mathbf{Z}(\theta^{\text{sca}} = \Theta, \varphi^{\text{sca}}; \theta^{\text{inc}} = 0, \varphi^{\text{inc}} = \varphi^{\text{sca}}; \Psi)\rangle_{\Psi}, \tag{228}$$

where $\theta \in [0,\pi]$ is the zenith (polar) angle, $\varphi \in [0, 2\pi)$ is the azimuth angle, and $\Theta$ is the angle between the incidence and scattering directions (i.e., the scattering angle); the ensemble average is taken over the uniform orientation distribution of the compound scatterer; and the normalization constant $C_{\text{sca}}$ is given by

$$C_{\text{sca}} = \int_{4\pi} \mathrm{d}\hat{\mathbf{n}}^{\text{sca}} \langle Z_{11}(\theta^{\text{sca}}, \varphi^{\text{sca}}; \theta^{\text{inc}} = 0, \varphi^{\text{inc}} = 0; \Psi)\rangle_{\Psi}. \tag{229}$$

Note that $\widetilde{\mathbf{F}}$ is independent of $\varphi^{\text{sca}}$ owing to the random orientation distribution of the scattering object, while $C_{\text{sca}}$ represents the ensemble-averaged scattering cross section $\langle\langle \overline{W}^{\text{sca}}(t)\rangle\rangle \big/ \langle\langle I^{\text{inc}}(t)\rangle\rangle$ for the case of unpolarized incident plane-wave field (cf. Eq. (209)). It is easily seen that the $(1,1)$ element of the scattering matrix $\widetilde{\mathbf{F}}(\Theta)$ (often called the phase function) is normalized according to

$$\frac{1}{2}\int_0^{\pi} \mathrm{d}\Theta\, \widetilde{F}_{11}(\Theta)\, \sin\Theta = 1. \tag{230}$$

The results of our computations are tabulated in Table 2 and visualized in Figs. 15–18. Table 2 gives the corresponding extinction,

$$Q_{\mathrm{ext}} = \frac{C_{\mathrm{ext}}}{\pi R^2}, \tag{231}$$

and scattering,

$$Q_{\mathrm{sca}} = \frac{C_{\mathrm{sca}}}{\pi R^2}, \tag{232}$$

efficiency factors, where

$$C_{\mathrm{ext}} = \langle K_{11}(\hat{\mathbf{n}}^{\mathrm{inc}};\Psi)\rangle_{\Psi} \tag{233}$$

is the $\hat{\mathbf{n}}^{\mathrm{inc}}$-independent extinction cross section $\langle\langle \overline{W}^{\mathrm{ext}}(t)\rangle\rangle / \langle\langle I^{\mathrm{inc}}(t)\rangle\rangle$ for unpolarized incident light (cf. Eq. (208)). Also tabulated are the absorption efficiency factor

$$Q_{\mathrm{abs}} = Q_{\mathrm{ext}} - Q_{\mathrm{sca}}, \tag{234}$$

the single-scattering albedo

$$\varpi = \frac{Q_{\mathrm{sca}}}{Q_{\mathrm{ext}}}, \tag{235}$$

and the asymmetry parameter

$$\langle\cos\Theta\rangle = \frac{1}{2}\int_0^{\pi} \mathrm{d}\Theta\, \widetilde{F}_{11}(\Theta)\, \sin\Theta\, \cos\Theta. \tag{236}$$

Unlike the case with the DDA, FDTDM, and PSTDM, the averaging over orientations by the STMM and II-TMM computer programs is performed analytically so that the accuracy of computations is unaffected by simulating the uniform orientation distribution of the compound object by a limited set of discrete orientations. This analytical procedure also made the STMM and II-STM computations for the randomly oriented composite object much faster.

The DDA simulations were performed with the code ADDA 1.2 on the computer cluster of the supercomputing center of the Novosibirsk State University. We used the default parameters of the code while controlling the discretization level by the number $n_x$ of so-called "dipoles" along the particle diameter. Five values ranging from 64 to 128 were considered, corresponding to dipole sizes from $\lambda/20$ to $\lambda/40$, where $\lambda = 2\pi/k_1$ is the wavelength in the infinite surrounding medium. The orientation averaging was performed with a built-in adaptive procedure which adjusts the number of simulated orientations of the compound object to keep the relative uncertainty in $C_{\mathrm{ext}}$ caused by averaging within $10^{-4}$ [228]. As a consequence, the final numerical uncertainty is controlled mostly by $n_x$. To further improve the accuracy, we applied the extrapolation to the zero dipole size, as described in [338]. This procedure also provides an internal error estimate which, for the majority of computed values, was adequate, i.e., was within the actual differences from the STMM results. Fig. 15 shows both the "raw" DDA results for $n_x = 64$ and 128 and the "extrapolated" ones. One can see that the quantitative agreement between the latter and the internally converged STMM results is quite good (the corresponding phase functions $\widetilde{F}_{11}(\Theta)$ typically differ by less than 1%).

Figs. 16–18 demonstrate a similarly impressive agreement between the STMM results and those obtained with the II-TMM, FDTDM, and PSTDM computer programs. Note that unlike the STMM, the II-TMM is based on an alternative approach to calculate the object's $T$ matrix which is more general, but can make it somewhat more cumbersome to obtain the same benchmark precision. The FDTDM and PSTDM results have been calculated with $\lambda/40$ and $\lambda/50$ spatial grid sizes and exhibit expected convergence towards the STMM curves.

This quantitative comparison of completely independent direct computer solvers of the MMEs obviously certifies that these five techniques can be used in reliable far-field calculations of electromagnetic scattering by DRMs.

Note that when running the STMM computer program, we increasingly tightened all numerical accuracy parameters until the final results converged internally to a very high accuracy. While this "subjective" convergence of the STMM results does not guarantee the same "objective" convergence, it is still expected to be a good indicator of the actual accuracy of the final numbers. Given the virtual absence of such benchmark numerical data in the published literature, we tabulate the converged STMM scattering-matrix results in Appendix A.

## 8. Direct computer modeling of electromagnetic scattering by Type-1 discrete random media

In this section we will discuss the results of representative calculations of electromagnetic scattering by Type-1 DRMs based on direct computer solutions of the MMEs. In most cases we will use the model of a DRM in the form of a cluster of $N$ identical small spherical particles randomly and uniformly distributed throughout an imaginary spherical volume $V$ with a radius $R$, as shown in Fig. 19a (after [339]).

### *8.1. Far-field speckle and its suppression*

Let us first consider far-field scattering of a quasi-monochromatic plane-wave field by two different fixed clusters of $N = 80$ identical spherical particles distributed throughout an imaginary spherical volume with a size parameter of $k_1 R = 40$. The size parameter of the constituent spherical particles is $k_1 r = 4$ and their relative refractive index is $m = 1.32$. The coordinates of the particles forming either cluster were chosen using a random number generator, but otherwise they are fixed. The laboratory spherical coordinate system used to describe far-field scattering by either cluster is shown in Fig. 14b where, as before, the unit vectors $\hat{\mathbf{n}}^{\mathrm{inc}}$ and $\hat{\mathbf{n}}^{\mathrm{sca}}$ specify the directions of incidence and scattering. The zenith and azimuth angles of the incidence direction are assumed to be $\theta^{\mathrm{inc}} = 0°$ and $\varphi^{\mathrm{inc}} = 0°$, respectively. The incident plane-wave field is assumed to be circularly polarized in the counter-clockwise sense when looking in the direction of propagation, which implies that $\langle\langle V^{\mathrm{inc}}(t)\rangle\rangle = \langle\langle I^{\mathrm{inc}}(t)\rangle\rangle$ and $\langle\langle Q^{\mathrm{inc}}(t)\rangle\rangle = \langle\langle U^{\mathrm{inc}}(t)\rangle\rangle = 0$; the double angular brackets denote averaging over a time interval $T \gg T_{\mathrm{f}}$.

The two panels of Fig. 20a show the corresponding time-independent far-field angular dis-tributions of the intensity $\langle\langle I^{\text{sca}}(r\hat{\mathbf{n}}^{\text{sca}},t)\rangle\rangle$ scattered in the backward hemisphere. These intensity distributions were calculated using the STMM computer program described in [243] and reveal typical random speckle patterns. Fig. 20b shows the result obtained by averaging the scattered intensity over the uniform orientation distribution of the multi-particle configuration used to create the top panel of Fig. 20a. This orientation averaging replaces averaging over a time interval $T \gg T_{\text{v}}$ and is intended to simulate averaging over uniformly random positions of all 80 particles by taking advantage of the efficient analytical procedure afforded by the STMM, as discussed in Section 7. Predictably, the average intensity pattern is rotationally symmetric with respect to the incidence direction and is fairly featureless, the strong and narrow backscattering peak being the only notable exception.

To interpret the results of these computations, we will invoke the mathematical concept of ordered multi-particle sequences representing the various terms on the right-hand side of Eq. (148). Fig. 21a shows schematically two such sequences depicted using the blue and yellow colors. To make the discussion even more physically appealing, we will assign a cumulative phase to each multi-particle sequence by assuming that each particle of the sequence resides in the far zone of the preceding particle. For example, particle 4 of the blue sequence in Fig. 21a is in the far zone of particle 3, particle 3 is in the far zone of particle 2, etc. In other words, we will use the far-field version of the Neumann expansion (148):

$$\begin{aligned}\widetilde{\mathbf{E}}(\mathbf{r},t) &= \widetilde{\mathbf{E}}^{\text{inc}}(\mathbf{r},t) + \widetilde{\mathbf{E}}^{\text{sca}}(\mathbf{r},t)\\ &= \widetilde{\mathbf{E}}^{\text{inc}}(\mathbf{r},t) + \sum_{i=1}^{N} g(r_i)\overset{\leftrightarrow}{A}_i(\hat{\mathbf{r}}_i,\hat{\mathbf{n}}^{\text{inc}})\cdot\widetilde{\mathbf{E}}^{\text{inc}}(\mathbf{R}_i,t)\\ &\quad+\sum_{i=1}^{N}\sum_{j(\neq i)=1}^{N} g(r_i)g(R_{ij})\overset{\leftrightarrow}{A}_i(\hat{\mathbf{r}}_i,\hat{\mathbf{R}}_{ij})\cdot\overset{\leftrightarrow}{A}_j(\hat{\mathbf{R}}_{ij},\hat{\mathbf{n}}^{\text{inc}})\cdot\widetilde{\mathbf{E}}^{\text{inc}}(\mathbf{R}_j,t)\\ &\quad+\sum_{i=1}^{N}\sum_{j(\neq i)=1}^{N}\sum_{l(\neq j)=1}^{N} g(r_i)g(R_{ij})g(R_{jl})\overset{\leftrightarrow}{A}_i(\hat{\mathbf{r}}_i,\hat{\mathbf{R}}_{ij})\cdot\overset{\leftrightarrow}{A}_j(\hat{\mathbf{R}}_{ij},\hat{\mathbf{R}}_{jl})\cdot\overset{\leftrightarrow}{A}_l(\hat{\mathbf{R}}_{jl},\hat{\mathbf{n}}^{\text{inc}})\cdot\widetilde{\mathbf{E}}^{\text{inc}}(\mathbf{R}_l,t)\\ &\quad+\cdots,\end{aligned} \tag{237}$$

where we imply the notation of Fig. 12 and indicate explicitly the temporal dependence of the macroscopic electric field vector of the incident quasi-monochromatic plane-wave field. It is then easily seen that the expression for the partial electric field contributed by the blue four-particle sequence in Fig. 21a at the observation point includes the complex exponential factor $\exp[\mathrm{i}k_1(r_4 + R_{43} + R_{32} + R_{21} + \hat{\mathbf{n}}^{\text{inc}}\cdot\mathbf{R}_1)]$. Thus the corresponding cumulative phase of the blue four-particle sequence is

$$\delta_{\text{blue}} = k_1(r_4 + R_{43} + R_{32} + R_{21} + \hat{\mathbf{n}}^{\text{inc}}\cdot\mathbf{R}_1). \tag{238}$$

The cumulative phases of other multi-particle sequences are determined analogously. For example, that of the yellow three-particle sequence is given by

$$\delta_{\text{yellow}} = k_1(r_{3'} + R_{3'2'} + R_{2'1'} + \hat{\mathbf{n}}^{\text{inc}}\cdot\mathbf{R}_{1'}). \tag{239}$$

It is important to recognize that the very concept of the cumulative phase becomes questionable if at least one particle of a sequence is located in the near zone of the preceding particle, which obviously happens in densely packed DRMs (e.g., Figs. 4i and 5). We will see however that qualitative interpretations of STMM results based on the notion of the cumulative phase can be qualitatively instructive even in the case of random particulate volumes with substantial packing densities.

The origin of the far-field speckles in the two panels of Fig. 20a can now be understood by recognizing that in the far zone of the entire cluster the partial field due to any multi-particle sequence is an outgoing transverse spherical wavelet centered at the origin of the last particle of the sequence. Since the distance to the far-zone observation point is much greater than the radius $R$ of the imaginary particulate volume $V$, all such partial wavelets at the observation point propagate in essentially the same direction given by the unit vector $\hat{\mathbf{n}}^{\rm sca}$ (Fig. 21a). The four-element column $\langle\langle \mathbf{I}^{\rm sca}(r\hat{\mathbf{n}}^{\rm sca},t)\rangle\rangle$ in Eq. (177) at the observation point can be directly expressed in terms of the elements of the scattering coherency dyadic $\overleftrightarrow{\rho}^{\rm sca}(r\hat{\mathbf{n}}^{\rm sca}) = \langle\langle \widetilde{\mathbf{E}}^{\rm sca}(r\hat{\mathbf{n}}^{\rm sca},t) \otimes [\widetilde{\mathbf{E}}^{\rm sca}(r\hat{\mathbf{n}}^{\rm sca},t)]^* \rangle\rangle$ according to

$$\langle\langle \mathbf{I}^{\rm sca}(r\hat{\mathbf{n}}^{\rm sca},t)\rangle\rangle = \frac{1}{2}\sqrt{\frac{\varepsilon_1}{\mu_0}} \begin{bmatrix} \hat{\boldsymbol{\theta}}^{\rm sca}\cdot\overleftrightarrow{\rho}^{\rm sca}(r\hat{\mathbf{n}}^{\rm sca})\cdot\hat{\boldsymbol{\theta}}^{\rm sca} + \hat{\boldsymbol{\varphi}}^{\rm sca}\cdot\overleftrightarrow{\rho}^{\rm sca}(r\hat{\mathbf{n}}^{\rm sca})\cdot\hat{\boldsymbol{\varphi}}^{\rm sca} \\ \hat{\boldsymbol{\theta}}^{\rm sca}\cdot\overleftrightarrow{\rho}^{\rm sca}(r\hat{\mathbf{n}}^{\rm sca})\cdot\hat{\boldsymbol{\theta}}^{\rm sca} + \hat{\boldsymbol{\varphi}}^{\rm sca}\cdot\overleftrightarrow{\rho}^{\rm sca}(r\hat{\mathbf{n}}^{\rm sca})\cdot\hat{\boldsymbol{\varphi}}^{\rm sca} \\ -\hat{\boldsymbol{\theta}}^{\rm sca}\cdot\overleftrightarrow{\rho}^{\rm sca}(r\hat{\mathbf{n}}^{\rm sca})\cdot\hat{\boldsymbol{\varphi}}^{\rm sca} - \hat{\boldsymbol{\varphi}}^{\rm sca}\cdot\overleftrightarrow{\rho}^{\rm sca}(r\hat{\mathbf{n}}^{\rm sca})\cdot\hat{\boldsymbol{\theta}}^{\rm sca} \\ {\rm i}\,(\hat{\boldsymbol{\varphi}}^{\rm sca}\cdot\overleftrightarrow{\rho}^{\rm sca}(r\hat{\mathbf{n}}^{\rm sca})\cdot\hat{\boldsymbol{\theta}}^{\rm sca} - \hat{\boldsymbol{\theta}}^{\rm sca}\cdot\overleftrightarrow{\rho}^{\rm sca}(r\hat{\mathbf{n}}^{\rm sca})\cdot\hat{\boldsymbol{\varphi}}^{\rm sca}) \end{bmatrix}, \tag{240}$$

where $\hat{\boldsymbol{\theta}}^{\rm sca}$ and $\hat{\boldsymbol{\varphi}}^{\rm sca}$ are the polar-angle and azimuth-angle unit vectors of the scattering direction such that $\hat{\mathbf{n}}^{\rm sca} = \hat{\mathbf{r}} = \hat{\boldsymbol{\theta}}^{\rm sca} \times \hat{\boldsymbol{\varphi}}^{\rm sca}$. According to Eq. (237), the dyadic product $\widetilde{\mathbf{E}}^{\rm sca}(r\hat{\mathbf{n}}^{\rm sca},t) \otimes [\widetilde{\mathbf{E}}^{\rm sca}(r\hat{\mathbf{n}}^{\rm sca},t)]^*$ at any moment in time is the sum of an infinite number of terms, each describing the result of interference of two spherical wavelets centered at the end particles of two particle sequences.

Fig. 21a exemplifies one such pair. If the interference of the corresponding pair of spherical wavelets at the observation point is constructive (destructive) then it serves to increase (decrease) the total intensity scattered in the direction $\hat{\mathbf{n}}^{\rm sca}$. The result of the interference depends largely on the phase difference $\Delta = \delta_{\rm blue} - \delta_{\rm yellow}$ given by

$$\Delta = k_1(r_4 + R_{43} + R_{32} + R_{21} + \hat{\mathbf{n}}^{\rm inc}\cdot\mathbf{R}_1 - r_{3'} - R_{3'2'} - R_{2'1'} - \hat{\mathbf{n}}^{\rm inc}\cdot\mathbf{R}_{1'}). \tag{241}$$

The total scattered intensity in the far zone of the particulate volume is the sum of the interference results contributed by all possible pairs of particle sequences. The minimal angular width of such interference maxima and minima is proportional to $1/k_1R$, while their number grows rapidly with $N$. These two factors explain the typical spotty appearance of the scattering patterns in Fig. 20a.

It is sometimes asserted that a speckle pattern can be caused only by monochromatic incident light, for example by that generated by a continuous laser. In actuality, however, all one needs in order to observe speckles is a fixed scattering object illuminated by a quasi-monochromatic plane-wave field.

The two panels of Fig. 20a exemplify the variability of the quasi-instantaneous speckle patterns that can be expected of a temporally changing DRM. After the quasi-instantaneous speckle patterns have been computed or measured for a representative set of evolving states of a DRM, one can choose to

- analyze the statistical information content of differences between the individual speckle patterns; or
- apply an averaging procedure, thereby isolating the static component of the speckle patterns.

We have already mentioned that these two approaches are known as dynamic and static light scattering.

### *8.2. Static scattering by Type-1 discrete random media*

In what follows, we simulate ensemble-averaged light-scattering characteristics of an imaginary spherical volume randomly and uniformly filled with identical particles by creating only one random $N$-particle configuration and then averaging over all possible orientations of this configuration with respect to the laboratory coordinate system. The fidelity of this approach will be analyzed later in this subsection.

We have already seen in Fig. 20 that averaging over the equiprobable orientation distribution of an 80-particle configuration effectively eliminates the speckle pattern and yields the combination of a smooth background and a notable backscattering peak. It turns out that the existence of both features can be explained qualitatively by using the notion of the cumulative phase of a multi-particle sequence introduced above. Specifically, each far-field speckle element can be thought of as being the result of constructive or destructive interference of two wavelets contributed by specific multi-particle sequences, such as those shown in Fig. 21a. The phase difference (241) evaluated at the far-zone observation point changes randomly as the particles move, so that the average result of the interference is zero. However, we will demonstrate below that certain classes of wavelet pairs interfere constructively irrespective of particle positions and thereby are responsible for the residual scattering pattern.

Let us make a simplifying assumption that $\varphi^{\text{sca}} = \varphi^{\text{inc}}$ and define the scattering direction in terms of the scattering angle $\Theta = \theta^{\text{sca}}$. Then scattering in the far zone can be conveniently described in terms of the dimensionless $4 \times 4$ scattering matrix (228). Numerous STMM computations have demonstrated that the elements populating the upper right and lower left $2 \times 2$ blocks of this matrix are negligibly small compared to the other elements, which is an expected result of averaging over the equiprobable orientation distribution of a multi-particle group coupled with sufficient uniformity of the initial particle positions throughout the scattering volume (cf. Table A.1). Specifically, the scattering matrix has the following typical structure:

$$\widetilde{\mathbf{F}}(\Theta) \approx \begin{bmatrix} \widetilde{F}_{11}(\Theta) & \widetilde{F}_{21}(\Theta) & 0 & 0 \\ \widetilde{F}_{21}(\Theta) & \widetilde{F}_{22}(\Theta) & 0 & 0 \\ 0 & 0 & \widetilde{F}_{33}(\Theta) & \widetilde{F}_{34}(\Theta) \\ 0 & 0 & -\widetilde{F}_{34}(\Theta) & \widetilde{F}_{44}(\Theta) \end{bmatrix}, \tag{242}$$

where the scattering matrix elements denoted by a zero are at least an order of magnitude smaller than the smallest nonzero element (in the absolute-value sense). Note that the relations $\widetilde{F}_{12}(\Theta) = \widetilde{F}_{21}(\Theta)$ and $\widetilde{F}_{43}(\Theta) = -\widetilde{F}_{34}(\Theta)$ are caused by the uniform orientation distribution of a multi-particle cluster. In all examples discussed below, the size parameter of the imaginary spherical volume filled with particles is fixed at $k_1R = 50$, while the size parameter and relative refractive index of the particles are fixed at $k_1r = 4$ and $m = 1.32$.

The plot of the phase function $\widetilde{F}_{11}(\Theta)$ in Fig. 22 reveals several fundamental consequences of increasing the number of particles $N$ in the volume. First of all, there is a strong and narrow forward-scattering enhancement owing to the systematically constructive interference of the wavelets singly scattered by the constituent particles in the exact forward direction. This feature is detailed in the top left-hand panel of Fig. 23 and, according to Fig. 21b, can be called forward-scattering localization of electromagnetic waves [287]. Indeed, the left-hand panel of Fig. 21b shows that the exact forward-scattering direction is unique in that the phases of the wavelets forward-scattered by all the individual particles in the DRM are precisely the same, irrespective of the specific instantaneous particle coordinates [1]. It is straightforward to show that if there were no multi-particle sequences, the constructive interference of these single-particle wavelets would cause an increase of the forward-scattering phase function $\widetilde{F}_{11}(0)$ by a factor of $N$. The top left-hand panel of Fig. 23 shows that this increase does occur for $N = 2$, 5, and 20, but eventually the $\widetilde{F}_{11}(0)$ value saturates. This behavior can be explained qualitatively by referring to a multi-particle interaction effect whereby particle 3 in the right-hand panel of Fig. 21b "shadows" particle 2 by attenuating the incident field exciting particle 2.

The second remarkable consequence of increasing $N$ is that the phase function at backscattering angles starts to develop a narrow peak with a maximum at $\Theta = 180°$ (see the top right-hand panel of Fig. 23). The qualitative explanation of this so-called weak localization of electromagnetic waves[9] (otherwise known as the coherent backscattering effect) is illustrated in Fig. 21c. The blue and yellow outgoing wavelets are contributed by the same chain of $n$ particles but sequenced in opposite order. The opposite sequencing is largely inconsequential owing to the reciprocity relation for the scattering dyadic (93a). Therefore, the two conjugate wavelets interfere at the observation point constructively or destructively mostly depending on the resulting phase difference between the blue and yellow sequences given by

$$\Delta = k_1(\mathbf{R}_1 - \mathbf{R}_n) \cdot (\hat{\mathbf{n}}^{\text{inc}} + \hat{\mathbf{n}}^{\text{sca}}). \tag{243}$$

If the observation direction $\hat{\mathbf{n}}^{\text{sca}}$ is far from the exact backscattering direction $-\hat{\mathbf{n}}^{\text{inc}}$ then the average effect of this interference is zero owing to randomly varying positions of particles 1 and $n$. However, at exactly the backscattering direction the differential phase $\Delta$ vanishes identically for any $n$-particle chain, thereby causing the interference to be always constructive and create a backscattering intensity peak.

The third obvious consequence of increasing the number of particles in the DRM is the progressively smooth and featureless profile of the phase function at scattering angles $30° \le \Theta \le 170°$. This effect manifests itself as the "diffuse" intensity background in Fig. 20b and is mostly caused by another class of wavelet pairs illustrated in Fig. 21d. In this case the wavelet caused by the yellow sequence of $n$ particles is the same as that caused by the blue sequence and thus "interferes with itself." Since the self-interference is always constructive irrespective of the specific chain of particles owing to the identity $\Delta \equiv 0$, the positive contribution of this class of wavelet pairs survives the ensemble averaging for any incidence and scattering directions. The qualitative explanation of the progressive smoothness of the phase-function curves with increasing $N$ in Fig. 22 is that the side-scattered intensity is averaged over the contri-butions from the rapidly increasing number of multi-particle chains.

[9] Note that the frequently used term "weak localization of photons" is thoroughly inappropriate since it refers to an interference phenomenon that is purely classical and has nothing to do with QED photons.

The bottom left-hand panel of Fig. 22 shows that the most prominent effect of increasing $N$ on the ratio $-\widetilde{F}_{21}(\Theta)/\widetilde{F}_{11}(\Theta)$ is to smooth out the low-frequency oscillations in the single-sphere curve and, on average, to make this ratio more neutral. This implies that the main contribution to the second Stokes parameter of the scattered light, $\langle\langle Q^{\rm sca}(r\hat{\mathbf{n}}^{\rm inc}, t)\rangle\rangle$, comes from single-particle chains, whereas the contributions from many-particle chains are largely randomized.

A fundamental property of the ratio $\widetilde{F}_{22}(\Theta)/\widetilde{F}_{11}(\Theta)$ is that it is identically equal to unity for scattering by a single sphere [1,34]. Therefore, the rapidly increasing deviation of this ratio from unity for $N \geq 5$ in Fig. 22 can also be attributed to multi-particle chains. Similarly, $\widetilde{F}_{33}(\Theta) \equiv \widetilde{F}_{44}(\Theta)$ and $\widetilde{F}_{33}(180°)/\widetilde{F}_{11}(180°) \equiv 1$ for scattering by a single spherical particle, but the cumulative contribution from multi-particle chains in particulate volumes with $N \geq 5$ cause rapidly growing violations of these identities.

If the incident plane-wave field is polarized linearly in the $xz$-plane then the angular distribution of the corresponding cross-polarized scattered intensity is defined by $\frac{1}{2}[\widetilde{F}_{11}(\Theta) - \widetilde{F}_{22}(\Theta)]$. This quantity is plotted in Fig. 23 along with the quantity $\frac{1}{2}[\widetilde{F}_{11}(\Theta) + \widetilde{F}_{44}(\Theta)]$ defining the same-helicity scattered intensity for the case of the incident plane-wave field polarized circularly in the counterclockwise direction when looking in the direction of the unit vector $\hat{\mathbf{n}}^{\rm inc}$ Both quantities provide the most definitive demonstration of the onset of weak localization with increasing $N$. Indeed, the corresponding single-particle curves show no backscattering enhancement whatsoever, so the backscattering peaks that develop with increasing $N$ (and thus with growing contributions from multi-particle chains) can be attributed unequivocally to weak localization.

Fig. 23 also depicts the angular profiles of the linear and circular polarization ratios defined as

$$\mu_{\rm L}(\Theta) = \frac{\widetilde{F}_{11}(\Theta) - \widetilde{F}_{22}(\Theta)}{\widetilde{F}_{11}(\Theta) + 2\widetilde{F}_{12}(\Theta) + \widetilde{F}_{22}(\Theta)} \tag{244}$$

and

$$\mu_{\rm C}(\Theta) = \frac{\widetilde{F}_{11}(\Theta) + \widetilde{F}_{44}(\Theta)}{\widetilde{F}_{11}(\Theta) - \widetilde{F}_{44}(\Theta)}, \tag{245}$$

respectively. The first quantity pertains to the case of a linearly polarized plane-wave incident field and is the ratio of the cross-polarized and co-polarized scattered intensities. The second quantity is relevant to the case of a circularly polarized plane-wave incident field and is the ratio of the same-helicity and opposite-helicity scattered intensities [34]. Fig. 23 demonstrates that the contribution from multi-particle chains serves to increase significantly the background deviations of both polarization ratios from zero, while weak localization causes pronounced backscattering peaks in the $\mu_{\rm L}$ and $\mu_{\rm C}$ angular profiles.

Let us now examine whether it was indeed appropriate to calculate each ensemble-averaged scattering pattern in Figs. 22 and 23 by averaging over orientations of only one quasi-random $N$-particle configuration. We essentially assumed that the results thus obtained would be statistically representative of the average over all possible realizations of the $N$-particle group, at least for large $N$. The correctness of this assumption is confirmed by Fig. 24 computed for two different realizations of a random 200-particle group populating a $k_1R = 50$ imaginary spherical volume. The refractive index of the identical $k_1r = 4$ particles is again 1.32. The reader can see that although the two sets of initial coordinates of the 200 particles were quite different, averaging over all orientations of each configuration yielded virtually indistinguishable results.

A more subtle and less ubiquitous manifestation of coherent backscattering can be exhibited by a DRM populated by quasi-Rayleigh particles with sizes significantly smaller than the wavelength [296]. Fig. 25 depicts the ratio $-\widetilde{F}_{21}(\Theta)/\widetilde{F}_{11}(\Theta)$ for a spherical particulate volume with $k_1R = 31$ populated by $N = 1, \ldots, 1875$ identical spherical particles with $k_1r = 2$ and $m = 1.31$ [243]. It can be seen that unlike the $-\widetilde{F}_{21}(\Theta)/\widetilde{F}_{11}(\Theta)$ trend in Fig. 22, the increase of $N$ first to 75 and then to 750 causes the onset and swift growth of a new feature not exhibited by the $N = 1$ curve. This narrow asymmetric minimum at backscattering angles was called the polarization opposition effect [340].

Like other manifestations of coherent backscattering, the polarization opposition effect is caused by pairs of multi-particle sequences exemplified by Fig. 21c. A qualitative interpretation of this specific feature is shown in Fig. 21g using simple two-particle sequences [341]. Particles 1–4 lie in a plane normal to the incidence direction and are assumed to have sizes significantly smaller than the wavelength. Particles 1 and 2 lie in the scattering plane (defined again as the plane through the illumination and observation directions), while the line through particles 3 and 4 is normal to this plane. If the incident quasi-monochromatic plane-wave field is unpolarized then both magenta sequences contribute scattered light polarized negatively with respect to the scattering plane (i.e., having positive values of the Stokes parameter $Q$), whereas both blue sequences contribute positively polarized scattered light (i.e., having negative values of the Stokes parameter $Q$). The phase difference between the conjugate magenta sequences is identically equal to zero, while that between the blue sequences is zero when the angle $\alpha = 180° - \Theta$ (traditionally called the phase angle) is zero, but oscillates rapidly with increasing $\alpha$. Therefore, on average, weak localization will enhance the negatively polarized scattering contributions over a wider range of phase angles than the positively polarized contributions. The result is the polarization opposition effect in the form of a negative polarization minimum at a small $\alpha$ comparable to the angular width of the coherent phase-function peak.

Despite its subtlety,[10] the polarization opposition effect was observed in the laboratory much earlier than the more ubiquitous backscattering intensity peak. Fig. 26 shows polarization measurements by Lyot [342] for a particulate surface obtained by burning a tape of magnesium under a glass plate until the deposit on the plate was completely opaque. Lyot described the observed phase curve of polarization as "puzzling" and tentatively attributed it to the very small size of magnesia grains. Lyot's results were recently reproduced and supplemented by photometric measurements [343] (see Fig. 27). The latter revealed an equally narrow backscattering intensity peak, thereby confirming that the backscattering intensity and polarization features have weak localization as their common cause. The polarization opposition effect with its typically asymmetric angular profile was not formally identified as a manifestation of weak localization until 1993 [340]. However, its physical origin is precisely the same as that of the so-called azimuthal asymmetry of the coherent backscattering cone observed in the late 1980s [298,344,345].

[10] For example, we have already pointed out that the bottom left-hand panel of Fig. 22 shows no signs of a sharp polarization minimum at backscattering angles emerging with increasing $N$.

It appears that Oetking [346] was the first to observe weak localization in the form of a narrow intensity peak centered at the exact backscattering direction. However, neither Lyot nor Oetking offered a correct theoretical explanation of their laboratory results. The first theoretical prediction of weak localization was made by Watson [347] with a reference to a private communication from R. Ruffine. The first deliberate laboratory demonstrations of coherent backscattering accompanied by a correct theoretical interpretation should be credited to Kuga and Ishimaru [348], Tsang and Ishimaru [349], Van Albada and Lagendijk [350], and Wolf and Maret [351]. Further references can be found in [25,38,352–354]. Remarkable manifestations of weak localization in planetary astrophysics are discussed in [31,296,306,317,339,340,355–358].

Qualitatively, the effect of increasing the number of particles $N$ in a DRM can be expected to be twofold. On one hand, it serves to increase the number of multi-particle sequences and thereby enhances such corollaries of the far-field Neumann expansion (237) as the smoothness of the scattered intensity at side-scattering directions and the various weak localization features at backscattering directions. On the other hand, it eventually yields packing density values so high that they cause features in the scattering patterns not implied by the far-field Neumann expansion. Therefore, the above qualitative interpretation of numerically exact STMM results can become partly or completely inadequate [304,305,318]. Fig. 25 shows that this is indeed the case: the black solid curve reveals a high-frequency ripple reminiscent of a homogeneous spherical particle with a size parameter comparable to that of the entire particulate volume. The corresponding packing density of 50% is so high that the expansion (237) along with the assumptions of randomness and statistical uniformity of particle positions become inapplicable.

Despite this conclusion, the direct solutions of the MMEs displayed in Figs. 22 and 23 do demonstrate that the classical corollaries of the low-density limit can survive – at least in a semi-quantitative sense – volume packing densities reaching 30%. Such values are typical of particle suspensions and many particulate surfaces.

Extensive STMM results reported in [287,296,298,304] have shown that the coherent backscattering peaks such as those in Fig. 23 are rounded at $\Theta = 180°$ owing to the finite size of the respective DRMs. The angular widths of the backscattering peaks and of the polarization opposition minimum are inversely proportional to $k_1R$ and are independent of $N$ until the effects of packing density start to dominate. For the same $k_1R$, the angular widths of the backscattering peaks (but not their amplitudes!) are weakly dependent on the particle size parameter and refractive index. Mixtures of spherical particles with different size parameters or different refractive indices also reveal all typical manifestations of weak localization, thereby further corroborating the universal interference nature of this phenomenon [307].

In [301], the conventional orientation-averaging procedure developed in the framework of the STMM was generalized to include the case of illumination by a finite Gaussian beam. Extensive computations demonstrated that all scattering patterns observed in the far zone of a random multisphere object and their evolution with decreasing width of the incident beam can still be interpreted in terms of forward-scattering interference, coherent backscattering, and diffuse background. It was shown in particular that the increasing violation of electromagnetic reciprocity with decreasing beam width suppresses and eventually eradicates all observable manifestations of weak localization and strongly suppresses the forward-scattering interference, while doing virtually nothing to the angular profiles of intensity and polarization at intermediate scattering angles.

To conclude this subsection, let us discuss the applicability of the effective-object methodology introduced in Section 6 to Type-1 DRMs. Specifically, we consider the result of substitut-ing an imaginary spherical volume filled with a large number of identical particles ("inclusions") by a homogeneous spherical object of the same radius, as shown in Fig. 28. Obviously, this substitution belongs to the category of semi-stochastic EOAs. In Fig. 29, the thick gray curves depict the orientation-averaged far-field STMM results for an imaginary $k_1R = 10$ spherical volume populated by $N = 15000$ identical spherical inclusions, each having a size parameter of $k_1r = 0.2$ and a refractive index of $m = 1.2$. For comparison, the thin black curves show the Lorenz–Mie results for the effective-medium counterpart of this imaginary spherical volume in the form of a homogeneous spherical particle with $k_1R = 10$ and $m_{\mathrm{eff}} = 1.023115$. Note that this effective refractive-index value follows from the Maxwell-Garnett effective-medium rule (EMR) [77] for the resulting 12% volume fraction.

It is patently obvious from Fig. 29 that despite the extremely small size parameter of the inclusions and their very large number, the Maxwell-Garnett EMR fails to reproduce the far-field dimensionless scattering matrix of the Type-1 particulate volume at side- and backscattering angles. In fact, the results of extensive Lorenz–Mie computations for effective refractive indices other than 1.023115 (not shown) revealed even worse agreement with the STMM curves. The likely qualitative explanation of this failure is the "bumpiness" effect wherein the discrete inclusions do not reproduce sufficiently well the perfectly smooth spherical surface of the effective Maxwell-Garnett scatterer responsible for the large-amplitude maxima and minima in the Lorenz–Mie curves. Not surprisingly, the Maxwell-Garnett EMR reproduces the STMM extinction cross section and asymmetry parameter much more accurately, the corresponding ratios being $C_{\mathrm{ext}}^{\mathrm{STMM}} / C_{\mathrm{ext}}^{\mathrm{MG}} = 1.0375$ and $\langle \cos \Theta \rangle^{\mathrm{STMM}} / \langle \cos \Theta \rangle^{\mathrm{MG}} = 0.9976$.

## 9. Direct computer modeling of static scattering by Type-2 discrete random media

In this section, we discuss the results of representative far-field STMM calculations for Type-2 DRMs to analyze how well they can be replicated by the effective-object methodology (see also [258,284,321–324,327,328]). For the purposes of our analysis, a heterogeneous object is modeled as an actual spherical body randomly filled with $N$ identical small spherical inclusions, as shown in Fig. 30a. Following the approach outlined in the preceding section, the statistical randomness and uniformity of the object's interior is simulated in two steps. First, we use a random-number generator to create a fixed yet quasi-random and quasi-uniform configuration of the $N$ inclusions, while making sure that the volumes of the inclusions do not cross the object's boundary and do not overlap. Second, we average all far-zone optical observables over the equiprobable orientation distribution of the resulting heterogeneous object using the STMM code described in [244].

The STMM results shown in Fig. 31 are obtained by assuming that the size parameter of the spherical host is fixed at $k_1R = 12$, while that of the inclusions takes on values $k_1r = 0.3$ and 1. The respective numbers of the inclusions are $N = 12800$ and 346, both implying the same $\rho = 20\%$ volume fraction. The refractive indices of the host and the inclusions are fixed at $m_{\mathrm{host}} = 1.33$ and $m_{\mathrm{incl}} = 1.55$, respectively. For comparison, we also show the results of Lorenz–Mie computations for a homogeneous spherical object with the size parameter $k_1R = 12$ and the refractive index $m_{\mathrm{LM}} = 1.372$. This refractive index provides the best fit of the Lorenz–Mie scattering matrix to that calculated for the heterogeneous object with $N = 12800$ inclusions and, in fact, is very close to the value $m_{\mathrm{MG}} = 1.3728$ predicted by the Maxwell-Garnett EMR for the given host and inclusion refractive indices and the inclusion volume fraction. Again, this EMR is predicated on the replacement of the heterogeneous target by an equidimensional homogeneous object with the same outer boundary, as exemplified by Figs. 30a,b, and belongs to the category of semi-stochastic EOAs. Since the Maxwell-Garnett effective refractive index is independent of $k_1 r$, the thick gray curves in Fig. 31 represent the EMR substitution for both heterogeneous objects.

It is obvious that if the boundary of the host body is perfectly spherical then the Maxwell-Garnett EMR must reproduce the well-known Lorenz–Mie identity $\widetilde{F}_{22}(\Theta)/\widetilde{F}_{11}(\Theta) \equiv 1$. Therefore, a deviation of the ratio $\widetilde{F}_{22}(\Theta)/\widetilde{F}_{11}(\Theta)$ for a heterogeneous spherical object from 100% is the most direct and unequivocal indicator of the numerical inaccuracy of the effective-medium methodology. Fig. 31 shows that the inclusion size parameter $k_1 r = 0.3$ yields $\widetilde{F}_{22}(\Theta)/\widetilde{F}_{11}(\Theta)$ values hardly distinguishable from 100%, whereas the inclusion size parameter $k_1 r = 1$ causes an obvious failure of the EMR.

Comparison of Figs. 29 and 31 reveals that the performance of the Maxwell-Garnett EMR is markedly better in the case of the Type-2 DRM, probably owing to the absence of the bumpiness effect. In fact, the nearly perfect agreement between the STMM curves for $k_1 r = 0.3$ and the Lorenz–Mie curves in Fig. 31 provides a convincing *numerical* validation of the effective-object hypothesis underlying the Maxwell-Garnett rule for Type-2 DRMs and should motivate efforts to derive this rule analytically from the MMEs. Still the STMM results for $k_1 r = 1$ in Fig. 31 show that the range of applicability of the EMR in terms of the maximal permissible inclusion size parameter can be quite limited. This result should also be explained by the analytical derivation.

In general, the optical cross sections and the asymmetry parameter are known to be less sensitive functions of the object's morphology than the elements of the scattering matrix. One can therefore expect a somewhat better accuracy of the Maxwell-Garnett prediction of the integral radiometric characteristics than that of the angular scattering-matrix profiles even for relatively large inclusions. This is indeed the case, the corresponding ratios being very close to unity for both inclusion size parameters: $C_{\text{ext}}^{\text{STMM}}/C_{\text{ext}}^{\text{MG}} = 1.0066$ and $\langle\cos\Theta\rangle^{\text{STMM}}/\langle\cos\Theta\rangle^{\text{MG}} =$ 0.9975 for $k_1 r = 0.3$ and $C_{\text{ext}}^{\text{STMM}}/C_{\text{ext}}^{\text{MG}} = 1.0209$ and $\langle\cos\Theta\rangle^{\text{STMM}}/\langle\cos\Theta\rangle^{\text{MG}} = 0.9779$ for $k_1 r = 1$.

To further substantiate the effective-medium hypothesis, in Fig. 32 we show the results of $T$-matrix computations for a spherical host with $k_1 R = 10$ and $m_{\text{host}} = 1.33$ randomly populated by two kinds of $k_1 r = 0.3$ inclusions having refractive indices $m_{\text{incl},1} = 1.45$ and $m_{\text{incl},2} = 1.6$. The number of each kind of inclusions is 4000. It is seen that the $T$-matrix results can be reproduced nearly perfectly by the Lorenz–Mie results for a homogeneous spherical object with $k_1 R = 10$ and $m_{\text{LM}} = 1.37$. Interestingly, almost the same refractive index ($m_{\text{EMR}} = 1.3696$) follows from the $n$-component effective-mixing rule [77]. The agreement between the respective extinction cross sections and asymmetry parameters is also excellent: $C_{\text{ext}}^{\text{STMM}}/C_{\text{ext}}^{\text{LM}} = 0.9895$ and $\langle\cos\Theta\rangle^{\text{STMM}}/\langle\cos\Theta\rangle^{\text{LM}} = 0.9943$.

Finally, in Figs. 33 and 34 we display the $T$-matrix results for two cases when the refractive index of the host exceeds that of the inclusions. Specifically, $m_{\text{host}} = 1.4$ in Fig. 33 and $m_{\text{host}} = 1.6$ in Fig. 34, while the inclusions are spherical voids with $m_{\text{incl}} = 1$. The other pa-rameters of both heterogeneous spherical objects are as follows: $k_1 R = 10$, $k_1 r = 0.3$, and $N = 8000$. It is seen that in these two cases, the diviations of the STMM curves for the scattering matrix elements other than the phase function from their best-fit Lorenz–Mie counterparts (corresponding to $m_{\mathrm{LM}} = 1.32$ in Fig. 33 and $m_{\mathrm{LM}} = 1.472$ in Fig. 34) are more noticeable than before, while the $\widetilde{F}_{22}(\Theta)/\widetilde{F}_{11}(\Theta)$ STMM curve in Fig. 34 signals significant problems with the very EMA methodology. Furthemore, the corresponding Maxwell-Garnett refractive indices ($m_{\mathrm{MG}} = 1.3123$ and 1.4694, respectively) differ substantially from their best-fit Lorenz–Mie values. Yet Table 3 shows that the Maxwell-Garnett refractive indices yield more accurate predictions of the extinction cross section and asymmetry parameter than the Lorenz–Mie refractive indices inferred by best-fitting the STMM scattering-matrix results in Figs. 33 and 34. Again, the still-to-be-developed analytical theory of the macroscopic effective-medium regime will need to explain all these numerically exact findings.

## 10. First-order-scattering approximation

Although using a numerically exact computer solver of the MMEs is the preferred way of quantifying electromagnetic scattering by a DRM, the applicability of this direct approach is still limited in terms of the number of constituent particles and the overall size of the particulate volume relative to the wavelength. However, there are two well-defined and often-encountered kinds of Type-1 DRM which allow for an explicit use of the far-field Foldy equations discussed in Subsection 4.10. As a result, one can derive analytically rather simple expressions or equations for key optical observables which provide for much more efficient computations by bypassing the calculation of the electromagnetic field itself. The particles forming either kind of DRM are sparsely and randomly distributed, but their number $N$ must be sufficiently small for the first kind or tend to infinity for the second kind. In either case the far-field conditions (89)–(91) do not apply to the whole DRM, which makes it necessary to first compute the ensemble-averaged Poynting–Stokes tensor and then use it to quantify the energy budget of the DRM and the reading of a near-field WCR [34].

Let us first consider the first kind of Type-1 DRM by assuming that:

- $N$ is sufficiently small and the average interparticle distance is sufficiently large that in the framework of the Foldy equations each particle can be considered as being "excited" only by the incident field;
- the $N$-particle DRM is observed from a distance $r$ much greater than any linear dimension $L$ of the imaginary volume $V$ circumscribing the DRM:

$$r \gg L; \tag{246}$$

- the observation point is allowed to be in the near zone of the entire DRM but is assumed to be distant enough to reside in the far zone of any of the $N$ particles constituting the DRM;
- all $N$ particles are moving randomly and independently of each other throughout the imaginary volume $V$;
- the physical states of the $N$ particles change randomly and independently of each other as well as independently of the particle positions, where, as before, the physical state of a particle includes all its physical characteristics except coordinates.

These requirements are often satisfied in laboratory and in situ measurements of light scattering by tenuous collections of small particles such as those discussed in [176,359–366].

According to the above assumptions, the second term on the right-hand side of Eq. (141) can be neglected in comparison with the first term. Let us choose the origin $O$ of the laboratory coordinate system close to the geometrical center of the $N$-particle DRM and assume that the observation point resides close enough to be in the near zone of the entire object yet sufficiently far to be in the far zone of any of the $N$ constituent particles (Fig. 35). Eqs. (75), (85), (86), and (140) then imply that

$$\widetilde{\mathbf{E}}^{\mathrm{sca}}(\mathbf{r}) = \sum_{i=1}^{N} \exp(\mathrm{i}k_1\hat{\mathbf{n}}^{\mathrm{inc}} \cdot \mathbf{R}_i) \frac{\exp(\mathrm{i}k_1 r_i)}{r_i} \overleftrightarrow{A}_i(\hat{\mathbf{r}}_i, \hat{\mathbf{n}}^{\mathrm{inc}}) \cdot \widetilde{\mathbf{E}}_0^{\mathrm{inc}}. \tag{247}$$

Let us now assume that the $N$-particle DRM is ergodic so that we can use Eq. (219b). Also, all particle positions $\mathbf{R}_i$, as well as all particle physical states $\xi_i$ (and thus the corresponding particle-centered scattering dyadics $\overleftrightarrow{A}_i(\hat{\mathbf{r}}_i, \hat{\mathbf{n}}^{\mathrm{inc}})$) as functions of time are considered to be independent random processes. This implies that averaging over all the individual-particle physical states and over all the individual-particle coordinates can be performed independently:

$$\langle \ldots \rangle_{\Psi} = \langle \langle \ldots \rangle_{\mathbf{R}} \rangle_{\xi}. \tag{248}$$

To average over the individual particle coordinates, we assume that the corresponding coordinate probability density functions are given by

$$p_{\mathbf{R}}(\mathbf{R}_i) = \begin{cases} 1/V & \text{if } \mathbf{R}_i \in V, \\ 0 & \text{if } \mathbf{R}_i \notin V \end{cases} \quad \text{for any } i = 1, \ldots, N. \tag{249}$$

This means that the individual positions of all the $N$ particles throughout the entire volume $V$ are mutually independent and statistically equiprobable. This is consistent with the assumption that the average particle packing density is sufficiently small. Finally, we assume that that the angular dependence of the individual particle-centered scattering dyadics is weak enough that at the large distance $r$ from the DRM,

$$\overleftrightarrow{A}_i(\hat{\mathbf{r}}_i, \hat{\mathbf{s}}) \approx \overleftrightarrow{A}_i(\hat{\mathbf{r}}, \hat{\mathbf{s}}) \quad \text{for any } i, \tag{250}$$

where $\hat{\mathbf{r}}$ is the unit vector originating at $O$ and pointing in the direction of the observation point $\mathbf{r}$ (Fig. 35).

Let us first quantify the energy budget of the entire $N$-particle DRM. This entails surrounding the volume $V$ by an imaginary sphere $S$ with a radius $r$ much greater than the volume's typical linear dimension $L$, as sketched in Fig. 35, and evaluating the integral

$$\langle\langle \overline{W}^{\mathrm{abs}}(t) \rangle\rangle = -\mathrm{Re} \int_S \mathrm{d}^2\mathbf{r} \langle\langle \overline{\mathbf{S}}(\mathbf{r}, t) \rangle\rangle \cdot \hat{\mathbf{r}}. \tag{251}$$

The explicit derivation detailed in [34] requires two more assumptions. First, the size parameter of the volume $V$ must be much greater than unity:

$$k_1 L \gg 1. \tag{252}$$

Second, the sum of the individual extinction cross sections of the $N$ particles forming the DRM must be much smaller than the geometrical cross section of the volume $V$. The final result, formulated here for the general case of quasi-monochromatic scattering, is as follows:

$$\langle\langle \overline{W}^{\mathrm{abs}}(t) \rangle\rangle = \langle\langle \overline{W}^{\mathrm{ext}}(t) \rangle\rangle - \langle\langle \overline{W}^{\mathrm{sca}}(t) \rangle\rangle, \tag{253}$$

where

$$\langle\langle \overline{W}^{\text{ext}}(t)\rangle\rangle = \sum_{i=1}^{N}[\langle K_{11}(\hat{\mathbf{n}}^{\text{inc}};\xi_i)\rangle_{\xi_i}\langle\langle I^{\text{inc}}(t)\rangle\rangle + \langle K_{12}(\hat{\mathbf{n}}^{\text{inc}};\xi_i)\rangle_{\xi_i}\langle\langle Q^{\text{inc}}(t)\rangle\rangle$$
$$+ \langle K_{13}(\hat{\mathbf{n}}^{\text{inc}};\xi_i)\rangle_{\xi_i}\langle\langle U^{\text{inc}}(t)\rangle\rangle + \langle K_{14}(\hat{\mathbf{n}}^{\text{inc}};\xi_i)\rangle_{\xi_i}\langle\langle V^{\text{inc}}(t)\rangle\rangle], \tag{254}$$

$$\langle\langle \overline{W}^{\text{sca}}(t)\rangle\rangle = \int_{4\pi} d\hat{\mathbf{r}} \sum_{i=1}^{N}[\langle Z_{11}(\hat{\mathbf{r}},\hat{\mathbf{n}}^{\text{inc}};\xi_i)\rangle_{\xi_i}\langle\langle I^{\text{inc}}(t)\rangle\rangle + \langle Z_{12}(\hat{\mathbf{r}},\hat{\mathbf{n}}^{\text{inc}};\xi_i)\rangle_{\xi_i}\langle\langle Q^{\text{inc}}(t)\rangle\rangle$$
$$+ \langle Z_{13}(\hat{\mathbf{r}},\hat{\mathbf{n}}^{\text{inc}};\xi_i)\rangle_{\xi_i}\langle\langle U^{\text{inc}}(t)\rangle\rangle + \langle Z_{14}(\hat{\mathbf{r}},\hat{\mathbf{n}}^{\text{inc}};\xi_i)\rangle_{\xi_i}\langle\langle V^{\text{inc}}(t)\rangle\rangle]. \tag{255}$$

In the above formulas, $\mathbf{K}(\hat{\mathbf{n}}^{\text{inc}};\xi_i)$ and $\mathbf{Z}(\hat{\mathbf{r}},\hat{\mathbf{n}}^{\text{inc}};\xi_i)$ are the particle-centered extinction and phase matrices of particle $i$, respectively.

Let us now consider the electromagnetic response of the two distant polarimetric WCRs shown in Fig. 36, each having its optical axis centered at the volume element $V$. Both instruments are located in the near zone of the DRM yet sufficiently far from it so that each partial wavelet contributing to the right-hand side of Eq. (247) becomes locally flat by the time it reaches a WCR. Furthermore, although the acceptance solid angle $\Delta\Omega$ of either WCR is very small, its distance $r$ from the center of the DRM is large enough that the solid angle subtended by $V$, as viewed from the WCR, is smaller than $\Delta\Omega$. As a result, either WCR captures all $N$ partial wavelets irrespective of particles' locations within $V$, while WCR 2 also captures the incident plane wave.

According to Subsection 4.5, WCR 1 integrates over its objective lens the time-averaged Stokes column vector of the superposition of the $N$ quasi-plane wavelets propagating in essentially the same direction $\hat{\mathbf{r}}_1$. Since WCR 1 does not capture the incident plane wavefront, it can be shown [34] that the quasi-monochromatic response of WCR 1 averaged over a sufficiently long period of time is given by

$$\langle\langle \overline{\textbf{Signal 1}}(t)\rangle\rangle = \frac{S_{\text{ol}}}{r^2}\sum_{i=1}^{N}\langle \mathbf{Z}(\hat{\mathbf{r}}_1,\hat{\mathbf{n}}^{\text{inc}};\xi_i)\rangle_{\xi_i}\langle\langle \mathbf{I}^{\text{inc}}(t)\rangle\rangle. \tag{256}$$

The {objective lens, diaphragm} filter of WCR 2 passes the incident plane wave in addition to the $N$ partial quasi-plane wavelets. As a consequence, the integration of the resulting Stokes column vector over the entrance pupil of WCR 2 yields [34]:

$$\langle\langle \overline{\textbf{Signal 2}}(t)\rangle\rangle = S_{\text{ol}}\langle\langle \mathbf{I}^{\text{inc}}(t)\rangle\rangle - \sum_{i=1}^{N}\langle \mathbf{K}(\hat{\mathbf{n}}^{\text{inc}};\xi_i)\rangle_{\xi_i}\langle\langle \mathbf{I}^{\text{inc}}(t)\rangle\rangle + \frac{S_{\text{ol}}}{r^2}\sum_{i=1}^{N}\langle \mathbf{Z}(\hat{\mathbf{n}}^{\text{inc}},\hat{\mathbf{n}}^{\text{inc}};\xi_i)\rangle_{\xi_i}\langle\langle \mathbf{I}^{\text{inc}}(t)\rangle\rangle. \tag{257}$$

Eqs. (253)–(257) represent the so-called first-order-scattering approximation for the Type-1 DRM in the form of a small ergodic group of sparsely distributed particles. Comparison of these formulas with their far-field counterparts (205)–(209) shows that the reading of a near-zone yet sufficiently distant WCR can be quantified by summing up the corresponding single-particle far-field readings.

A fundamental consequence of the additivity of the extinction and phase matrices in Eqs.

(254)–(257) is that the actual $N$-particle DRM is optically indistinguishable from that consisting of $N$ statistically identical particles, each having the same average extinction and phase matrices given by

$$\begin{cases} \langle \mathbf{K}(\hat{\mathbf{n}}^{\text{inc}};\xi)\rangle_\xi = \dfrac{1}{N}\displaystyle\sum_{i=1}^{N} \langle \mathbf{K}(\hat{\mathbf{n}}^{\text{inc}};\xi_i)\rangle_{\xi_i}, \\ \langle \mathbf{Z}(\hat{\mathbf{r}},\hat{\mathbf{n}}^{\text{inc}};\xi)\rangle_\xi = \dfrac{1}{N}\displaystyle\sum_{i=1}^{N} \langle \mathbf{Z}(\hat{\mathbf{r}},\hat{\mathbf{n}}^{\text{inc}};\xi_i)\rangle_{\xi_i}. \end{cases} \tag{258}$$

The matrices $\langle \mathbf{K}(\hat{\mathbf{n}}^{\text{inc}};\xi)\rangle_\xi$ and $\langle \mathbf{Z}(\hat{\mathbf{r}},\hat{\mathbf{n}}^{\text{inc}};\xi)\rangle_\xi$ can be thought of as being averaged over a synthetic distribution of physical states of one particle $p_\xi(\xi)$ derived from the $N$ individual-particle distributions $p_{\xi_i}(\xi_i)$. Then Eqs. (254)–(257) take the following simplified form:

$$\begin{aligned} \langle\langle \overline{W}^{\text{ext}}(t)\rangle\rangle = N[&\langle K_{11}(\hat{\mathbf{n}}^{\text{inc}};\xi)\rangle_\xi \langle\langle I^{\text{inc}}(t)\rangle\rangle + \langle K_{12}(\hat{\mathbf{n}}^{\text{inc}};\xi)\rangle_\xi \langle\langle Q^{\text{inc}}(t)\rangle\rangle + \langle K_{13}(\hat{\mathbf{n}}^{\text{inc}};\xi)\rangle_\xi \langle\langle U^{\text{inc}}(t)\rangle\rangle \\ &+ \langle K_{14}(\hat{\mathbf{n}}^{\text{inc}};\xi)\rangle_\xi \langle\langle V^{\text{inc}}(t)\rangle\rangle], \end{aligned} \tag{259}$$

$$\begin{aligned} \langle\langle \overline{W}^{\text{sca}}(t)\rangle\rangle = N\int_{4\pi} \mathrm{d}\hat{\mathbf{r}}\, [&\langle Z_{11}(\hat{\mathbf{r}},\hat{\mathbf{n}}^{\text{inc}};\xi)\rangle_\xi \langle\langle I^{\text{inc}}(t)\rangle\rangle + \langle Z_{12}(\hat{\mathbf{r}},\hat{\mathbf{n}}^{\text{inc}};\xi)\rangle_\xi \langle\langle Q^{\text{inc}}(t)\rangle\rangle \\ &+ \langle Z_{13}(\hat{\mathbf{r}},\hat{\mathbf{n}}^{\text{inc}};\xi)\rangle_\xi \langle\langle U^{\text{inc}}(t)\rangle\rangle + \langle Z_{14}(\hat{\mathbf{r}},\hat{\mathbf{n}}^{\text{inc}};\xi)\rangle_\xi \langle\langle V^{\text{inc}}(t)\rangle\rangle], \end{aligned} \tag{260}$$

$$\langle\langle \overline{\textbf{Signal 1}}(t)\rangle\rangle = \frac{S_{\text{ol}}}{r^2} N \langle \mathbf{Z}(\hat{\mathbf{r}}_1,\hat{\mathbf{n}}^{\text{inc}};\xi)\rangle_\xi \langle\langle \mathbf{I}^{\text{inc}}(t)\rangle\rangle, \tag{261}$$

$$\langle\langle \overline{\textbf{Signal 2}}(t)\rangle\rangle = S_{\text{ol}} \langle\langle \mathbf{I}^{\text{inc}}(t)\rangle\rangle - N\langle \mathbf{K}(\hat{\mathbf{n}}^{\text{inc}};\xi)\rangle_\xi \langle\langle \mathbf{I}^{\text{inc}}(t)\rangle\rangle + \frac{S_{\text{ol}}}{r^2} N\langle \mathbf{Z}(\hat{\mathbf{n}}^{\text{inc}},\hat{\mathbf{n}}^{\text{inc}};\xi)\rangle_\xi \langle\langle \mathbf{I}^{\text{inc}}(t)\rangle\rangle. \tag{262}$$

The principal advantage of the first-order-scattering approximation is that it obviates the need to explicitly solve the MMEs for a statistically representative set of sparse $N$-particle configurations and replaces this complicated task by the much simpler task of finding the far-field solution of the MMEs for one isolated particle followed by averaging this solution over a representative distribution of particle physical states. Furthermore, there is no need to satisfy the most challenging requirement of the far-field approximation, viz., the inequality (91), by applying it to the entire volume $V$.

The analytical derivation of the first-order-scattering approximation does not involve an explicit requirement that the $N$ constituent particles be in the far-zones of each other. Instead, the most important explicit requirement leading to Eqs. (259)–(262) is that the second term on the right-hand side of Eq. (141) be much smaller than the first term. However, this requirement does imply that the average separation between the particles must be appropriately large and their total number $N$ must be sufficiently small. These qualitative criteria were analyzed using numerically exact STMM results in [288]. Further insight can be gained from recalling that Eq. (138) is valid in the near zone as well as in the far zone of a DRM. Therefore, far-field STMM computations based on this formula should be a good test of the accuracy of Eq. (259). Table 4 shows the values of the ratio $\langle\langle \overline{W}^{\text{ext}}(t)\rangle\rangle^{\text{FOSA}} / \langle\langle \overline{W}^{\text{ext}}(t)\rangle\rangle^{\text{STMM}}$ for an imaginary spherical volume with a size parameter $k_1 R = 50$ randomly filled with $N$ identical spherical particles having a size parameter of $k_1 r = 4$ and a refractive index of $m = 1.32$. The incident field is assumed to be quasi-monochromatic and unpolarized. Also shown are the corresponding values of the packing density $\rho = N(r/R)^3$. It is obvious that only packing densities of one percent or less can ensure high numerical accuracy of the first-order-scattering approximation.

The main difference between the far-zone formula (98) and the near-zone formula (261) is that the latter completely ignores the forward-scattering interference explained in Fig. 21b and discussed in Subsection 8.2. Yet at side- and back-scattering angles both formulas should give similar results provided that the main requirements of the first-order-scattering approximation are met. In particular, the ratios of the elements of the phase matrix must become $N$-independent. Fig. 22 shows that this is the case only when $N$ is smaller than 20. According to Table 4, this again implies that the packing density must be less than one percent.

## 11. Radiative transfer and coherent backscattering

### *11.1. Radiative transfer theory*

Another analytical approach directly derivable from the MMEs is what is traditionally called the radiative transfer theory. In this case it is assumed that:

- the $N$ particles forming the Type-1 DRM (Fig. 3a) are separated widely enough that each of them is located in the far zones of all the other particles;
- the observation point is located in the far zone of any particle in the group (but, in general, in the near zone of the entire group);
- $N$ is very large: $N \to \infty$.

The first assumption implies the applicability of the algebraic far-field FEs (149) and (151). According to the second assumption, the total field at any observation point located sufficiently far from any particle in the sparse DRM is the superposition of the incident plane wave and $N$ partial spherical wavelets contributed by the $N$ particles. The observation point does not have to be in the far zone of the entire group and can be anywhere in space, including inside the DRM, as long as it resides in the far zones of all the $N$ particles constituting the DRM (see Subsection 4.10).

The third assumption implies that we can replace the full far-field Neumann expansion (237) by the much simpler so-called Twersky expansion. Indeed, the terms with $j = i$ and $l = j$ in the triple summation on the right-hand side of Eq. (237) are excluded, but the terms with $l = i$ are retained. Therefore, we can decompose this summation as follows:

$$\sum_{i=1}^{N} \sum_{\substack{j=1 \\ j \neq i}}^{N} \sum_{\substack{l=1 \\ l \neq j}}^{N} \cdots = \sum_{i=1}^{N} \sum_{\substack{j=1 \\ j \neq i}}^{N} \sum_{\substack{l=1 \\ l \neq i \\ l \neq j}}^{N} \cdots + \sum_{i=1}^{N} \sum_{\substack{j=1 \\ j \neq i}}^{N} \sum_{l=1}^{N} \delta_{jl} \times \cdots , \tag{263}$$

where $\delta_{jl}$ is the Kronecker delta. Higher-order summations in Eq. (237) can be decomposed similarly. The first group of terms on the right-hand side of Eq. (263) is contributed by "self-avoiding" sequences of particles, whereas the second group includes contributions from sequences that involve a particle more than once. The approximation introduced by Twersky [367] helps simplify Eq. (237) by retaining only the terms contributed by all self-avoiding multi-particle sequences. In the limit $N \to \infty$ the Twersky approximation accounts for the overwhelming majority of multi-particle sequences and thus can be expected to yield asymptotically accurate results.

Since we are dealing with a near-field problem, the solution must be based on the calculation of the time-averaged Poynting–Stokes tensor or, more generally, the time-averaged dyadic correlation function (152). Using the Twersky approximation of the Neumann expansion (237), the Twersky approximation for the dyadic correlation function can be formulated diagrammatically according to Fig. 37. The different terms entering the expanded expression inside the angular brackets on the right-hand side of this equation can be classified using the notation introduced in Fig. 38a. In this particular case, the upper and lower multi-particle sequences involve different particles. However, the two multi-particle sequences can involve one or more common particles, as indicated in Figs. 38c–f by the dashed connectors. Moreover, if the number of common particles in a diagram is two or more then they can enter the upper and lower sequences in the same order, as in Fig. 38d, or in the reverse order, as in Fig. 38e. The diagrams without crossing connectors are called ladder diagrams. Two such diagrams are exemplified by Figs. 21d,e. Fig. 38f gives an example of a mixed diagram wherein two common particles appear in the same order while two other common particles appear in the reverse order. By the very nature of the Twersky approximation, no particle can appear in either the upper or the lower sequence more than once.

According to the preceding discussion, the assumption of full ergodicity of the DRM allows us to replace the calculation of the time average $\langle\langle \overleftrightarrow{C}(\mathbf{r}', \mathbf{r}; t)\rangle\rangle$ by the calculation of the ensemble average $\langle \overleftrightarrow{C}(\mathbf{r}', \mathbf{r}; \Psi)\rangle_{\Psi} = \langle \overleftrightarrow{C}(\mathbf{r}', \mathbf{r}; \mathbf{R}, \xi)\rangle_{\mathbf{R},\xi}$, where $\mathbf{R}$ denotes the complete set of particle coordinates and $\xi$ denotes the complete set of particle physical states. This problem is still very complex in general, but becomes more manageable if we further assume that:

- The position and physical state of each particle are statistically independent of each other and of those of all the other particles.
- The physical states of all the particles have the same statistical characteristics.
- The spatial distribution of the particles throughout the medium is completely random and statistically uniform.
- All diagrams with crossing connectors in the diagrammatic expansion of the dyadic correlation function can be ignored. This is the gist of the ladder approximation [368].

The subsequent analytical derivation is detailed in [34] (see also [25,187]) and is not dwelled upon in this Report since it contains no new concepts and is a straightforward mathematical exercise. An important intermediate step is the emergence of the following matrix integro-differential equation:

$$\hat{\mathbf{q}} \cdot \nabla \widetilde{\mathbf{I}}(\mathbf{r}, \hat{\mathbf{q}}) = -n_0 \langle \mathbf{K}(\hat{\mathbf{q}}; \xi)\rangle_{\xi} \widetilde{\mathbf{I}}(\mathbf{r}, \hat{\mathbf{q}}) + n_0 \int_{4\pi} \mathrm{d}\hat{\mathbf{q}}' \langle \mathbf{Z}(\hat{\mathbf{q}}, \hat{\mathbf{q}}'; \xi)\rangle_{\xi} \widetilde{\mathbf{I}}(\mathbf{r}, \hat{\mathbf{q}}') \tag{264}$$

traditionally called the radiative transfer equation (RTE). Here, $n_0 = N/V$ is the average number of particles per unit volume; $\langle \mathbf{K}(\hat{\mathbf{q}}; \xi)\rangle_{\xi}$ is the single-particle extinction matrix averaged over the physical states of all the $N$ particles; $\langle \mathbf{Z}(\hat{\mathbf{q}}, \hat{\mathbf{q}}'; \xi)\rangle_{\xi}$ is the single-particle phase matrix, also averaged over the physical states of all the $N$ particles constituting the DRM; and

$$\widetilde{\mathbf{I}}(\mathbf{r},\hat{\mathbf{q}}) = \begin{bmatrix} \widetilde{I}(\mathbf{r},\hat{\mathbf{q}}) \\ \widetilde{Q}(\mathbf{r},\hat{\mathbf{q}}) \\ \widetilde{U}(\mathbf{r},\hat{\mathbf{q}}) \\ \widetilde{V}(\mathbf{r},\hat{\mathbf{q}}) \end{bmatrix} \tag{265}$$

is the real-valued so-called specific intensity column vector. The RTE is supplemented by the boundary condition

$$\widetilde{\mathbf{I}}(\mathbf{r},\hat{\mathbf{q}}_{\leftarrow})\big|_{\mathbf{r}\in S} = \delta(\hat{\mathbf{n}}^{\mathrm{inc}} - \hat{\mathbf{q}}_{\leftarrow})\langle\langle \mathbf{I}^{\mathrm{inc}}(t)\rangle\rangle, \tag{266}$$

where $S$ is the boundary of the Type-1 DRM (Fig. 3a), $\hat{\mathbf{q}}_{\leftarrow}$ is any unit vector directed *into* the volume $V$, and $\delta(\hat{\mathbf{s}})$ is the solid-angle delta function. Note that Eqs. (264)–(266) are valid in the general case of the quasi-monochromatic plane-wave incident field (168)–(169).

It is convenient to decompose the total specific intensity column vector into so-called coherent (subscript "c") and diffuse (subscript "d") components:

$$\widetilde{\mathbf{I}}(\mathbf{r},\hat{\mathbf{q}}) = \delta(\hat{\mathbf{n}}^{\mathrm{inc}} - \hat{\mathbf{q}})\mathbf{I}_{\mathrm{c}}(\mathbf{r}) + \widetilde{\mathbf{I}}_{\mathrm{d}}(\mathbf{r},\hat{\mathbf{q}}). \tag{267}$$

It is easily seen that these quantities are solutions of the following boundary-value problems:

$$\hat{\mathbf{n}}^{\mathrm{inc}} \cdot \nabla \mathbf{I}_{\mathrm{c}}(\mathbf{r}) = -n_0 \langle \mathbf{K}(\hat{\mathbf{n}}^{\mathrm{inc}};\xi)\rangle_{\xi} \mathbf{I}_{\mathrm{c}}(\mathbf{r}), \tag{268}$$

$$\mathbf{I}_{\mathrm{c}}(\mathbf{r})\big|_{\mathbf{r}\in S_{\mathrm{ill}}} = \langle\langle \mathbf{I}^{\mathrm{inc}}(t)\rangle\rangle, \tag{269}$$

$$\hat{\mathbf{q}} \cdot \nabla \widetilde{\mathbf{I}}_{\mathrm{d}}(\mathbf{r},\hat{\mathbf{q}}) = -n_0 \langle \mathbf{K}(\hat{\mathbf{q}};\xi)\rangle_{\xi} \widetilde{\mathbf{I}}_{\mathrm{d}}(\mathbf{r},\hat{\mathbf{q}}) + n_0 \int_{4\pi} \mathrm{d}\hat{\mathbf{q}}' \langle \mathbf{Z}(\hat{\mathbf{q}},\hat{\mathbf{q}}';\xi)\rangle_{\xi} \widetilde{\mathbf{I}}_{\mathrm{d}}(\mathbf{r},\hat{\mathbf{q}}')$$

$$+ n_0 \langle \mathbf{Z}(\hat{\mathbf{q}},\hat{\mathbf{n}}^{\mathrm{inc}};\xi)\rangle_{\xi} \mathbf{I}_{\mathrm{c}}(\mathbf{r}), \tag{270}$$

$$\widetilde{\mathbf{I}}_{\mathrm{d}}(\mathbf{r},\hat{\mathbf{q}}_{\leftarrow})\big|_{\mathbf{r}\in S} = \mathbf{0}, \tag{271}$$

where $S_{\mathrm{ill}}$ the illuminated part of the boundary $S$ and $\mathbf{0}$ is a zero four-component column. The obvious solution of Eq. (268) is the straightforward matrix generalization of the famous Bouguer exponential attenuation law [93,94,100]:

$$\mathbf{I}_{\mathrm{c}}(\mathbf{r}) = \exp[-n_0 s \langle \mathbf{K}(\hat{\mathbf{n}}^{\mathrm{inc}};\xi)\rangle_{\xi}]\langle\langle \mathbf{I}^{\mathrm{inc}}(t)\rangle\rangle, \tag{272}$$

where $s$ is the distance between the observation point $\mathbf{r}$ and $S_{\mathrm{ill}}$ along the straight line parallel to $\hat{\mathbf{n}}^{\mathrm{inc}}$.

The solution of the RTE can be directly used to compute relevant near-field optical observables. For example, the energy-budget problem is solved by using the following formula for the time-averaged local Poynting vector:

$$\langle\langle \overline{\mathbf{S}}(\mathbf{r},t)\rangle\rangle = \int_{4\pi} \mathrm{d}\hat{\mathbf{q}}\,\hat{\mathbf{q}}\,\widetilde{I}(\mathbf{r},\hat{\mathbf{q}}), \tag{273}$$

where $\widetilde{I}(\mathbf{r},\hat{\mathbf{q}})$, traditionally called the specific intensity, is the first element of the specific intensity column vector (265). The reading of a polarization-sensitive WCR centered around the

“propagation direction” $\hat{\mathbf{q}}$ per unit time is given by

$$\langle\langle \overline{\mathbf{Signal}}(\mathbf{r},\hat{\mathbf{q}};t)\rangle\rangle = S_{\mathrm{ol}}\int_{\Delta\Omega_{\hat{\mathrm{q}}}} \mathrm{d}\hat{\mathbf{q}}'\widetilde{\mathbf{I}}(\mathbf{r},\hat{\mathbf{q}}') \approx \begin{cases} S_{\mathrm{ol}}\mathbf{I}_{\mathrm{c}}(\mathbf{r}) + S_{\mathrm{ol}}\Delta\Omega\widetilde{\mathbf{I}}_{\mathrm{d}}(\mathbf{r},\hat{\mathbf{n}}^{\mathrm{inc}}) \text{ if } \hat{\mathbf{q}} = \hat{\mathbf{n}}^{\mathrm{inc}}, \\ S_{\mathrm{ol}}\Delta\Omega\widetilde{\mathbf{I}}_{\mathrm{d}}(\mathbf{r},\hat{\mathbf{q}}) \text{ if } \hat{\mathbf{q}} \neq \hat{\mathbf{n}}^{\mathrm{inc}}, \end{cases} \tag{274}$$

where it is assumed that the WCR is placed inside the DRM (Fig. 39) and, as before, $\Delta\Omega$ is the WCR’s acceptance solid angle.

The implications of the derivation of Eqs. (264)–(274) directly from the MMEs are quite profound and are discussed in [34], while the genesis of these formulas is traced in [100]. There are several efficient computer solvers of the RTE [25,57,59,60,64,66,369–372] which make it much easier to deal with the RTE than with the MMEs. The fact that the reading of the WCR can be modeled theoretically by solving the RTE often makes the {WCR, RTE} combination a useful optical-characterization tool. Moreover, comparison of Eqs. (273) and (274) shows that a WCR can be used to measure the local time-averaged Poynting vector by integrating its signal over the entire range $\hat{\mathbf{q}} \in 4\pi$ and thereby solve the energy-budget problem experimentally. Needless to say, to enable such optical-characterization and energy-budget applications based on the radiative transfer theory, the DRM must possess the specific macro- and microphysical properties discussed in the beginning of this subsection.

### *11.2. The Tyndall effect*

It is easily seen that in the absence of the integral term on the right-hand side of Eq. (264), the solution $\widetilde{\mathbf{I}}(\mathbf{r},\hat{\mathbf{q}})$ of the RTE subject to the boundary condition (266) would reduce to

$$\widetilde{\mathbf{I}}(\mathbf{r},\hat{\mathbf{q}}) = \delta(\hat{\mathbf{n}}^{\mathrm{inc}} - \hat{\mathbf{q}})\mathbf{I}_{\mathrm{c}}(\mathbf{r}). \tag{275}$$

This is equivalent to using the coherent-field approximation, i.e., to keeping only the first term on the right-hand side of Eq. (218b). As a consequence, the reading of the WCR in Fig. 39 would be nonzero only if the inward optical axis of the instrument was perfectly aligned with the incidence direction (cf. Eq. (274)). The fact that a WCR immersed in or looking at a turbid medium and having its axis not aligned with the incidence direction can generate a nonzero signal is explained by the presence of the integral term in the RTE causing a non-zero diffuse specific intensity column vector $\widetilde{\mathbf{I}}_{\mathrm{d}}(\mathbf{r},\hat{\mathbf{q}})$ and the resulting inadequacy of the coherent-field approximation. This optical phenomenon was first identified by John Tyndall [373,374] and is often called the Tyndall effect. Its physical origin can be traced all the way back to the inequalities (187) and (188).

Typical manifestations of the Tyndall effect primarily caused by the last term on the right-hand side of Eq. (270) are shown in Figs. 30c–e. In Fig. 30c the laser beam is “invisible” when it passes through the glass containing pure water but becomes “visible” (i.e., causes a nonzero reaction of the photographic camera) when it passes through a colloidal suspension. Similarly, the “solar rays” become “visible” upon scattering by haze or fog particles in Figs. 30d,e.

### *11.3. Weak localization*

We have seen that a major approximation in deriving Eqs. (264)–(274) was keeping only the ladder component of the dyadic correlation function. An improvement could be the computation of the so-called “cyclical” component caused by pairs of multi-particle sequences exempli-fied by Figs. 21c,f. Indeed, let us again consider the scattering by a Type-1 DRM as shown schematically in Fig. 40. The DRM is illuminated by a quasi-monochromatic plane-wave field. It is straightforward to show that upon statistical averaging, the contribution to the total Poynting–Stokes tensor of all the diagrams of the type illustrated in Fig. 41 must vanish at near-field observation points located either inside (observation point 1) or outside (observation point 2) the object. However, as discussed in Subsection 8.2, there is an exception corresponding to the situation when the observation point is in the far zone of the entire DRM and is located within its "back-shadow" (observation point 3 in Fig. 40). Then the class of diagrams illustrated by Figs. 21c,f and 41c–e makes a nonzero contribution that causes the coherent backscattering effect. These diagrams are called maximally crossed or cyclical [375] because they can be drawn in such a way that all connectors cross at one point.

The inclusion of the cyclical diagrams makes the computation of the total Poynting–Stokes tensor much more involved [25] and limits the range of problems that can be solved analytically [305,318]. A fully analytical solution has so far been derived only for a semi-infinite layer composed of nonabsorbing Rayleigh scatterers [376]. In general, no closed-form analytical equation similar to the RTE has been derived for the computation of the coherent component of the total Poynting–Stokes tensor. As a consequence, this cyclical component is often computed using the direct Monte Carlo summation of the cyclical diagrams [339,377].

### *11.4. Validation of the analytical theory of radiative transfer and weak localization*

By virtue of being a direct corollary of the MMEs, the radiative transfer–weak localization (RT–WL) theory contains no adjustable parameters inherent in semi-empirical and phenomenological approaches. As such, it can unambiguously be compared with computer solutions of the MMEs and results of controlled laboratory experiments. This is very important, since some of the assumptions made earlier in this section are semi-qualitative and thus need to be clarified quantitatively. Indeed, the RT–WL theory is fundamentally based on the asymptotic requirements $\rho \ll 1$ and $N \gg 1$, where, as before, $\rho$ is the particle packing density. The first inequality ensures that particle positions inside the volume are random, mutually independent, and statistically uniform. Furthermore, in the case of particles with sizes comparable to and greater than the wavelength, it ensures that each particle is located in the far zones of all the other particles constituting the DRM. The second inequality allows one to ignore non-self-avoiding diagrams in the far-field Neumann expansion (237). The combination of these inequalities implies that the overall size parameter of the DRM must be much greater than unity. While these inequalities are essential in the derivation of the RT–WL theory from the MMEs, the derivation in and of itself does not yield specific numerical estimates of the largest allowable packing density and the smallest allowable number of particles. Such estimates can only be derived from quantitative comparisons of the approximate RT–WL results with numerical data obtained by either directly solving the MMEs or performing a detailed optical experiment on a fully characterized DRM.

An important consequence of the analytical derivation summarized in Subsection 11.1 is that although Eq. (274) has been obtained while assuming that the observation point $\mathbf{r}$ is located in the near zone of the DRM, the entire volume starts to behave like a single far-field scatterer as $r \to \infty$ [34]. This makes it possible to validate the RT theory (alone and in combination with the WL theory) using far-field STMM computations and the Monte Carlo computer simulator described in [339,377]. Some results of this validation [308] (see also [295,310,311]) are shown in Fig. 42. The computations were carried out for two models of a spherical Type-1 DRM with a size parameter of $k_1R = 40$. All constituent spherical particles are identical and have the refrac-tive index $m = 1.31$ and the size parameter $k_1 r = 2$. The number of particles and the corresponding packing density are $N = 250$, $\rho = 3.125\%$ in the left-hand column and $N = 500$, $\rho = 6.25\%$ in the right-hand column, where, as before, $\rho = Nr^3 / R^3$. Fig. 42 displays separately the RT-only and the combined RT–WL results.

The comparison in Fig. 42 leads to the following instructive conclusions:

- Although the DRMs studied contain modest numbers of particles, the packing density deviates from zero significantly, and the size parameter of the DRMs is moderate, the quantitative agreement between the exact STMM and approximate RT–WL results is quite evident. Overall, this comparison confirms the mesoscopic rooting of the RT–WL theory in the MMEs traced in Subsections 8.2 and 11.1.

- A scattering-angle range where the STMM and RT results disagree fundamentally is that corresponding to forward-scattering directions. This result can be explained by different ways of treating the effect of forward-scattering interference. Indeed, in the framework of the expressly near-field RT theory, this effect is incorporated mathematically in the computation of the exponential attenuation rate inside the particulate volume [25,34], whereas, in the framework of far-field STMM computations it causes the strong and narrow interference peak discussed in Subsection 8.2.[11]

- Outside a relatively narrow range of backscattering angles, the RT-only and the full RT–WL results are very close. This is consistent with the physical interpretation of weak localization as a backscattering interference phenomenon.

- The RT-only results do not reproduce the backscattering peaks in the phase function $\widetilde{F}_{11}(\Theta)$ and in the linear and circular polarization ratios defined by Eqs. (244) and (245), as well as the asymmetric minimum in the ratio $-\widetilde{F}_{12}(\Theta)/\widetilde{F}_{11}(\Theta)$ at backscattering angles exhibited by the STMM results. The inclusion of the cyclical diagrams serves to reproduce these backscattering features very closely, which is again indicative of their weak-localization nature.

- The residual differences between the RT–WL and the STMM results at side- and backscattering angles decrease with decreasing packing density, which is an expected result. However, they persist even at packing densities as small as ~3%, possibly in part because the reduction of $\rho$ is achieved by decreasing $N$ and thus violating more significantly the requisite inequality $N \gg 1$.

In another recent paper [378], the RT theory was tested against the results of a controlled laboratory experiment. Specifically, the results of high-accuracy measurements of the Stokes reflection matrix for fully-characterized suspensions of submicrometer-sized latex particles in water were compared with the results of a numerically exact computer solution of the RTE based on the so-called adding method [57,66]. The quantitative performance of the RTE was monitored by increasing the volume packing density of the latex particles from 2% to 10%. The results of this study indicate that the RTE can be applied safely to DRMs with packing densities up to ~2%. Radiative-transfer results for packing densities of the order of 5% should be taken with great caution, while the polarized bidirectional reflectivity of suspensions with larger packing densities cannot be accurately predicted. These conclusions are generally consistent with the results of [308].

[11] Note that in [312] the RT exponential extinction law was reproduced by near-field STMM computations.

## 12. Fixed particulate media

We have seen in Section 8 that the diffuse speckle-free regime naturally develops from the speckle regime upon averaging optical observables over changing particle positions. Furthermore, we have seen in Subsection 11.1 that it is the averaging over random particle coordinates that effectively leads to the RTE (264). In the case of a fixed particulate medium such as a powder surface, a sheet of paper, or a layer of paint, the speckle regime caused by scattering of a collimated monochromatic or quasi-monochromatic beam persists and is easily detectable with a WCR having a sufficiently fine angular resolution [44,45,379].

As discussed in Section 5.3 of [238], the speckle regime can get suppressed in many practical applications owing to the use of polychromatic sources of light, uncollimated illumination, and/or detectors of light integrating over a wide solid angle of scattering directions. In particular, it is the non-detection of speckle in such applications that has led to the widespread belief that the RT theory or its *ad hoc* modifications can be used to describe electromagnetic scattering by fixed particulate layers.

It is important to recognize however that the RTE has never been derived directly from the MMEs by averaging optical observables over a range of incidence and/or scattering directions or over a finite spectral range instead of averaging over varying particle positions. Therefore, the only way to verify quantitatively whether a fixed particulate medium can behave optically as a DRM is to analyze the results of direct computer solutions of the MMEs.

Fig. 43 shows the results of STMM computations of the dimensionless scattering matrix for two objects. The first one is a fixed configuration of $N = 200$ particles with $k_1 r = 4$ and $m = 1.32$ quasi-randomly and quasi-uniformly populating an imaginary $k_1 R = 50$ spherical volume and yielding a 10% packing density. In this case the scattering matrix is defined according to

$$\widetilde{\mathbf{F}}(\Theta) = \frac{4\pi}{C_{\text{sca}}} \mathbf{Z}(\theta^{\text{sca}} = \Theta, \varphi^{\text{sca}} = 0; \theta^{\text{inc}} = 0, \varphi^{\text{inc}} = 0), \tag{276}$$

where $C_{\text{sca}}$ is given by

$$C_{\text{sca}} = \int_{4\pi} d\hat{\mathbf{n}}^{\text{sca}} Z_{11}(\theta^{\text{sca}}, \varphi^{\text{sca}}; \theta^{\text{inc}} = 0, \varphi^{\text{inc}} = 0), \tag{277}$$

and is depicted by thin black curves. The second object is a DRM modeled by assuming a uniform orientation distribution of the first object. In this case the scattering matrix is defined by Eq. (228) and is depicted by thick gray curves. Consistent with the discussion in Subsection 8.1, the sharp large-amplitude oscillations exhibited by the thin black curves represent speckles typical of a fixed multi-particle configuration, whereas the smooth thick gray curves are representative of a DRM.

Fig. 44 is analogous to Fig. 43, but now the scattering matrix (276) of the fixed multi-particle configuration computed at a single wavelength is replaced by the average over a range of wavelengths:

$$\widetilde{\mathbf{F}}(\Theta) = \frac{4\pi}{\langle C_{\text{sca}} \rangle_{\Delta\lambda}} \langle \mathbf{Z}(\theta^{\text{sca}} = \Theta, \varphi^{\text{sca}} = 0; \theta^{\text{inc}} = 0, \varphi^{\text{inc}} = 0) \rangle_{\Delta\lambda}. \tag{278}$$

It is assumed that (i) the incident field is a polychromatic parallel beam with quasi-monochromatic components, and (ii) all quasi-monochromatic components have the same Stokes parameters (see Section 13.6 of [34]). The spectral range $\Delta\lambda$ is equal to $1/10$ of the central wavelength, which implies that $k_1 R$ ranges from 47.5 to 52.5 and $k_1 r$ ranges from 3.8 to 4.2. The numerical integration over $\Delta\lambda$ was performed using a Gaussian quadrature formula with 100 division points.

The comparison of Figs. 43 and 44 is quite revealing. First of all, it confirms that averaging the scattering matrix over a finite spectral range serves as an extremely efficient suppressor of speckles generated by a fixed multi-particle configuration. Second of all, it demonstrates that as a consequence of spectral averaging the scattering properties of the fixed multi-particle configuration become very similar to those of the "morphologically-equivalent" DRM. This result [380] is qualitatively consistent with Eq. (241) which shows that the phase difference between two multi-particle sequences can be randomized not only by changing particle positions but also by varying the wavelength.

Although these conclusions should be viewed as preliminary and should be corroborated by further research, they appear to support the conventional belief that depending on specific measurement settings (e.g., polychromatic illumination), the notion of a DRM can often be broadened to encompass fixed particulate media.

## 13. Concluding remarks

The overall objective of this Report was to outline the first-principles physical framework of the discipline of electromagnetic scattering by a (slowly varying) DRM, formulate the resulting physical and mathematical problems in maximally rigorous terms, and discuss the most robust and well-characterized ways of addressing these problems. We intentionally focused on numerically exact computer solutions of the MMEs as the most reliable way of obtaining profound physical insights unavailable with phenomenological and heuristic theories. We also discussed how the first-order-scattering approximation, the radiative transfer theory, and the theory of weak localization of electromagnetic waves can be derived directly from the Maxwell equations for very specific and well-defined kinds of particulate medium.[12] The main advantage of these numerical and analytical corollaries of the MMEs is that they obviate the need to introduce fictitious tunable parameters and poorly defined notions such as dependent, independent, and incoherent scattering; elementary volume elements; incoherent light rays; photons as particles of light or blobs of electromagnetic energy without phases; and collective scattering effects. The whole evolution of physics has been in the direction of replacing phenomenological and heuristic approaches with first-principles ones. A major objective of this Report was to summarize recent contributions to this process.

Consistent with this objective, we stayed away from discussing phenomenological and semi-empirical theories of light scattering by particulate media other than the effective-medium approach. As explained in [91,92], facile theories such as those described in [84–90] are inherently flawed in that they are typically devoid of primordial physical parameters of a DRM involved in the solution of the Maxwell equations and instead feature numerous artificial adjustable parameters. As a consequence, they represent little more than a conglomerate of contrived yet enticingly simple formulas intended to provide a back-of-an-envelope solution of the profoundly complex scattering problem. The use of freely tunable *ad hoc* parameters makes these models a flexible interpolation tool capable of fitting almost any data. The price one has to pay for this interpolation capability is that the best-fit model usually has little (if any) physical meaning.

[12] A more detailed discussion of the phenomenological origin of the radiative transfer theory and its recent transformation into a legitimate branch of statistical electromagnetics can be found in [100].

We hope that this Report serves as a convincing demonstration of substantial recent progress that has made the discipline of electromagnetic scattering by a DRM a full-fledged branch of physical optics (or, to use a catchy term, of "disordered photonics" [381]). In particular, direct computer solutions of the MMEs discussed in Section 8 and straightforward analytical derivations reviewed in Section 11 have fully confirmed the purportedly mesoscopic origin of the theory of radiative transfer and weak localization [353,354,382–384]. Indeed, they clearly demonstrate how the "macroscopic" regime of this theory emerges from the "microscopic" particle-level regime of Maxwell's electromagnetics upon averaging over random realizations of a large sparse multi-particle group. Both theoretical and experimental studies discussed in Subsection 11.4 (see also [385]) have revealed the inevitable breakdown of the RT–WL regime when the particle packing density exceeds a certain threshold. This emphasizes the importance of efficient computer solvers of the MMEs which have no intrinsic limitations on packing density and, in combination with the ever growing power of computer clusters, should eventually facilitate the solution of outstanding problems of unprecedented complexity.

Still the range of scattering problems that can be solved exactly remains limited. As a consequence, approximate theories of light scattering by DRMs will still be practiced in the foreseeable future to handle full-scale "real-life" problems. It is therefore imperative to use advanced computer solvers of the MMEs as well as controlled laboratory experiments to quantify numerical errors of approximate approaches and understand their origin. Although further research is still needed to better validate such popular modeling tools as the first-order-scattering approximation, the radiative transfer equation, the theory of weak localization, and the effective-medium approach, significant progress has already been achieved, as discussed in Subsections 8.2 and 11.4 and Sections 9 and 10.

The main subject of this Report can be characterized as the direct scattering problem, i.e., the calculation of electromagnetic scattering by a known, well-defined system. We have not discussed how to solve the inverse scattering problem, i.e., determine the physical characteristics of a particulate object by analyzing its measured scattering and absorption properties. The vastness of this applied discipline obviously necessitates a separate review. Similarly left out are the countless specific applications of electromagnetic scattering by particulate media in various branches of science and technology.

In this Report we focused on isolated particulate media. Yet there is an urgent need to consider even more complex problems involving different combinations of volume and/or surface scattering. Good examples would be a densely packed particulate layer bounded from below by a plane interface and a layer of continuous fluctuating medium hosting randomly positioned discrete particles and bounded by random rough interfaces. It is safe to say that the first-principles treatment of such problems is still at an early stage of development [305,318,386–389].

Finally we note that an essential assumption made at the very outset of this Report is that the infinite host medium surrounding the particles is nonabsorbing. A preliminary first-principles analysis of the general case of an absorbing host can be found in [179,390,391].

## Acknowledgments

We thank an anonymous reviewer for providing helpful comments on the initial version of this Report. We appreciate numerous insightful discussions with Yuri Barabanenkov, Matthew Berg, Anatoli Borovoi, Oleg Bugaenko, Petr Chýlek, Helmut Domke, Joop Hovenier, Vsevolod Ivanov, Michael Kahnert, George Kattawar, Nikolai Khlebtsov, Nikolai Kiselev, Thomas Kulp, Andrew Lacis, K.-N. Liou, Pavel Litvinov, James Lock, Daniel Mackowski, M. Pinar Mengüç, Karri Muinonen, Antti Penttilä, Thomas Reichardt, Vera Rosenbush, Yuri Shkuratov, Viktor Tishkovets, Cornelis van der Mee, Gorden Videen, and Edgard Yanovitskij. We thank Alexandra Ivanova for providing Fig. 14a, Antti Penttilä for providing Figs. 19a and 30a, and Guanglin Tang and Jianing Zhang for help with FDTDM and PSTDM computations displayed in Figs. 17 and 18. M.I.M., L.L., B.C., and P.Y. acknowledge continued support from the NASA Remote Sensing Theory Project managed by Lucia Tsaoussi, the NASA Radiation Sciences Program managed by Hal Maring, and the NASA ACE Project. Some of the numerical results reported in this paper were obtained with the "Discover" supercomputer at the NASA Center for Climate Simulation. J.M.D. was supported by the National Academy of Sciences of Ukraine under the Main Astronomical Observatory GRAPE/GPU/GRID Computing Cluster Project. M.A.Yu. was supported by the Russian Science Foundation grant No. 14-15-00155.

## Appendix A. Benchmark STMM results

Owing to the equiprobable orientation distribution, the dimensionless scattering matrix (232) has the following symmetric structure [1,34]:

$$\tilde{\mathbf{F}}(\Theta) = \begin{bmatrix} \tilde{F}_{11}(\Theta) & \tilde{F}_{21}(\Theta) & \tilde{F}_{13}(\Theta) & \tilde{F}_{14}(\Theta) \\ \tilde{F}_{21}(\Theta) & \tilde{F}_{22}(\Theta) & \tilde{F}_{23}(\Theta) & \tilde{F}_{24}(\Theta) \\ -\tilde{F}_{13}(\Theta) & -\tilde{F}_{23}(\Theta) & \tilde{F}_{33}(\Theta) & \tilde{F}_{34}(\Theta) \\ \tilde{F}_{14}(\Theta) & \tilde{F}_{24}(\Theta) & -\tilde{F}_{34}(\Theta) & \tilde{F}_{44}(\Theta) \end{bmatrix}. \tag{A.1}$$

Table A.1 is a tabulation of the 10 independent elements of the scattering matrix computed with the STMM program [244] for the randomly oriented compound object shown in Fig. 14a and specified in Section 7. Note that this table well exemplifies Eq. (246). In Table A.2, we also tabulate the coefficients appearing in the expansions of the numerically most significant scattering matrix elements in Wigner $d$-functions $d^s_{mn}(\Theta)$ or, equivalently, in generalized spherical functions $P^n_{mm'}(\cos\Theta) = \mathrm{i}^{m-m'} d^n_{mm'}(\Theta)$ [20,25,34,66,392,393]:

$$\tilde{F}_{11}(\Theta) = \sum_{n=0}^{n_{\max}} \alpha_1^n P_{00}^n(\cos\Theta) = \sum_{n=0}^{n_{\max}} \alpha_1^n d_{00}^n(\Theta), \tag{A.2}$$

$$\tilde{F}_{22}(\Theta) + \tilde{F}_{33}(\Theta) = \sum_{n=0}^{n_{\max}} (\alpha_2^n + \alpha_3^n) P_{22}^n(\cos\Theta) = \sum_{n=0}^{n_{\max}} (\alpha_2^n + \alpha_3^n) d_{22}^n(\Theta), \tag{A.3}$$

$$\tilde{F}_{22}(\Theta) - \tilde{F}_{33}(\Theta) = \sum_{n=0}^{n_{\max}} (\alpha_2^n - \alpha_3^n) P_{2,-2}^n(\cos\Theta) = \sum_{n=0}^{n_{\max}} (\alpha_2^n - \alpha_3^n) d_{2,-2}^n(\Theta), \tag{A.4}$$

$$\tilde{F}_{44}(\Theta) = \sum_{n=0}^{n_{\max}} \alpha_4^n P_{00}^n(\cos\Theta) = \sum_{n=0}^{n_{\max}} \alpha_4^n d_{00}^n(\Theta), \tag{A.5}$$

$$\widetilde{F}_{21}(\Theta) = \sum_{n=0}^{n_{\max}} \beta_1^n P_{02}^n(\cos\Theta) = -\sum_{n=0}^{n_{\max}} \beta_1^n d_{02}^n(\Theta), \tag{A.6}$$

$$\widetilde{F}_{34}(\Theta) = \sum_{n=0}^{n_{\max}} \beta_2^n P_{02}^n(\cos\Theta) = -\sum_{n=0}^{n_{\max}} \beta_2^n d_{02}^n(\Theta). \tag{A.7}$$

Note that

$$\langle\cos\Theta\rangle = \frac{1}{3}\alpha_1^1. \tag{A.8}$$

The number of nonzero terms in the expansions (A.2)–(A.7) is, strictly speaking, infinite. In practice, however, a finite upper summation limit $n_{\max}$ is chosen such that the corresponding truncated sums differ from the respective scattering matrix elements within the requisite numerical accuracy on the entire interval $\Theta \in [0, \pi]$ of scattering angles. All numerical accuracy parameters in the STMM program were increasingly tightened until the numbers in Tables A.1 and A.2 converged to within plus/minus a few units in the last decimals given.

## Appendix B. List of acronyms

| | |
|---|---|
| DDA | discrete-dipole approximation |
| DRM | discrete random medium |
| EMA | effective-medium approximation |
| EMR | effective-medium rule |
| EOA | effective-object approximation |
| FDTDM | finite-difference time-domain method |
| FEs | Foldy equations |
| II-TMM | invariant-imbedding *T*-matrix method |
| MMEs | macroscopic Maxwell equations |
| PSTDM | pseudo-spectral time-domain method |
| QED | quantum electrodynamics |
| RT | radiative transfer |
| RTE | radiative transfer equation |
| STMM | superposition *T*-matrix method |
| TMM | *T*-matrix method |
| VIE | volume integral equation |
| WCR | well-collimated radiometer |
| WL | weak localization |

**Table 1**
Cartesian coordinates of 10 spherical inclusions.

| $n$ | $x_n$ | $y_n$ | $z_n$ |
|---|---|---|---|
| 1 | –0.215062 | 6.479603 | 0.616824 |
| 2 | –3.756010 | –2.754431 | 5.549602 |
| 3 | 0.650697 | 0.515307 | –0.017826 |
| 4 | –2.364920 | 0.805033 | –4.337800 |
| 5 | 5.008396 | –4.096047 | 1.241592 |
| 6 | –4.504373 | –4.444519 | –0.820851 |
| 7 | 7.303638 | 0.831435 | –0.230329 |
| 8 | 4.725006 | 5.314130 | 1.544096 |
| 9 | –0.219794 | –7.116933 | –0.691158 |
| 10 | 3.957806 | 1.528642 | –4.454259 |

**Table 2**
Integral optical characteristics.

| Technique | $Q_{\text{ext}}$ | $Q_{\text{sca}}$ | $Q_{\text{abs}}$ | $\varpi$ | $\langle\cos\Theta\rangle$ |
|---|---|---|---|---|---|
| STMM[1] | 1.9104 | 1.8839 | 0.02652 | 0.98612 | 0.62184 |
| II-TMM | 1.9105 | 1.8839 | 0.02657 | 0.98609 | 0.62186 |
| DDA (extrapolated) | 1.9129 | 1.8865 | 0.02634 | 0.98623 | 0.62199 |
| FDTDM ($\lambda/50$) | 1.9034 | 1.8769 | 0.02650 | 0.98608 | – |
| PSTDM ($\lambda/50$) | 1.9129 | 1.8864 | 0.02652 | 0.98614 | – |

[1]The STMM results are expected to be accurate to plus/minus one unit in the last digits shown.

**Table 3**
Extinction cross-section and asymmetry parameter ratios.

| $m_{\text{host}}$ | $\dfrac{C_{\text{ext}}^{\text{STMM}}}{C_{\text{ext}}^{\text{LM}}}$ | $\dfrac{C_{\text{ext}}^{\text{STMM}}}{C_{\text{ext}}^{\text{MG}}}$ | $\dfrac{\langle\cos\Theta\rangle^{\text{STMM}}}{\langle\cos\Theta\rangle^{\text{LM}}}$ | $\dfrac{\langle\cos\Theta\rangle^{\text{STMM}}}{\langle\cos\Theta\rangle^{\text{MG}}}$ |
|---|---|---|---|---|
| 1.4 | 1.0755 | 1.0235 | 1.0348 | 1.0121 |
| 1.6 | 0.9061 | 0.9903 | 0.9035 | 0.9926 |

**Table 4**
Comparison of STMM and first-order-scattering approximation results.

| $N$ | $\rho$ | $\frac{\langle\langle \overline{W}^{\text{ext}}(t)\rangle\rangle^{\text{FOSA}}}{\langle\langle \overline{W}^{\text{ext}}(t)\rangle\rangle^{\text{STMM}}}$ |
|---|---|---|
| 1 | 0.0005 | 1 |
| 2 | 0.0010 | 1.0001 |
| 5 | 0.0026 | 1.0059 |
| 20 | 0.0102 | 1.0336 |
| 50 | 0.0256 | 1.4451 |
| 100 | 0.0512 | 1.6952 |
| 200 | 0.1024 | 2.4506 |
| 400 | 0.2048 | 4.7936 |
| 600 | 0.3072 | 7.0776 |

**Table A.1**
Elements of the normalized scattering matrix calculated with the STMM computer program.

| $\Theta$ (deg) | $\tilde{F}_{11}$ | $\tilde{F}_{21}$ | $\tilde{F}_{13}$ | $\tilde{F}_{14}$ | $\tilde{F}_{22}$ | $\tilde{F}_{23}$ | $\tilde{F}_{24}$ | $\tilde{F}_{33}$ | $\tilde{F}_{34}$ | $\tilde{F}_{44}$ |
|---|---|---|---|---|---|---|---|---|---|---|
| 0 | 49.71733 | 0.00000 | 0.00000 | –0.00014 | 49.70581 | 0.00322 | 0.00000 | 49.70581 | 0.00000 | 49.69506 |
| 10 | 12.42281 | –0.38719 | 0.00002 | –0.00000 | 12.41798 | 0.00155 | –0.00002 | 12.31709 | 1.48848 | 12.31480 |
| 20 | 3.82084 | –0.65502 | 0.00002 | 0.00037 | 3.81143 | 0.00028 | 0.00001 | 3.66047 | –0.77437 | 3.65477 |
| 30 | 2.46558 | 0.32314 | –0.00004 | –0.00033 | 2.45656 | –0.00048 | –0.00001 | 2.41573 | 0.25041 | 2.41197 |
| 40 | 2.96876 | –0.02893 | –0.00002 | 0.00015 | 2.95631 | –0.00032 | 0.00002 | 2.89553 | –0.57369 | 2.88959 |
| 50 | 1.00871 | 0.22695 | –0.00003 | –0.00019 | 0.99662 | 0.00000 | –0.00016 | 0.89948 | 0.28267 | 0.89468 |
| 60 | 1.04317 | –0.09226 | 0.00002 | –0.00008 | 1.03129 | 0.00054 | –0.00002 | 0.98187 | –0.26947 | 0.97817 |
| 70 | 0.40002 | 0.19768 | 0.00001 | –0.00012 | 0.39089 | 0.00027 | 0.00003 | 0.29308 | 0.03631 | 0.29193 |
| 80 | 0.53819 | –0.09156 | 0.00001 | 0.00004 | 0.52842 | 0.00018 | –0.00003 | 0.49913 | –0.11133 | 0.49892 |
| 90 | 0.24845 | 0.13979 | –0.00013 | 0.00018 | 0.24058 | –0.00015 | 0.00013 | 0.15711 | –0.03949 | 0.15781 |
| 100 | 0.16012 | –0.03083 | 0.00000 | 0.00010 | 0.14943 | –0.00039 | –0.00008 | 0.12980 | –0.01340 | 0.13269 |
| 110 | 0.19072 | 0.09828 | –0.00024 | –0.00043 | 0.17921 | 0.00032 | –0.00024 | 0.10879 | –0.04054 | 0.11226 |
| 120 | 0.10523 | 0.01397 | –0.00007 | 0.00000 | 0.08762 | 0.00003 | –0.00023 | 0.02627 | –0.06017 | 0.03571 |
| 130 | 0.17251 | 0.04556 | 0.00023 | –0.00022 | 0.14978 | 0.00007 | 0.00001 | 0.10869 | –0.00973 | 0.12077 |
| 140 | 0.32879 | 0.00751 | 0.00023 | –0.00009 | 0.29739 | –0.00018 | –0.00050 | 0.04533 | –0.27640 | 0.06741 |
| 150 | 0.14684 | 0.02090 | 0.00016 | 0.00075 | 0.09466 | 0.00010 | 0.00066 | 0.03747 | –0.04800 | 0.07547 |
| 160 | 0.76055 | 0.01776 | –0.00121 | 0.00079 | 0.70377 | –0.00073 | 0.00117 | 0.39784 | –0.52574 | 0.44526 |
| 170 | 0.40934 | 0.22432 | 0.00055 | –0.00063 | 0.37337 | –0.00021 | 0.00003 | 0.24795 | –0.04579 | 0.27032 |
| 180 | 0.38852 | 0.00000 | 0.00000 | –0.00581 | 0.24250 | 0.00000 | 0.00000 | –0.24250 | 0.00000 | –0.09647 |

**Table A.2**
Expansion coefficients calculated with the STMM computer program.

| $n$ | $\alpha_1^n$ | $\alpha_2^n$ | $\alpha_3^n$ | $\alpha_4^n$ | $\beta_1^n$ | $\beta_2^n$ |
|---|---|---|---|---|---|---|
| 0 | 1.00000 | 0.00000 | 0.00000 | 0.90630 | 0.00000 | 0.00000 |
| 1 | 1.86553 | 0.00000 | 0.00000 | 1.89662 | 0.00000 | 0.00000 |
| 2 | 2.29661 | 3.61509 | 3.43653 | 2.21923 | –0.09381 | 0.13309 |
| 3 | 1.86558 | 2.44923 | 2.45022 | 1.91094 | –0.02746 | 0.00610 |
| 4 | 1.94148 | 2.23947 | 2.13993 | 1.92623 | –0.04024 | 0.18993 |
| 5 | 1.82032 | 1.92938 | 1.86526 | 1.79708 | 0.00112 | –0.05645 |
| 6 | 1.96397 | 2.10089 | 2.06841 | 1.96043 | 0.05146 | 0.21263 |
| 7 | 2.17566 | 2.09321 | 2.04990 | 2.15775 | 0.12290 | –0.02730 |
| 8 | 2.51398 | 2.61833 | 2.58468 | 2.50124 | 0.13264 | 0.12198 |
| 9 | 2.79420 | 2.70146 | 2.69891 | 2.82529 | 0.23022 | –0.11972 |
| 10 | 2.80854 | 3.03972 | 2.98293 | 2.79412 | 0.21611 | –0.04008 |
| 11 | 2.83598 | 2.74434 | 2.74189 | 2.88242 | 0.29442 | –0.20541 |
| 12 | 2.56817 | 2.85635 | 2.82965 | 2.61048 | 0.21075 | –0.07325 |
| 13 | 2.73807 | 2.57889 | 2.48558 | 2.70521 | 0.26578 | –0.20495 |
| 14 | 2.55909 | 2.82990 | 2.84695 | 2.63705 | 0.10429 | 0.00955 |
| 15 | 3.05048 | 2.82332 | 2.72507 | 3.01411 | 0.13466 | –0.19354 |
| 16 | 3.04930 | 3.32459 | 3.32160 | 3.10138 | –0.02963 | 0.03529 |
| 17 | 3.36077 | 3.16079 | 3.22556 | 3.51389 | –0.27596 | –0.19771 |
| 18 | 2.99934 | 3.37988 | 3.27850 | 3.00480 | –0.46426 | –0.79836 |
| 19 | 2.08446 | 2.05418 | 2.00679 | 2.13122 | –0.04756 | –0.78261 |
| 20 | 1.23452 | 1.52855 | 1.26855 | 1.03331 | 0.17120 | –0.84357 |
| 21 | 0.03859 | 0.01828 | 0.00073 | 0.02990 | 0.22586 | –0.01528 |
| 22 | 0.10947 | 0.12174 | 0.10523 | 0.09718 | 0.03328 | –0.02163 |
| 23 | 0.03300 | 0.03695 | 0.03270 | 0.02979 | 0.01255 | –0.00690 |
| 24 | 0.00812 | 0.00913 | 0.00803 | 0.00728 | 0.00372 | –0.00150 |
| 25 | 0.00171 | 0.00192 | 0.00166 | 0.00150 | 0.00091 | –0.00026 |
| 26 | 0.00032 | 0.00036 | 0.00030 | 0.00027 | 0.00019 | –0.00004 |
| 27 | 0.00005 | 0.00006 | 0.00005 | 0.00004 | 0.00004 | 0.00000 |
| 28 | 0.00001 | 0.00001 | 0.00001 | 0.00001 | 0.00001 | 0.00000 |
| 29 | 0.00000 | 0.00000 | 0.00000 | 0.00000 | 0.00000 | 0.00000 |

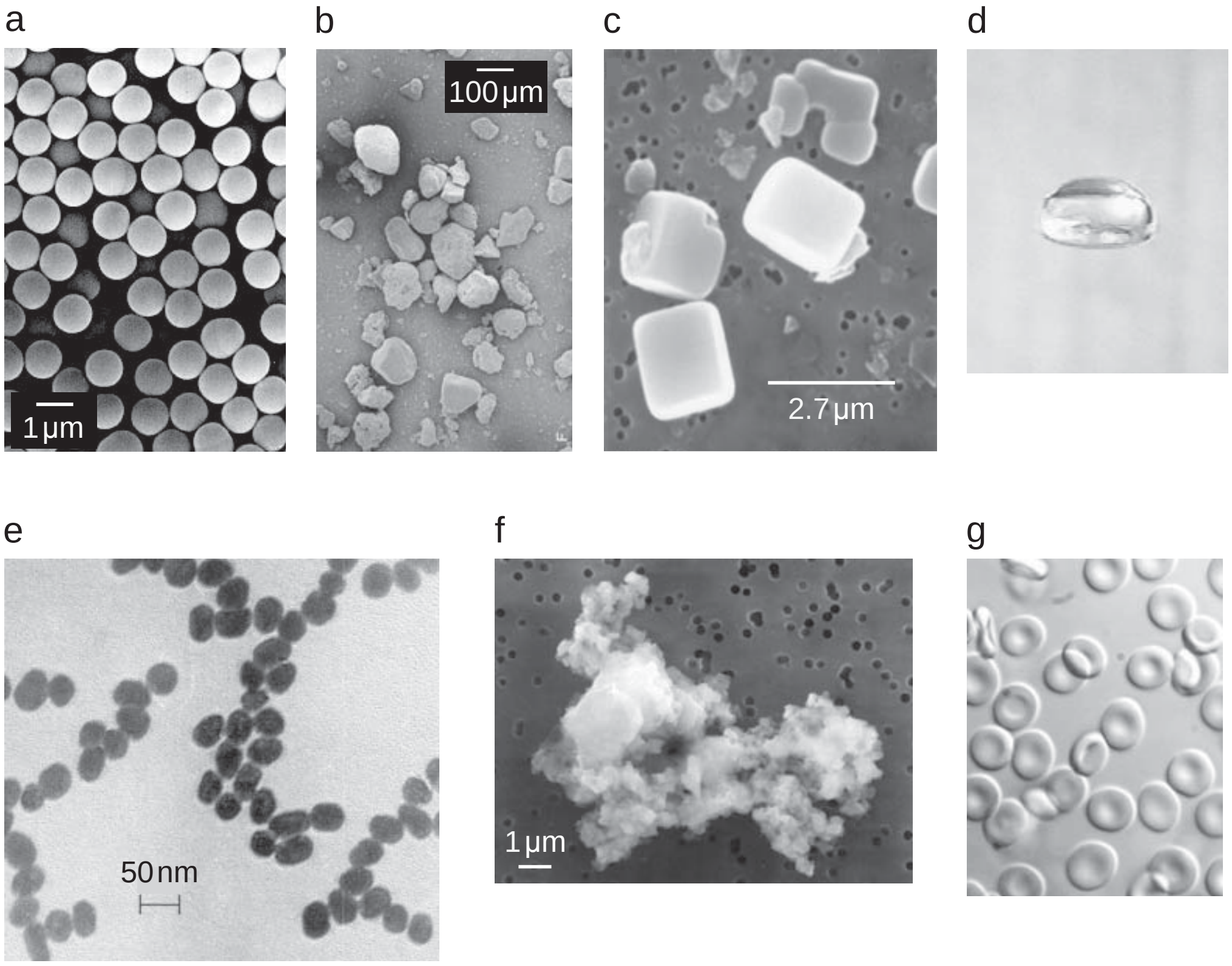

**Fig. 1.** Examples of manmade and natural small particles. (a) Commercial glass spheres (after [37]). (b) Sahara desert sand (after [38]). (c) Dry sea-salt particles (after [39]). (d) A 6-mm-diameter falling raindrop. (e) 40-nm-diameter gold particles (after [40]). (f) Interplanetary dust particle U2012C11 collected by a NASA U2 aircraft. (g) Red blood cells.

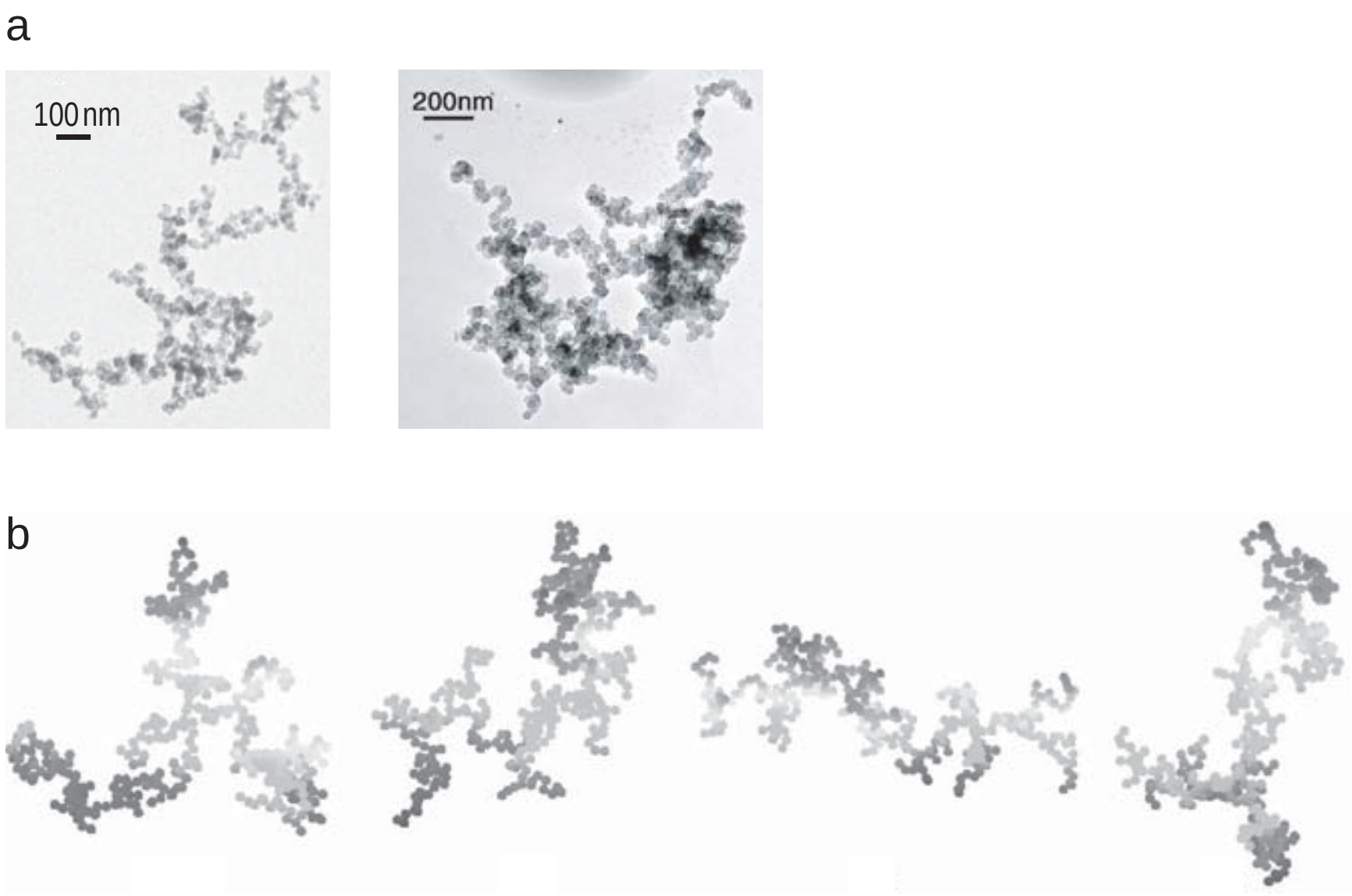

**Fig. 2.** (a) Natural and (b) modeled soot fractals (after [41–43]).

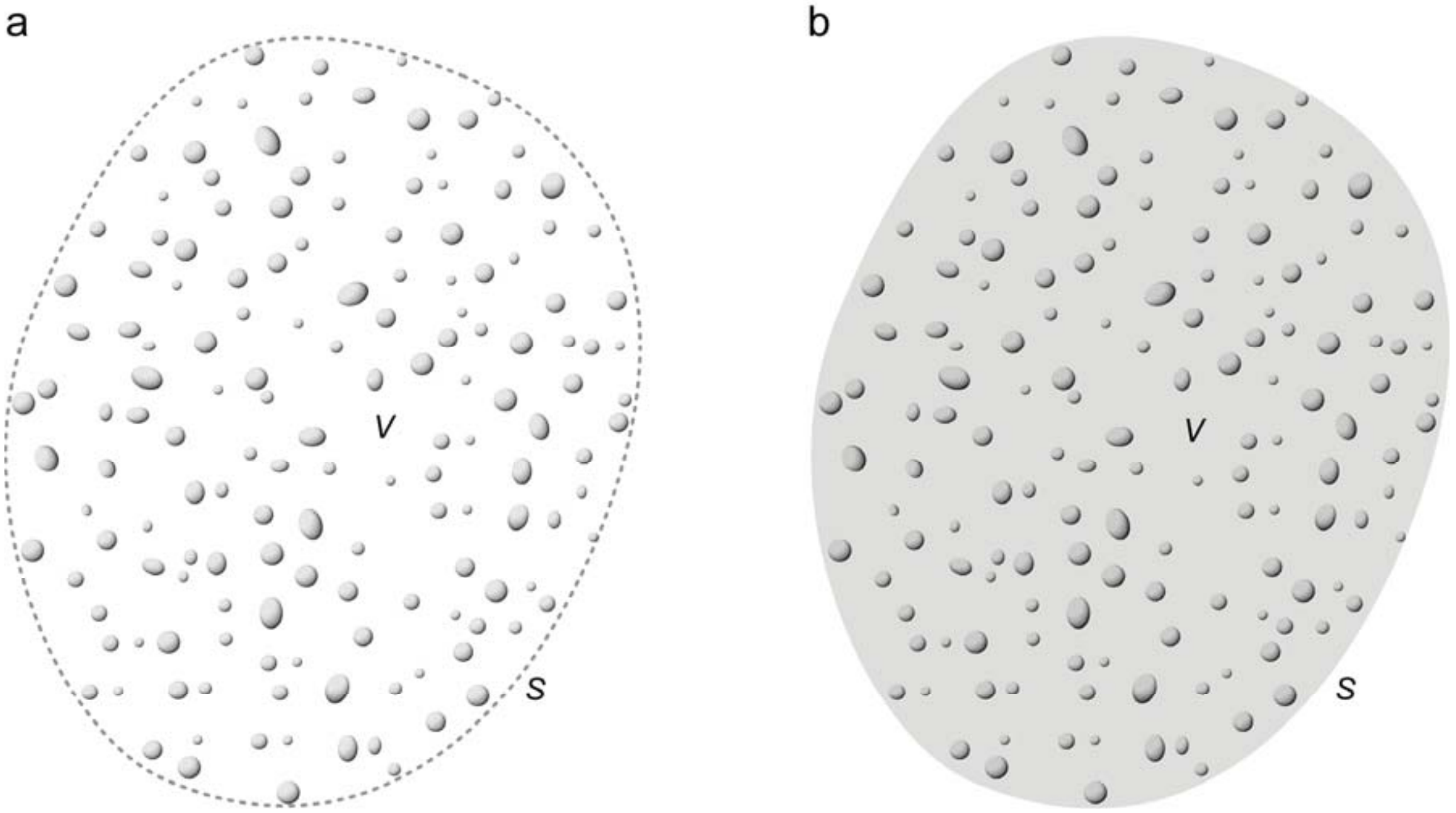

**Fig. 3.** Two types of discrete random medium. (a) Type 1: particles are randomly distributed throughout an imaginary volume $V$. (b) Type 2: particles are randomly distributed throughout a host volume $V$ having a refractive index different from that of the surrounding infinite space.

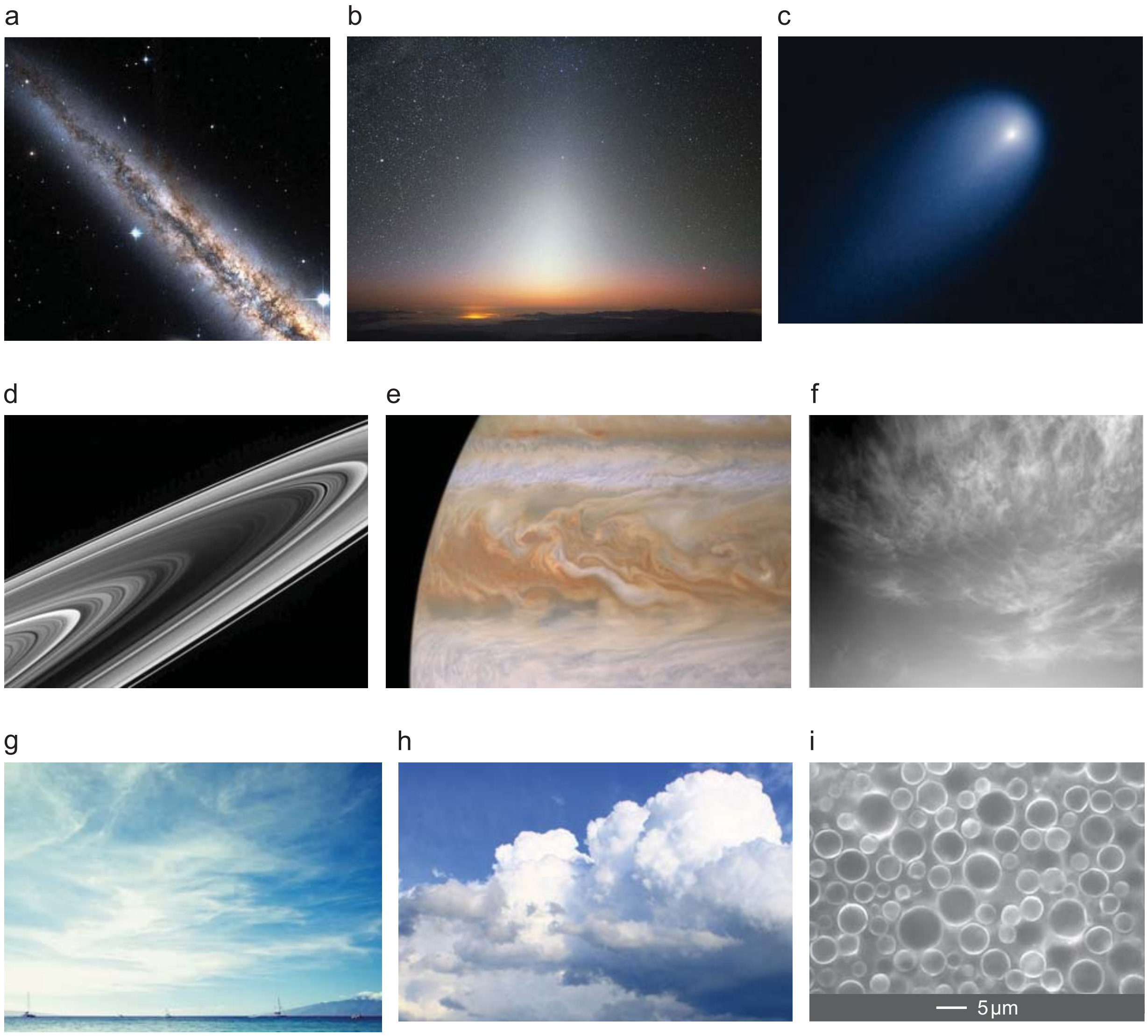

**Fig. 4.** Examples of natural discrete random media. (a) Clouds of interstellar dust, arranged in huge patches and tentacles, appears dark when they are silhoutted against the stars in the mid-plane of the galaxy NGC 891. Image taken with NASA's Hubble Space Telescope. (b) Ghostly glow caused by the scattering of sunlight by the interplanetary dust cloud. (c) The dusty atmosphere of the comet ISON photographed on 10 April 2013 with NASA's Hubble Space Telescope. (d) Particulate Saturn's rings photographed from NASA's Cassini spacecraft. (e) Jovian clouds photographed from NASA's Cassini spacecraft. (f) Thin diffuse clouds in the atmosphere of Mars photographed from NASA's Opportunity rover. Cirrus (g) and liquid-water (h) clouds in the Earth's atmosphere. (i) Raw milk.

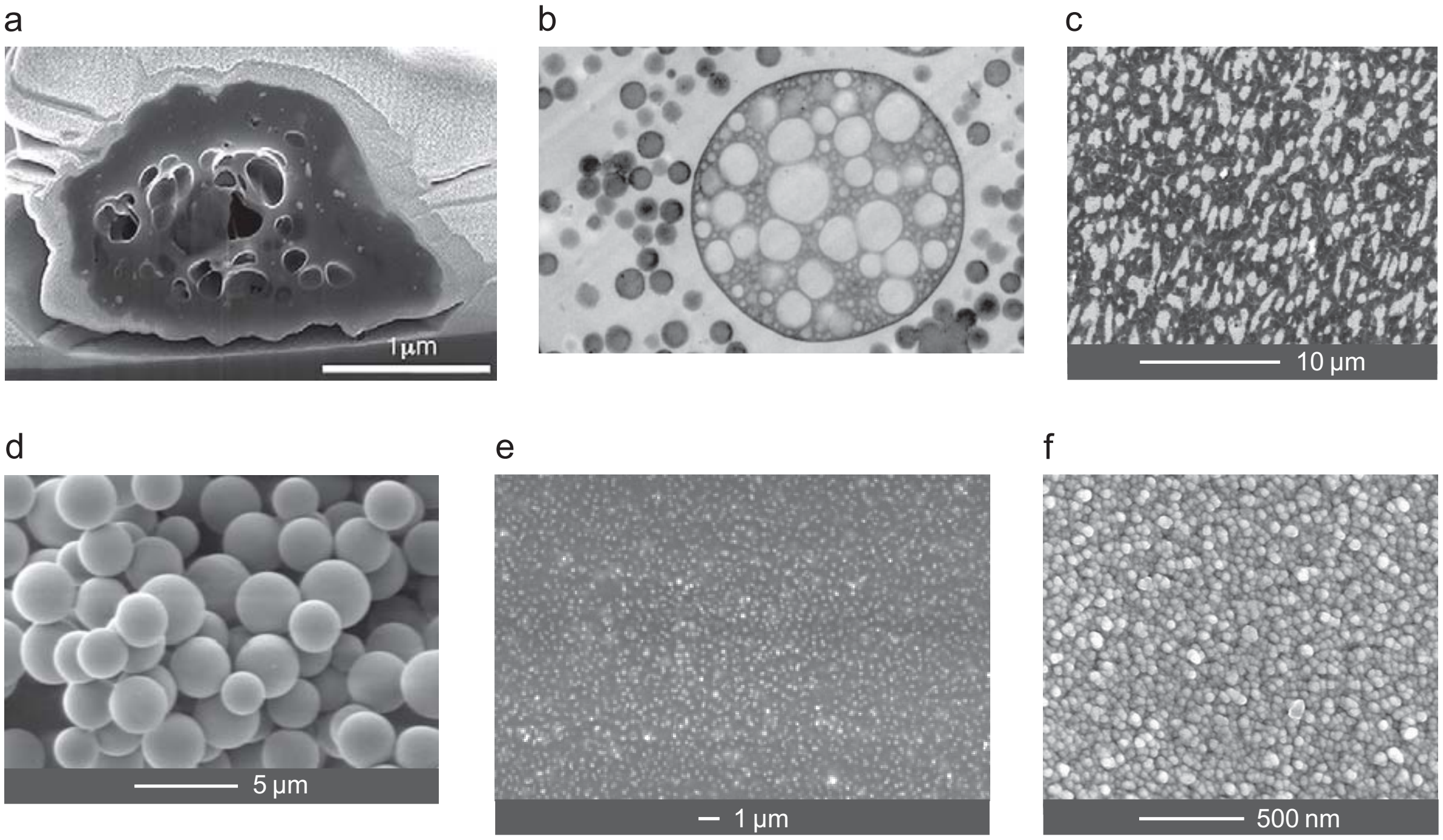

**Fig. 5.** Examples of natural and manmade discrete random media. (a) Cross-section of a ~2.2-μm highly porous natural organic-matter aerosol particle (after [51]). (b) Transmission electron micrograph of a high-impact polystyrene sample cut with an oscillating diamond knife. The large composite particle has a diameter of ~3 μm (after [52]). (c) Backscattered electron micrograph of the cross section of an olefin polymer blend polished using an oscillating diamond knife at room temperature (after [53]). (d) Particulate surface composed of glass microspheres. (e) Electron micrograph of a paint film formed by $TiO_2$ particles immersed in a binder. (f) Dense coating formed by 30-nm $Y_2O_3$ crystals.

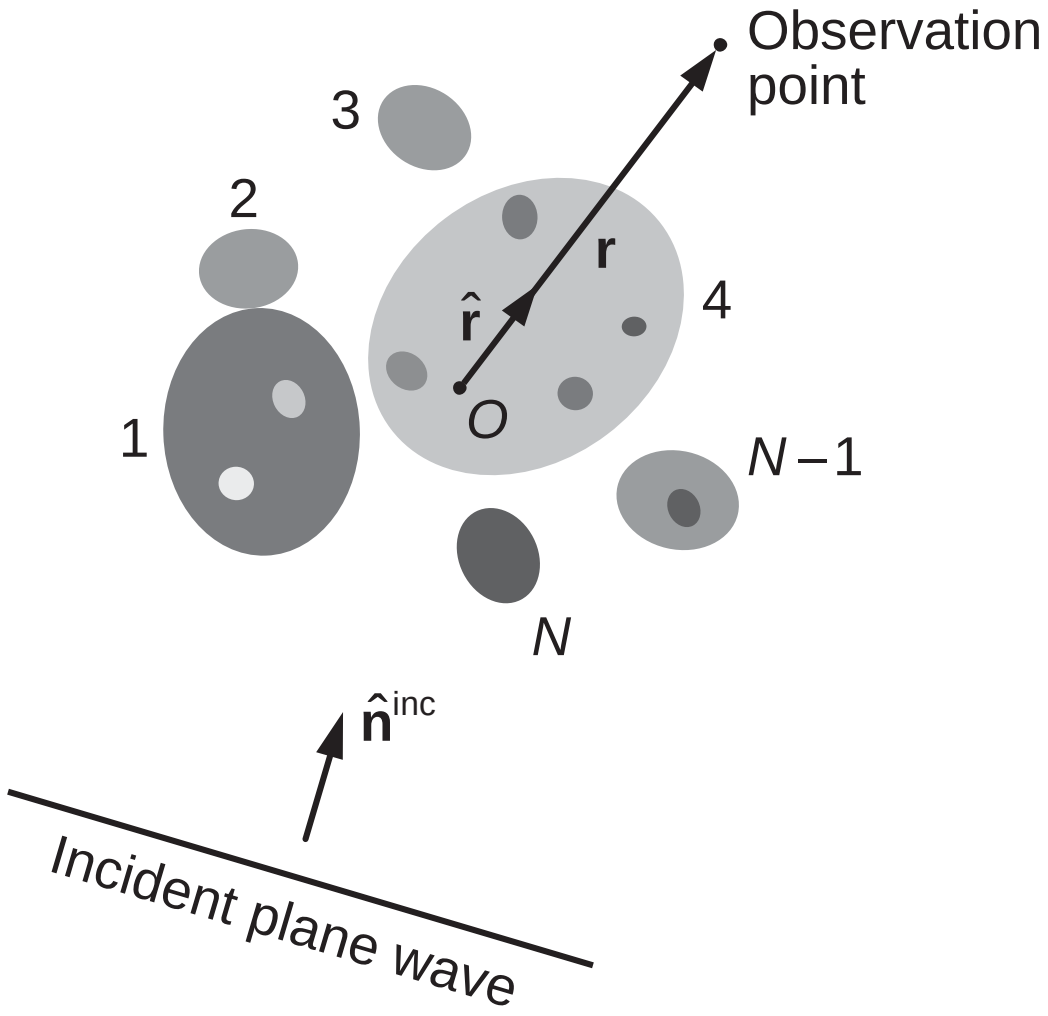

**Fig. 6.** Standard electromagnetic scattering problem. The fixed finite scattering object consists of $N$ distinct and potentially inhomogeneous components. The shaded areas collectively represent the interior region $V_{\mathrm{INT}}$, while the unshaded exterior region $V_{\mathrm{EXT}}$ is unbounded in all directions.

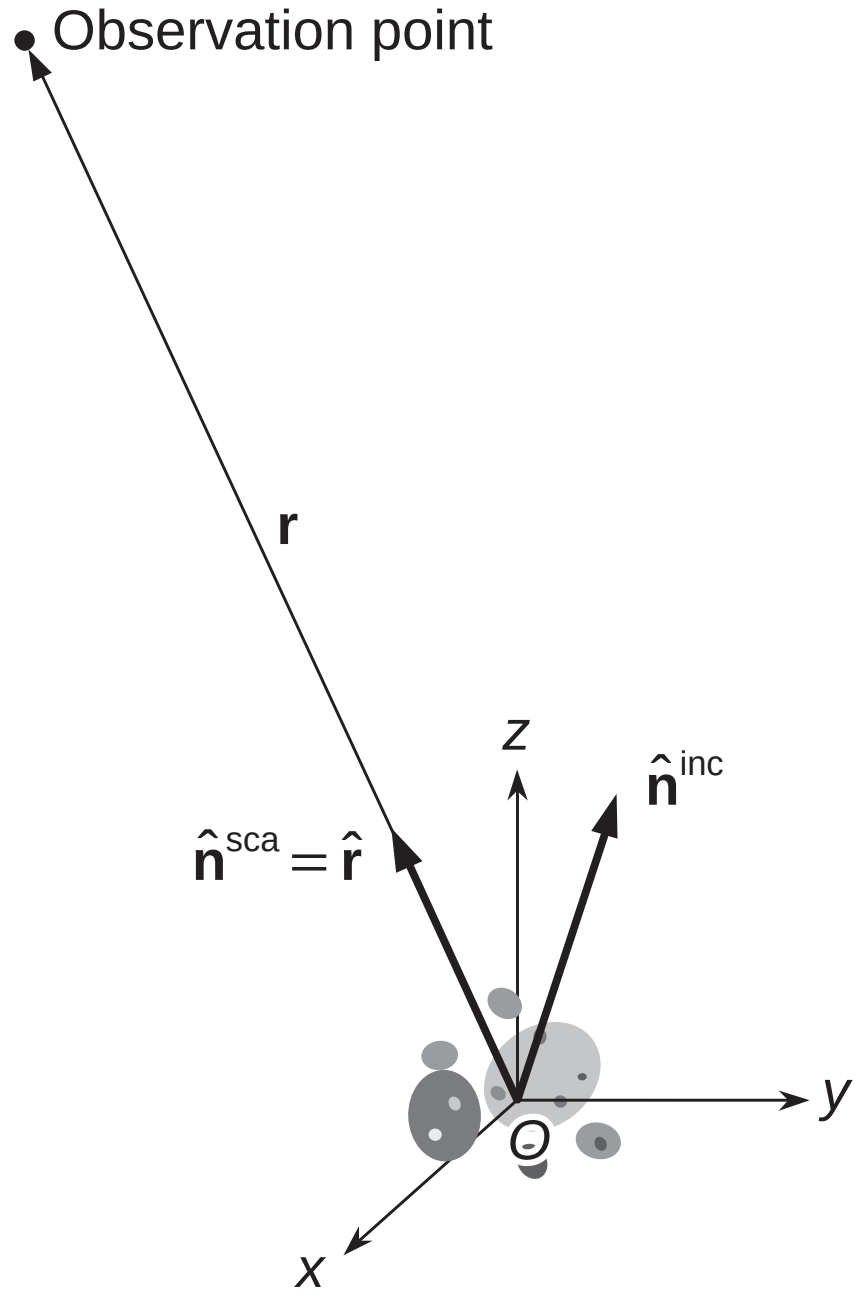

**Fig. 7.** Scattering in the far zone of the object.

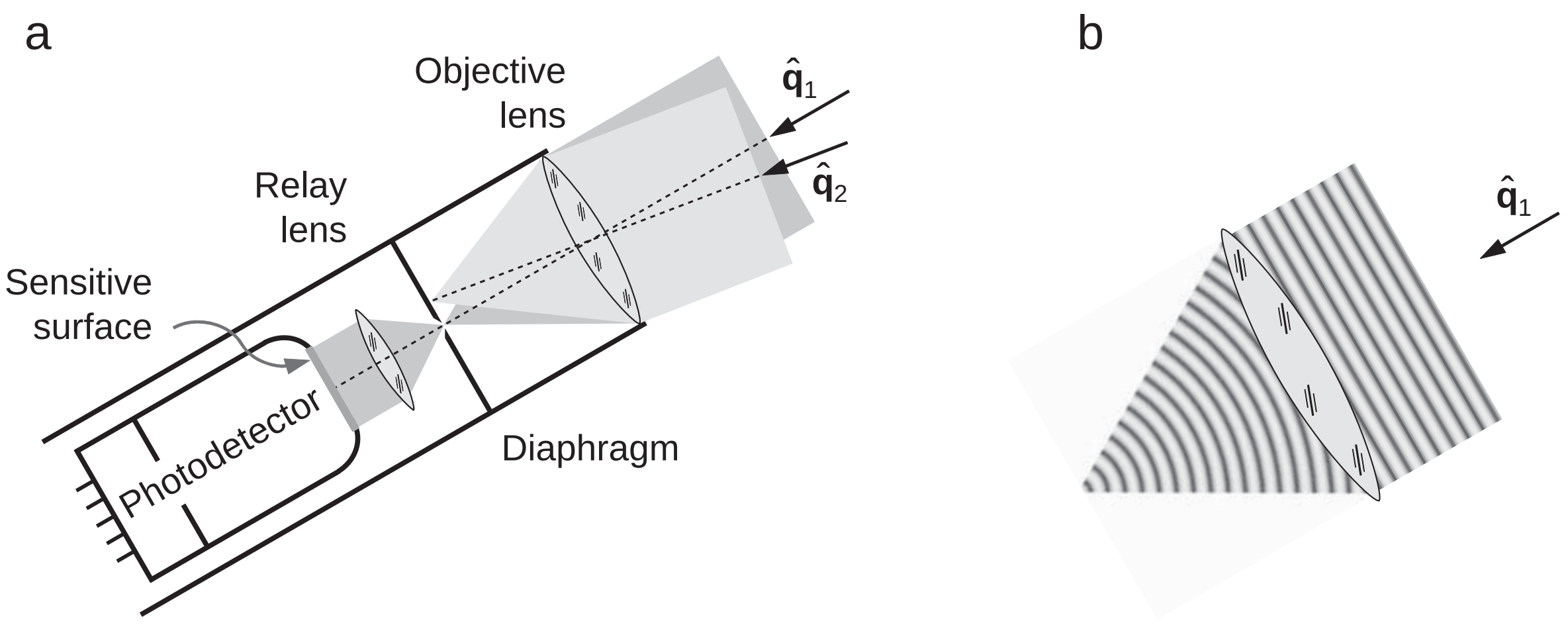

**Fig. 8.** Optical scheme of a well-collimated radiometer.

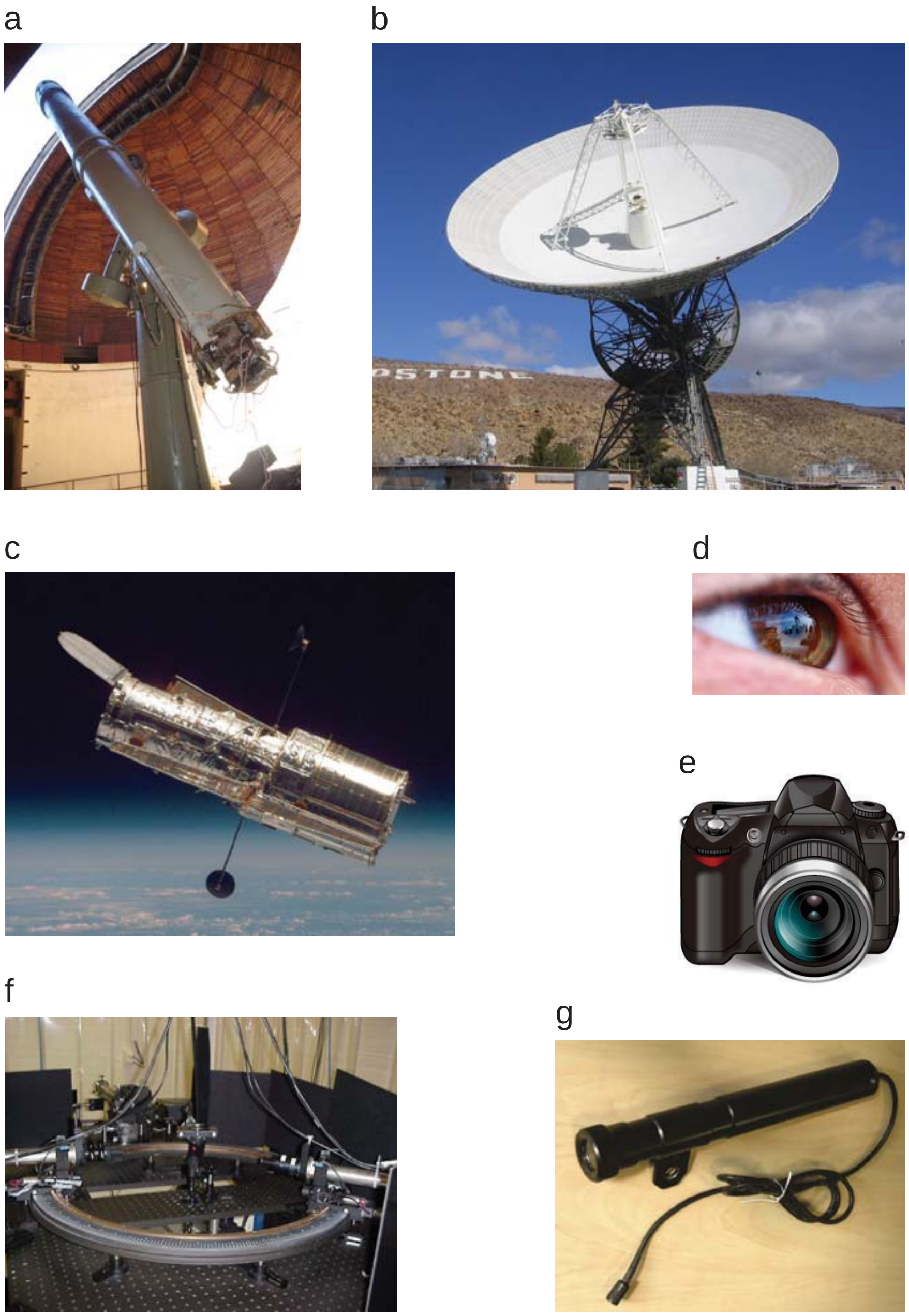

**Fig. 9.** Examples of well-collimated radiometers. (a) 26-in refractor of the Pulkovo Observatory. (b) NASA's 34-m Goldstone radio telescope. (c) NASA's Hubble Space Telescope. (d) Human eye. (e) Digital photographic camera. (f) Light scattering setup built at the University of Amsterdam (after [176]). (g) Gershun tube (after [177]).

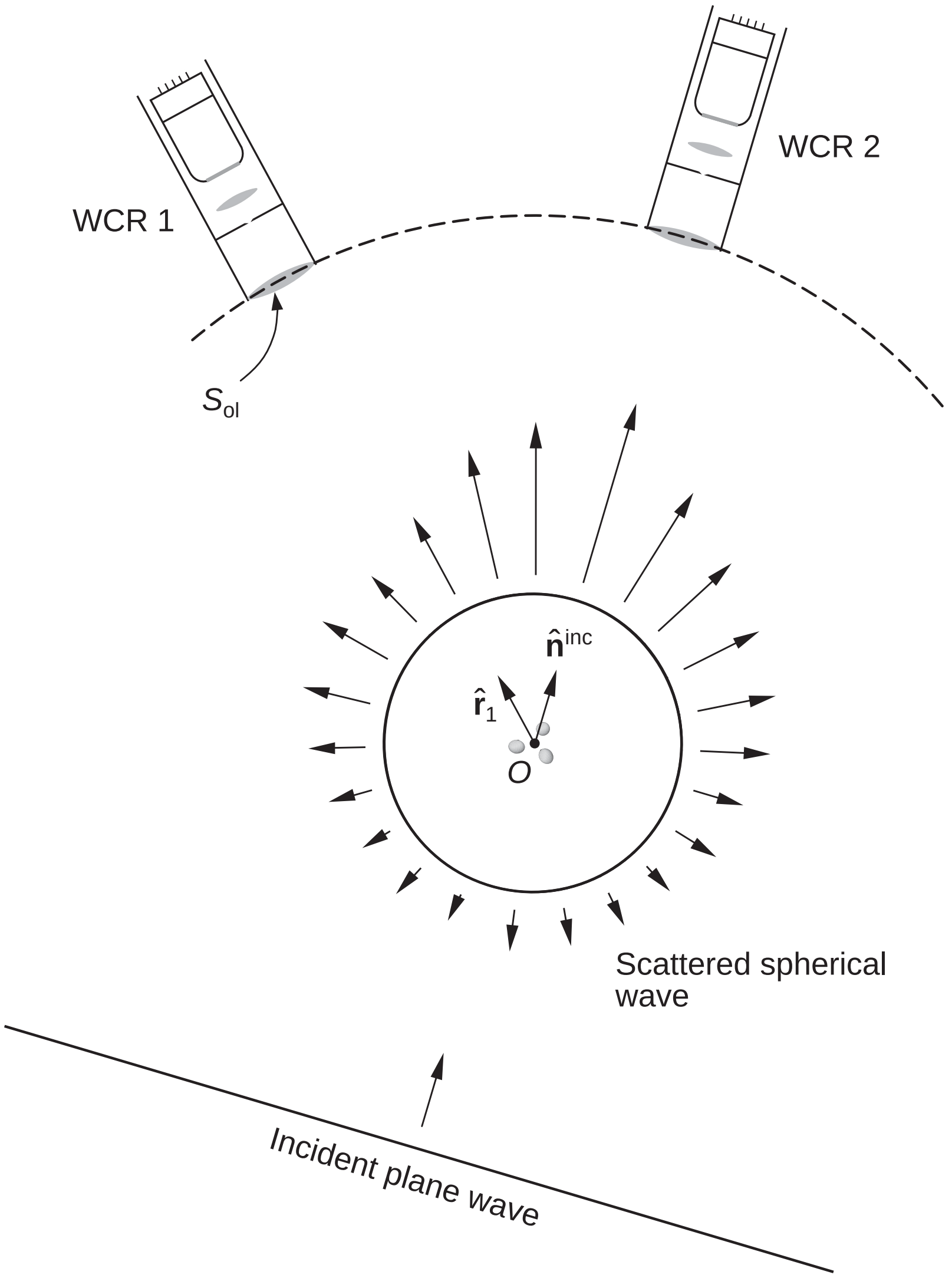

**Fig. 10.** The response of a polarization-sensitive well-collimated radiometer depends on the line of sight.

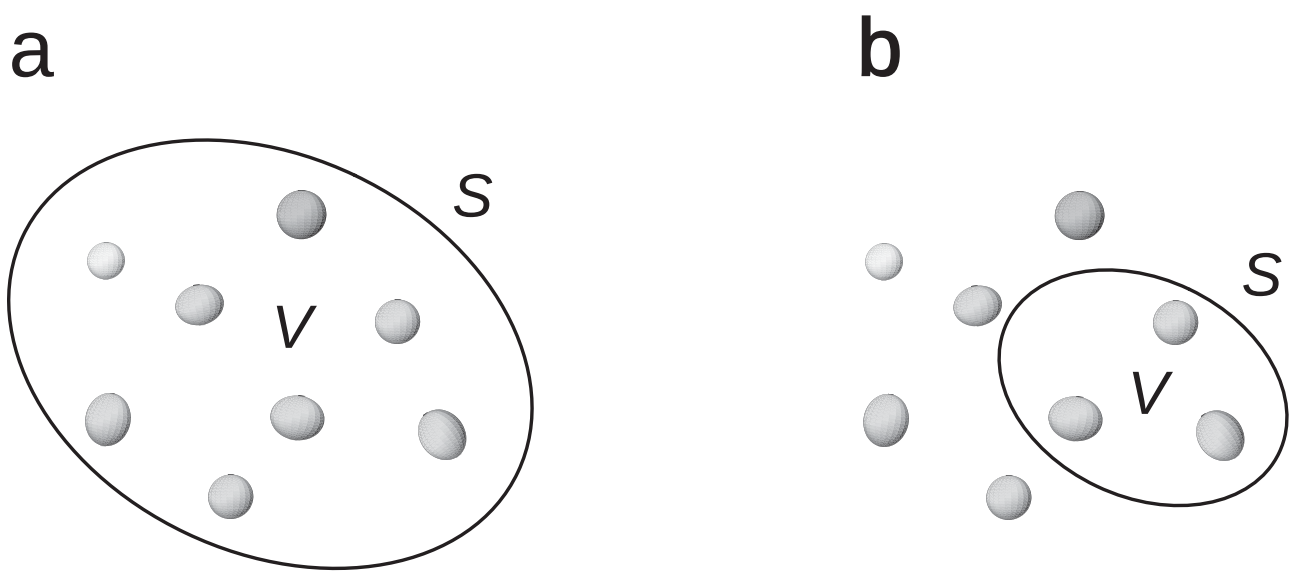

**Fig. 11.** Energy budget of a finite volume enclosing (a) the entire scattering object or (b) a part of the object.

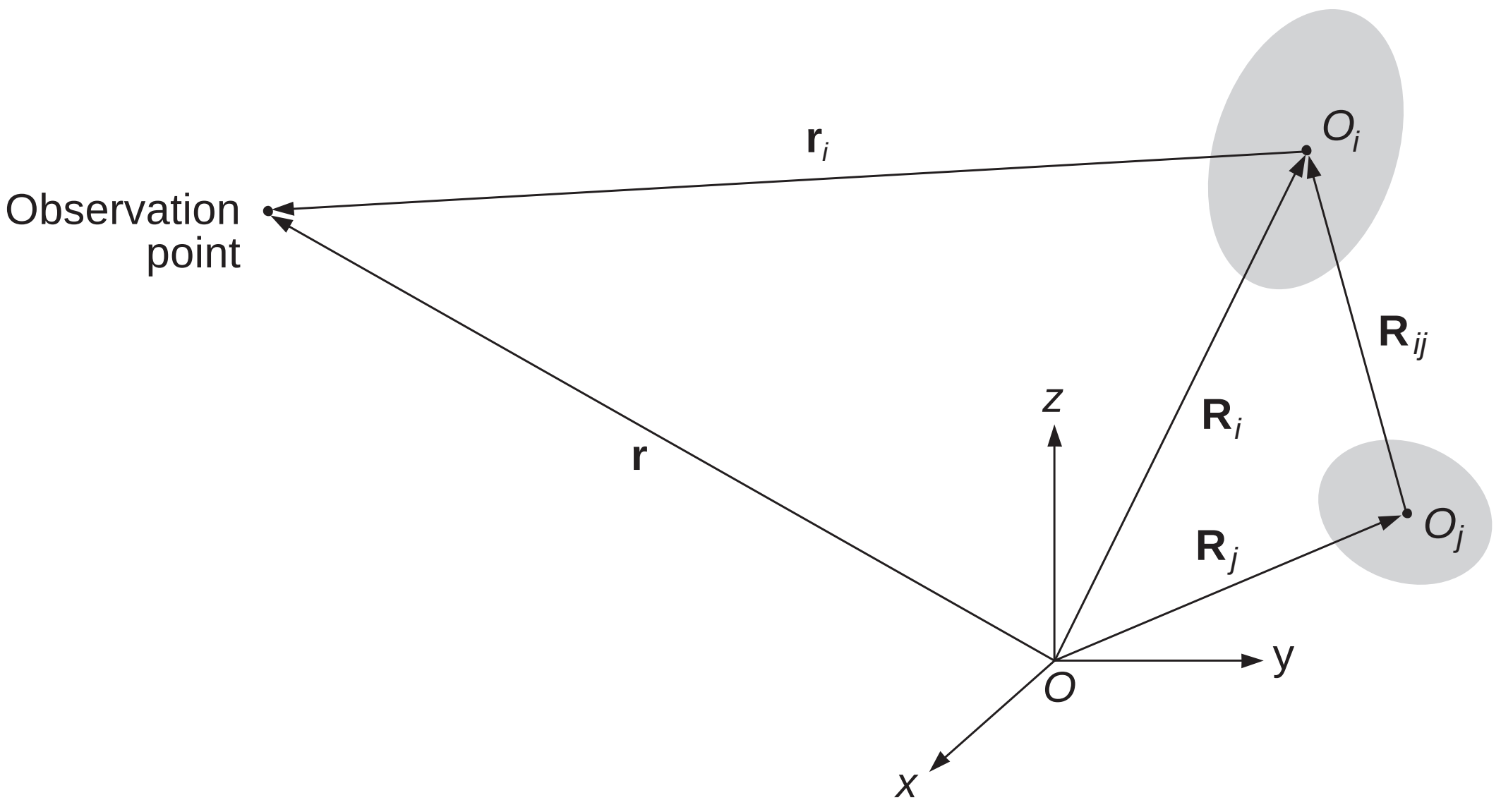

**Fig. 12.** Vector notation used in the far-field Foldy equations.

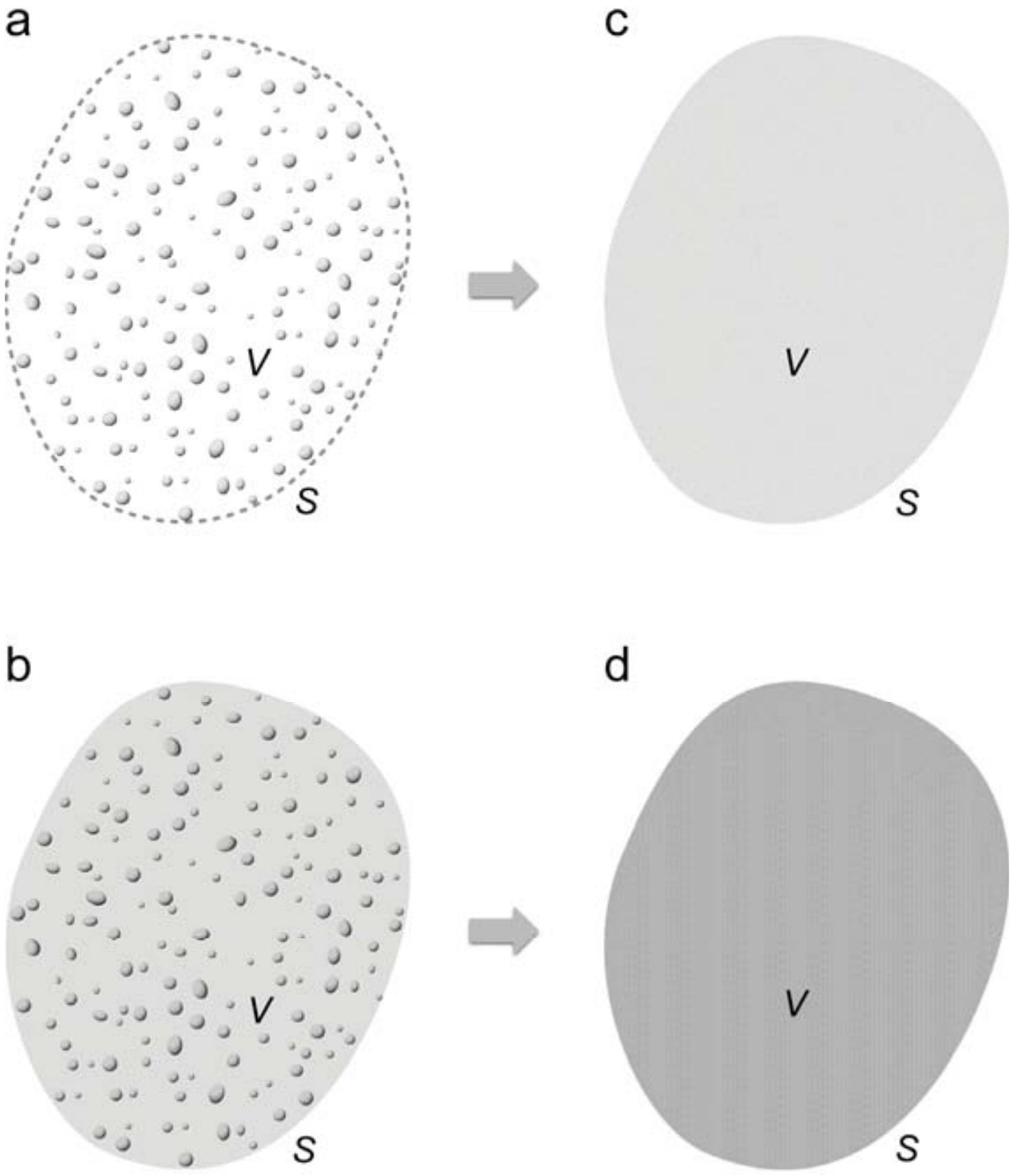

**Fig. 13.** Effective-medium methodology.

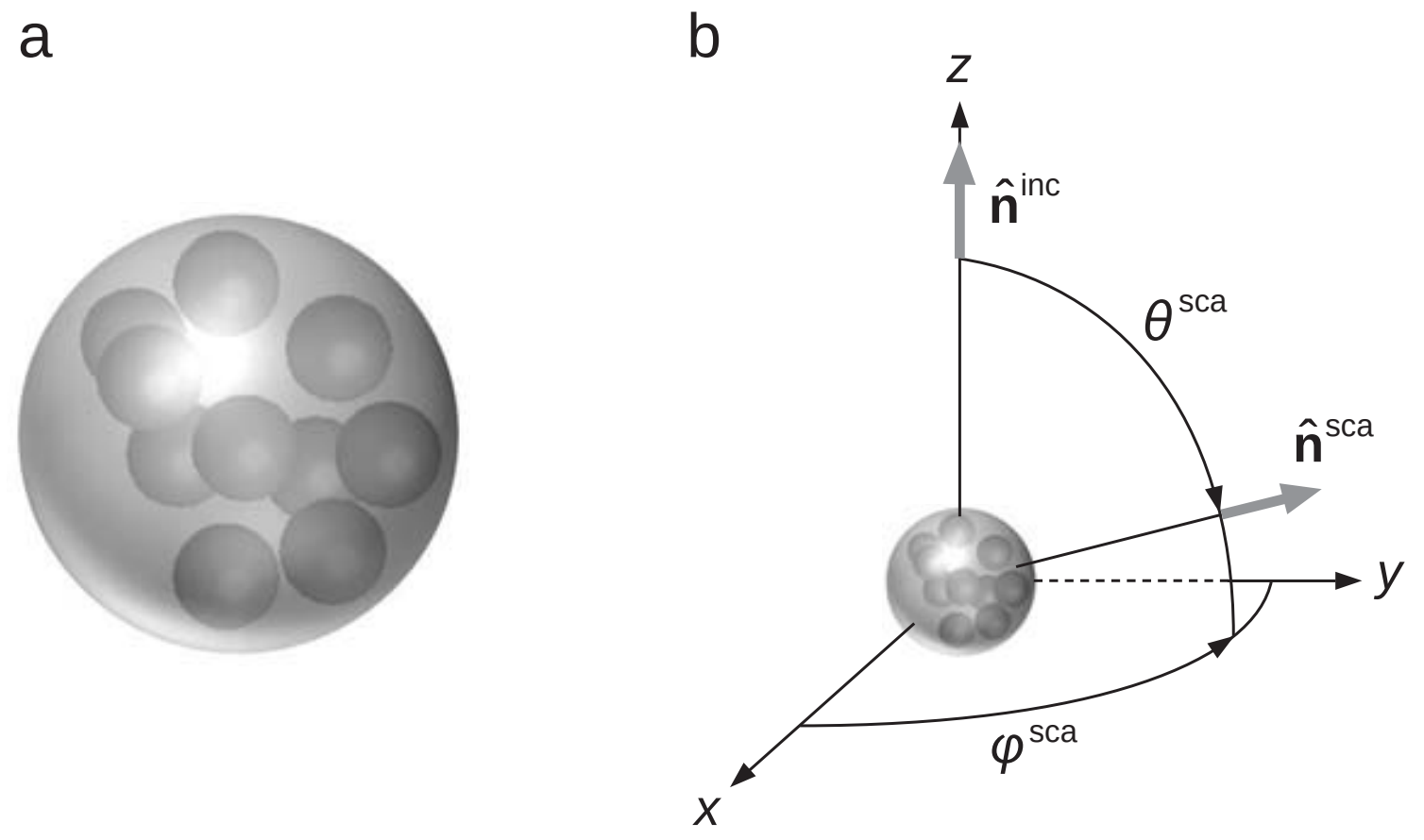

**Fig. 14.** (a) Model compound scatterer. (b) Scattering geometry.

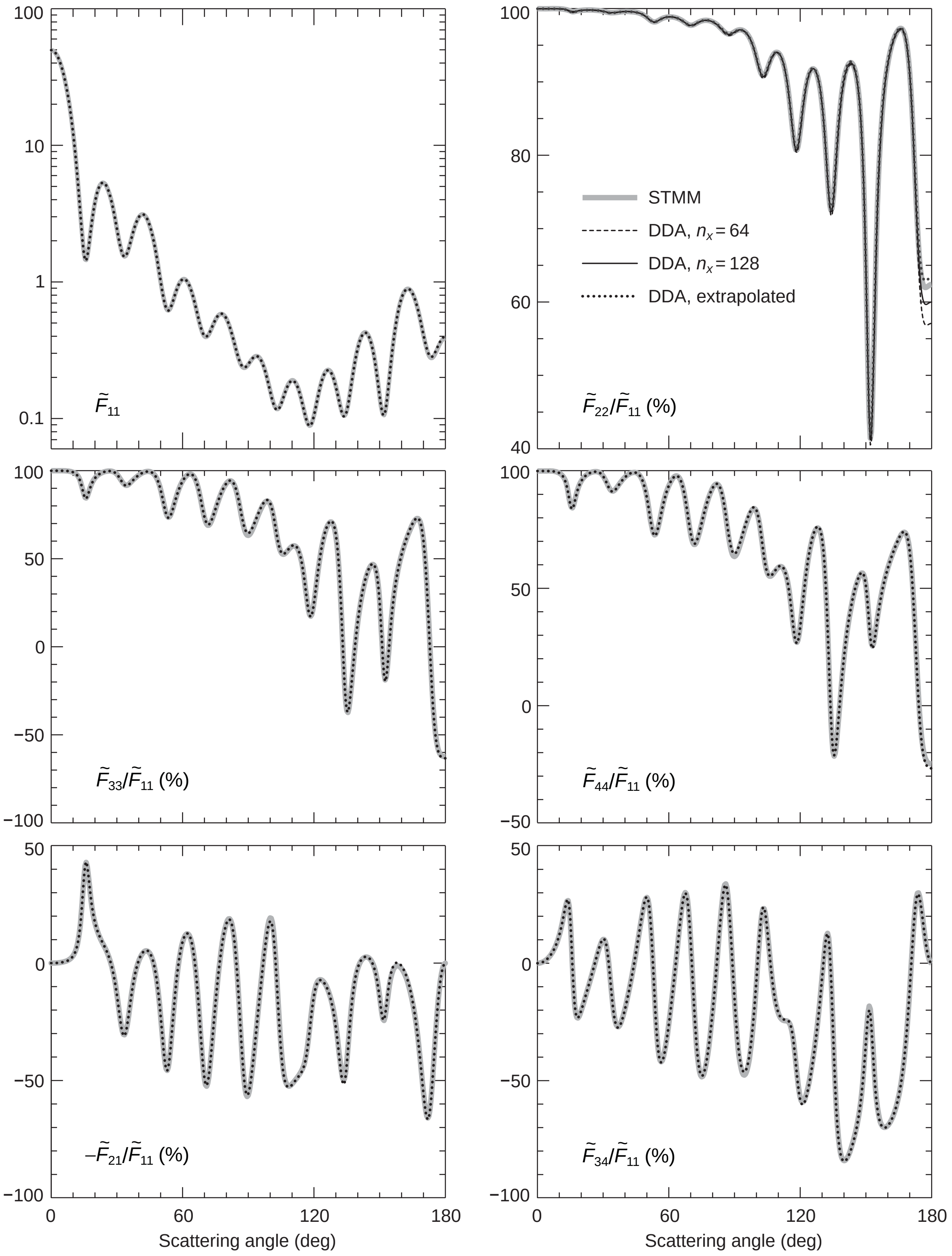

**Fig. 15.** Elements of the dimensionless scattering matrix computed using the STMM and DDA for the randomly oriented composite object shown in Fig. 14a. The $n_x = 64$ and $n_x = 128$ DDA results are shown only in the $\tilde{F}_{22}/\tilde{F}_{11}$ panel.

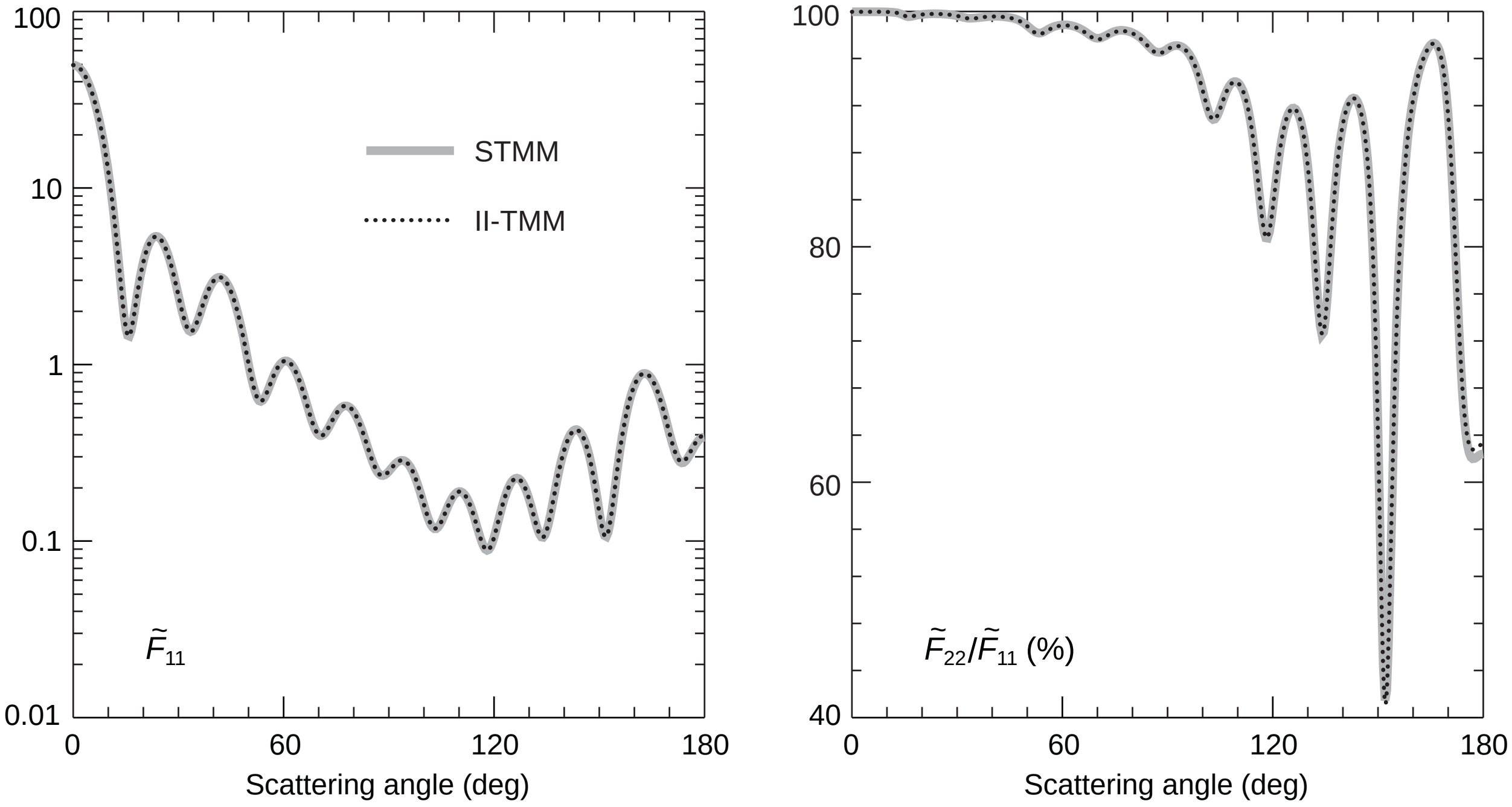

**Fig. 16.** As in Fig. 15, but for STMM vs. II-TMM results.

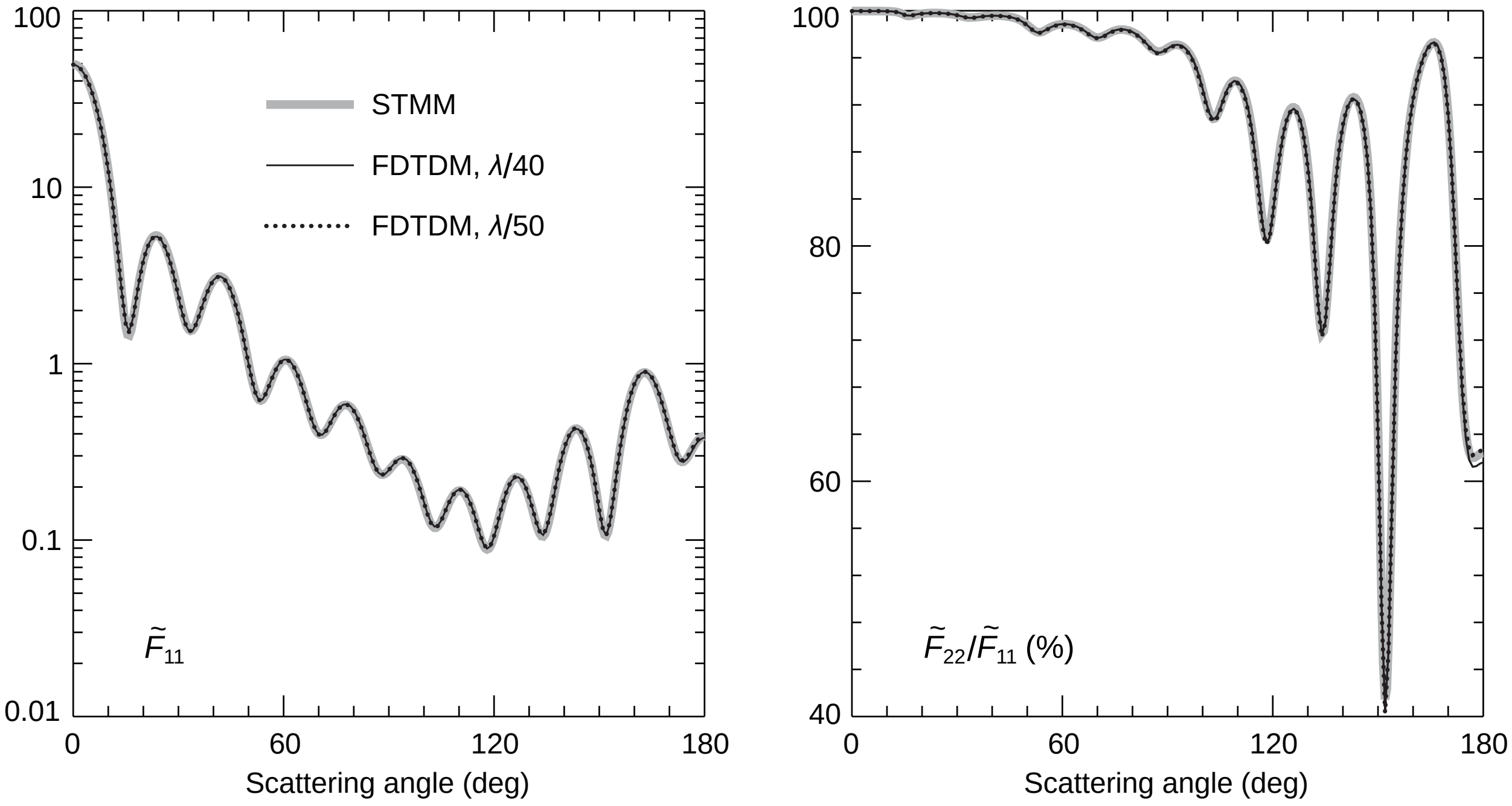

**Fig. 17.** As in Fig. 15, but for STMM vs. FDTDM results.

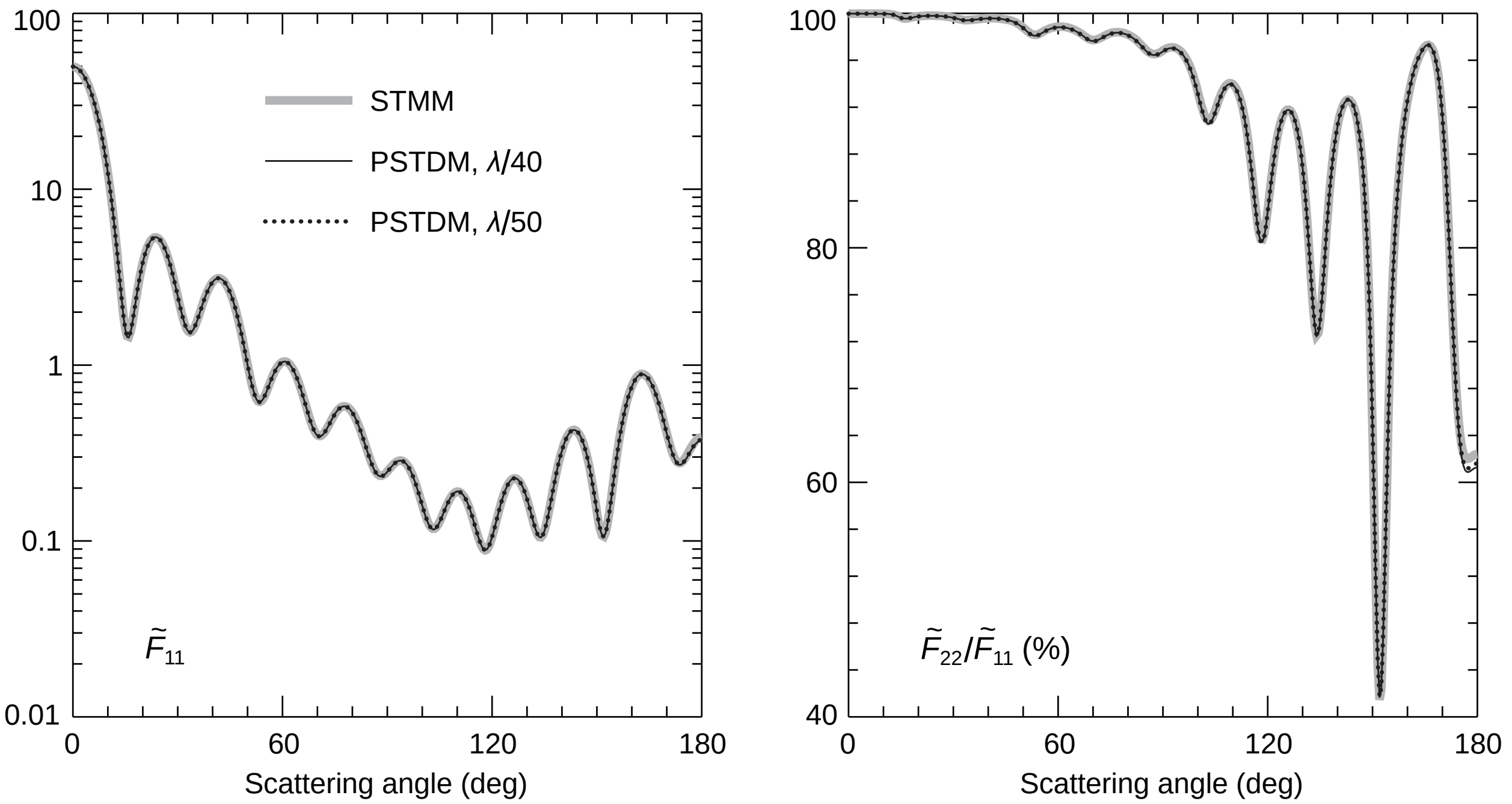

**Fig. 18.** As in Fig. 15, but for STMM vs. PSTDM results.

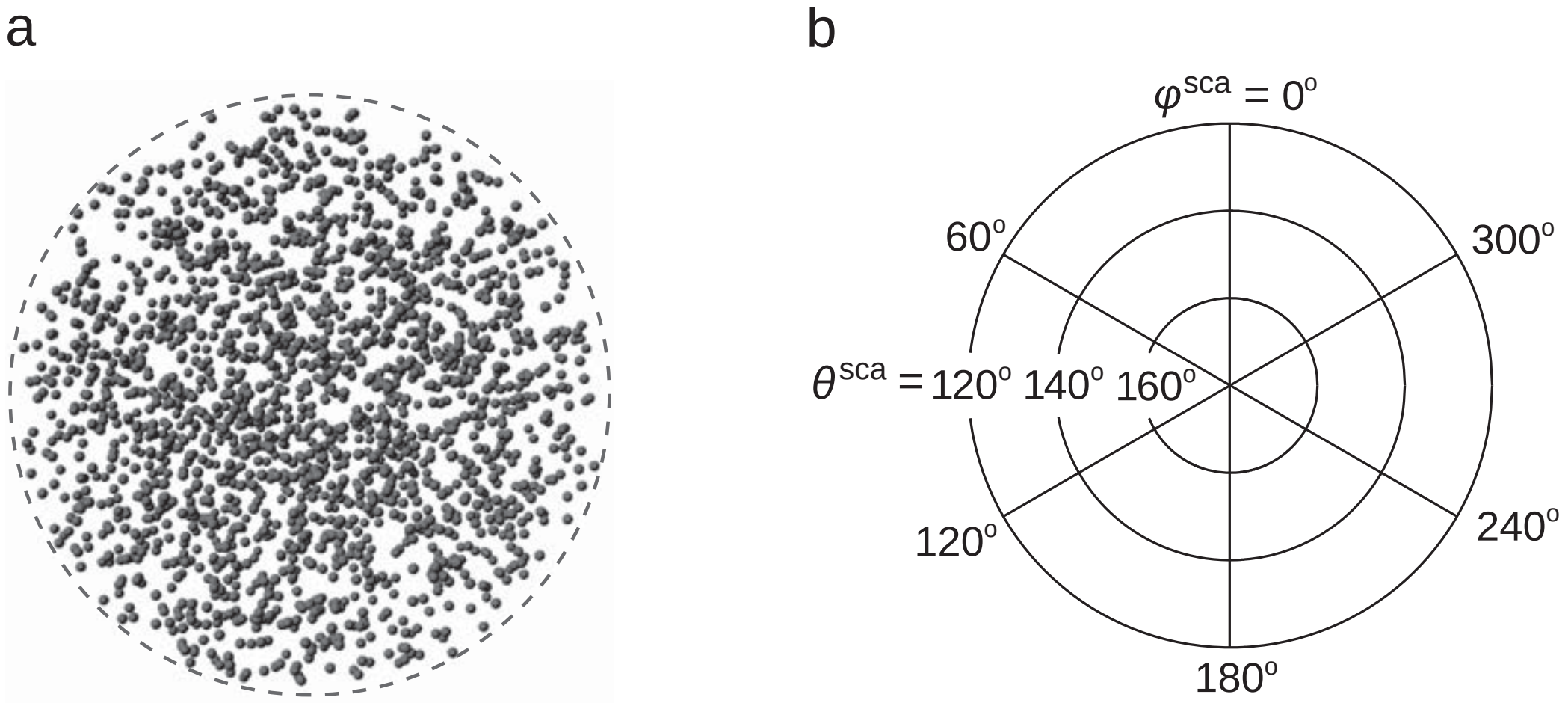

**Fig. 19.** (a) An imaginary spherical volume populated by randomly positioned spherical particles. (b) Angular coordinates used in Fig. 20.

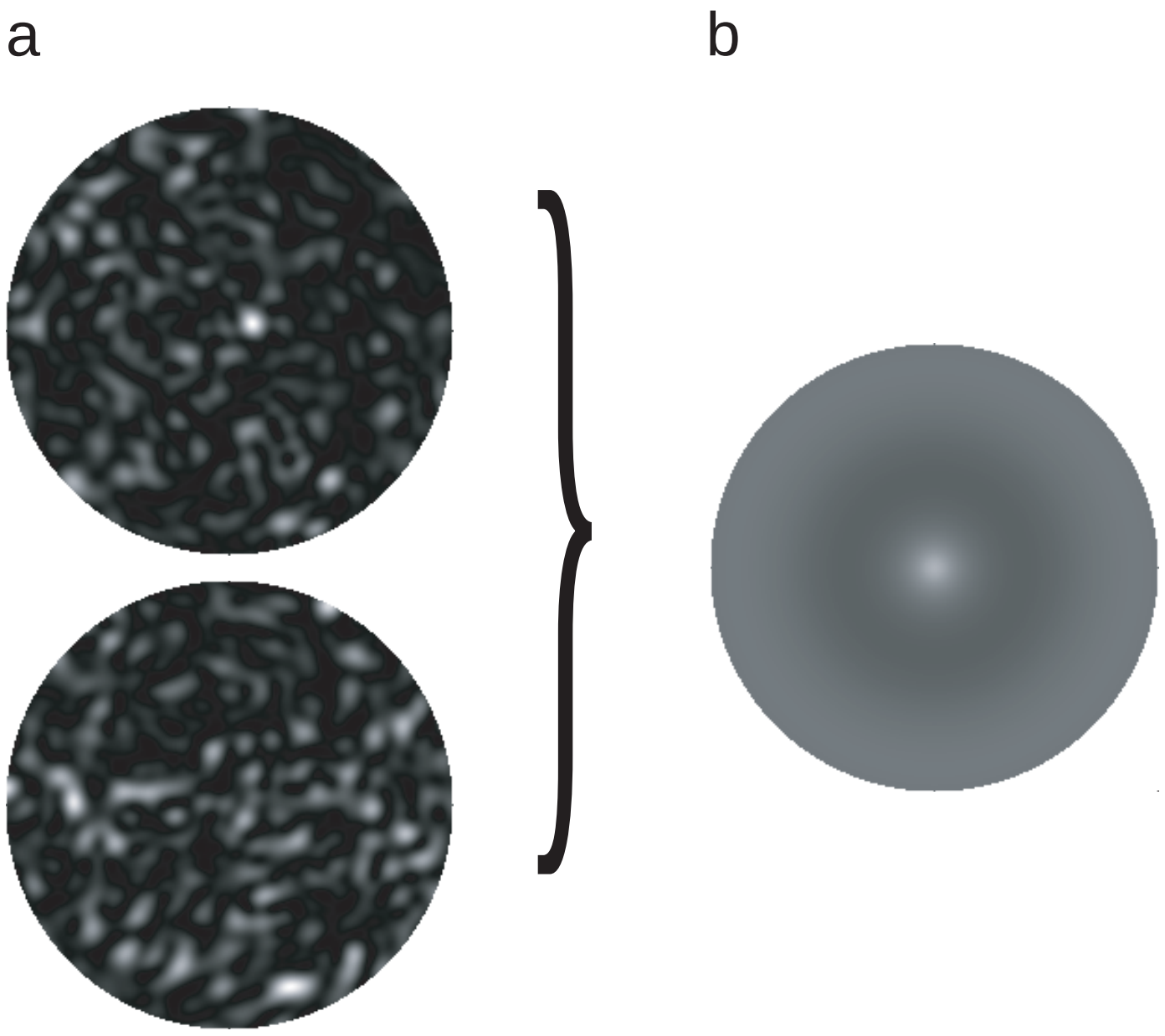

**Fig. 20.** (a) Angular distributions of the scattered intensity for two fixed spherical particulate volumes. (b) As in panel (a), but averaged over random particle positions. The gray scale is individually adjusted in order to maximally reveal the fine structure of each scattering pattern. Fig. 19b shows the angular coordinates used for all three panels.

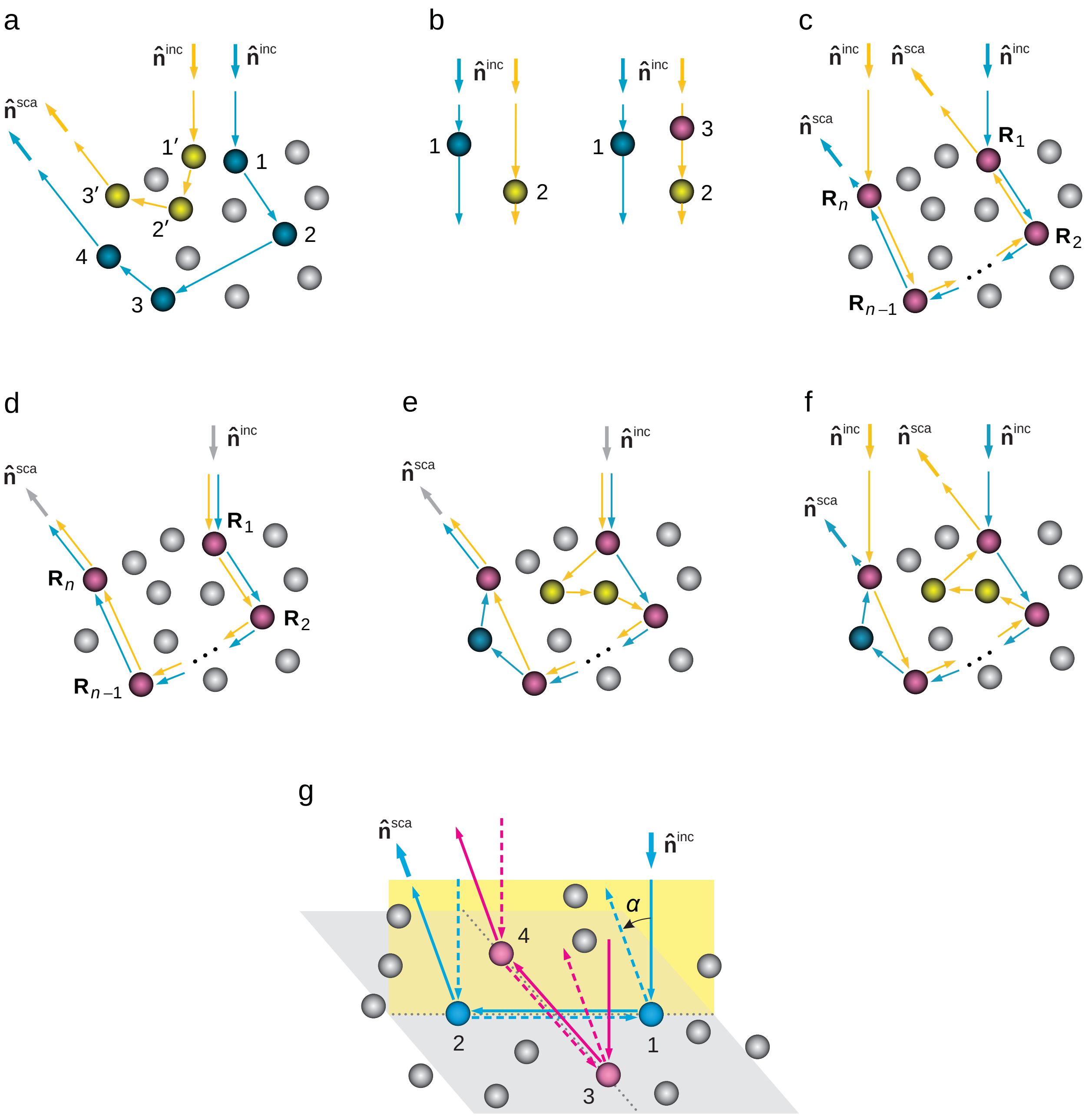

**Fig. 21.** (a) Interference origin of speckle. (b) Forward-scattering interference. (c) Interference origin of weak localization. (d) Interference origin of the diffuse background. (e) A pair of particle sequences contributing to the time-averaged diffuse background. (f) A pair of particle sequences contributing to time-averaged weak localization. (g) Interference origin of the polarization opposition effect.

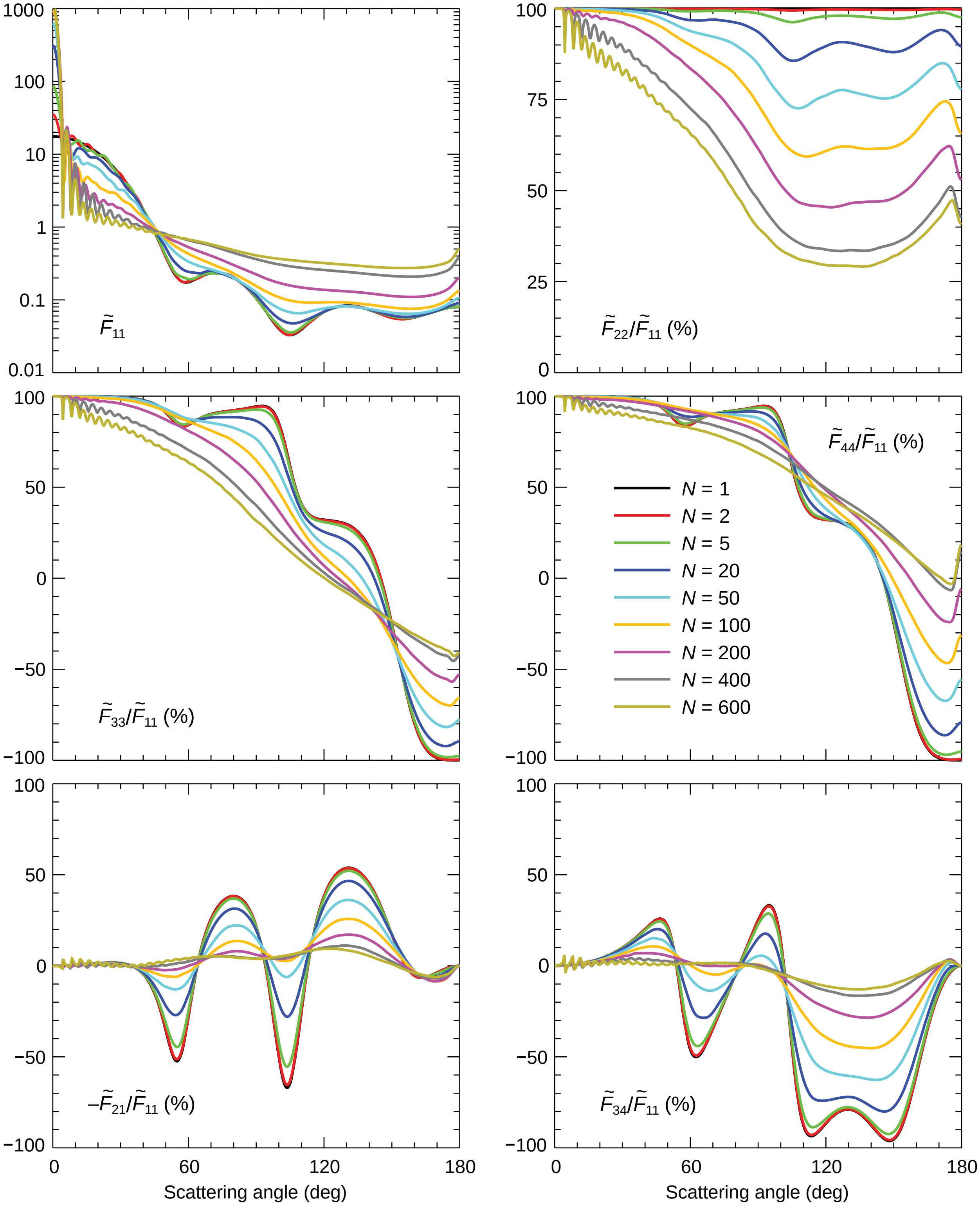

**Fig. 22.** Elements of the dimensionless scattering matrix computed for an imaginary $k_1R = 50$ spherical volume of discrete random medium uniformly populated by $N = 1, 2, \ldots, 600$ particles with $k_1r = 4$ and $m = 1.32$.

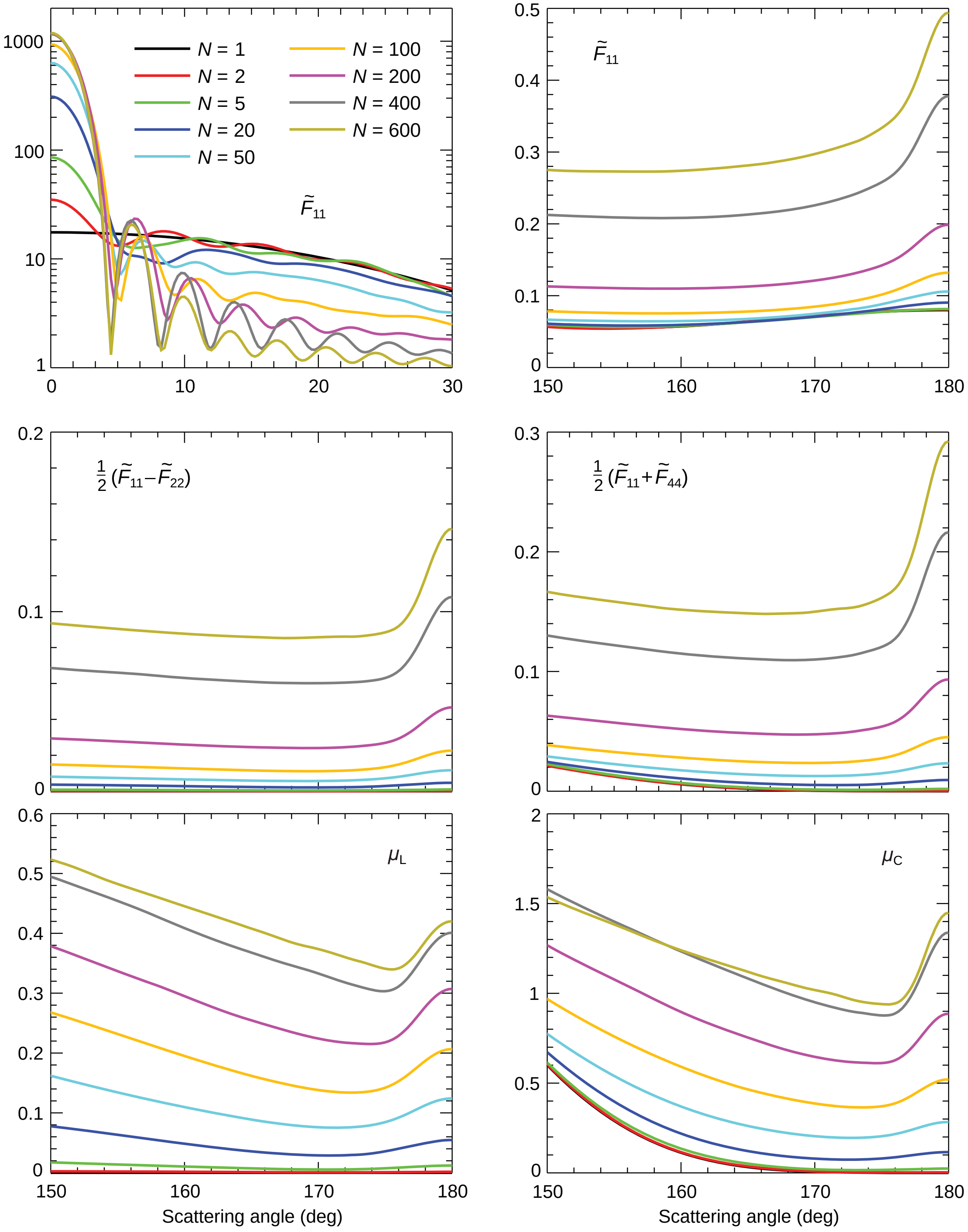

**Fig. 23.** Elements of the dimensionless scattering matrix and polarization ratios computed for an imaginary $k_1R = 50$ spherical volume of discrete random medium uniformly populated by $N =$ 1, 2, …, 600 particles with $k_1r = 4$ and $m = 1.32$.

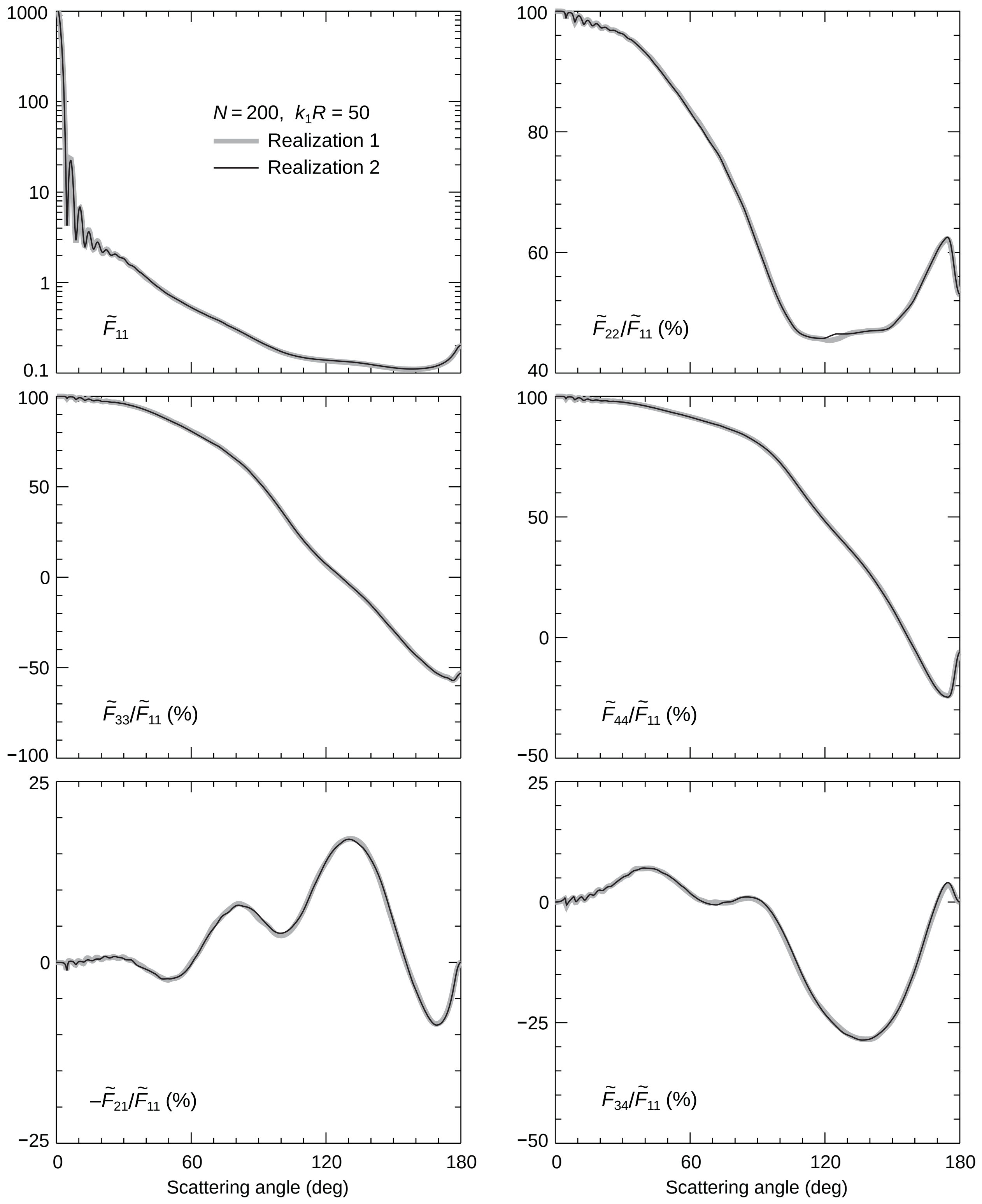

**Fig. 24.** Elements of the dimensionless scattering matrix for two realizations of an imaginary spherical volume of discrete random medium with $k_1R = 50$, $N = 200$, $k_1r = 4$, and $m = 1.32$.

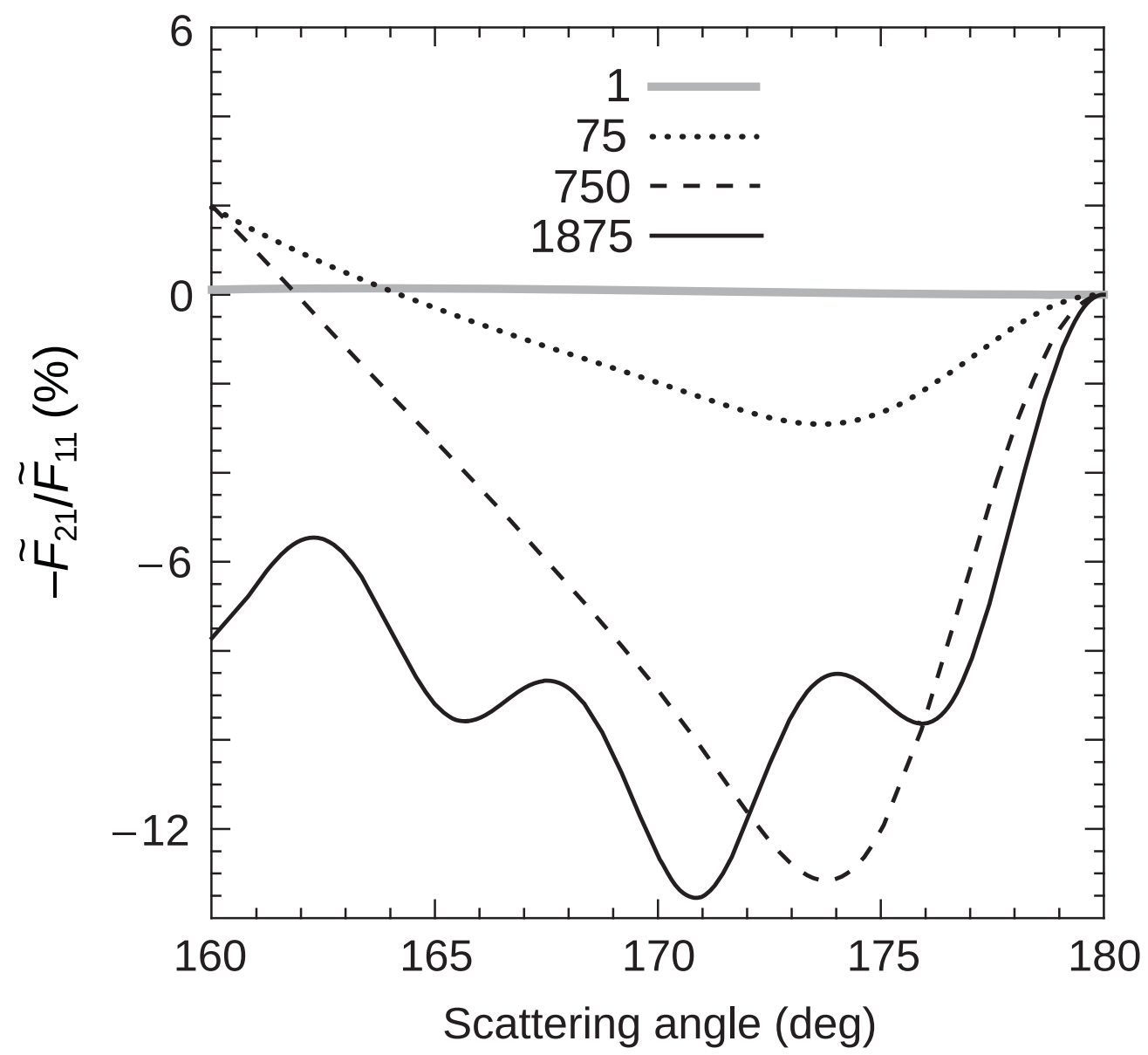

**Fig. 25.** Polarization opposition effects.

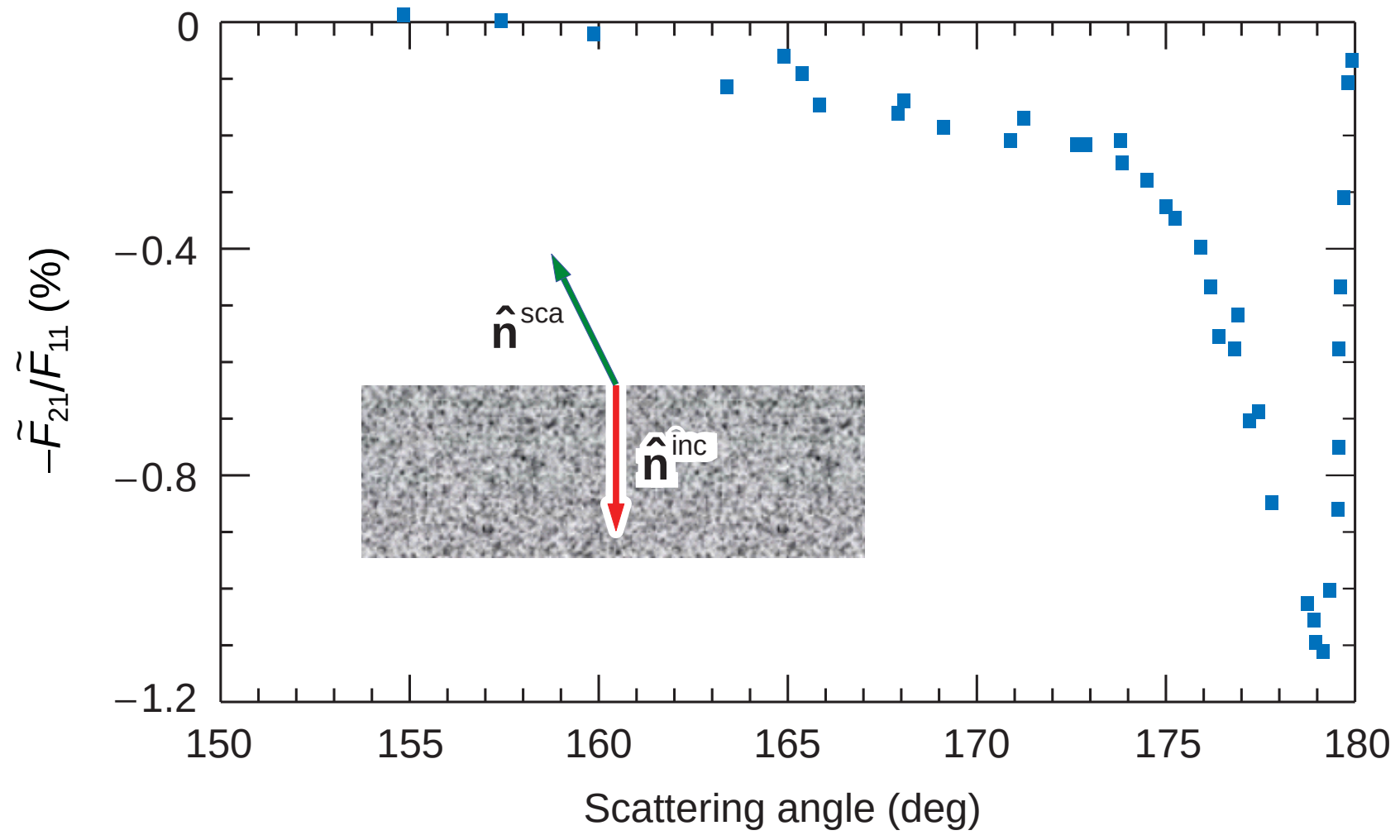

**Fig. 26.** Polarization measurements for a particulate surface composed of small magnesia particles.

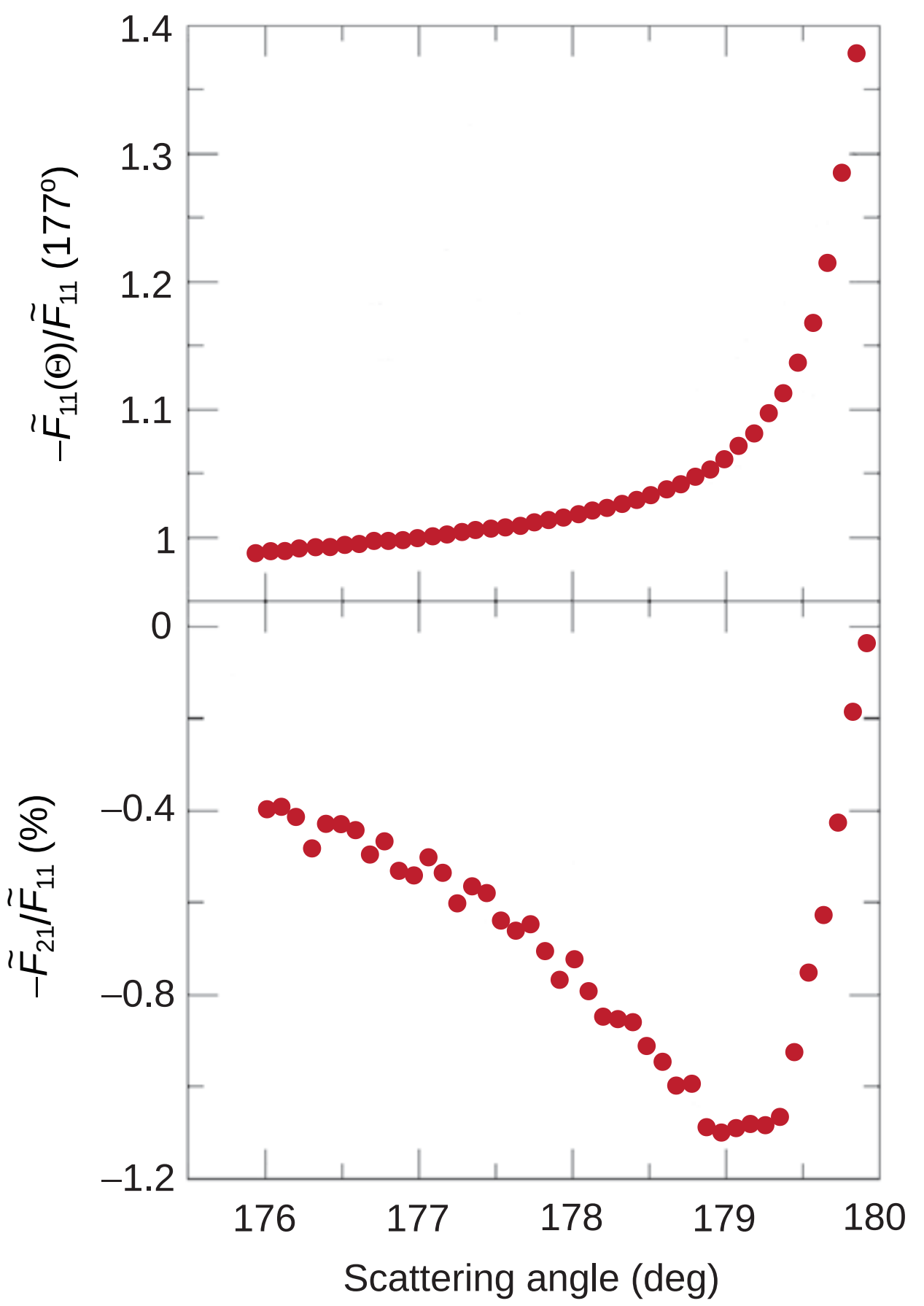

**Fig. 27.** Measurements of intensity and polarization of light backscattered by a particulate surface composed of small magnesia particles.

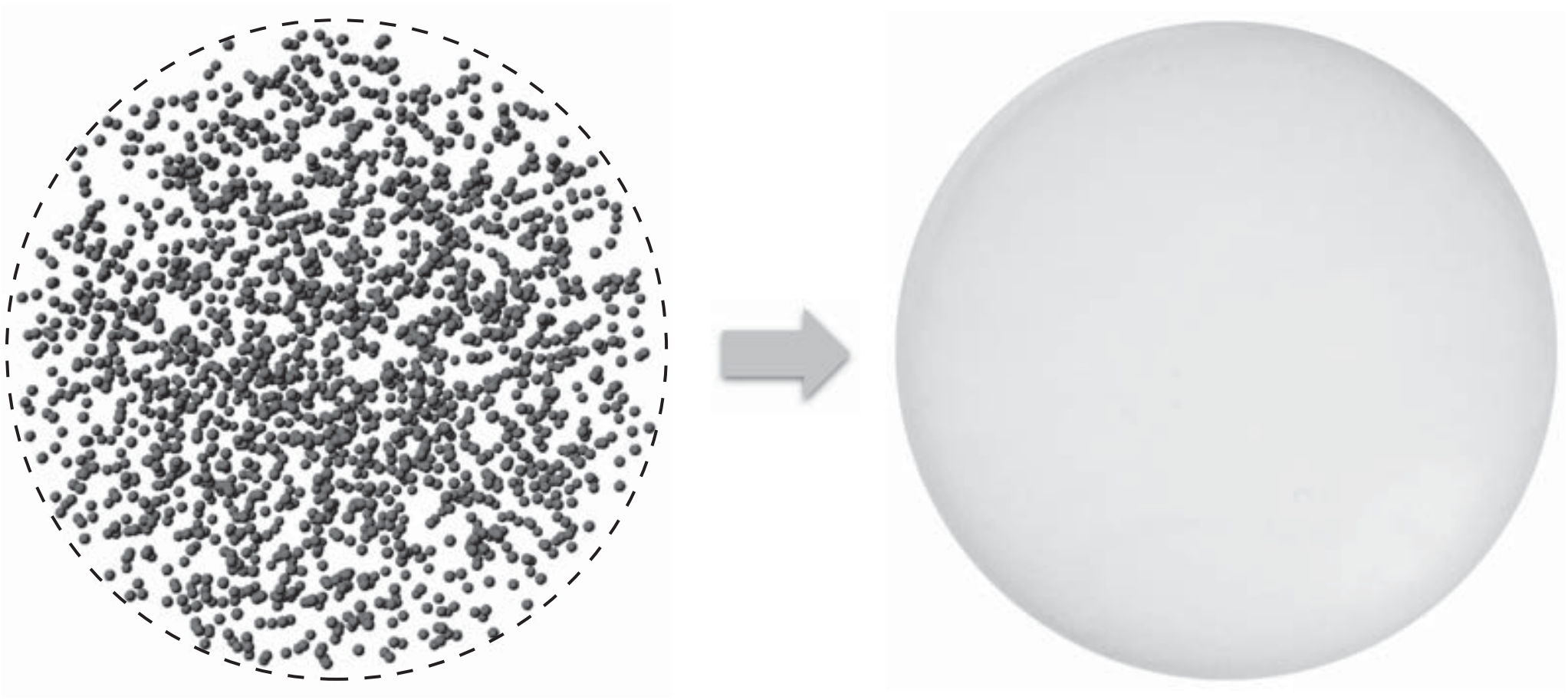

**Fig. 28.** An equidimensional homogeneous spherical particle replaces the imaginary spherical volume filled with a large number of identical inclusions.

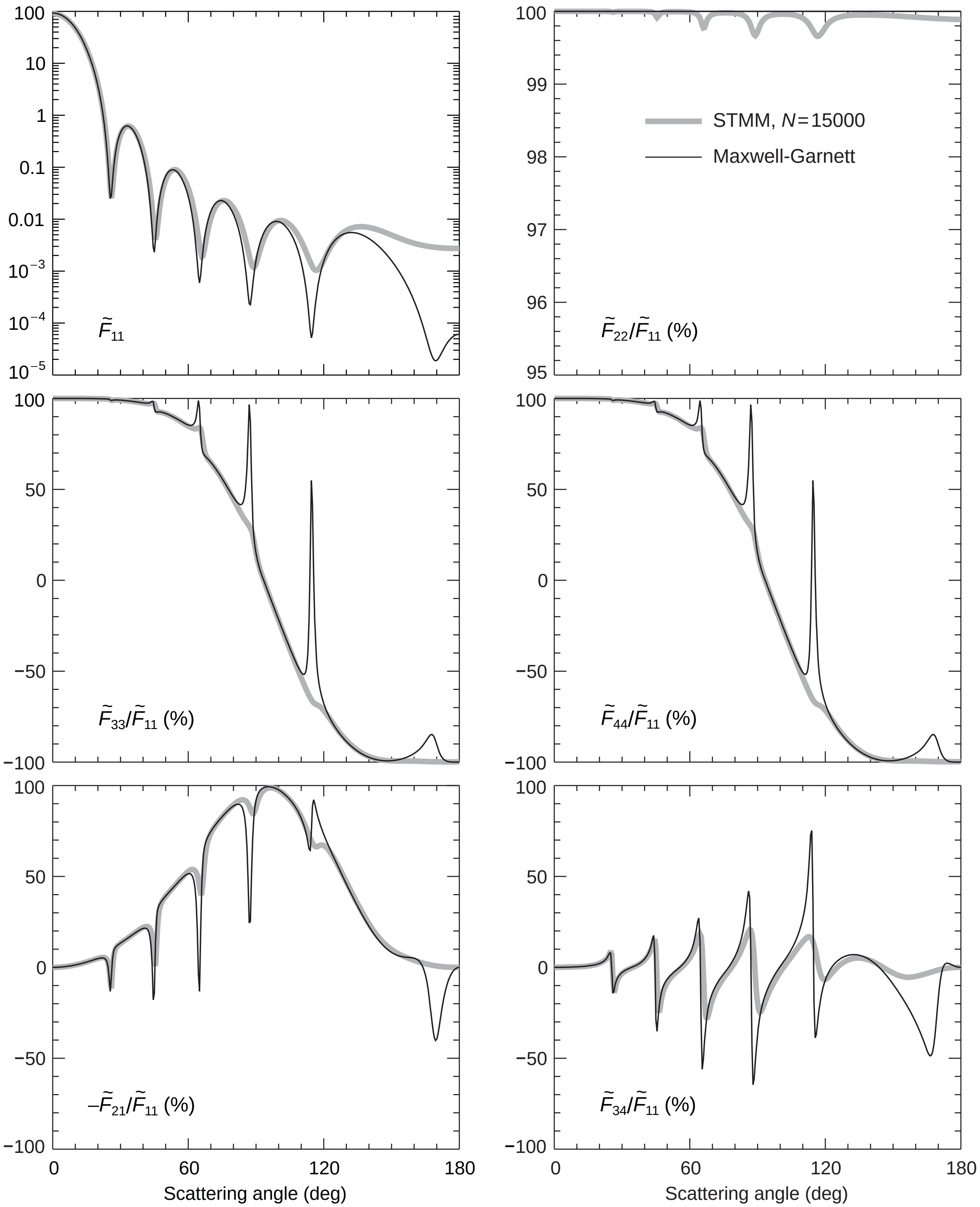

**Fig. 29.** Orientation-averaged elements of the dimensionless scattering matrix for an imaginary spherical volume of discrete random medium with $k_1R = 10$, $N = 15000$, $k_1r = 0.2$, and $m = 1.2$. The thin black curves show the result of using the Maxwell-Garnett approximation.

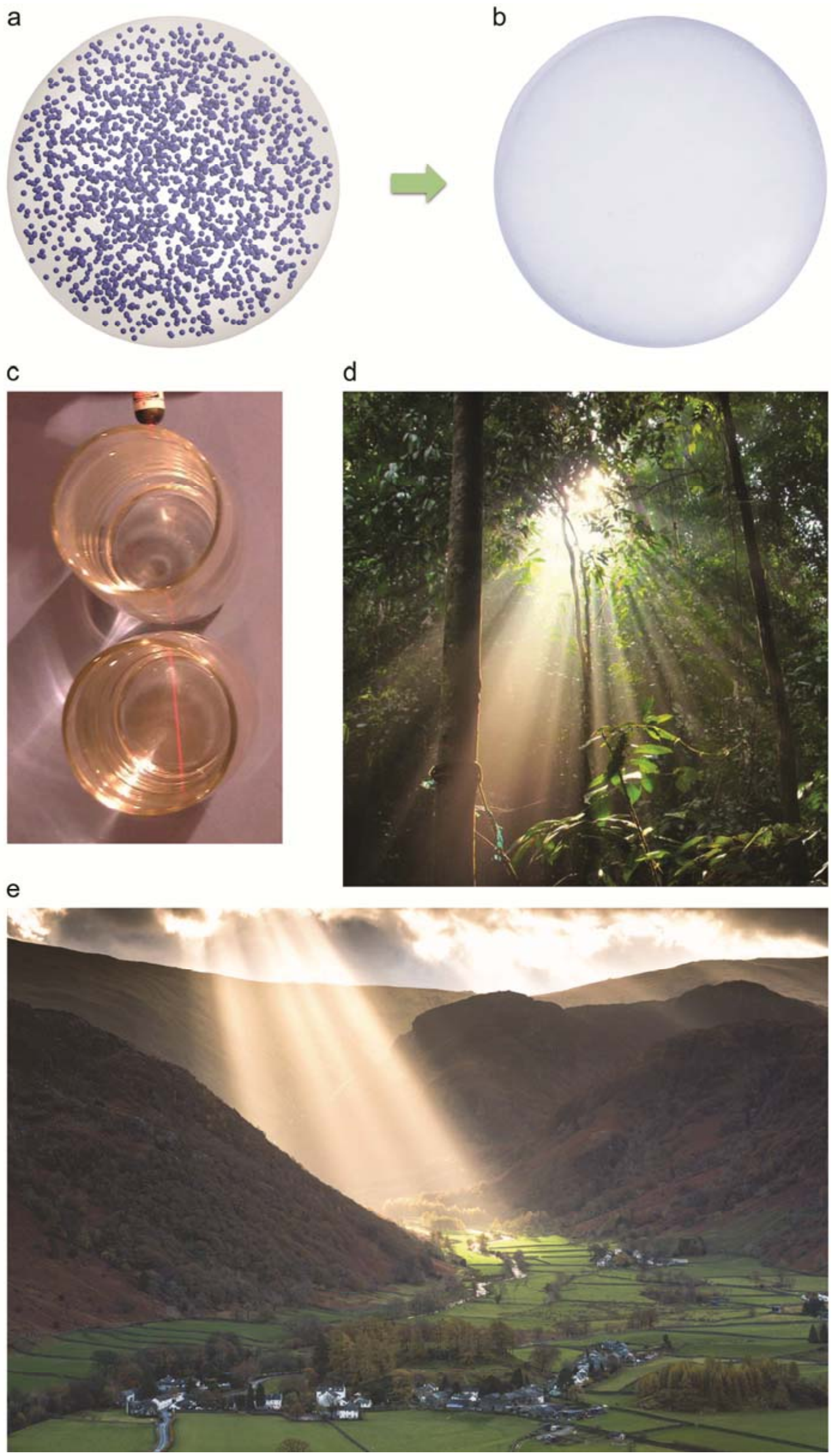

**Fig. 30.** (a,b) Heterogeneous spherical target and its effective-medium counterpart. (c–e) Manifestations of the Tyndall effect.

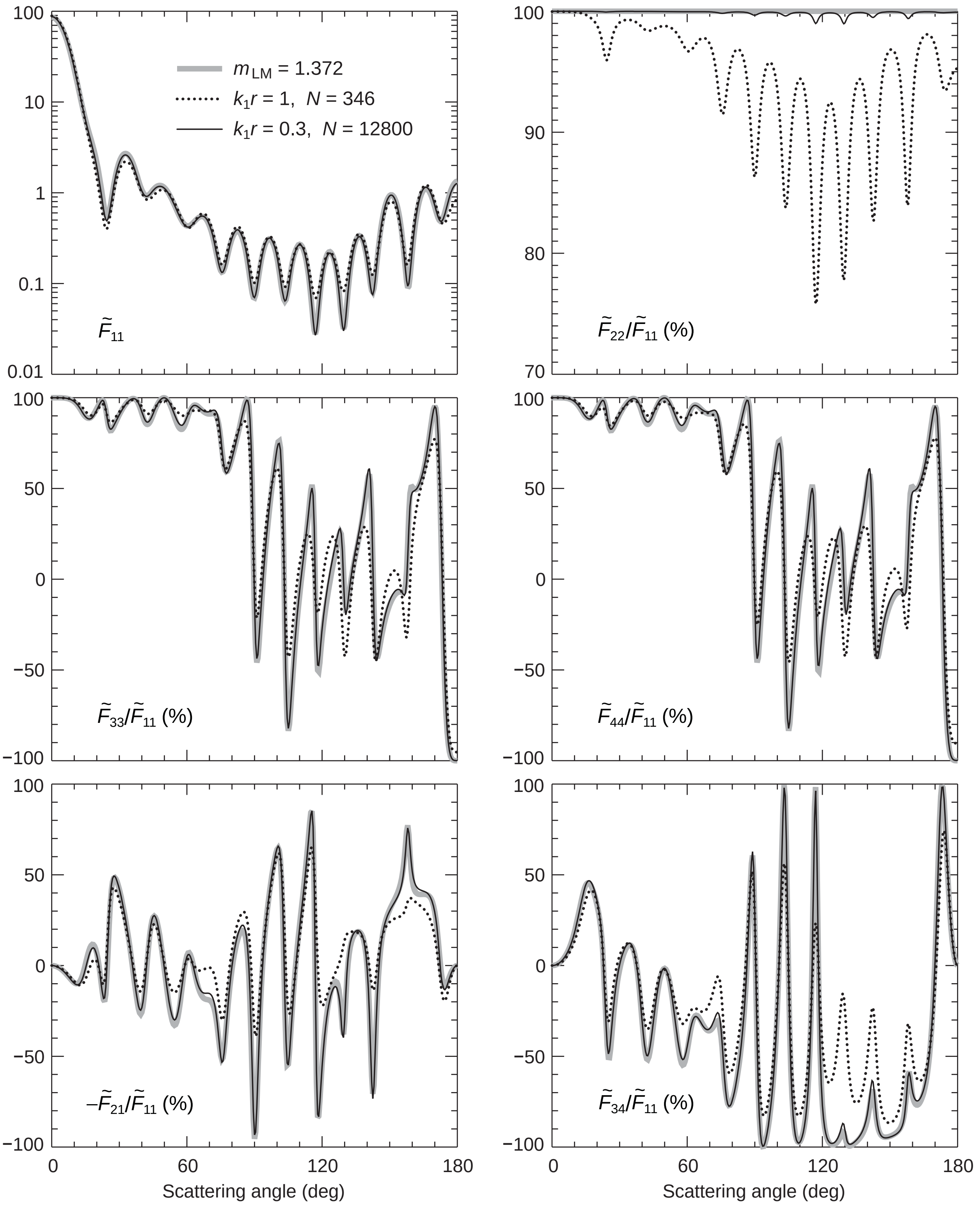

**Fig. 31.** Elements of the dimensionless scattering matrix for randomly heterogeneous and homogeneous spherical objects with a fixed size parameter $k_1R = 12$ (see text).

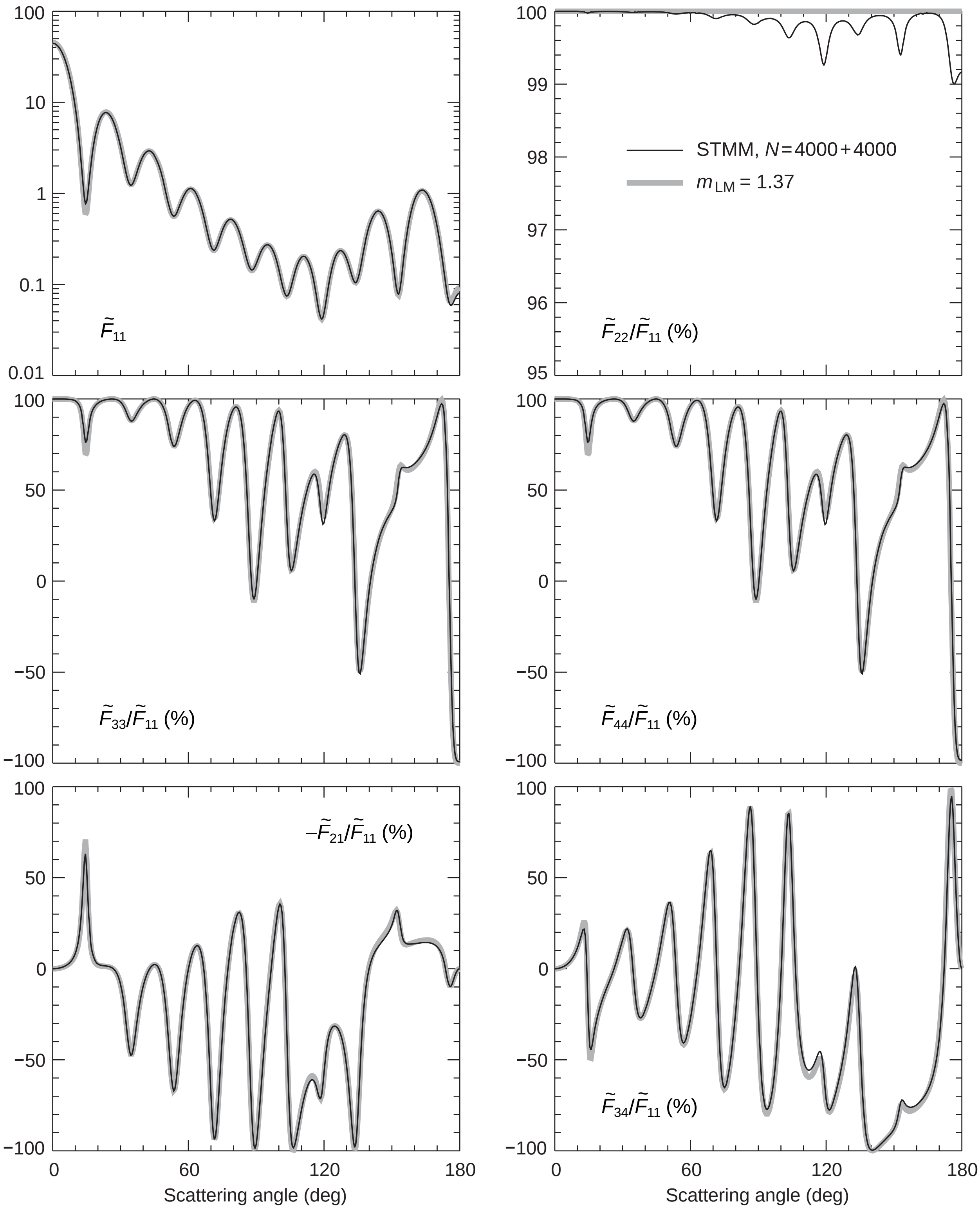

**Fig. 32.** Elements of the dimensionless scattering matrix for randomly heterogeneous and homogeneous spherical objects with a fixed size parameter $k_1R = 10$ (see text).

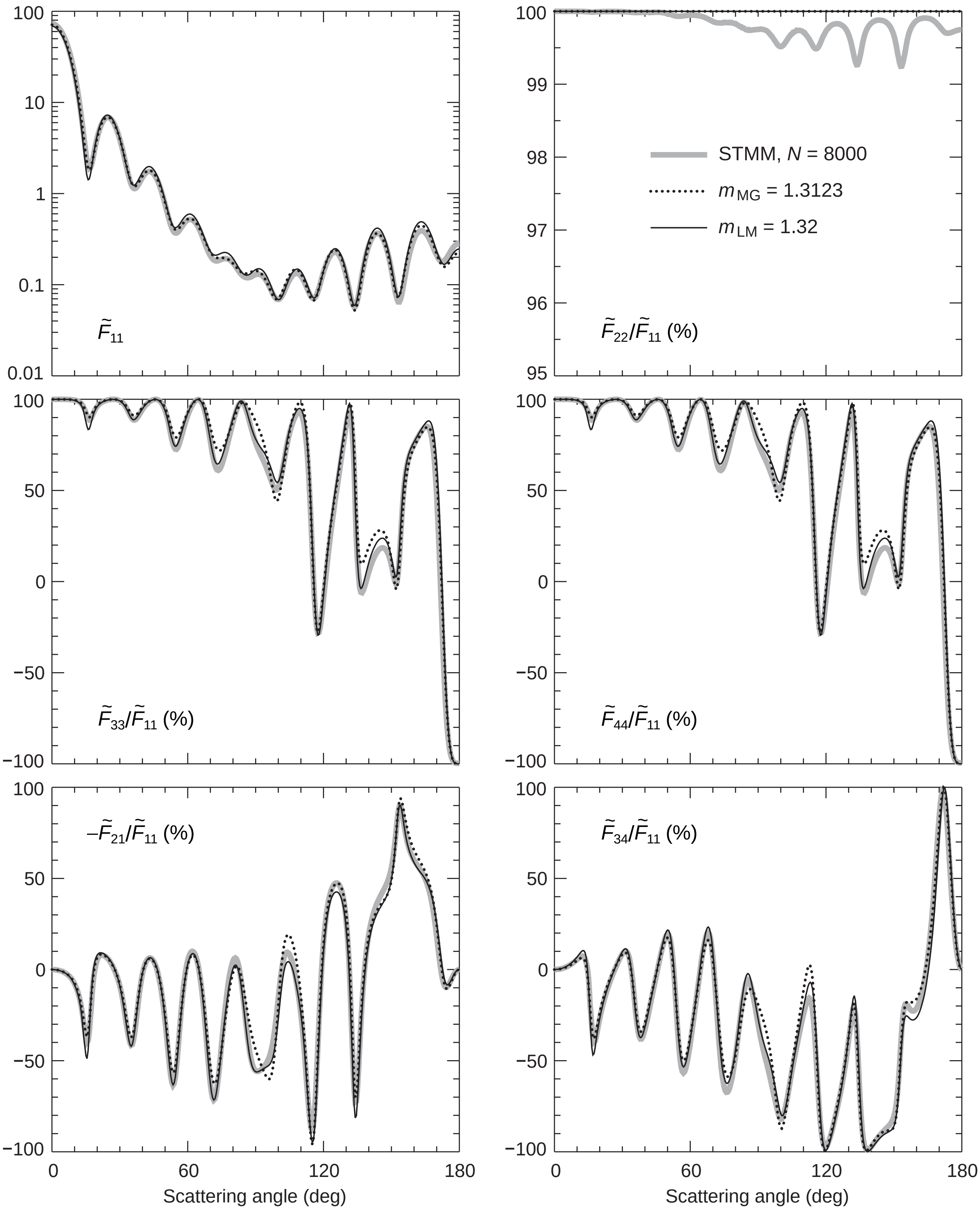

**Fig. 33.** Elements of the dimensionless scattering matrix for randomly heterogeneous and homogeneous spherical objects with a fixed size parameter $k_1R = 10$ (see text).

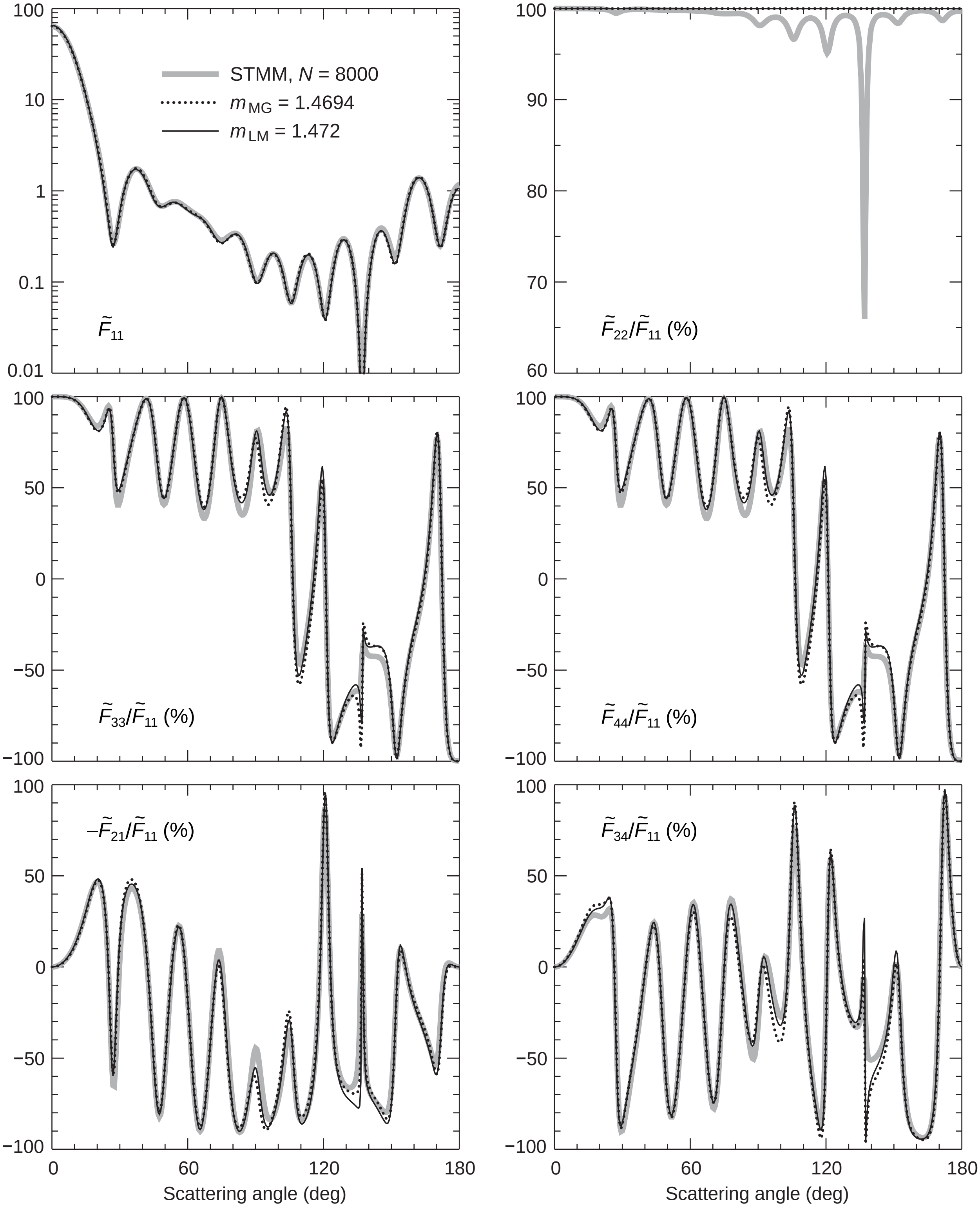

**Fig. 34.** Elements of the dimensionless scattering matrix for randomly heterogeneous and homogeneous spherical objects with a fixed size parameter $k_1 R = 10$ (see text).

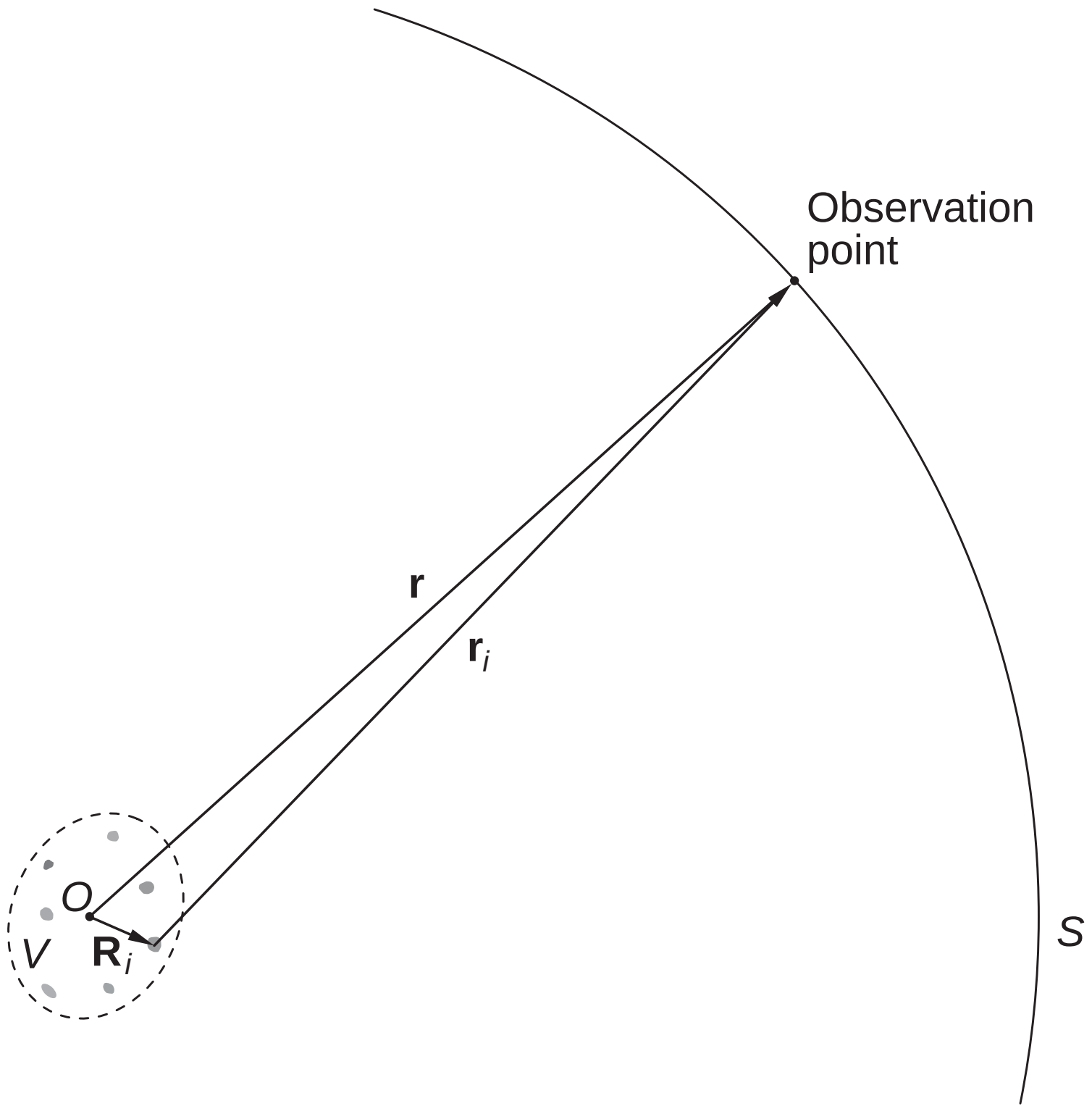

**Fig. 35.** The Type-1 DRM is composed of a small number of particles sparsely populating an imaginary volume $V$ and is observed from a sufficiently large distance $r$.

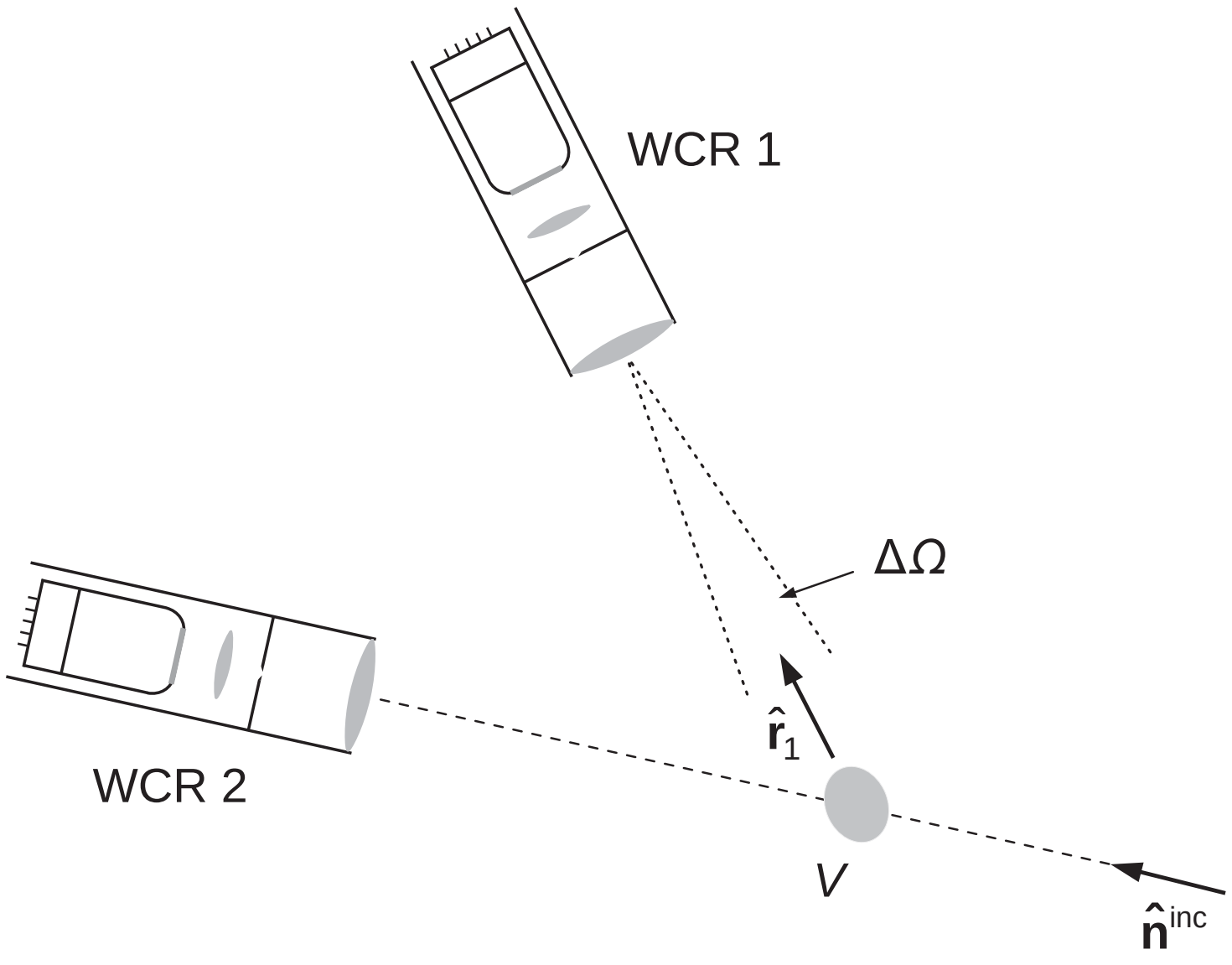

**Fig. 36.** Near-field measurements of electromagnetic scattering by a small sparse DRM.

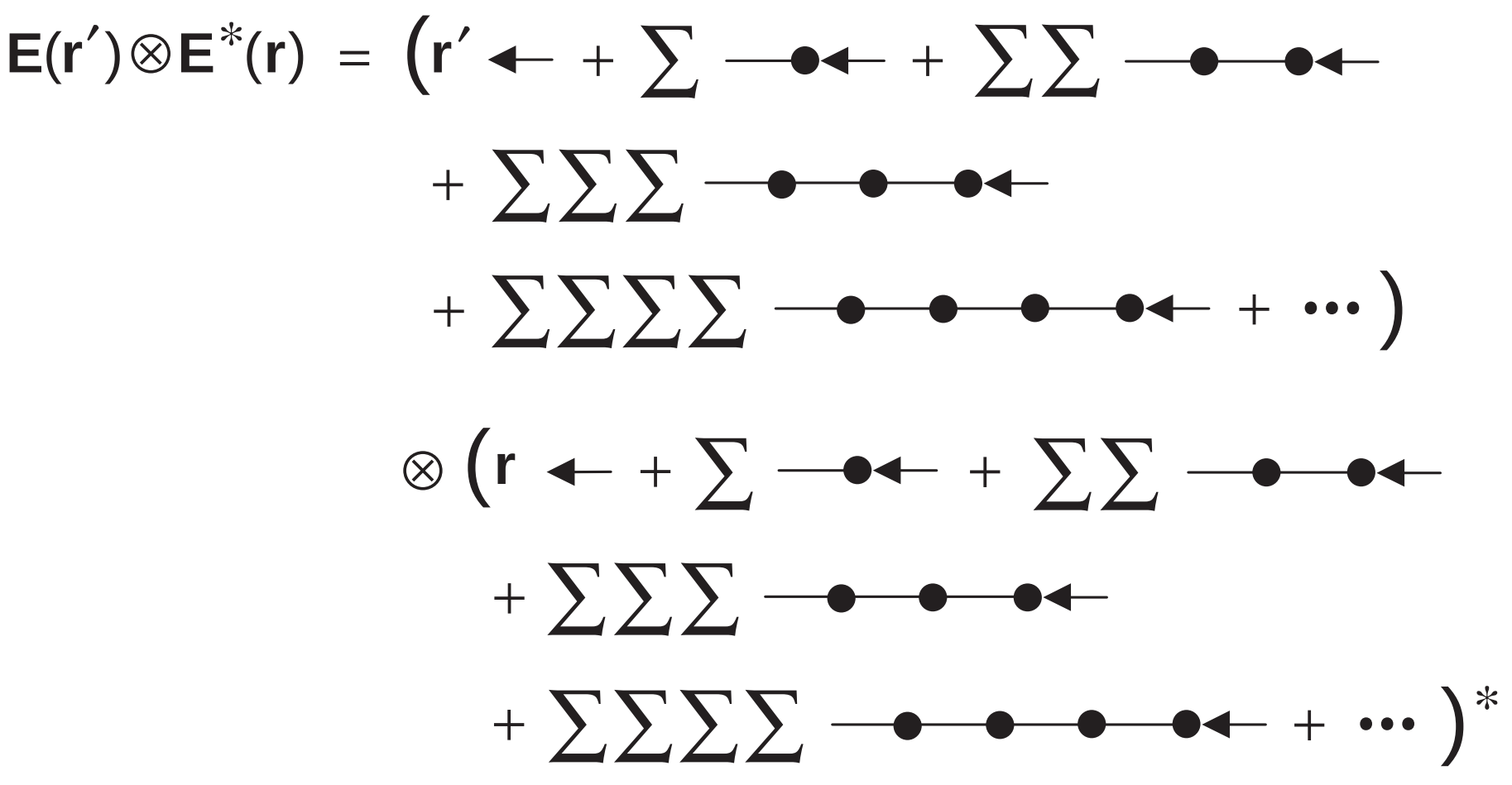

**Fig. 37.** The Twersky approximation for the dyadic correlation function. Each arrow denotes the local incident field; each dot denotes the left-multiplication by the corresponding scattering dyadic; and each horizontal line denotes multiplication by the corresponding $g$-function (150).

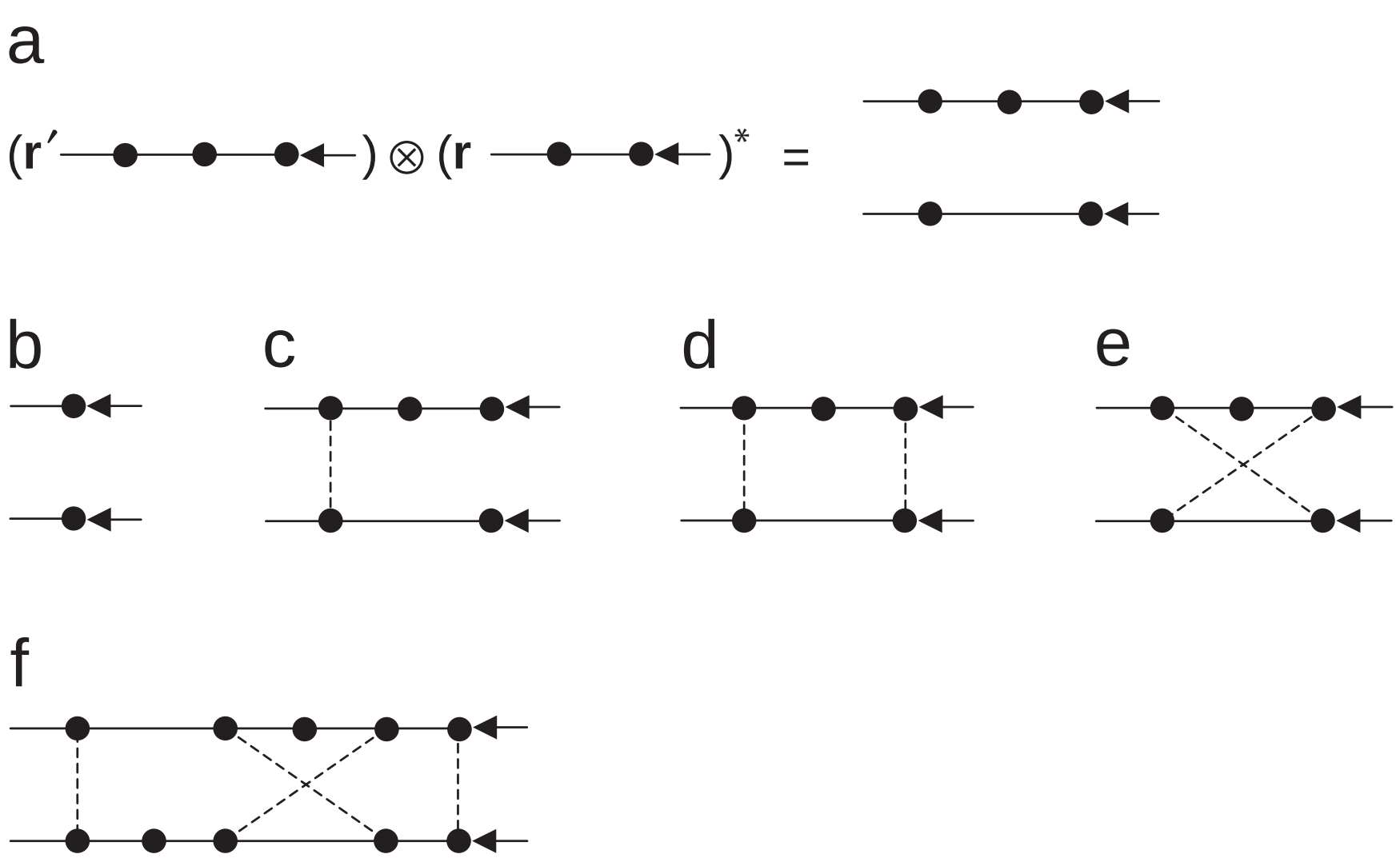

**Fig. 38.** Classification of various terms entering the expanded Twersky approximation for the dyadic correlation function.

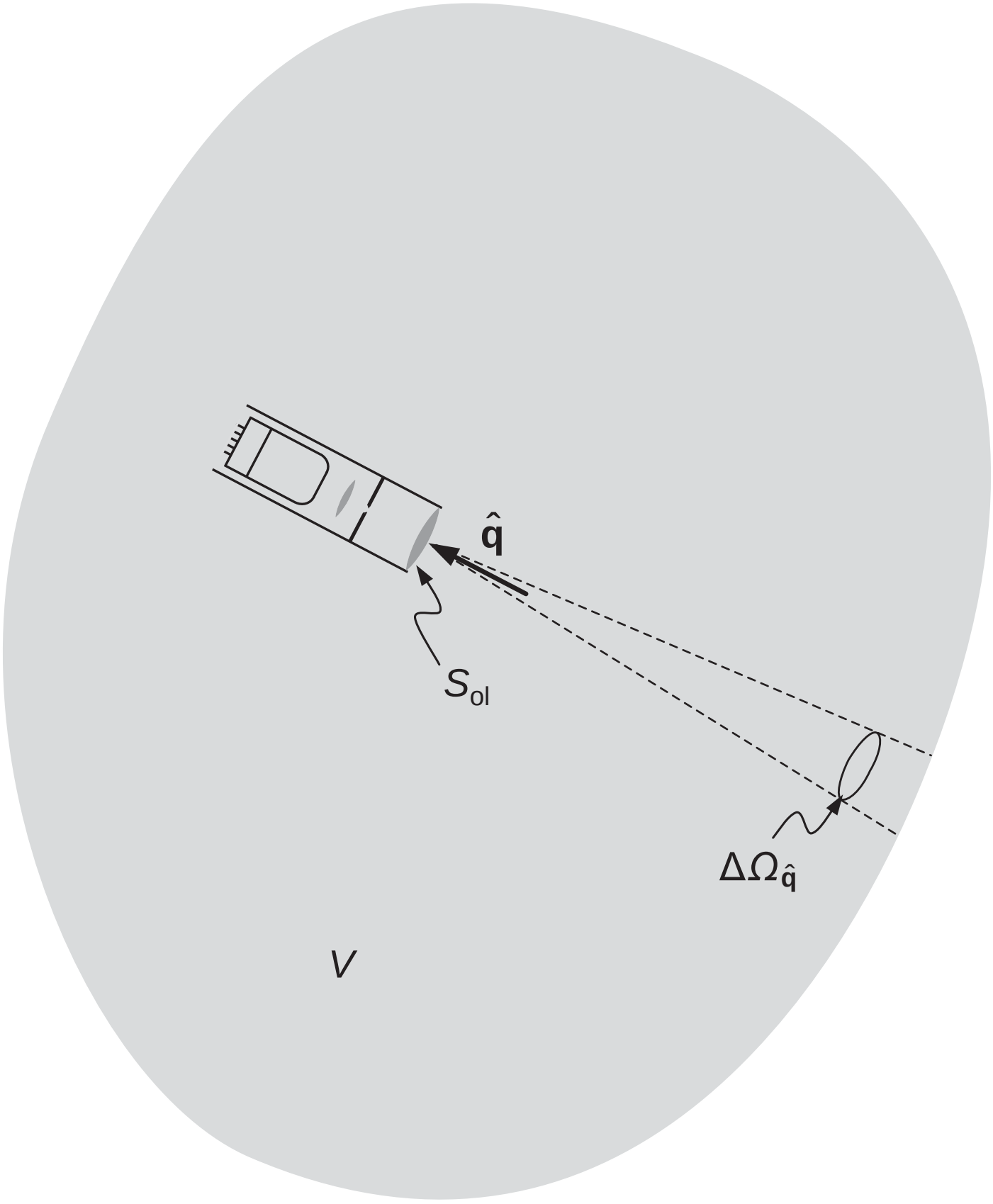

**Fig. 39.** A WCR placed inside the DRM. The size of the WCR is exaggerated relative to that of the DRM for demonstration purposes. The uniform shading is intended to emphasize that the constituent particles move randomly throughout the volume *V* during the measurement.

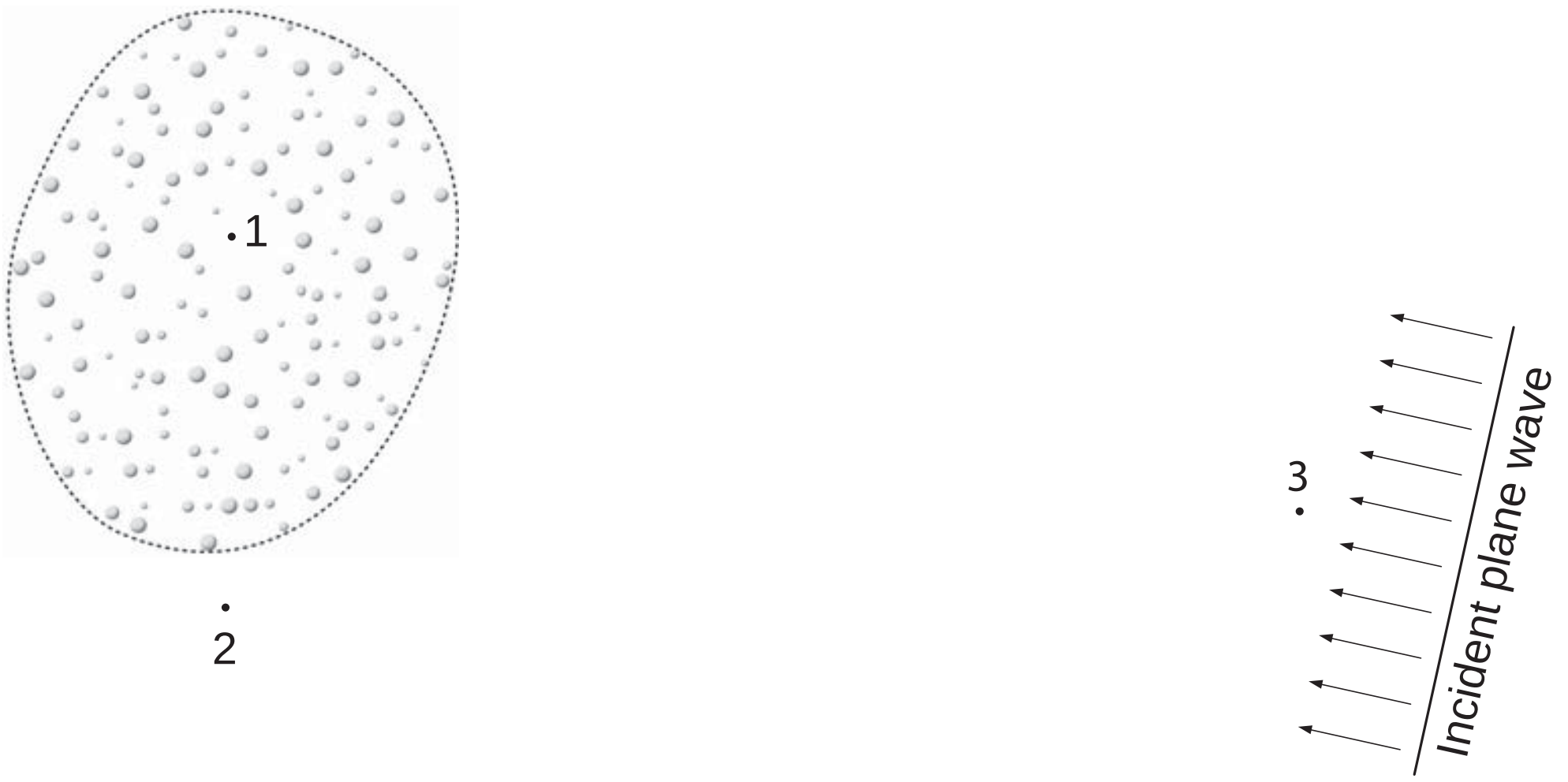

**Fig. 40.** Electromagnetic scattering by a sparse Type-1 DRM. The size of the DRM is exaggerated relative to its distance from observation point 3 for demonstration purposes.

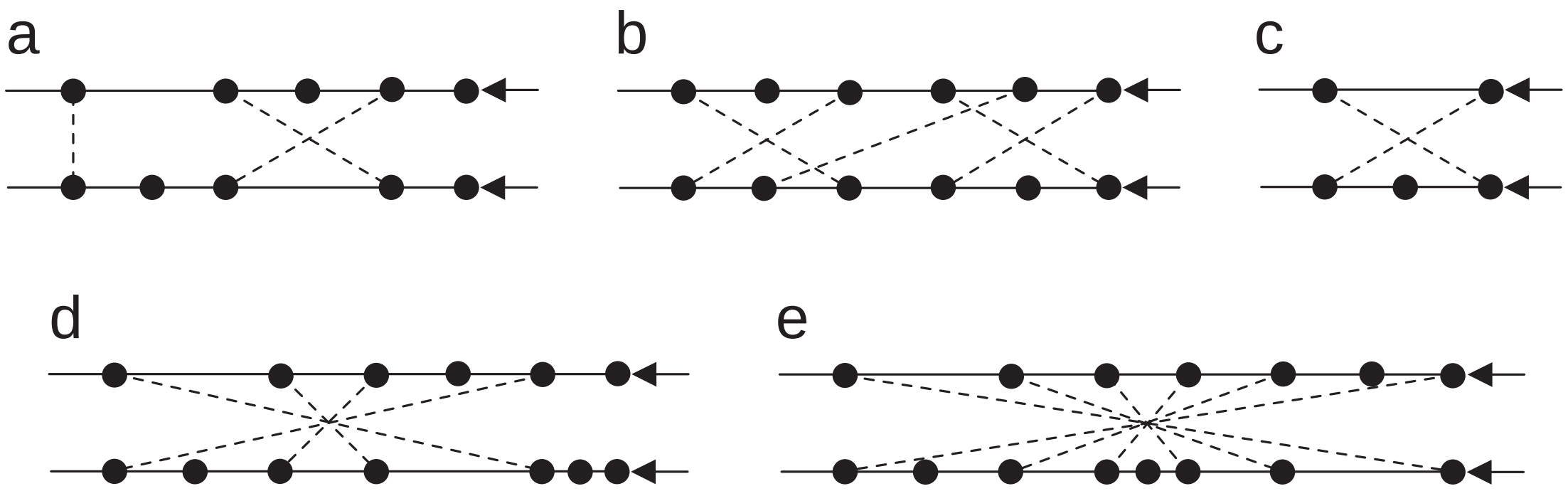

**Fig. 41.** Diagrams with crossing connectors.

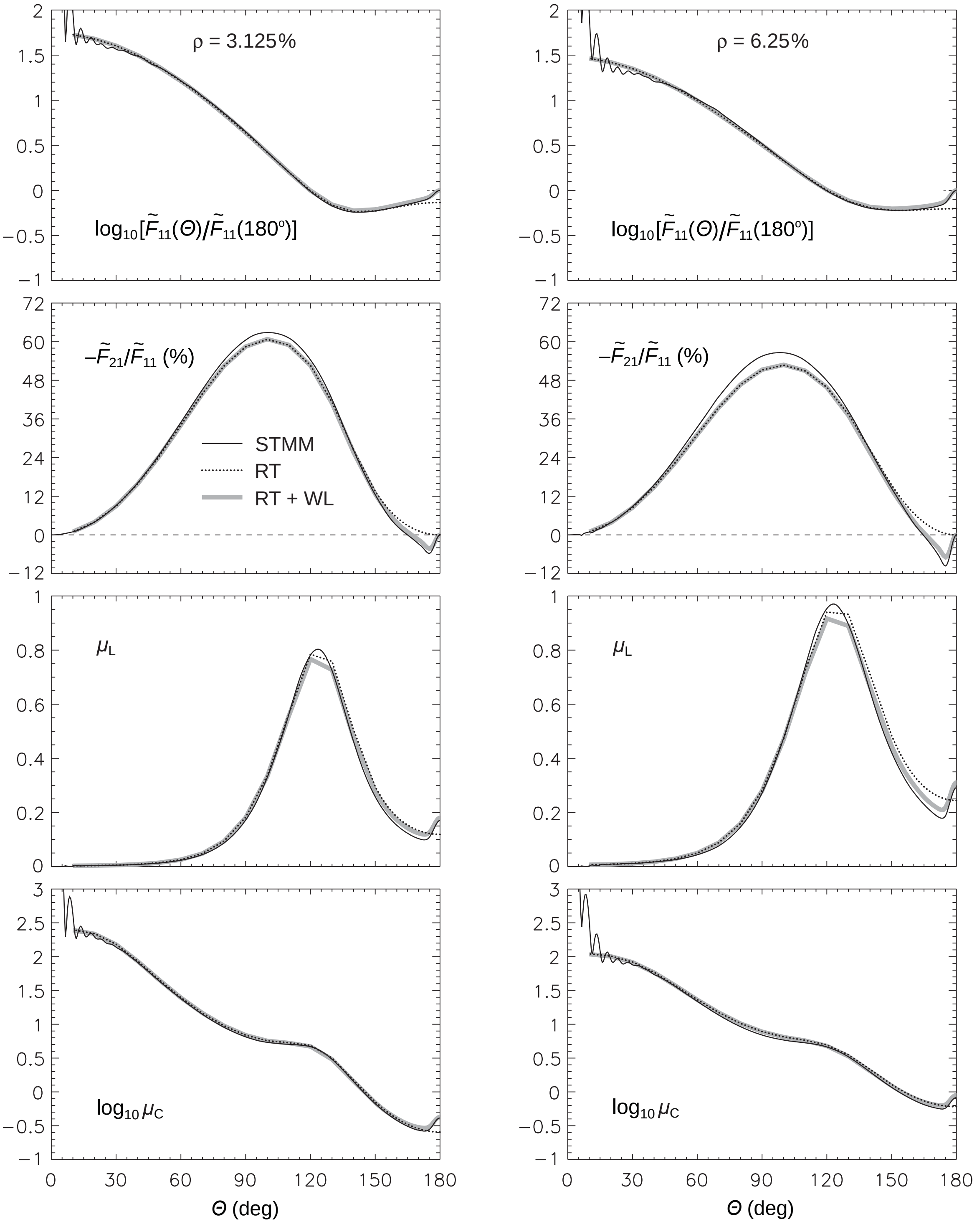

**Fig. 42.** Scattering by a spherical Type-1 DRM with a size parameter of $k_1R = 40$ and packing densities of $\rho = 3.125\%$ and 6.250%, populated with identical spherical particles with a size parameter of $k_1r = 2$ and a refractive index of $m = 1.31$. The solid, dotted, and thick gray curves depict the STMM, RT-only, and RT–WL results, respectively. The RT phase functions are shifted downward to match the RT–WL phase functions at $\Theta = 150°$.

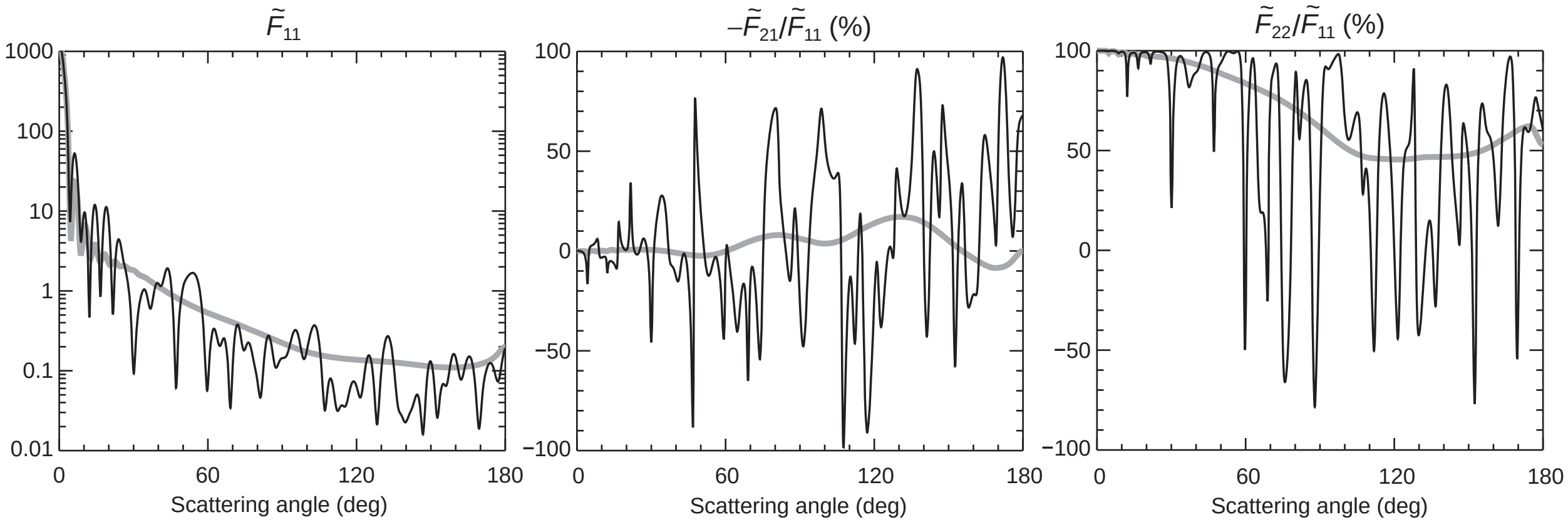

**Fig. 43.** Elements of the dimensionless scattering matrix computed for an imaginary $k_1R = 50$ spherical volume populated by $N = 200$ particles with $k_1r = 4$ and $m = 1.32$. Black curves: the multi-particle configuration is fixed. Gray curves: the results are averaged over the uniform orientation distribution of the multi-particle configuration.

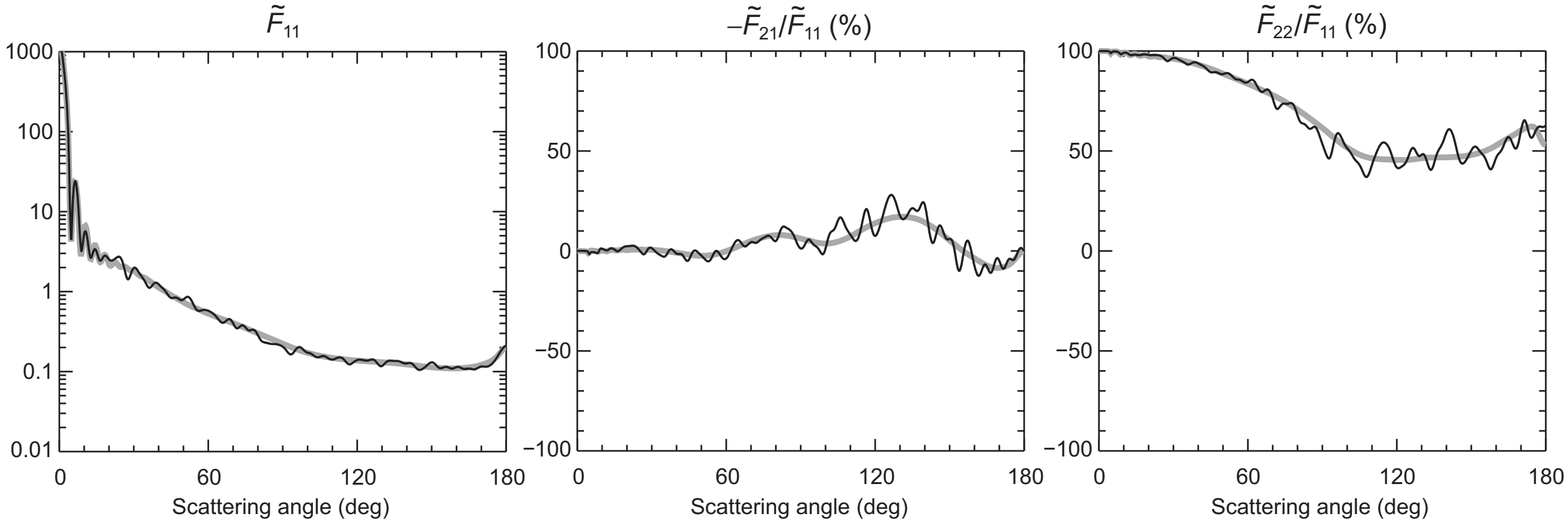

**Fig. 44.** Elements of the dimensionless scattering matrix computed for an imaginary spherical volume populated by $N = 200$ particles with $m = 1.32$. Black curves: the multi-particle configuration is fixed and the results are averaged over a range of wavelengths such that $k_1R$ varies from 47.5 to 52.5 and $k_1r$ varies from 3.8 to 4.2. Gray curves: the results are averaged over the uniform orientation distribution of the multi-particle configuration at a single wavelength such that $k_1R = 50$ and $k_1r = 4$.